\RequirePackage{fix-cm}
\documentclass[natbib,smallextended]{svjour3}       
\usepackage{graphicx}
\usepackage{aas_macros}
\usepackage{amssymb}
\usepackage{color}
\usepackage[colorlinks=true,citecolor=blue,urlcolor=blue,breaklinks]{hyperref}
\usepackage[dvipsnames,table,xcdraw]{xcolor}
\usepackage{tikz}
\usetikzlibrary{mindmap, backgrounds, calc}
\def\sllnewer#1{#1}
\def\sllnew#1{#1}

\def\CP#1{#1}
\def\CPi#1{#1}
\def\CPn#1{#1}

\def\schrodinger{Schr\"{o}dinger}
\journalname{Living Rev. Relativ.}
\begin{document}

\title{Dynamical boson stars}

\author{Steven L.\ Liebling \and Carlos Palenzuela}

\institute{S.~L.\ Liebling%
\at
Long Island University\\
Brookville, NY 11548, U.S.A.\\
\email{steve.liebling@liu.edu}
\and
C.\ Palenzuela%
\at
Universitat de les Illes Balears\\
Palma de Mallorca, 07122 Baleares, Spain\\
\email{carlos.palenzuela@uib.es}
}

\date{Received: date / Accepted: date}

\maketitle

\begin{abstract}
The idea of stable, localized bundles of energy has strong appeal as a
model for particles. In the 1950s, John Wheeler envisioned such
bundles as smooth configurations of electromagnetic energy that he
called \emph{geons}, but none were found. Instead, particle-like
solutions were found in the late 1960s with the addition of a scalar
field, and these were given the name \emph{boson stars}.  Since then,
boson stars find use in a wide variety of models as sources of dark
matter, as black hole mimickers, in simple models of binary systems,
and as a tool in finding black holes in higher dimensions with only a
single Killing vector.  We discuss important varieties of boson stars,
their dynamic properties, and some of their uses, concentrating on
recent efforts.
\end{abstract}

\setcounter{tocdepth}{3}
\tableofcontents


{\small
\section*{What's New}
\addcontentsline{toc}{section}{What's New}

\paragraph{\textbf{Second revision (2022):}}
We have added a diagram of the varieties of different bosonic stars at the beginning of
Sect.~\ref{section:varieties}. We clarified the differences between multi-state
and multi-field stars and added discussion of $\ell$-boson stars in Sect.~\ref{sec:multi}.
We have added Sect.~\ref{section:opensoftware} with links to open-source software
related to boson stars.
We discuss the significant progress made in evolving: (i)~rotating boson stars in the new
Sect.~\ref{sec:stability_w_angmom} and (ii)~boson star binaries in Sect.~\ref{dynamics_binary_bs}.
Since the first revision, the LIGO-Virgo-Kagra collaboration has progressed significantly
with a large number of observations while the Event Horizon Telescope has produced images
of Sgr~A* and M87, and this progress informs some of the new discussion.
Figures~\ref{fig:rotating_rprofile}, 
\ref{fig:lboson_massomega}, 
\ref{fig:hybrid_lboson},
\ref{fig:rotatingBS_NAI}, 
\ref{fig:rotatingBS_NAI_evolution},
\ref{fig:q1bbs},
\ref{fig:qne1bbs},
\ref{fig:atoms},
\ref{fig:sgraoptions},
and
\ref{fig:gw190521}
have been added, and
the number of references has increased from 367 to 484.

~\\

\paragraph{\textbf{First revision (2017):}}
We have added Sects.~\ref{sec:proca} (Proca stars) and \ref{scalarclouds} (Kerr black holes with scalar hair and superradiance).
We have revised and extended Sect.~\ref{section:astro}, extended Sect.~\ref{section:math}. Figures~\ref{fig:fermionboson}, \ref{fig:proca}, \ref{fig:hairyBH}, \ref{fig:bbs_headon}, \ref{fig:bbs_orbital:MJN}, \ref{fig:eht}, and \ref{fig:bbs_headon_psi4}
have been added, and 
the number of references has increased from 223 to 367.
}

\section{Introduction}
\label{section:introduction}

Particle-like objects have a very long and broad history in science, arising long
before Newton's corpuscles of light, and spanning the range from fundamental to astronomical.
In the mid-1950s, John Wheeler sought to construct stable, particle-like solutions from
only the smooth, classical fields of electromagnetism coupled to general
relativity \citep{Wheeler:1955zz,Power:1957zz}. Such solutions would
represent something of a ``gravitational atom'', but the solutions Wheeler found, which he
called \emph{geons}, were unstable.
However,
in the following decade, Kaup replaced electromagnetism\footnote{But see the discussion 
\CP{of oscillatons with a real vector field in Sect.~\ref{subsection:varieties_oscillatons}} and
of geons in AdS in Sect.~\ref{sec:other}.}
with a complex scalar
field \citep{Kaup:1968zz}, and found \emph{Klein--Gordon geons} that, in all their guises,
have become well-known as today's \emph{boson stars} (see Sect.~II of \citealt{Schunck:2003kk}
for a discussion of the naming history of boson stars).

As compact, stationary configurations of scalar field bound
by gravity, boson stars are called upon to fill a number of different
roles. Most obviously, could such solutions actually represent
astrophysical objects, either observed directly or indirectly through its gravity? Instead, if
constructed larger than a galaxy, could a boson star serve as the dark
matter halo that explains the flat rotation curve observed for most galaxies?

The equations describing boson stars are relatively simple,
and so even if they do not exist in nature, they 
still serve as a simple and important model for compact objects, ranging from
particles to stars and galaxies. In all these cases,
boson stars represent a balance between the dispersive nature of
the scalar field  and the attraction of gravity holding 
it together.

This review is organized as follows. The rest of this section 
describes some general features about boson stars. The system of equations describing
the evolution of the scalar field and gravity (i.e., the Einstein--Klein--Gordon
equations) are presented in Sect.~\ref{section:solving}. These equations are
restricted to the spherical symmetric case (with a harmonic ansatz for the complex
scalar field and a simple massive potential) to obtain a boson-star family of
solutions. To accommodate all their possible uses, a large variety of boson-star types
have come into existence, many of which are described in more detail in
Sect.~\ref{section:varieties}. For example, one can vary the form of the
scalar field potential to achieve a larger range of masses and compactnesses 
than with just a mass term in the potential. Certain types of potential admit 
soliton-like solutions even in the absence of gravity, leading to so-called
Q-stars. One can adopt Newtonian gravity instead of general relativity, or construct 
solutions from a real scalar field instead of a complex one. It is also possible to find
solutions coupled  to an electromagnetic field or a perfect fluid,
leading respectively to charged boson stars and fermion-boson stars. 
Rotating boson stars are found to have an angular momentum which is not arbitrary, but instead quantized, and can even coexist with a Kerr black hole. \CP{Multi-field} boson
stars with more than one complex scalar field are also considered.
Recently, stars made of a massive vector field 
have been constructed 
which more closely match the original geon proposal
because such a field has the same unit spin as Maxwell.

We discuss the dynamics of boson stars in Sect.~\ref{section:dynamics}.
Arguably, the most important property of boson-star dynamics concerns their stability.
Approaches to analyzing their stability include
linear perturbation analysis, catastrophe theory, and fully non-linear, numerical evolutions. 
The
latter option allows for the study of the final state of perturbed stars.
Possible endstates include dispersion to infinity of the scalar field, migration from unstable
to stable configurations, and collapse to a black hole. There is also the
question of formation of boson stars. Full numerical evolutions in 3D allow for the
merger of binary boson stars, which display a large range of different behaviors as well
producing distinct gravitational-wave signatures.

Finally, we review the impact of boson stars in astronomy in Sect.~\ref{section:astro}
(as astrophysical objects, black hole mimickers, gravitational-wave sources, and sources of dark
matter) and in mathematics in Sect.~\ref{section:math} (appearing in critical behavior,
the Hoop conjecture, other dimensions and anti-de~Sitter spacetimes, and gravitational analogs). We conclude with some remarks
and future directions. 

\subsection{The nature of a boson star}
\label{sec:origins}

Boson stars (BS) are constructed with a complex scalar field coupled to gravity
(as described in Sect.~\ref{section:solving}). A complex scalar field $\phi(t,\mathbf{r})$ can
be decomposed into two real scalar fields $\phi_{\mathrm{R}}$ and $\phi_{\mathrm{I}}$ mapping every spacetime event 
to the complex plane
\begin{equation}
\phi(t,\mathbf{r}) \equiv \phi_{\mathrm{R}}(t,\mathbf{r}) + i \phi_{\mathrm{I}}(t,\mathbf{r}) \,.
\end{equation}

Such a field possesses energy because of its spatial gradients and time derivatives, and
this energy gravitates holding the star together. Less clear is what supports the star
against the force of gravity. Its constituent scalar field obeys a Klein--Gordon wave equation
which tends to disperse fields. This is the same dispersion which underlies the Heisenberg
uncertainty principle. Indeed, Kaup's original work \citep{Kaup:1968zz}
found energy eigenstates for a semi-classical, complex scalar field, discovering that gravitational collapse was not inevitable. \cite{PhysRev.187.1767}
followed up on this work by quantizing a real scalar field 
representing some number of bosons and they found the same field equations.

\begin{figure}[htbp]
\centerline{\includegraphics[width=9.0cm]{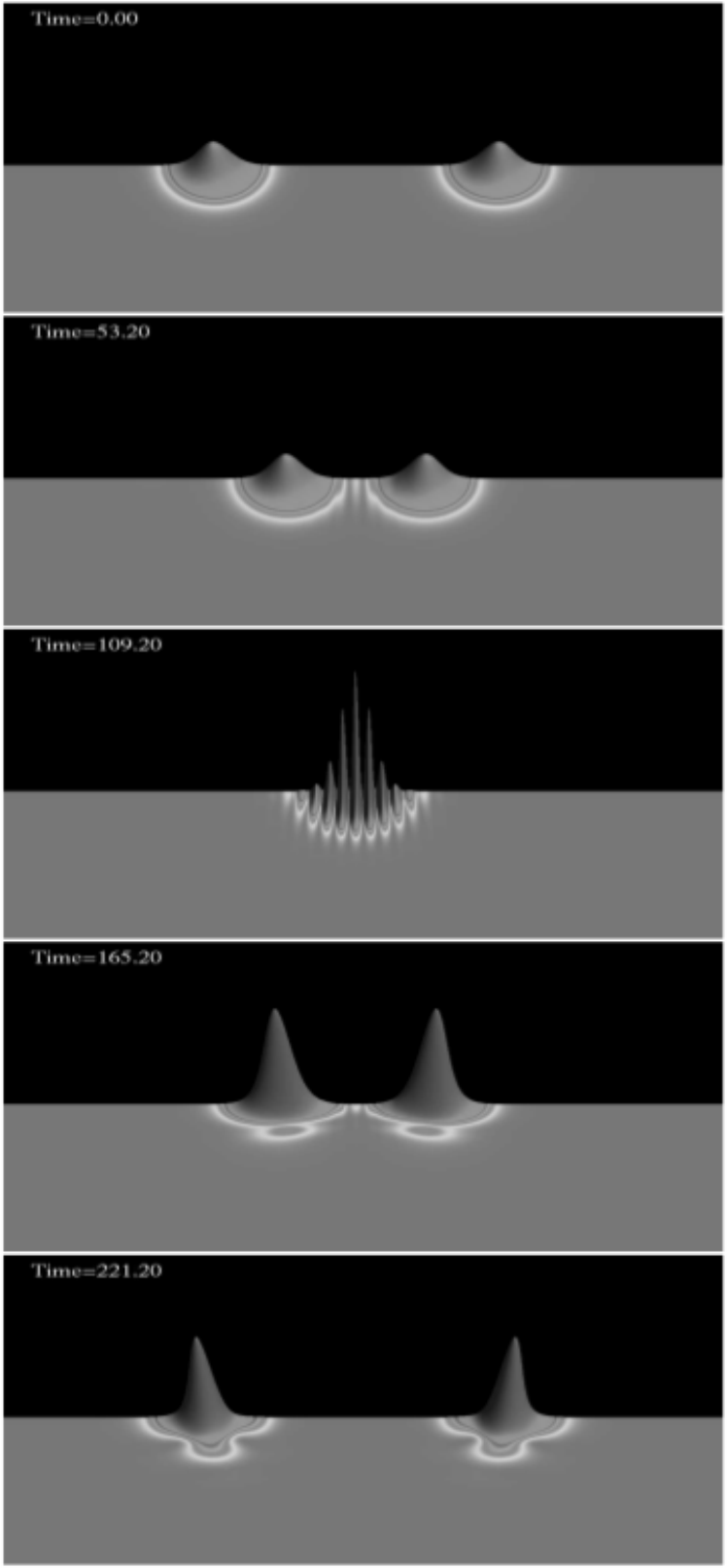}}
\caption{Demonstration of the solitonic nature of the (mini-)boson star. Shown are snapshots of
the magnitude squared of the complex scalar field for a head-on collision of two
identical mini-boson stars. The interacting stars display an interference pattern as they
pass through each other, recovering their individual identities after the collision.
However, note that the BSs have a larger amplitude after their interaction and
so are not true solitons. The collision can therefore
be considered \emph{inelastic}.
Reproduced with permission from \cite{bosonunpub}.
See also \cite{2005PhDT.........2L} (e.g., Figure~5.12).}
\label{fig:soliton}
\end{figure}

None of this guarantees that such solutions balancing dispersion against gravitational
attraction exist. In fact, a
widely known theorem,
\emph{Derrick's theorem} \citep{Derrick:1964ww} (see also \citealp{1966JMP.....7.2066R} and its extension to the case of a general non-canonical scalar field \citealt{2013PhRvD..88f7302D}), uses a clever scaling
argument to show that no regular, static, nontopological localized scalar field
solutions are stable in three (spatial) dimensional flat space. This constraint is avoided by
adopting a harmonic ansatz for the complex scalar field
\begin{equation}
  \phi( \mathbf{r},t) = \phi_0(\mathbf{r}) e^{i \omega t}
\end{equation}
and by working with gravity.
Although the field is no longer static, as shown in Sect.~\ref{section:solving} the spacetime
remains static. The star itself is a stationary, soliton-like solution as demonstrated
in Fig.~\ref{fig:soliton}.

There are, of course, many other soliton and soliton-like solutions in three dimensions finding a
variety of ways to evade Derrick's theorem. For example, the field-theory monopole
of 't~Hooft and Polyakov is a localized solution of a properly gauged triplet scalar field. Such a solution
is a topological soliton because the monopole possesses false vacuum energy which is
topologically trapped. The monopole is one among a number of different topological defects
that requires an infinite amount of energy to ``unwind'' the potential energy trapped within
(see \citealt{vilenkinbook} for a general introduction to defects and the introduction
of \citealt{ryderbook} for a discussion of relevant classical field theory concepts).

\sllnew{Derrick's Theorem is technically limited to flat space, leaving open the possibility that general relativity could allow for the formation of regular, static solutions. Efforts to exclude this possibility include work by Hod with
certain assumptions about the scalar
potential (including monotonically increasing) \citep{Hod:2018dij,Hod:2019yfl}. A more general result was presented
by \cite{Carloni:2019cyo}.
Further work considered AdS \citep{Peng:2019uzw} and other couplings \citep{Liu:2022iyj}.
}

In Sect.~\ref{section:solving}, we present the underlying equations and mathematical solutions, but here
we are concerned with the physical nature of these boson stars.  When searching for an actual boson star, we 
look not for a quantized wave function or even a semiclassical one. 
Instead, we look to a fundamental scalar to provide the bosonic material of the star. 
\sllnew{Only in the last decade}
has a scalar particle been experimentally found with the discovery by the Large Hadron Collider (LHC) of the standard model Higgs boson with a mass roughly 
$125\mathrm{\ GeV}/c^2$ \citep{Aad:2012tfa,Chatrchyan:2012xdj,2015EPJC...75..212K}.
Of course, other proposed bosonic candidates remain, such as the axion particle.

Boson stars are then either a collection of stable fundamental bosonic particles bound by gravity,
or else a collection of unstable particles that, with the gravitational binding, have an inverse
process efficient enough to reach an equilibrium.
They can thus be considered a Bose--Einstein condensate (BEC), 
although boson stars can also exist in an excited state as well.

Indeed, applying the uncertainty principle to a boson star by assuming it to be a macroscopic quantum state results in
an excellent estimate for the maximum mass of a BS. One begins with the
Heisenberg uncertainty principle of quantum mechanics
\begin{equation}
\Delta p\, \Delta x \ge \hbar
\end{equation}
and assumes the BS is confined within some radius $\Delta x = R$ with a maximum momentum of $\Delta p = mc$
where $m$ is the mass of the constituent particle
\begin{equation}
m c R \ge \hbar \,.
\end{equation}
This inequality is consistent with the star being described by a
Compton wavelength of $\lambda_{\mathrm{C}} = h/\left( mc \right)$.
We look for the maximum possible mass $M_{\max}$ for the boson star which will saturate the uncertainty bound
and drive the radius of the star towards its Schwarzschild radius $R_{\mathrm{S}} \equiv 2\,GM_{\max}/c^2$.
Substituting yields
\begin{equation}
\frac{2\,G m\,M_{\max}}{c} = \hbar \,,
\end{equation}
which gives an expression for the maximum mass
\begin{equation}
M_{\max} = \frac{1}{2} \frac{\hbar c}{Gm} \,.
\end{equation}
Recognizing the Planck mass $M_{\mathrm{Planck}} \equiv \sqrt{\hbar c / G}$,
we obtain the estimate
of $M_{\max} = 0.5\,M^2_{\mathrm{Planck}}/m$ \sllnew{(see \cite{Herdeiro:2022gzp} for a discussion of the conditions under which a boson star is the legitimate classical limit of a quantum field)}.
This simple estimate indicates that the maximum mass of the BS is inversely related to
the mass of the constituent scalar field. We will see below in Sect.~\ref{section:solving} that
this inverse relationship continues to hold with the explicit solution of the differential equations
for a simple mass term in the potential,
but can vary with the addition of self-interaction terms.
Indeed
depending on the strength of the coupling $m$ and the other parameters of the 
self-interaction potential, 
the size and mass of the boson stars can vary
from atomic to astrophysical scales.

Despite their connection to fundamental physics, one can also view boson stars in
analogy with models of neutron stars. In particular, as we discuss in the following
sections, both types of star demonstrate somewhat similar mass versus radius
curves for their solutions, with a transition in stability at the local maxima of the mass.
There is also a correspondence between (massless) scalar fields and a stiff, perfect
fluid (see Sect.~2.1 and Appendix~A of \citealt{Brady:2002iz}), but the correspondence
does not mean that the two are equivalent \citep{Faraoni:2012hn}.
More than just an analogy, boson stars can serve as a very useful model
of a compact star, having certain advantages over a fluid neutron star model:
  (i)~the equations governing its dynamics avoid developing discontinuities, 
      in particular there is no sharp stellar surface,
 (ii)~there is no concern about resolving turbulence,
and
(iii)~one avoids uncertainties in the equation of state \sllnew{(at the cost of having to choose a potential for the bosonic field)}.

%
%
\subsection{Other reviews}
\label{sec:otherreviews}

%
%

A number of other reviews of boson stars have appeared. 
\cite{Schunck:2003kk} concentrate on the possibility of detecting BSs,
extending their previous reviews \citep{Mielke:1997re,Mielke:2000im}.
In 1992, a number of reviews appeared:
\cite{Jetzer:1991jr} concentrates on the astrophysical relevance of BS
(in particular their relevance for explaining dark matter)
while \cite{1992IJMPD...1..101L} focus on their formation.
Other reviews include \cite{Straumann:1991pt,1992PhR...221..251L}.
\cite{Mielke:2016war} reviewed rotating boson stars, while \cite{Herdeiro:2015gia} reviewed Kerr black holes with scalar hair.

\sllnew{
\cite{Braaten:2019knj} have published an extensive review focused on axion stars, while \cite{Visinelli:2021uve} wrote a broad review of boson stars.
Most recently, \cite{Shnir:2022lba} presented a short review focusing on multipolar scalar configurations. 
}


\section{Solving for boson stars}
\label{section:solving}

In this section, we present the equations governing boson-star solutions, namely
the Einstein equations for the geometry description and the Klein--Gordon
equation to represent the (complex) scalar field. We refer to this coupled
system as the Einstein--Klein--Gordon (EKG) equations.

The covariant equations describing boson stars are presented in
Sect.~\ref{subsection:lagrangian}, which is followed by choosing
particular coordinates consistent with a 3+1 decomposition in
Sect.~\ref{subsection:3+1_decomposition}. A form for the potential of
the scalar field is then chosen and solutions are presented in
Sect.~\ref{subsection:spherical}.

\subsection{Conventions}
\label{section:conventions}

Throughout this review, Roman letters from the beginning of the
alphabet $a,b,c,\dots$ denote spacetime indices ranging from 0 to 3,
while letters near the middle $i,j,k,\dots$ range from 1 to 3,
denoting spatial indices. Unless otherwise stated, we use units such 
that $\hbar=c=1$ so that the Planck mass becomes 
$M_{\mathrm{Planck}} = G^{-1/2}$. We also use
the signature convention $(-,+,+,+)$ for the metric.

\subsection{The Lagrangian, evolution equations and conserved quantities}
\label{subsection:lagrangian}

The EKG evolution equations can be derived from the action \cite{1984ucp..book.....W}
\begin{equation}\label{action}
  {\cal S} =  \int \left( \frac{1}{16 \pi G} R + {\cal L_M} \right)
    \sqrt{-g}\, d^4x
\end{equation}
where $R$ is the Ricci scalar of the spacetime represented by the metric $g_{ab}$,
and its determinant $\sqrt{-g}$.
The term ${\cal L_M}$ describes the matter, which here is that
of a complex scalar field, $\phi$
\begin{equation}\label{Lagrangian1}
  {\cal L_M} = 
   - \frac{1}{2} \left[ g^{ab} \nabla_a \bar \phi\, \nabla_b \phi
   + V\left( \left|\phi \right|^2\right) \right] \,,
\end{equation}
where $\bar \phi$ is the complex conjugate of the field
and $V(|\phi|^2)$ is the potential depending only on 
the magnitude of the scalar field, consistent with the $U(1)$ invariance of the field in the complex plane.

Variation of the action in Eq.~(\ref{action}) with respect to the metric $g^{ab}$
leads to the well-known Einstein equations 
\begin{eqnarray}\label{EE1}
   R_{ab} &-& \frac{R}{2} g_{ab} = 8 \pi G T_{ab} \\
\label{stress-energy}
   T_{ab} &=& \frac{1}{2} \left[\nabla_a \bar \phi \, \nabla_b \phi
                        + \nabla_a \phi \, \nabla_b \bar \phi \right]
            - \frac{1}{2} g_{ab} \left[g^{cd} \nabla_c \bar \phi \, \nabla_d \phi
                                      + V\left(|\phi|^2\right)\right] \,,
\end{eqnarray}
where $R_{ab}$ is the Ricci tensor and $T_{ab}$ is the real stress-energy tensor.
Eqs.~(\ref{EE1}) form a system of 10 non-linear partial differential equations for the
spacetime metric components $g_{ab}$ coupled to the scalar field
via the stress-energy tensor given in Eq.~(\ref{stress-energy}).

On the other hand, the variation of the action in Eq.~(\ref{action})
with respect to the scalar field $\phi$, leads to the
Klein--Gordon (KG) equation
\begin{equation}\label{KG1}
  g^{ab} \nabla_a \nabla_b \phi = \frac{d V}{d |\phi|^2} \phi \, .
\end{equation}
An equivalent equation is obtained when varying the action with respect
to the complex conjugate $\bar \phi$. The simplest potential leading to
boson stars is the so-called free field case, where the potential takes the form
\begin{equation}\label{potential}
  V( |\phi|^2 ) = m^2~ |\phi|^2 \,,
\end{equation}
with $m$ a parameter that can be identified with the bare
mass of the field theory.

According to Noether's theorem, the invariance of the
Klein--Gordon Lagrangian in Eq.~(\ref{Lagrangian1}) under global
$U(1)$ transformations $\phi \rightarrow \phi e^{i \varphi}$
(such that  $\delta \phi = i \phi$)
implies the existence of a conserved current
\begin{eqnarray}\label{Noether_current}
  J^{a} = \frac{\partial {\cal L_M}}{\partial (\nabla_{a} \phi)} \delta \phi
        + \frac{\partial {\cal L_M}}{\partial (\nabla_{a} \bar \phi)} \delta \bar \phi
        = \frac{i}{2} g^{ab} \left(\bar \phi \, \nabla_{b} \phi-
                             \phi \, \nabla_{b} \bar \phi \right) \,,
\end{eqnarray}
satisfying the conservation law
\begin{equation}
  \nabla_a J^{a} = \frac{1}{\sqrt{-g}} \partial_a 
         \left( \sqrt{-g}\, J^a  \right) = 0 \,.
\end{equation}
The spatial integral of the time component %
of this current
defines the conserved Noether charge, given by
\begin{equation}\label{Noether_charge}
   N = \int J^0 \sqrt{-g} \, dx^3 \,,
\end{equation}
which can be associated with the total number of bosonic
particles \citep{PhysRev.187.1767}.
If one neglects the binding energy of the star, then the total mass can be expressed simply in terms of the bare mass as $mN$.

\subsection{The 3+1 decomposition of the spacetime}
\label{subsection:3+1_decomposition}

Although the spacetime description of general relativity is
very elegant, the covariant form of Einstein equations is not
suitable to describe how an initial configuration evolves towards the future.
It is, therefore, more intuitive to instead consider a succession of spacetime
geometries, where the evolution of a given slice is given by the
Einstein equations (for more detailed treatments see \citealt{alcubierre2008introduction, baumgarte2010numerical, bona2009elements, Gourgoulhon:2007ue}).
In order to convert the four-dimensional,
covariant Einstein equations to a more intuitive ``space+time''
or 3+1 decomposition, the following steps are taken:
\begin{itemize}
\item \emph{specify the choice of coordinates.}
The spacetime is foliated by a family of spacelike hypersurfaces,
which are crossed by a congruence of time lines that will determine
our observers (i.e., coordinates). This congruence is described by
the vector field $t^a = \alpha n^a +\beta^a$, where $\alpha$ is the
lapse function which measures the proper time of the observers,
$\beta^a$ is the shift vector that measures the displacement of the
observers between consecutive hypersurfaces and $n^a$ is the timelike
unit vector normal to the spacelike hypersurfaces. 

\item \emph{decompose every 4D object into its 3+1 components.}
The choice of coordinates allows for the definition of a projection
tensor ${\gamma^a}_b \equiv \delta^a_b + n^a\, n_b$. Any four-dimensional tensor can
be decomposed into 3+1 pieces using the spatial projector to obtain
the spatial components, or contracting with $n^a$ for the time
components. For instance, the line element can be written in a general form as
\begin{equation}\label{line_element}
   ds^2 = - \alpha^2\, dt^2 + \gamma_{ij} (dx^i + \beta^i dt)\,
                                          (dx^j + \beta^j dt) \,.
\end{equation}
The stress-energy tensor can then be decomposed into its various components as
\begin{equation}\label{3+1_stress_energy}
  \tau \equiv T^{ab}\, n_a\, n_b \,, \qquad S_i \equiv T_{ab}\, n^a
  \,{\gamma^a}_i \,, \qquad S_{ij} \equiv T_{ab}\, {\gamma^a}_i\, {\gamma^b}_j \,. 
\end{equation}

\item \emph{write down the field equations in terms of the 3+1 components.}
Within the framework outlined here, the induced 
(or equivalently, the spatial 3D) metric $\gamma_{ij}$ and the scalar field $\phi$ 
are as yet still unknown (remember that
the lapse and the shift just describe our choice of coordinates).
In the original 3+1 decomposition (ADM formulation \citealt{adm1962}) 
an additional geometrical tensor $K_{ij} \equiv -\left(1/2 \right)
{\cal L}_{\mathbf{n}} \gamma_{ij} = -1/\left(2\alpha\right) \left( \partial_t-{\cal L}_\beta \right)\gamma_{ij}$
is introduced to describe the change of
the induced metric along the congruence of observers. Loosely speaking,
one can view the determination of $\gamma_{ij}$ and $K_{ij}$ as akin
to the specification of a position and velocity for projectile motion.
In terms
of the extrinsic curvature and its trace, $\mathrm{trK} \equiv {K_i}^i$, the Einstein equations can be written as
\begin{equation}
 \label{energyconst}
     {R_i}^i + \left(\mathrm{trK}\right)^2 - {K_i}^j\, {K_j}^i 
      = 16\, \pi\, G\, \tau\,
\end{equation}
\begin{equation}
  \label{momconst}
     \nabla_j\;\left({K_i}^j - \mathrm{trK}\;{\delta_i}^j \right) 
      = 8\, \pi\, G\, S_i 
\end{equation}
\begin{equation}
 \label{Kevol}
       \left(\partial_t - \mathcal{L}_\beta\right) K_{ij}
   = - \nabla_i \nabla_j  \alpha
    + \alpha \left( R_{ij} -2{K_i}^k {K_{jk}} + \mathrm{trK}\,K_{ij} 
     - 8\pi G \left[S_{ij}-{\frac{\gamma_{ij}}{2}}\left(\mathrm{trS} - \tau\right)\right] \right)
\end{equation}
In a similar fashion, one can introduce a quantity $Q \equiv - {\cal L}_{\mathbf{n}} \phi$ 
for the Klein--Gordon equation which reduces it to an equation
first order in time, second order in space
\begin{equation}
  \label{evolKG}
   \partial_t (\sqrt{\gamma}\, Q) 
 - \partial_i (\beta^i \sqrt{\gamma} Q) 
  + \partial_i (\alpha\, \sqrt{\gamma}\, \gamma^{ij}\, \partial_j \phi)
   = \alpha\, \sqrt{\gamma}\, \frac{d V}{d |\phi|^2} \phi\,.
\end{equation}

\item \emph{enforce any assumed symmetries.} Although the boson star is found by a harmonic ansatz for the time dependence,
here we choose to retain the full time-dependence. However, a considerably simplification is provided by
assuming that the spacetime is spherically symmetric.
Following \cite{2005PhDT.........2L}, the most general metric in this
case can be written in terms of spherical coordinates as
\begin{equation}\label{spherical_timedependent_metric}
ds^2 = \left(- \alpha^2 + a^2\, \beta^2 \right) dt^2
    + 2\,a^2\,\beta \,dt\,dr + a^2\, dr^2 + r^2\, b^2\, d\Omega^2 \,,
\end{equation}
where $\alpha(t,r)$ is the lapse function, $\beta(t,r)$ is the radial component of the
shift vector and $a(t,r),b(t,r)$ represent components of the spatial metric,
with $d\Omega^2$ the metric of a unit two-sphere. With this metric, the extrinsic
curvature
only has two independent components
$K^i_j = {\mathrm{diag}} \left( {K^r}_r,
{K^{\theta}}_{\theta}, {K^{\theta}}_{\theta} \right)$.
The constraint equations, Eqs.~(\ref{energyconst}) and~(\ref{momconst}), can now be written as
\begin{eqnarray}
 \label{energyconst_rt}
  -\frac{2}{a r b} \left\{ \partial_r \left[ \frac{\partial_r (r b)}{a} \right]
     + \frac{1}{r b} \left[ \partial_r \left( \frac{r b}{a} \partial_r \left(r b\right) \right) - a \right] 
    \right\} &+& 4 {K^r}_r\, {K^{\theta}}_{\theta} + 2 {K^{\theta}}_{\theta}\, {K^{\theta}}_{\theta}
  \\
 \label{momconst_rt}
   &=& \frac{8 \pi G}{a^2} \left[ \left|\Phi\right|^2 + \left|\Pi\right|^2 + a^2 V\left(|\phi|^2\right) \right]
 \nonumber \\
   \partial_r {K^{\theta}}_{\theta} + \frac{\partial_r\left(r b\right)}{r b} ({K^{\theta}}_{\theta} - {K^{r}}_{r})
   &=& \frac{2 \pi G}{a} \left( \bar \Pi \Phi + \Pi \bar \Phi \right) \,,
\end{eqnarray}
where we have defined the auxiliary scalar-field variables
\begin{equation}
    \Phi \equiv \partial_r \phi \,, \qquad 
    \Pi \equiv \frac{a}{\alpha} \left( \partial_t \phi - \beta \partial_r \phi \right) \,.
\end{equation}
The evolution equations for the metric and extrinsic curvature
components reduce to
\begin{eqnarray}
\label{gevolrr_rt}
   \partial_t a &=& \partial_r (a \beta) - \alpha a {K^{r}}_{r} 
  \\
\label{gevoltt_rt}
   \partial_t b &=& \frac{\beta}{r} \partial_r (r b) - \alpha b {K^{\theta}}_{\theta} 
  \\ 
 \label{Kevolrr_rt}
   \partial_t {K^{r}}_{r} - \beta \partial_r {K^{r}}_{r} 
  &=& - \frac{1}{a} \partial_r \left( \frac{\partial_r \alpha}{a} \right)\nonumber \\
   && + \alpha \left\{ - \frac{2}{a r b} \partial_r \left[ \frac{\partial_r (r b)}{a} \right]
     + \mathrm{trK}\, {K^r}_r - \frac{4\pi\,G}{a^2} \left[ 2 |\Phi|^2 + a^2 V(|\phi|^2) \right]
 \right\}
  \nonumber \\
 \label{Kevoltt_rt}
    \partial_t {K^{\theta}}_{\theta} - \beta \partial_r {K^{\theta}}_{\theta} 
  &=& \frac{\alpha}{(r b)^2} 
     - \frac{1}{a (r b)^2} \partial_r \left[ \frac{\alpha r b}{a} \partial_r (r b) \right] 
     + \alpha \left[ \mathrm{trK}\, {K^{\theta}}_{\theta} - 4\pi\,G V(|\phi|^2) \right] \,.
\end{eqnarray}

Similarly, the reduction of the Klein--Gordon equation to first order
in time and space leads to the following set of evolution equations
\begin{eqnarray}
  \label{evolKG1_a}
  \partial_t \phi &=& \beta \Phi + \frac{\alpha}{a} \Pi 
  \\
\label{evolKG1_b}
  \partial_t \Phi &=& \partial_r \left( \beta \Phi + \frac{\alpha}{a} \Pi \right)
  \\
  \partial_t \Pi &=& \frac{1}{(r b)^2} 
       \partial_r \left[ (r b)^2 \left( \beta \Pi + \frac{\alpha}{a} \Phi \right) \right]
\nonumber \\
& &
       + 2 \left[ \alpha {K^{\theta}}_{\theta} - \beta \frac{\partial_r (r b)}{r b} \right] \Pi
       - \alpha a \frac{d V}{d |\phi|^2} \phi \,.
\label{evolKG1_c}
\end{eqnarray}
This set of equations,
Eqs.~(\ref{energyconst_rt})\,--\,(\ref{evolKG1_c}), describes
general, time-dependent, spherically symmetric solutions of a
gravitationally-coupled complex scalar field. In the next section, we
proceed to solve for the specific case of a boson star.

\end{itemize}

\subsection{Mini-boson stars}
\label{subsection:spherical}

The concept of a star entails a configuration of matter
which remains localized. One, therefore, looks for a localized and
time-independent 
matter configuration such that the gravitational field is stationary
and regular everywhere. As shown in \cite{1987PhRvD..35.3640F}, %
such a configuration does not exist for a real scalar field. But since
the stress-energy tensor %
depends only on the modulus of the scalar field and its gradients, one can relax the
assumption of time-independence of the scalar field while retaining a time-independent
gravitational field.
The key is to assume
a harmonic ansatz for the scalar field
\begin{equation}\label{harmonic}
  \phi( \mathbf{r},t) = \phi_0(\mathbf{r}) e^{i \omega t} \,,
\end{equation}
where $\phi_0$ is a real scalar which is the profile of the star
and $\omega$ is a real constant denoting the angular frequency of the
phase of the field in the complex plane.

We consider spherically symmetric, equilibrium configurations
corresponding to minimal energy solutions while requiring 
the spacetime to be static. In Schwarzschild-like coordinates, the 
general, spherically symmetric, static
metric can be written as
\begin{equation}\label{spherical_metric}
ds^2 = - \alpha\left(r\right)^2 dt^2 + a\left(r\right)^2 dr^2 + r^2 d\Omega^2 \,,
\end{equation}
in terms of two real metric functions, $\alpha$ and $a$. 
The coordinate $r$ is an areal radius such that spheres of
constant $r$ have surface area $4\pi r^2$. For this reason, these coordinates are often
called polar-areal coordinates.

The equilibrium equations are obtained by substituting 
the metric of Eq.~(\ref{spherical_metric}) and the harmonic ansatz of
Eq.~(\ref{harmonic}) into the spherically symmetric EKG system of
Eqs.~(\ref{gevolrr_rt}\,--\,\ref{evolKG1_c}) with $\beta=0,b=1$, resulting in 
three first order partial differential equations~(PDEs)
\begin{eqnarray}
\label{ekg2a}
\partial_r a &=& -\frac{a}{2r} \left(a^2 -1\right) + 4\pi\,G r a^3 \tau 
\\
\label{ekg2b} 
\partial_r \alpha &=& \frac{\alpha}{2r} \left(a^2 -1\right) + 4\pi\,G r \alpha a^2 {S^r}_r 
\\
\label{ekg2c}
\partial_r \Phi &=& - \left[ 1 + a^2 + 4\pi\,G r^2 a^2 \left( S^r_r - \tau\right) \right] \frac{\Phi}{r} 
                - \left( \frac{ \omega^2 }{\alpha^2} 
                          - \frac{d V}{d |\phi|^2} \right) a^2\, \phi_0 \,.
\end{eqnarray}
Notice that these equations hold for any stress-energy contributions and
for a generic type of self-potential $V(|\phi|^2)$.
In order to close the system of Eqs.~(\ref{ekg2a}\,--\,\ref{ekg2c}), we still have to prescribe 
this potential. The simplest case admitting localized solutions is the free
field case of Eq.~(\ref{potential}) for which the potential describes a field with mass $m$ and
for which the equations can be written as
\begin{eqnarray}
\label{ekg3a}
\partial_r a &=& \frac{a}{2} \left\{- \frac{a^2-1}{r} \right.
       \left. + 4\pi\,G r \left[ \left(\frac{\omega^2}{\alpha^2} + m^2\right)
                a^2 \phi_0^2 + \Phi^2 \right] \right\} \,, 
\\
\label{ekg3b}
\partial_r \alpha &=& \frac{\alpha}{2}\left\{\ \frac{a^2-1}{r} \right.
     \left. + 4\pi\,G r \left[ \left(\frac{\omega^2}{\alpha^2}-m^2\right)
               a^2 \phi_0^2 + \Phi^2 \right] \right\} \,,
\\
\label{ekg3c}
\partial_r \Phi &=& - \left\{ 1 + a^2 - 4\pi\,G r^2 a^2 m^2 \phi_0^2 \right\} \frac{\Phi}{r} 
                - \left( \frac{\omega^2}{\alpha^2} - m^2 \right) \phi_0\, a^2 \,.\label{ekg2}
\end{eqnarray}
In order to obtain a
physical solution of this system, we have to impose the following
boundary conditions,
\begin{eqnarray}
\phi_0\left(0 \right) &=& \phi_{c}, \label{BC1}\\
\Phi\left(0 \right) &=& 0,\label{BC2}\\
a\left(0 \right) &=& 1,\label{BC3} \\
\lim_{r\rightarrow \infty}\phi_0 \left(r \right) &=& 0, \label{BC4}\\
\lim_{r\rightarrow \infty}\alpha\left(r \right) &=& \lim_{r\rightarrow \infty} \frac{1}{a(r)}\,, \label{BC5}
\end{eqnarray}
which guarantee regularity at the origin and asymptotic flatness.
For a given central value of the field $\{\phi_{c}\}$, we need
only to adjust the eigenvalue
$\{\omega\}$ 
to find a solution which matches the
asymptotic behavior of Eqs.~(\ref{BC4}\,--\,\ref{BC5}).
This system can be solved as a shooting problem 
by integrating from $r=0$ towards the outer boundary $r=r_{\mathrm{out}}$
(see \citealt{Dias:2015nua} for a review on numerical methods to find
stationary gravitational solutions). 
Equation~(\ref{ekg3b}) is linear and homogeneous in $\alpha$ and one is therefore
able to rescale the field consistent with Eq.~(\ref{BC5}). We can get
rid of the constants in the equations by re-scaling the variables
in the following manner
\begin{equation}
  {\tilde \phi_0} \equiv \sqrt{4\pi\,G} \phi_0 \,, \qquad
  {\tilde r} \equiv m\,r \,, \qquad 
  {\tilde t} \equiv \omega\, t \,, \qquad
  {\tilde \alpha} \equiv (m/\omega) \alpha \,.
\end{equation}

Notice that the form of the metric in Eq.~(\ref{spherical_metric}) resembles
Schwarzschild allowing the association $a^2 \equiv (1 - 2\, M/r)^{-1}$,
where $M$ is the ADM mass of the spacetime.
This allows us to define a more general mass aspect function
\begin{equation}\label{aspect_mass}
   M(r,t) = \frac{r}{2} \left( 1 - \frac{1}{a^2(r,t)}\right) \,,
\end{equation}
which measures the total mass contained in a coordinate sphere of radius $r$
at time $t$.

In isotropic coordinates, the spherically symmetric
metric can be written as
\begin{equation}\label{spherical_metric_isotropic}
ds^2 = - \alpha\left(R\right)^2 dt^2 + \psi\left(R\right)^4 
          \left(dR^2 + R^2 d\Omega^2 \right) \,,
\end{equation}
where $\psi$ is the conformal factor. A change of the radial coordinate
$R=R(r)$ can transform the solution obtained in Schwarzschild coordinates
into isotropic ones, in particular
\begin{eqnarray}
   R (r_{\max}) &=& \left[ \left( \frac{1 + \sqrt{a}}{2} \right)^2 \frac{r}{a} \right]_{r_{\max}} \\
   \frac{dR}{dr} &=& a \, \frac{R}{r} \,,
\end{eqnarray}
where the first condition is the initial value to integrate the second equation
backwards, obtained by imposing that far away from the
boson star the spacetime resembles the Schwarzschild solution. By comparing the 
angular metric coefficients, we also find that $\psi = \sqrt{r/R}$. Further details can
be found in Appendix~D of \cite{2005PhDT.........2L}.

As above, boson stars are spherically symmetric
solutions of the Eqs.~(\ref{ekg3a}\,--\,\ref{ekg3c}) with asymptotic behavior
given by Eqs.~(\ref{BC1}\,--\,\ref{BC5}). For a given value of the central amplitude of the
scalar field $\phi_0(r=0) =\phi_c$, there exist configurations with some effective 
radius and a given mass satisfying the previous conditions for a different set of $n$ discrete
eigenvalues $\omega^{(n)}$. As $n$ increases, one obtains solutions with an
increasing number of nodes in $\phi_0$. 
The configuration without nodes is the \emph{ground state}, while all those with any
nodes are excited states.  As the number of nodes increases,
the distribution of the mass as a function of the radius becomes more homogeneous. 

As the amplitude $\phi_c$ increases, the stable configuration has a larger mass
while its effective radius decreases. This trend indicates that the compactness of the boson
star increases. However, at some point the mass instead decreases with increasing
central amplitude. Similar to models of neutron stars (see Sect.~4 of \citealt{cook_living_review}), this turnaround implies 
a maximum allowed mass for a boson star in the ground state, which
numerically was found to be $M_{\max} = 0.633\,M^2_{\mathrm{Planck}}/m$. The existence of a maximum
mass for boson stars is a relativistic effect, which is not present in the Newtonian limit, 
while the maximum of baryonic stars is an intrinsic property. 

\begin{figure}[htb]
\centerline{\includegraphics[width=9.0cm]{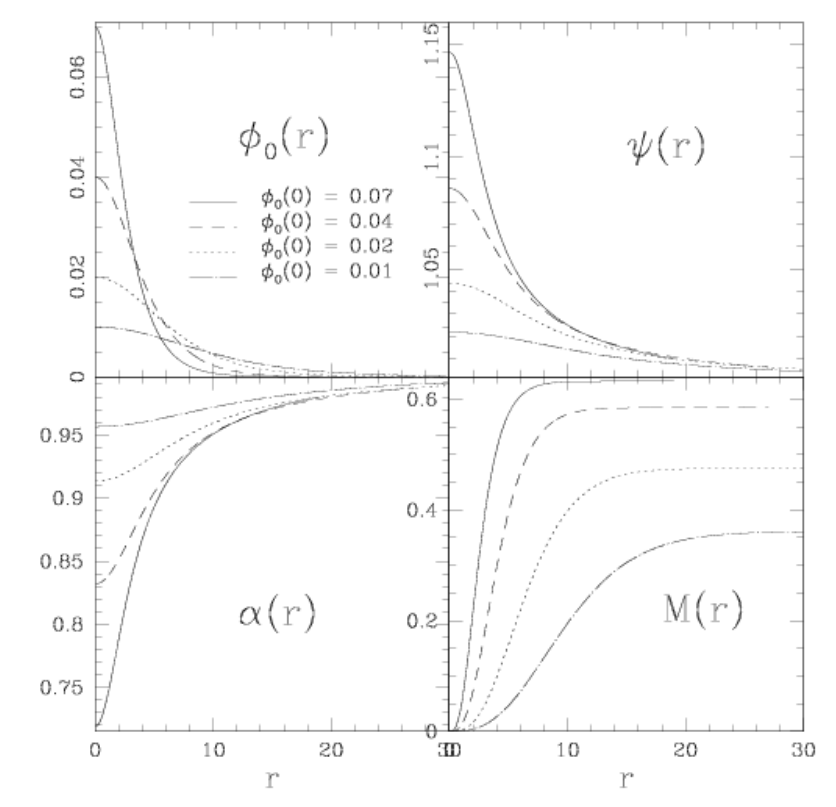}}
\caption{Profiles characterizing static, spherically symmetric boson
  stars with a few different values of the central scalar field (top
  left). Reproduced with permission from \cite{2005PhDT.........2L}.}
\label{fig:miniboson_id}
\end{figure}

Solutions for a few representative boson stars in the ground state are shown 
in Fig.~\ref{fig:miniboson_id} in isotropic coordinates. 
The boson stars becomes more compact for higher values of $\phi_c$, implying narrower
profiles for the scalar field, larger conformal factors, and smaller lapse functions,
as the total mass increases.

\section{Varieties of boson stars}
\label{section:varieties}

Quite a number of different flavours of boson
stars are present in the literature. They can have charge, a fermionic
component, or rotation. They can be constructed with various potentials
for the scalar field. The form of gravity which holds them together can
even be modified to, say, Newtonian gravity or even no gravity at all (Q-balls).
To a certain extent, such modifications are akin to varying the equation of
state of a normal, fermionic star.
Here we briefly review some of these variations, \CP{most of them represented in the diagram below}, paying particular attention
to recent work. 
%
%
\begin{figure*}[!ht]
\centerline{\includegraphics[width=\textwidth]{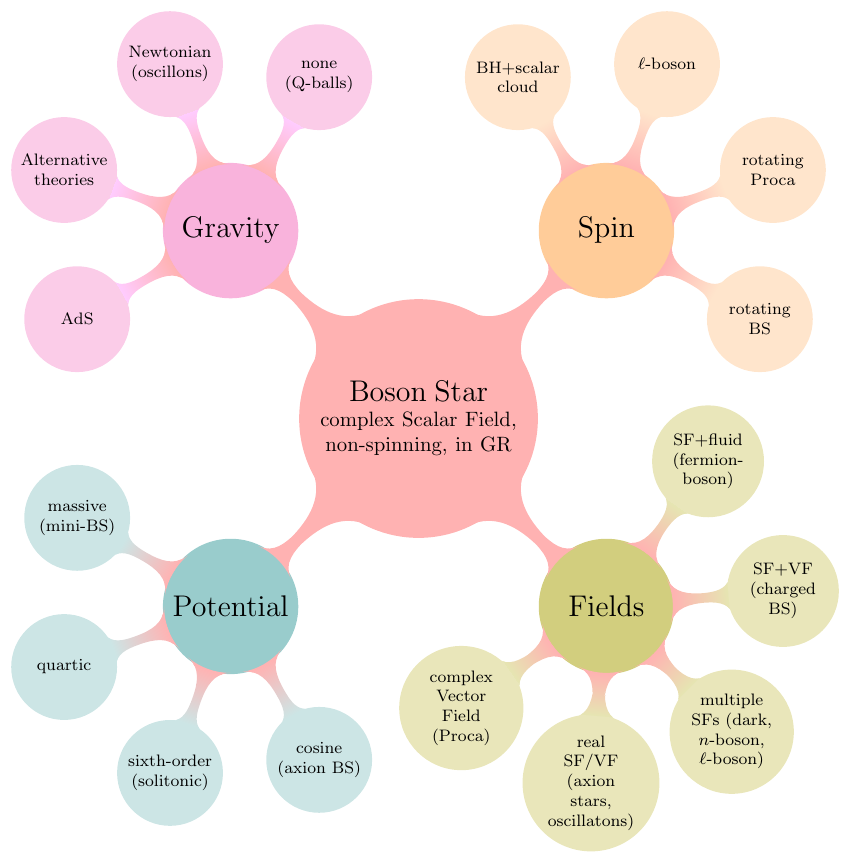}}
\end{figure*}
\subsection{Self-interaction potentials}
\label{subsection:varieties_selfinteraction}

Originally, boson stars were constructed with a free-field potential without any
kind of self-interaction, obtaining a maximum mass with a dependence
$M \approx M^2_{\mathrm{Planck}}/m$. This mass, for typical masses of bosonic particle candidates,
 is much smaller than the Chandrasekhar
mass $M_{\mathrm{Ch}} \approx M^3_{\mathrm{Planck}}/m^2$ obtained for fermionic
stars, and so they were known as mini-boson stars. In order to extend this limit and
reach astrophysical masses comparable to the Chandrasekhar mass, the potential
was generalized to include a self-interaction term that provided an extra
pressure against gravitational collapse.
To preserve the global $U(1)$ invariance, and hence to retain a conserved particle number, such a potential
should be a function of $|\phi|$.

\begin{figure}[htbp]
\centerline{\includegraphics[width=\textwidth]{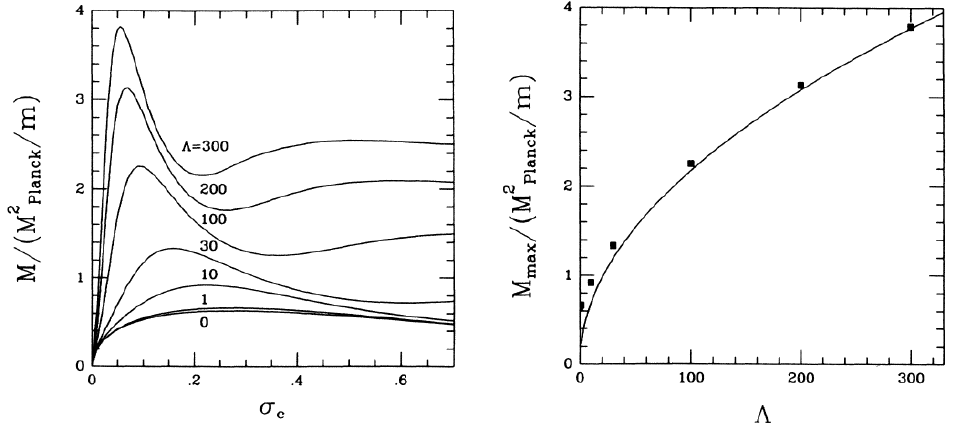}}
\caption{\emph{Left:} The mass of the boson star as a function of the
  central value of the scalar field in adimensional units $\sigma_c =
  \sqrt{4\pi\,G} \phi_c$. \emph{Right:} Maximum mass as a function of
  $\Lambda$ (squares) and the asymptotic $\Lambda \rightarrow \infty$
  relation of Eq.~(\ref{pot_quartic_maxmass}) (solid curve). Reproduced
  with permission from \cite{1986PhRvL..57.2485C}, copyright by APS.}
\label{fig:selfinteraction_id}
\end{figure}

\begin{figure}[htbp]
\centerline{\includegraphics[width=0.7\textwidth]{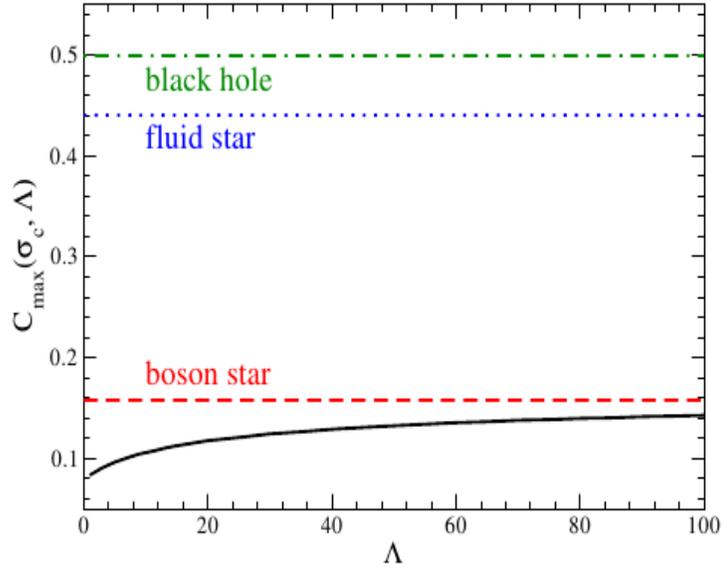}}
\caption{The compactness of a stable boson star (black solid line) as
  a function of the adimensional self-interaction parameter $\Lambda
  \equiv \lambda / \left(4\pi\,G m^2\right)$. The compactness is shown
  for the most massive stable star (the most compact BS is
  unstable). This compactness asymptotes for $\Lambda \rightarrow
  \infty$ to the value indicated by the red, dashed line. Also shown
  for comparison is the compactness of a Schwarzschild BH (green
  dot-dashed line), and the maximum compactness of a non-spinning
  neutron star (blue dotted line). Reproduced with permission
  from \cite{2010JCAP...11..002A}, copyright by IOP.}
\label{fig:selfinteraction_id2}
\end{figure}

Although the first expansion to nonlinear potentials was considered in \cite{Mielke:1981}
including fourth and sixth order $|\phi|$-terms, a deeper analysis was performed later considering a potential with only the quartic term
\cite{1986PhRvL..57.2485C}
\begin{equation}\label{pot_quartic}
    V\left( \left|\phi\right|^2 \right) = m^2 \left|\phi\right|^2 \,
                                         + \frac{\lambda}{2} \left|\phi\right|^4 ,
\end{equation}
with $\lambda$ a dimensionless coupling constant. Written in terms of a general potential,
the EKG equations remain the same.
The families of gravitational equilibrium can be
parametrized by the single dimensionless quantity $\Lambda \equiv \lambda / \left( 4\pi\,G m^2\right)$.
The potential of Eq.~(\ref{pot_quartic}) results in a 
maximum boson-star mass that now scales as
\begin{equation}\label{pot_quartic_maxmass}
    M_{\max} \approx 0.22 \Lambda^{1/2} M_{\mathrm{Planck}}/m 
                   = \left(0.1\mathrm{\ GeV}^2\right) M_{\odot} \lambda^{1/2}/m^2
\end{equation}
which is comparable to the Chandrasekhar mass for fermions with
mass $m_{\mathrm{fermion}} \sim m/\lambda^{1/4}$ \citep{1986PhRvL..57.2485C}. This 
self-interaction, therefore, allows
much larger masses than the mini-boson stars as long as $\Lambda \gg 1$, an 
inequality that may be satisfied even when $\lambda \ll 1$ for reasonable scalar boson
masses. The maximum mass as a function of the central value of the scalar field
is shown in Fig.~\ref{fig:selfinteraction_id} for different values of $\Lambda$.
The compactness of the most massive stable stars was studied in \cite{2010JCAP...11..002A},
finding an upper bound $M/R \lesssim 0.16$ for $\Lambda \gg 1$.
Figure~\ref{fig:selfinteraction_id2} displays
this compactness as a function of $\Lambda$ along with the compactness of
a Schwarzschild BH and non-spinning neutron star for comparison. The effect of repulsive ($\lambda>0$) and attractive ($\lambda<0$) quartic terms in the self-interaction potential have been studied in \cite{Eby:2015hsq}.

Many subsequent papers further analyze the EKG solutions with polynomial,
or even more general non-polynomial, potentials. One work in
particular \citep{2000IJMPD...9..601S} studied
the properties of the galactic dark matter halos modeled with these boson stars.
They found that a necessary condition to obtain stable, compact solutions with
an exponential decrease of the scalar field,
the series expansion of these potentials must contain the usual mass term $m^2|\phi|^2$.

More exotic ideas similarly try to include a 
pressure to increase the mass of BSs. \cite{2009PhRvD..79h4033A} consider
a form of repulsive self-interaction mediated by vector mesons within the mean-field approximation.
However, the authors leave the solution of the fully nonlinear system of
the Klein--Gordon and Proca equations to future work.

Other generalizations of the potential allow for the presence of \emph{nontopological soliton} solutions even in the absence of gravity,
with characteristics quite different than those of
the mini-boson stars. In order to obtain these solutions the potential must satisfy
two conditions. First, it must be a
function of $|\phi|^2$ to preserve the global $U(1)$ invariance. Second, the potential
should have an attractive term, bounded from below and positive for
$|\phi| \rightarrow \infty$. These conditions imply a potential of at least 
sixth order, a condition that is satisfied by the typical degenerate
vacuum form \citep{1987PhRvD..35.3637L,1987PhRvD..35.3658F}
\begin{equation}\label{pot_sixth}
    V\left( \left|\phi\right|^2 \right) = m^2~ \left|\phi\right|^2 \, 
                 \left( 1 - \frac{\left|\phi\right|^2}{ \phi^2_0}  \right)^2 \,,
\end{equation}
for which the potential has two degenerate minima at $\pm \phi_0$. The
case $|\phi|=0$ corresponds to the true vacuum state, while
$|\phi|=\phi_0$ represents the degenerate vacuum state.

The resulting soliton solution can be split into three
different regions. When gravity is negligible, the interior solution satisfies
$\phi \approx \phi_0$, followed by a shell of width $1/m$ over which $\phi$ changes
from $\phi_0$ to zero, and an exterior that is essentially vacuum.
This potential leads to a different scaling of the mass and radius
than that of the ground state \citep{1992PhR...221..251L}
\begin{equation}\label{pot_sixth_maxmass}
    M_{\max} \approx M^4_{\mathrm{Planck}}/ (m \, \phi_0^2) \,, \qquad
    R_{\max} \approx M^2_{\mathrm{Planck}}/ (m \, \phi_0^2).
\end{equation}

There is another type of non-topological \emph{soliton} star,
called Q-stars \citep{Lynn:1988rb},
which also admits soliton solutions in the absence of gravity (i.e., Q-balls \cite{Coleman:1985ki,1992PhR...221..251L}).
The potential, besides also being a function of $|\phi|$, must satisfy the following
conditions: it must behave like $\approx |\phi|^2$ near $\phi=0$, it has to be bounded
$< |\phi|^2$ in an intermediate region, and must be larger $> |\phi|^2$ for
$|\phi| \rightarrow \infty$. The Q-stars also have three regions; an interior solution
of radius $R \approx M_{\mathrm{Planck}}/\phi_0^2$,
(i.e., $\phi_0 \approx m$ is the free particle inverse Compton wavelength)
a very thin surface region of thickness $1/\phi_0$, and finally the exterior solution
without matter, which reduces to Schwarzschild in spherical symmetry. The mass of these
Q-stars scales
now as $M^3_{\mathrm{Planck}}/ \phi_0^2$, 
and for some choices of the sixth order self-interaction potential the compactness of the boson star (defined with the expected value of $R$ or $R^2$) can approach the black-hole limit \citep{Kleihaus:2011sx}.
The stability of these Q-stars has been studied
recently using catastrophe theory, such as \cite{2010PhRvD..81l4041T,Kleihaus:2011sx}. Rotating,
axisymmetric Q-balls were constructed in \cite{2005PhRvD..72f4002K,2008PhRvD..77f4025K}.
Related, rotating solutions in $2+1$ with the signum-Gordon equation instead of the KG equation
are found in \cite{2009PhRvD..80f7702A}.
Other interesting works have
studied the formation of Q-balls by the Affleck--Dine mechanism \citep{2000PhRvD..61d1301K}, their
dynamics in one, two and three spatial dimensions \citep{2000NuPhB.590..329B}, and
their viability as a self-interacting dark matter candidate \citep{2001PhRvL..87n1301K}.

\sllnew{
\textbf{Solitonic boson stars} represent solutions in the presence of gravity with the potential given by Eq.~(\ref{pot_sixth}). Very compact stable stars can be constructed numerically choosing small values of $\phi_0\ll 1$ \citep{2013PhRvD..88f4046M}.
Such solutions can be very compact with a very thin wall separating the inside from the outside \citep{Boskovic:2021nfs,Collodel:2022jly}.
Such stars are often studied within the context of mimicking black holes and serving as models of ultra compact objects which could be expected
to produce gravitational wave echoes \citep{Urbano:2018nrs,Cardoso:2021ehg}.
} 

It has been shown recently that very compact boson stars can also be found by using a V-shaped potential proportional to $|\phi|$ \citep{2012PhLB..714..120H}. The same V-shaped potential with an additional quadratic massive term has been considered in \cite{Kumar:2015sia}.
\sllnew{A recent study of ultra-compact objects \citep{Cardoso:2021ehg} found a general condition on the compactness of a} 
\sllnewer{non-rotating} 
\sllnew{boson star. In particular, for a general self-interaction potential that has a degenerate vacuum (here, precisely two minima), its most compact  boson stars can have a radius slightly smaller than the corresponding light ring,\footnote{A light ring is the location outside a gravitational well at which light will orbit, separating paths that fall inward and those that head outward. Only for very compact objects does a light ring exist outside the object.} but will have a maximum compactness of $C=M/R \approx 0.36$ \citep{Cardoso:2021ehg}.}

\CPi{Bose--Einstein condensates can arise also from periodic potentials. For instance, by using the potential Eq.~(\ref{eq:pot_axion}) associated with the axion field, one can  construct \emph{axion boson stars} made of a light complex scalar field \citep{Guerra:2019srj}. Similar solutions, using the semi-relativistic approach with two different periodic potentials, were already found almost a decade before in \cite{2011PhRvD..83d3525B}.  
\cite{Chan:2022bkz} study the evolution of such a star within an ambient axion gas. Further details on  axion stars are given in Sect.~\ref{subsection:varieties_oscillatons} .}

\cite{2009arXiv0910.1972B} consider a chemical potential to construct BSs, arguing that
the effect of the chemical potential is to reduce the parameter space of stable solutions.
Boson stars with a thermodynamically consistent equation of state, leading to an isotropic pressure, were considered in \cite{2011arXiv1108.3986C}. The solutions, obtained by integrating the TOV equations, reached compactnesses smaller (but comparable) to neutron stars. The extension to boson stars with finite temperature was considered in \cite{2014PhRvD..90l7501L}.

Related work modifies the kinetic term of the action instead of the potential.
\cite{2010GReGr..42.2663A} study the resulting BSs for a class of
\emph{K field theories}, finding solutions of two types: (i)~compact balls possessing
a naked singularity at their center and (ii)~compact shells with a singular
inner boundary which resemble black holes.
\cite{2008arXiv0804.3437A} consider coherent states of a scalar field instead
of a BS within \emph{k-essence} in the context of explaining dark matter. 
\cite{2008JHEP...07..094D} modify the kinetic term with just a minus sign
to convert the scalar field to a \emph{phantom field}. Although, a regular real scalar
field has no spherically symmetric, local static solutions, they find such solutions
with a real phantom scalar field.

\subsection{Newtonian boson stars}
\label{subsection:varieties_newtonian}

The Newtonian limit of the
Einstein--Klein--Gordon Eqs.~(\ref{EE1}\,--\,\ref{KG1}) can be derived by assuming
that the spacetime metric in the weak field approximation can 
be written as 
\begin{equation}\label{newtonian_limit}
    g_{00}= -(1+2\,V) \,, \qquad g_{ii} = 1 + 2\,V\,, \qquad %
                            g_{ij} = 0 \quad \mathrm{for} \quad i \neq j \,,
\end{equation}
where $V$ is the Newtonian gravitational potential. In this limit, the
Einstein equations reduce to the Poisson equation
\begin{equation}
    \nabla^2 V= 4\pi\,G T^{00} = 4\pi\,G m^2 \phi \bar \phi \,.
\end{equation}
Conversely, by assuming that
\begin{equation}
    \phi (x,t) \equiv \Phi(x,t) e^{i m t},
\end{equation}
in addition to the weak limit of Eq.~(\ref{newtonian_limit}), the
Klein--Gordon equation reduces to
\begin{equation}
\label{eq:SP}
    i \partial_t \Phi = -\frac{1}{2\,m} \nabla^2 \Phi + m\,V\,\Phi \,,
\end{equation}
which is just the \schrodinger~equation with $\hbar=1$.
Therefore, the EKG system is reduced in the Newtonian limit to the 
\schrodinger--Poisson~(SP) system \citep{1995PhDT........25G}.

The initial data is obtained by solving an eigenvalue problem
very similar to the one for boson stars, with similar assumptions
and boundary conditions. The solutions also share similar features
and display a similar behavior. A nice property of the Newtonian limit
is that all the solutions can be obtained by rescaling from
one known solution \citep{1995PhDT........25G},
\begin{equation}\label{reescaling_newtonian}
      \phi_2 = \phi_1 \left( \frac{N_2}{N_1}\right)^2 \,, \qquad 
      \omega_2 = \omega_1 \left( \frac{N_2}{N_1}\right)^2 \,, \qquad 
      r_2 = r_1 \left( \frac{N_1}{N_2} \right) \,,
\end{equation}
where $N\equiv m \int dx^3 \phi \bar \phi$ is the Newtonian
number of particles.

The possibility of including self-interaction terms in the potential
was considered in \cite{2006ApJ...645..814G}, studying also the gravitational
cooling (i.e., the relaxation and virialization through the emission of 
scalar field bursts) of spherical perturbations. Non-spherical
perturbations were further studied in \cite{2006PhRvD..74f3504B},
showing that the final state is a spherically symmetric configuration.
Single Newtonian boson stars were studied in \cite{1995PhDT........25G},
either when they are boosted with/without an external central potential.

\sllnew{Rotating stars in Newtonian gravity are discussed in the beginning of Sect.~\ref{subsection:varieties_rotating}.}
Numerical evolutions of binary boson stars in Newtonian gravity are discussed
in Sect.~\ref{dynamics_binary_bs}.

Recent work by Chavanis with Newtonian gravity
solves the Gross--Pitaevskii equation, a variant of Eq.~(\ref{eq:SP}) which involves a pseudo-potential for
a Bose--Einstein condensate, to model either dark matter or compact alternatives to
neutron stars \citep{2012AA...537A.127C, 2011PhRvD..84d3531C, 2011arXiv1108.3986C,Chavanis:2015zua,Chavanis:2016shp}.
However, see a rebuttal to some of this work \citep{Mukherjee:2014kqa}.

Much recent work considers boson stars from a quantum perspective as a Bose--Einstein condensate involving some number, $P$,
of scalar fields.
\cite{2012CMaPh.311..645M} study the
collapse of boson stars mathematically in the mean field limit in which $P \rightarrow \infty$.
\cite{2009JSP...137.1063K} argues for the existence of \emph{bosonic atoms} instead of stars.
\cite{2011JCoPh.230.5449B} use numerical methods to study the mean field dynamics of BSs.

%
%
%
\subsection{Charged boson stars}
\label{subsection:varieties_charged}

Charged boson stars result from the coupling of the bosonic field to the
electromagnetic field \citep{1989PhLB..227..341J}. The coupling between
gravity and a complex scalar field with a $U(1)$ charge arises
by considering the action of Eq.~\ref{action} with the following matter 
Lagrangian density
\begin{equation}\label{Lagrangian_EMKG}
  {\cal L_M} = 
   - \frac{1}{2} \left[ g^{ab} 
      \left(\nabla_a \bar\phi + i\,e\,A_a\,\bar\phi \right)
      \left(\nabla_b \phi - i\,e\,A_b\,\phi\right)
   + V\left( \left|\phi \right|^2\right) \right] 
   - \frac{1}{4} F_{ab} F^{ab} \,,
\end{equation}
where $e$ is the gauge coupling constant. The Maxwell tensor $F_{ab}$ can be decomposed
in terms of the vector potential $A_a$
\begin{equation}\label{potential_vector}
  F_{ab} = \nabla_a A_b - \nabla_b A_a \,.
\end{equation}
The system of equations obtained by performing the variations on the action forms
the Einstein--Maxwell--Klein--Gordon system, which contains the evolution equations
for the complex scalar field $\phi$, the vector potential $A_a$, and the spacetime metric $g_{ab}$ \citep{2006PhDT.........6P}.

Because a charged BS may be relevant for a variety of scenarios, 
we detail the resulting equations.
For example, \emph{cosmic strings} are also constructed from a charged, complex scalar field
and obeys these same equations. It is only when we choose the harmonic time dependence of the scalar
field that we distinguish from the harmonic azimuth of the cosmic string \citep{vilenkinbook}.
The evolution equations for the scalar field and for the Maxwell tensor are
\begin{eqnarray}\label{MEKG1}
 && g^{ab} \nabla_a \nabla_b \phi - 2\,i\,e A^a \nabla_a \phi
   - e^2\,\phi A_a A^a - i\, e\, \phi \, \nabla_a A^a = \frac{d V}{d |\phi|^2} \phi \\
\label{MEKG2}
 && \nabla_a F^{ab} = -J^b = i\,e\, (\bar \phi \nabla_b \phi - \phi \nabla_b \bar \phi) 
                  + 2\, e^2\, \phi\, \bar \phi A^b \,.
\end{eqnarray}
Notice that the vector potential is not unique; we can still add any curl-free
components without changing the Maxwell equations. The gauge freedom can be fixed by choosing,
for instance, the Lorentz gauge $\nabla_a A^a=0$. Within this choice, which sets
the first time derivative of the time component $A_0$, the Maxwell 
equations reduce to a set of wave equations in a curved background with
a non-linear current. This gauge choice resembles the \emph{harmonic gauge}
condition, which casts the Einstein equations as a system of non-linear,
wave equations \citep{1984ucp..book.....W}.

Either from Noether's theorem or by taking an additional covariant derivative
of Eq.~(\ref{MEKG2}),
one obtains that the electric current $J^a$ follows a conservation law. The spatial integral
of the time component of this current, which can be identified with the total
charge $Q$, is conserved. This charge is proportional to the number of particles,
$Q=e\,N$. The mass $M$ and the total charge $Q$ can be calculated by associating the
asymptotic behavior of the metric with that of Reissner--Nordstr\"{o}m metric,
\begin{equation}\label{RN_asymptotic}
    g_{rr} =\left(1 - \frac{2\,G\,M}{r} + \frac{G\,Q^2}{4\,\pi\,r^2} \right)^{-1} 
\quad \mathrm{for} \quad r \rightarrow \infty \,,
\end{equation}
which is the unique solution at large distances for a scalar field with compact support.

We look for a time independent metric by first assuming a harmonically varying scalar
field as in Eq.~(\ref{harmonic}). We work in spherical coordinates and assume spherical
symmetry.
With a proper  gauge choice, the vector potential takes a particularly simple form with only
a single,
non-trivial component $A_a=\left(A_0(r),0,0,0\right)$. This choice implies an everywhere vanishing
magnetic field so that the electromagnetic field
is purely electric. 
The boundary conditions
for the vector potential are obtained by requiring that the electric field vanishes
at the origin because of regularity, $\partial_r A_0 (r=0) = 0$.
Because the electromagnetic field depends only on derivatives of the potential, we can use this
freedom to set
$A_0 (\infty) = 0$ \citep{1989PhLB..227..341J}.

With these conditions, it is possible
to find numerical solutions in equilibrium as described in \cite{1989PhLB..227..341J}. It was shown that bound stable configurations exist only for values of the coupling constant less than or equal to a certain critical value, such that solutions are found for
${\tilde e}^2 \equiv e^2\,M^2_{\mathrm{Planck}}/(8\,\pi\,m^2)<1/2$. 
For ${\tilde e}^2 >1/2$ the repulsive Coulomb force is bigger than the gravitational attraction and
no solutions were found, although it has been 
shown recently that, due to the binding energy, solutions with ${\tilde e}^2 =1/2$ and even slightly higher are also allowed \citep{2013PhRvD..88b4053P}.
This bound on the BS charge in terms of its mass ensures that one cannot construct an \emph{
overcharged} BS, in analogy to the overcharged monopoles of \cite{Lue:2000qr}.
An overcharged monopole is one with more charge than mass and is therefore susceptible to
gravitational collapse by accreting sufficient (neutral) mass. However, because its charge is
higher than its mass, such collapse might lead to an extremal Reissner--Nordstr\"{o}m BH, but
BSs do not appear to allow for this possibility. Interestingly \cite{Sakai:2011wn} find
that if one removes gravity, the obtained Q-balls may have no limit on their charge.

\begin{figure}[htbp]
\centerline{\includegraphics[height=8.0cm,width=10.0cm]{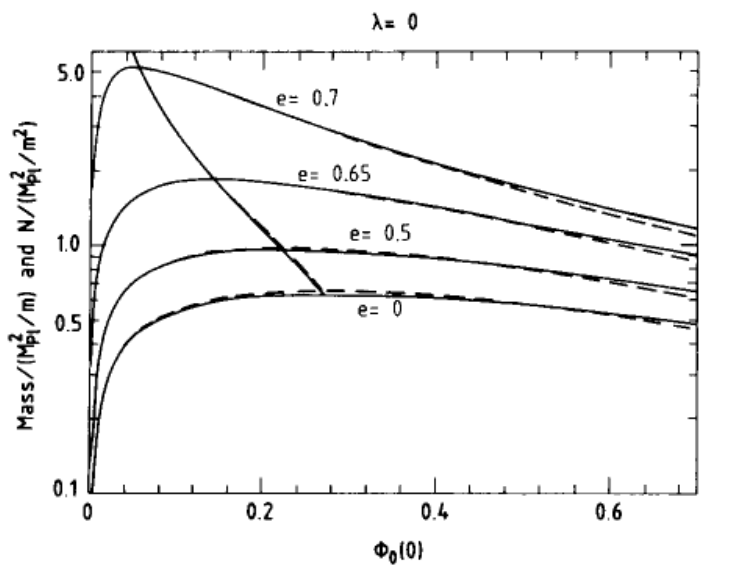}}
\caption{The mass (solid) and the number of particles (dashed) versus
  central scalar value for charged boson stars with four values of
  ${\tilde e}$ as defined in
  Sect.~\ref{subsection:varieties_charged}. The mostly-vertical
  lines crossing the four plots indicate the solution for each case
  with the maximum mass (solid) and maximum particle number
  (dashed). Reproduced with permission from \cite{1989PhLB..227..341J};
  copyright by Elsevier.}
\label{fig:charged_id}
\end{figure}

The mass and the number
of particles are plotted as a function of $\phi_c$ for different values of
${\tilde e}$ in Fig.~\ref{fig:charged_id}. Trivially, for ${\tilde e}=0$
the mini-boson stars of Sect.~\ref{subsection:spherical} are recovered. Excited solutions
with nodes are qualitatively similar \citep{1989PhLB..227..341J}. The stability of these
objects has been studied in \cite{1989PhLB..231..433J}, showing that the equilibrium
configurations with a mass larger than the critical mass are dynamically unstable, similar to uncharged BSs.

Work with charged BSs includes the publication of Maple\footnote{
   \url{http://www.maplesoft.com}}
routines to study boson nebulae
charge \citep{2010ChPhL..27a1101C,Murariu:2010zz,dariescu2008} and charged boson stars in the presence of a cosmological constant \citep{Kumar:2016sxx}. 

\sllnew{Because a charged black hole is subject to a superradiant instability~(see Sect.~\ref{scalarclouds}) which extracts energy from the hole, the confinement of such a black hole to a box is called a \emph{black hole bomb}.
The fate of this configuration is of interest as is the possibility that a hairy soliton is formed.
Within this context,
}
new regular solutions of  charged scalar fields in a cavity are presented in \cite{Ponglertsakul:2016wae}, which are stable only when the radius of the mirror is sufficiently large.

\CPn{The dynamical mechanism of the black hole bomb is studied numerically by perturbing a Reissner-Nordstrom black hole, either in Ads \cite{PhysRevLett.116.141102} or in a cavity \cite{PhysRevLett.116.141101} with a charged scalar field. Although these two studies are not finding soliton stars, they do find  remarkable agreement on the dynamical development of the superradiant instability, obtaining in both cases a stable hairy black hole as the final state.}
\sllnew{
Subsequently, \cite{Dias:2021acy} study the features of  hairy solitons inside a Minkowski box with Reissner--Nordstr\"{o}m describing the exterior.
Charged boson stars (and black holes) with wavy scalar hair are found in \cite{Brihaye:2021mqk}.
}

Other work generalizes the Q-balls and Q-shells found with a certain potential which
leads to the signum-Gordon equation for the scalar field \citep{2009PhLB..675..102K,2010PhRvD..82j4050K}.
In particular, shell solutions can be found with a black hole in its interior, which has implications
for black hole scalar hair (for a review of black hole uniqueness see \citealt{lrr-2012-7}).

One can also consider Q-balls coupled to an electromagnetic field, a regime appropriate
for particle physics. Within such a context, \cite{Eto:2010vi} study the
chiral magnetic effect arising from a Q-ball.
Other work in \cite{Brihaye:2009dx} studies charged, spinning Q-balls\sllnew{, and
\cite{Kunz:2021mbm} compare gauged Q-balls with a symmetry-breaking potential (the Friedberg--Lee--Sirlin model) with their corresponding gravitating charged boson stars.}

Charged BSs in anti-de~Sitter spacetimes have attracted some interest as noted at the end of
Sect.~\ref{sec:other}.

\subsection{Oscillatons and axion stars}
\label{subsection:varieties_oscillatons}

As mentioned earlier, it is not possible to find time-independent, spacetime solutions for
a real scalar field.
However, there are non-singular, time-dependent near-equilibrium configurations of self-gravitating
real scalar fields, which are known as \emph{oscillatons} \citep{1991PhRvL..66.1659S}.
These solutions
are similar to boson stars, with the exception that the spacetime must also have
a time dependence in order to avoid singularities.

In this case, the system is still described by the EKG Eqs.~(\ref{gevolrr_rt}\,--\,\ref{evolKG1_c}), with the
the additional simplification that the scalar field is strictly real,
$\phi = \bar \phi$. In order to find equilibrium configurations,
one expands both metric components $\{A(r,t) \equiv a^2,
 C(r,t) \equiv (a/\alpha)^2\}$ and the scalar field $\phi(r,t)$ as a truncated Fourier series
\begin{eqnarray}\label{oscillaton_expansion}
    \phi(r,t) &=& \sum_{j=1}^{j_{\max}} \phi_{2j-1}(r) \cos \left(\left[2\,j-1\right]\, \omega t \right) \,, \\
     A(r,t) &=& \sum_{j=0}^{j_{\max}} A_{2j}(r) \cos (2\, j\, \omega t) \,, \qquad 
     C(r,t) = \sum_{j=0}^{j_{\max}} C_{2j}(r) \cos (2\, j\, \omega t) \,, 
\end{eqnarray}
where $\omega$ is the fundamental frequency and $j_{\max}$
is the mode at which the Fourier series are truncated.
As noted in \cite{2002CQGra..19.6259U,2003CQGra..20.2883A}, the scalar field consists only
of odd components while the metric terms consist only of even ones.
Solutions
are obtained by substituting the expansions of Eq.~(\ref{oscillaton_expansion})
into the spherically symmetric Eqs.~(\ref{gevolrr_rt}\,--\,\ref{evolKG1_c}).
By matching terms of the same frequency,
the system of equations reduces to a set of coupled ODEs.
The boundary conditions are determined by requiring regularity at the origin and that
the fields become asymptotically flat at large radius. These form an eigenvalue problem for
the coefficients $\{ \phi_{2j-1}(r=0),A_{2j}(r=0),C_{2j}(r=0) \}$ corresponding to a given central value $\phi_1(r=0)$.
As pointed out in \cite{2002CQGra..19.6259U}, the frequency $\omega$ is determined by the
coefficient $C_0(\infty)$ and is therefore called an \emph{output value}.
Although the equations are non-linear, the Fourier series converges
rapidly, and so a small value of $j_{\max}$ usually suffices.

A careful analysis of the high frequency components of this construction reveals 
difficulties in avoiding infinite total energy while maintaining the asymptotically
flat boundary condition \citep{2004PhRvD..70b3002P}. Therefore, the truncated solutions constructed
above are not exactly time periodic. Indeed, very accurate numerical work has shown
that the oscillatons radiate scalar field on extremely long time scales while their
frequency increases \citep{fodor-2010-81,PhysRevD.84.065037}.
This work finds a mass loss rate of just one part in
$10^{12}$ per oscillation period, much too small for most numerical simulations to observe.
The solutions are, therefore, only near-equilibrium solutions
and can be extremely long-lived. \CP{Oscillatons have also been found in the context of dark matter for real vector fields, sharing many of the features of their scalar-field counterparts \citep{Brito:2015yga,Brito:2015yfh}. Because they are constructed from a real gauge vector field (albeit massive), these oscillatons would actually represent the closest realization to the electromagnetic geons sought by Wheeler.}%

\begin{figure}[htbp]
\centerline{\includegraphics[width=9.0cm]{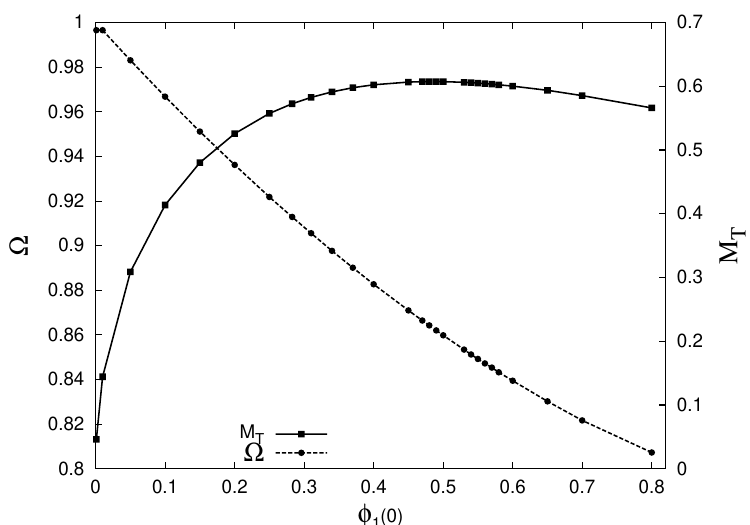}}
\centerline{\includegraphics[width=9.0cm]{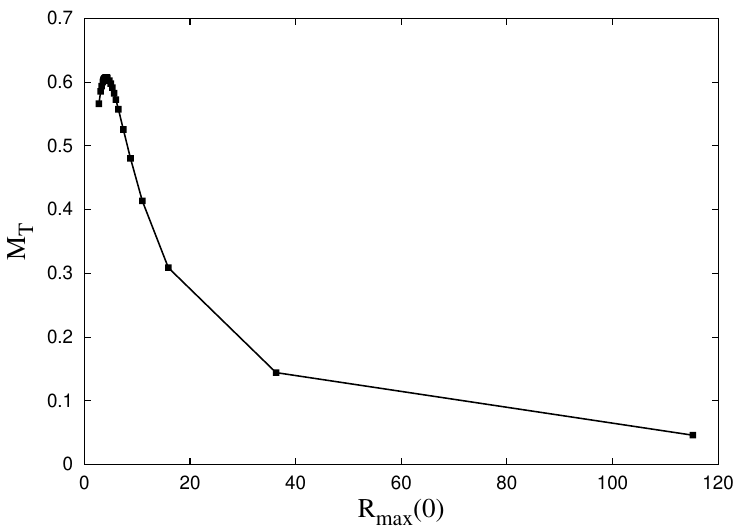}}
\caption{\emph{Top:} Total mass (in units of
  $M^2_{\mathrm{Planck}}/m$) and fundamental frequency of an
  oscillaton as a function of the central value of the scalar field
  $\phi_1(r=0)$. The maximum mass is
  $M_{\max}=0.607\,M^2_{\mathrm{Planck}}/m$. \emph{Bottom:} Plot of
  the total mass versus the radius at which $g_{rr}$ achieves its
  maximum. Reproduced with permission from \cite{2003CQGra..20.2883A};
  copyright by IOP.}
\label{fig:oscillaton_id}
\end{figure}

Although the geometry is oscillatory in nature, oscillatons
behave similarly to BSs.
In particular, they similarly transition from long-lived solutions to a dynamically
unstable branch separated at the maximum mass
$M_{\max} = 0.607\,M^2_{\mathrm{Planck}}/m$.
Figure~\ref{fig:oscillaton_id} displays the total mass curve, which shows
the mass as a function of central value.
Compact solutions can be found in the Newtonian framework when the weak
field limit is performed appropriately, reducing to the so-called Newtonian
oscillations \citep{2002CQGra..19.6259U}.
The dynamics produced by perturbations are also
qualitatively similar, including gravitational cooling, migration to more
dilute stars, and collapse to black holes \citep{2003CQGra..20.2883A}.
More recently, these studies have been
extended by considering the evolution in 3D of excited states \citep{2008PhRvD..77b4028B} 
and by including a quartic self-interaction potential \citep{ValdezAlvarado:2011dd}.
In \cite{2008CQGra..25x5004K}, a variational approach is used to construct oscillatons
in a reduced system similar to that of the sine-Gordon breather solution.
Such localized solutions have also been constructed in AdS
(see Sect.~\ref{sec:other}), and numerical evolutions suggest that they
are stable below some critical density \citep{Fodor:2015eia}.

Closely related, are \emph{oscillons} that exist in flatspace and that were first mentioned
as ``pulsons'' in \cite{1977ZhPmR..25..120B}. And so just as a Q-ball can be thought
of as a BS without gravity, an oscillon is an oscillaton in the absence of gravity.
Extensive
literature studies such solutions, many of which appear in \cite{Fodor:2008es}.
A series of papers establishes that oscillons similarly radiate on very long time
scales \citep{
Fodor:2008es,
Fodor:2008du,
Fodor:2009xw,
Fodor:2009kf}.
\CP{Recently, it has been demonstrated that oscillons also exist in the low-energy effective theory of an interacting massive vector field \citep{Zhang:2021xxa}. Interestingly, they found two types of vector oscillons, which despite having vanishing angular momentum and approximately spherically symmetric energy density, have a non-spherical field configuration (i.e., they are ``directional'' linearly polarized, with vanishing total intrinsic spin, and ``spinning'' circularly polarized oscillons, with a macroscopic intrinsic spin).}

An interesting numerical approach to evolving oscillons
adopts coordinates that blueshift and damp outgoing radiation of the
massive scalar field \citep{Honda:2000gv,Honda:2001xg}.
A detailed look at the long term dynamics of these solutions
suggests the existence of a fractal boundary in parameter space between oscillatons that
lead to expansion of a true-vacuum bubble and those that disperse \citep{Honda:2010gb}.
\cite{Dymnikova:2000dy} examine the collision of two of these bubbles in the context of a first order phase transition.
The reheating phase of inflationary cosmology generally feature oscillons
which may produce observable gravitational waves \citep{Antusch:2016con, Antusch:2015ziz}.

\CPi{
	The axion field is a real scalar field introduced by Peccei and Quinn as part of quantum chromodynamics~(QCD) to solve the CP problem \citep{Peccei:1977hh} and has since become a popular model for dark matter. 
	The  invariance of the axion Lagrangian under shift symmetry
	requires the axion potential to be a
	periodic function of $\phi$. 
	The simplest model for the axion potential, employed in most
	phenomenological studies, is the instanton potential
	\begin{equation}\label{eq:pot_axion}
		V\left( \phi \right) = (m_a f_a)^2~ \left| 1 - \cos(\phi/f_a)\right| \,,
	\end{equation}
	where $f_a$ is the axion decay constant and $m_a$ its mass.
	Bose--Einstein condensates of such a field are called \emph{axion stars}.
	As already mentioned, an extensive review of axion stars can be found in \cite{Braaten:2019knj}.
	}

\subsection{Rotating boson stars}
\label{subsection:varieties_rotating}
%

Boson stars with rotation were not explored until the mid-1990s
because of the lack of a strong astrophysical motivation and the technical
problems with the regularization along the axis of symmetry.
The first equilibrium solutions of rotating boson stars were obtained within 
Newtonian gravity~\sllnew{independently by two different groups \citep{Schupp:1995dy,1995PhRvD..52.5724S}}. 
\sllnew{
Approximate, analytic solutions for rotating boson stars were later found in four and five dimensions \citep{Kan:2016xkn}.
Recently, \cite{Kling:2020xjj} construct slowly rotating, Newtonian boson stars via the Gross--Pitaevskii--Poisson equation by perturbing the ground state boson star. 
}

In order to generate axisymmetric
time-independent solutions with angular momentum, one is naturally lead to the ansatz
\begin{equation}\label{harmonic2}
  \phi( \mathbf{r},t) = \phi_0(r,\theta) e^{i ( \omega t + k \varphi)} \,,
\end{equation}
where $\phi_0(r,\theta)$ is a real scalar representing the profile of the star,
$\omega$ is a real constant denoting the angular frequency of the field
and $k$ must be an integer so that the field $\phi$ is not multivalued in the
azimuthal coordinate $\varphi$. This integer, $k$,  is commonly known as the rotational
quantum number
\sllnew{(the letter adopted in the literature varies for this azimuthal winding number, sometimes calling it $s$, $\ell$, or $m$)}. 

\CP{General relativistic rotating boson stars were found
adopting the ansatz given by  Eq.~(\ref{harmonic2}) \citep{1996rscc.conf..138S,1997PhRvD..56..762Y}}. 
To obtain stationary axially symmetric solutions, two symmetries
were imposed on the spacetime described by two commuting Killing
vector fields $\xi = \partial_t$ and $\eta = \partial_{\varphi}$ in a
system of adapted (cylindrical) coordinates $\{t,r,\theta,\varphi \}$.
In these coordinates, the metric is independent 
of $t$ and $\varphi$ and can be expressed in isotropic coordinates in the
Lewis--Papapetrou form
\begin{equation}\label{axisymmetric_timedependent_metric}
ds^2 = -f dt^2 + \frac{l}{f} \left[ g \left(dr^2 + r^2\, d\theta^2\right)
     + r^2\, \sin^2 \theta \left( d\varphi - \frac{\Omega}{r} dt \right)^2 \right] \,,
\end{equation}
where $f,l,g$ and $\Omega$ are metric functions depending only on $r$ and $\theta$.
This means that we have to solve five coupled PDEs, four for the metric and 
one for the Klein--Gordon equation; these equations determine an elliptic quadratic
eigenvalue problem in two spatial dimensions. Near the axis, the scalar
field behaves as
\begin{equation}\label{regularity_condition}
  \lim_{r \rightarrow 0} \phi_0(r,\theta) = r^k\, h_k(\theta) + O(r^{r+2}) \,,
\end{equation}
so that for $k>0$ the field vanishes near the axis. Note that $h_k$ is some
arbitrary function different for different values of $k$ but no sum over $k$ is
implied in Eq.~(\ref{regularity_condition}).
This implies that the rotating 
star solutions have toroidal level surfaces instead of spheroidal ones as in
the spherically symmetric case $k=0$. In this case the metric coefficients are
simplified, namely $g=1$, $\Omega=0$ and $f=f(r)$, $l=l(r)$.

\begin{figure}[htbp]
\centerline{
  \includegraphics[width=0.49\textwidth]{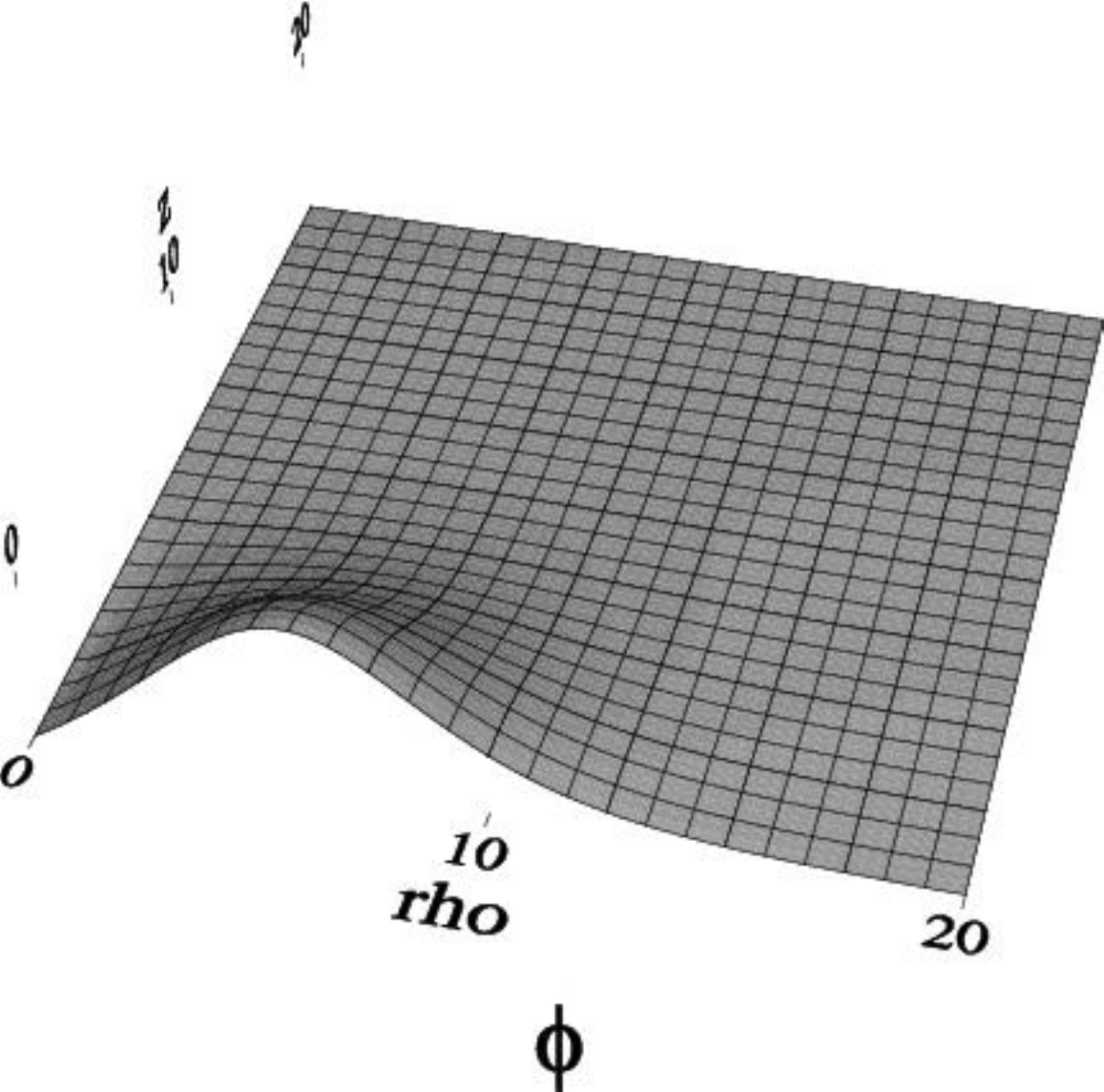}
  \includegraphics[width=0.49\textwidth]{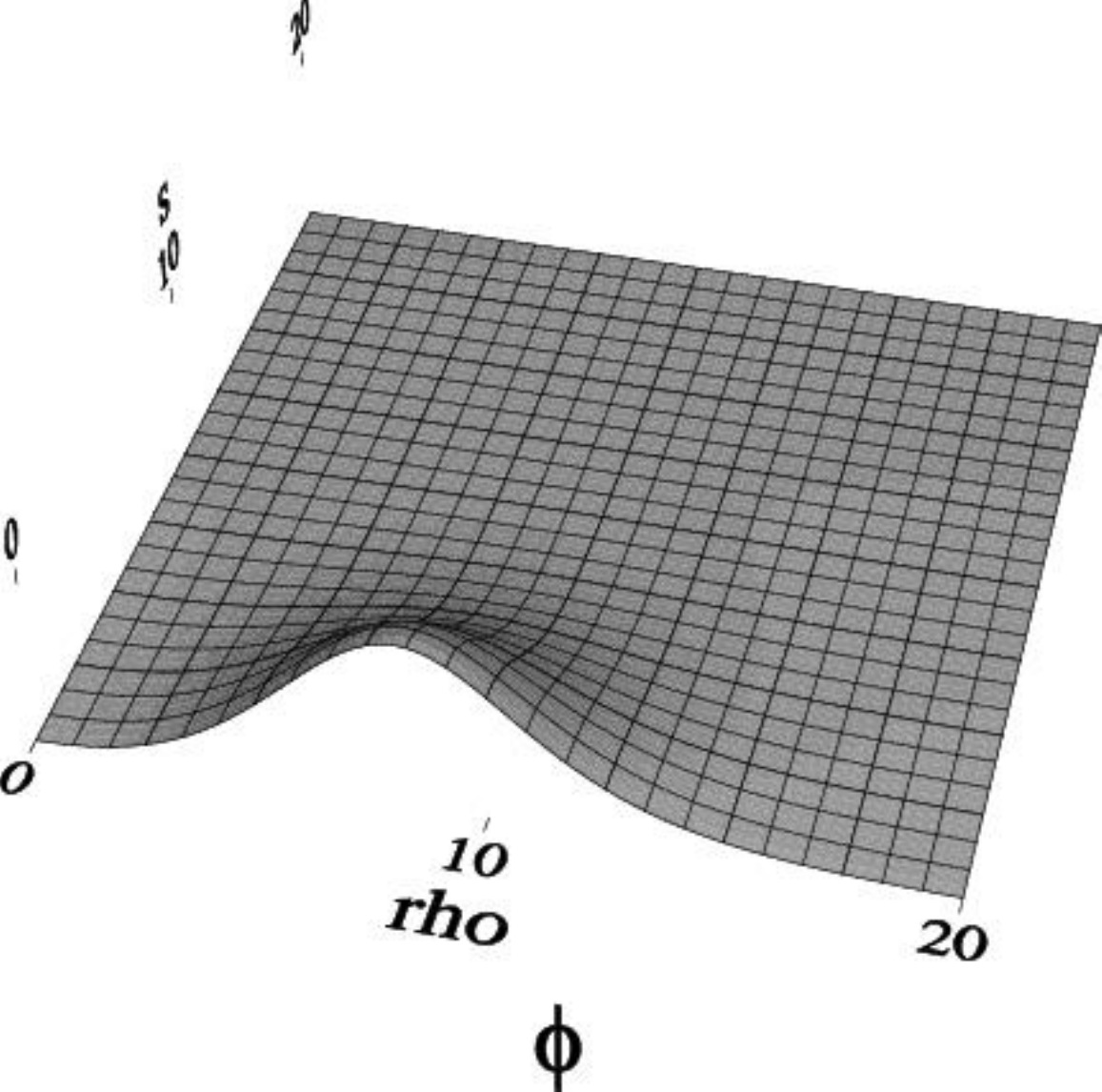}
}
\caption{The scalar field in cylindrical coordinates $\phi(\rho,z)$
  for two rotating boson-star solutions: (left) $k=1$ and (right)
  $k=2$. The two solutions have roughly comparable amplitudes in
  scalar field. Note the toroidal shape. Reproduced with permission
  from \cite{2005PhDT.........2L}.}
\label{fig:rotating}
\end{figure}

The entire family of solutions for $k=1$ and part of $k=2$ was computed using
the self-consistent field method \citep{1997PhRvD..56..762Y}, obtaining a maximum mass 
$M_{\max}=1.31\,M^2_{\mathrm{Planck}}/m$. Both families were
completely computed in \cite{2005PhDT.........2L} using faster
multigrid methods, although there were significant discrepancies in
the maximum mass, which indicates a problem with the regularity
condition on the \textit{z}-axis. The mass $M$ and angular momentum $J$ for
stationary asymptotically flat spacetimes can be obtained from their
respective Komar expressions. They can be read off from the asymptotic
expansion of the metric functions $f$ and $\Omega$ 
\begin{equation}
    f = 1 - \frac{2\,G M}{r} + {\rm O}\left(\frac{1}{r^2}\right) \,, \qquad
    \Omega = \frac{2\,J G}{r^2} + {\rm O}\left(\frac{1}{r^3}\right) \,.
\end{equation}
Alternatively, using the Tolman expressions for the angular momentum
and the Noether charge relation in Eq.~(\ref{Noether_charge}),
one obtains an important quantization
relation for the angular momentum \citep{1997PhRvD..56..762Y}
\begin{equation}\label{quantization_J}
    J = k\, N \,,
\end{equation}
for integer values of $k$. 
\sllnew{The quantization of angular momentum here contrasts with
the slowly-rotating Newtonian solutions in \cite{Kling:2020xjj} which, as noted above, connect continuously with the nonrotating solutions.}
This remarkable quantization condition for this classical solution also plays a role in the work of \cite{Dias:2011at} discussed in
Sect.~\ref{sec:other}. Also, \cite{Smolic:2015txa} discusses the quantization condition of rotating BSs in the context of symmetry.
Fig.~\ref{fig:rotating} shows the scalar field for
two different rotating BSs. Spinning BS solutions with a quartic self-interacting potential 
have been found too, as well as their Kerr BH limit \citep{Herdeiro:2016gxs}.

\begin{figure}[htbp]
		\centerline{\includegraphics[width=8.0cm]{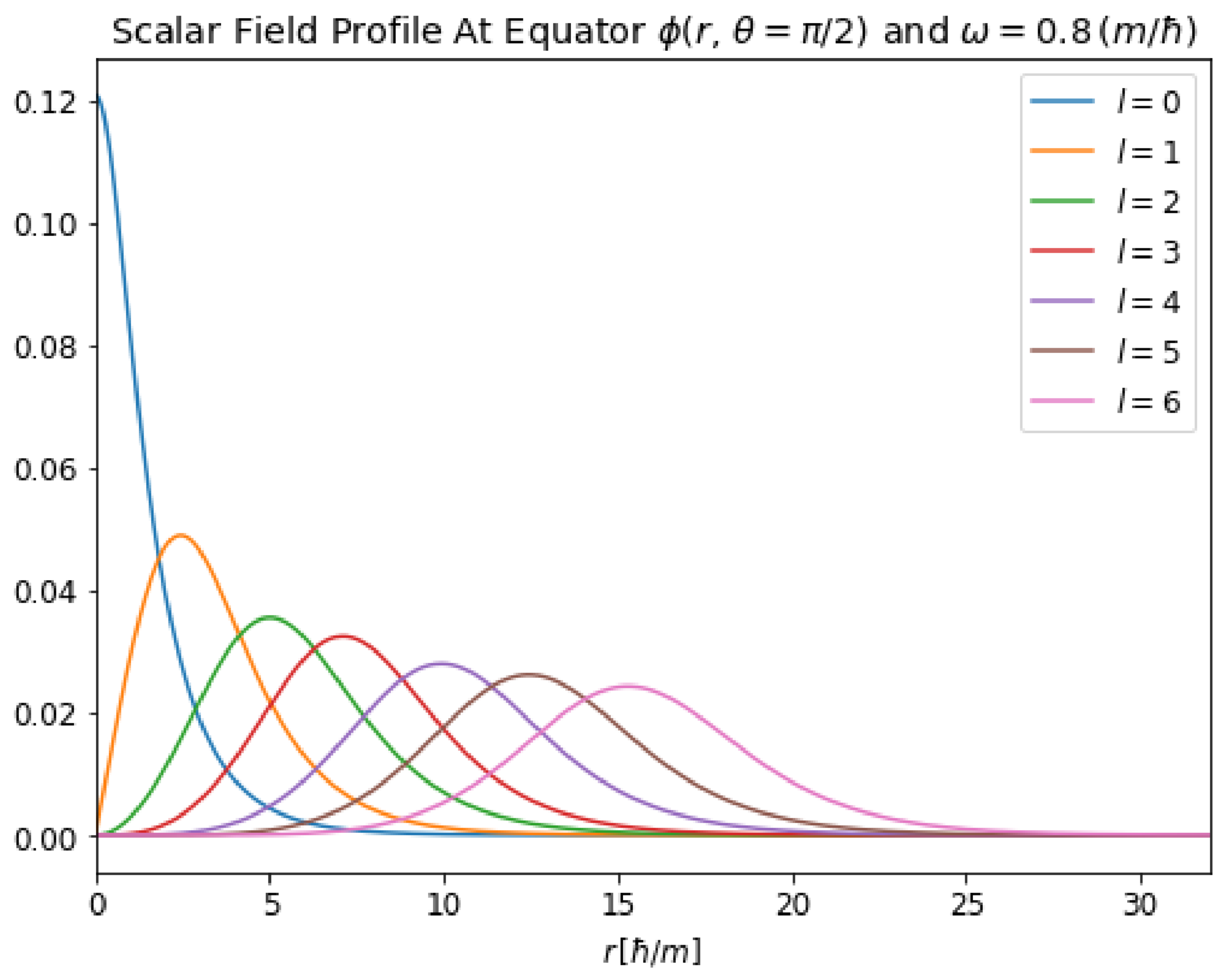}}
		\caption{Radial profiles along the equator of rotating boson stars
                        for fixed $\omega = 0.8 (m/\hbar)$ and $k \in [0, 6]$.
			Reproduced with permission from \cite{Ontanon:2021hbg}, copyright by IOP, which uses $l$ in place of $k$ to indicate the azimuthal quantum number.}
		\label{fig:rotating_rprofile}
\end{figure}

\sllnew{
More recently, quite a variety of rotating solutions have been found.
Most recently, \cite{Ontanon:2021hbg} construct rotating boson stars up to $k=6$ and determine the maximum masses and minimum frequencies. }\CPn{Their radial profiles are displayed in Fig.~\ref{fig:rotating_rprofile}, showing that they form a family of solutions with tori further away from the origin as the angular momentum increases} \sllnew{discretely}.
\sllnew{	
\cite{Herdeiro:2018wvd}  found rotating boson stars  with non-minimal scalar coupling.
The structure of charged, rotating boson stars were studied,
in particular the properties in terms of an effective description with an anisotropic fluid \citep{Collodel:2019ohy}.
\cite{Li:2019mlk} construct multi-state, multi-field rotating stars (see Sect.~\ref{sec:multi}), and 
\cite{Delgado:2020udb} construct rotating axion stars
(see Sect.~\ref{subsection:varieties_oscillatons} for a description of axion stars).
}

\sllnew{
\cite{Vaglio:2022flq} construct rotating boson stars in a regime in which they are expected to be stable and compares their multipole structure to Kerr.
\cite{Adam:2022nlq} construct rotating boson stars with various potentials to find a universal relation for the moment of inertia, the (dimensionless) angular momentum, and the quadrupole moment, that may help distinguish boson stars from compact neutron stars.
}

\sllnew{
Gauged rotating boson  and Dirac stars are studied in \cite{Herdeiro:2021jgc}.
Chains of rotating boson stars can be constructed by switching the sign of the amplitude
between adjacent stars \citep{Gervalle:2022fze}, generalizing the construction of
chains of non-rotating stars \citep{Herdeiro:2021mol}.
\cite{Collodel:2019uns} constructed rotating stars in massive tensor multi-scalar~(MTMS) theories of gravity.
}

Rotating boson stars have been shown to develop
a strong ergoregion instability when rapidly spinning on short characteristic
timescales (i.e., 0.1 seconds\,--\,1 week for objects with mass $M=1$--$10^6\,M_{\odot}$ 
and angular momentum $J> 0.4\,G M^2$), indicating that very compact objects
with large rotation are probably black holes \cite{2008PhRvD..77l4044C}.
The presence of light rings around rotating boson stars is studied in \cite{Grandclement:2016eng,Cunha:2022gde}, while geodesics on the spacetime of these solutions are studied in \cite{2014PhRvD..90b4068G}. \CPi{A detailed discussion of the stability of rotating BSs is deferred to Sect.~\ref{sec:stability_w_angmom}.}

A review by Mielke focuses  on rotating boson stars \citep{Mielke:2016war}.
Further discussion concerning the numerical methods and limitations of some of these approaches can also be found in \cite{2005PhDT.........2L}.

\subsection{Fermion-boson stars}
\label{subsection:varieties_fermionicbosonic}
%

The possibility of compact stellar objects made with a mixture of bosonic and
fermionic matter was studied in \cite{1989PhLB..233...99H,1990NuPhB.337..737H}.
In the simplest case, the bosonic component interacts with the fermionic component 
only via the gravitational field, although different couplings
were suggested in \cite{1990NuPhB.337..737H} and have been further explored 
in \cite{1998PhRvD..58l3003D,1996MPLA...11..647P}. Such a simple interaction is, at the very least,
consistent with models of a bosonic dark matter coupling only gravitationally
with visible matter, and the idea that such a bosonic component would become
gravitationally bound within fermionic stars is arguably a natural expectation.

One can consider a perfect fluid as the fermionic component such that the stress-energy tensor takes the standard form
\begin{eqnarray}\label{Tab_fermionboson}
   T^{\rm fluid}_{ab}  &=& (\mu + p) u_a\, u_b + p\, g_{ab}
\end{eqnarray}
where $\mu$ is the energy density, $p$ is the pressure of the fluid and
$u_a$ its four-velocity. Such a fluid requires an equation of state to close the system of 
equations (see \citealt{lrr-2008-7} for more about fluids in relativity). 
In \sllnew{much of the early work}
with fermion-boson stars, the fluid is described by a degenerate,
relativistic Fermi gas, so that the pressure is given by the parametric equation of state of
Chandrasekhar 
\begin{eqnarray}\label{EoS_chandra}
   \mu = K (\sinh t - t) \qquad
     p = \frac{K}{3} \left[ \sinh t - 8\, \sinh \left( \frac{t}{2} \right)
                             + 3\,t \right] \,,
\end{eqnarray}
where $K=m_n^4/(32\,\pi^2)$ and $m_n$ the mass of the fermion. The parameter $t$
is given by
\begin{eqnarray}\label{EoS_chandra_part}
     t(r) = 4 \log \left[ \frac{p_o}{m_n} 
             + \left( 1 + \left( \frac{p_o}{m_n} \right)^2 \right)^{1/2}\right] \,,
\end{eqnarray}
where $p_o$ is the maximum value of the momentum in the Fermi distribution
at radius $r$.

The perfect fluid obeys relativistic versions of the Euler equations, which account for the 
conservation of the fluid energy and momentum, plus the conservation of baryon
number (i.e., mass conservation). The complex scalar field representing the
bosonic component is once again
described by the Klein--Gordon equation. The spacetime is computed through 
the Einstein equations with a stress-energy tensor, which is a combination
of the complex scalar field and the perfect fluid
\begin{eqnarray}\label{Tab_fermion_boson}
     T_{ab} = T^{\phi}_{ab} + T^{\mathrm{fluid}}_{ab} \,.
\end{eqnarray}
After imposing the harmonic time dependence of Eq.~(\ref{harmonic}) on the complex scalar
field, assuming a static metric as in Eq.~(\ref{spherical_metric}) and the static
fluid $u_i=0$, one obtains the equations describing equilibrium fermion-boson configurations
\begin{eqnarray*}
\frac{da}{dr} &=& \frac{a}{2}\left\{\frac{1}{r}(1-a^{2})+4\pi\,G
  r\left[\left(\frac{\omega^{2}}{\alpha^{2}}+m^{2}\right)a^{2}\phi^{2}(r)+
    \Phi^{2}(r)+2a^{2}\mu\right]\right\} \\
\frac{d\alpha}{dr} &=& \frac{\alpha}{2}\left\{\frac{1}{r}(a^{2}-1)+4\pi\,G
  r\left[\left(\frac{\omega^{2}}{\alpha^{2}}-m^{2}\right)a^{2}\phi^{2}(r)+
  \Phi^{2}(r)+2a^{2}p\right]\right\} \\
\frac{d\phi}{dr} &=& \Phi(r) \\
\frac{d\Phi}{dr} &=& \left(m^{2}-\frac{\omega^{2}}{\alpha^{2}}\right)
a^{2}\phi -\left[1+a^{2}-4\pi\,Ga^{2}r^{2}\left(m^{2}\phi^{2}+
  \mu-p\right)\right] \frac{\Phi}{r} \\
\frac{dp}{dr} &=& -(\mu+p)\frac{\alpha^{'}}{\alpha} \,.
\end{eqnarray*}

These equations can be written in adimensional form by rescaling the variables and introducing the following quantities
\begin{eqnarray}
x\equiv mr \,, \quad \sigma(x)\equiv \sqrt{4\pi\,G} \phi(0,r) \,, \quad
\Omega\equiv \omega/m^{2} \,,  \nonumber \\ \bar{\mu}\equiv (4\pi\,G/m^{2})\mu
\,, \quad \bar{p}\equiv (4\pi\,G/m^{2})p \,. 
\end{eqnarray}
By varying the central value of the fermion energy density $\mu(r=0)$ and the scalar field $\phi(r=0)$, one finds stars dominated by either bosons
or fermions, with a continuous spectrum in between.
It was shown that the stability arguments made with boson stars 
can be generalized to these mixed objects \citep{1990PhLB..243...36J}.

\begin{figure}[htbp]
\centerline{\includegraphics[width=8.0cm]{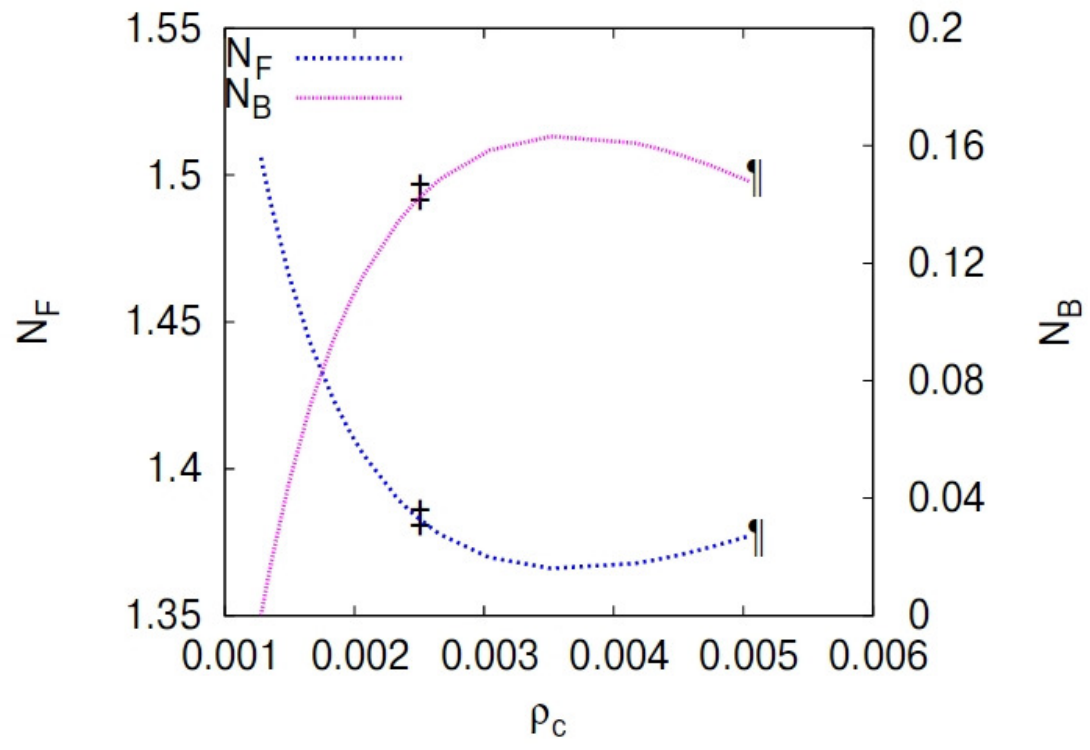}}
\caption{Initial data of a mixed fermion-boson star
with fixed total mass $M_T= 1.4$. The numbers of fermions, $N_F$, and bosons, $N$ (denoted $N_B$ in the figure, but just $N$ in this text), in terms of the central density, $\rho_c$, are plotted. The position
of the maximum of $N$ (and correspondingly the minimum of $N_F$) represents the critical point, with a maximum value $N/N_F = 12\%$,
which separates the stable and the unstable solutions. The
two configurations marked, one on each side of the maximum/minimum, correspond to $N/N_F \approx 10\% $. Reproduced with permission from \cite{2013PhRvD..87h4040V}, copyright by APS.}
\label{fig:fermionboson}
\end{figure}

More recently, neutron stars with a bosonic component, sourced by dark matter accretion, have also been considered \citep{2013PhRvD..87h4040V,Brito:2015yfh}. The fermionic matter for a cold star can be described easily by using simultaneously the polytropic and the ideal gas equation of state $P=K \rho^{\Gamma} = (\Gamma - 1) \rho \epsilon$, where $\rho$ is the rest-mass density, $\epsilon$ its internal energy, $K$ the polytropic constant, and $\Gamma$ the adiabatic index (i.e., the energy density can be written then as $\mu=\rho(1+\epsilon)$).
For standard masses of the neutron
star, stable configurations allow only about $N\approx 12\%  N_F$, where $N_F$ is the number of fermions.
Fig.~\ref{fig:fermionboson} displays both $N$ and $N_F$ for a fixed
total mass but with different central densities, $\rho_c$.
Similar studies have been performed by coupling fermion matter to oscillatons instead of boson stars \citep{Brito:2015yfh}.

\CP{Fermion-boson stars with a charged scalar field were studied in \cite{Kain:2021bwd}}.
The existence of slowly rotating fermion-boson stars was
shown in \cite{2001IJMPD..10..881D}, although no solutions were
found in previous attempts \citep{Kobayashi:1994qi}.
Also see \cite{Dzhunushaliev:2011ma} for unstable solutions consisting of a real
scalar field coupled to a perfect fluid with a polytropic equation of state.

\sllnewer{
A very different approach was taken by \cite{2022arXiv220910905G} deriving an equation of state for a particular bosonic dark matter field.
With this equation of state, they used a two-fluid formalism to construct
fermion-boson configurations of asymmetric bosonic dark matter with self-repulsion.
Considering either DM fully condensed in the core or distributed in a dilute halo around a neutron star, they found that while the former induces an effective softening of the equation of state, the latter mimics an apparent stiffening of strongly interacting matter.
}

%
\subsection{Multi-state, multi-field boson stars}
\label{sec:multi}
\sllnew{
Just as one can construct fermion-boson stars as a combination of a neutron star with a boson star, one can similarly combine multiple boson stars together. In this section, we first consider boson stars constructed from multiple states of the same field, which we call \emph{multi-state BSs}. Similarly, one can consider multiple fields, each possibly the superposition of multiple states, which we call \emph{multi-field BSs}.
}

\sllnew{
Motivated by the mode analysis conducted in the study of scalar collapse in AdS space \citep{Bizon:2011gg}, \cite{Choptuik:2019zji} construct \emph{multi-oscillator} boson stars by
explicitly promoting stable linear modes to auxiliary states. A full, nonlinear solve
for such a solution produces a multi-state boson star oscillating at two frequencies, and
this process can be applied to an arbitrary number of modes.
}

\sllnew{
\cite{Herdeiro:2020kvf} construct non-spherically symmetric, stationary BS solutions as the sum of $(N,l,m)$ states similar to hydrogen orbitals using the Einstein--De~Turck method,
a powerful approach in which an initial configuration is ``flowed'' to the stationary
one as in numerical approaches to Ricci flow \citep{Adam:2011dn,Garfinkle:2003an}.
}

It turns out that excited BSs, as dark matter halo candidates\sllnew{~(see Sect.~\ref{sec:darkmatter})}, provide for flatter, and
hence more realistic, galactic rotation curves than ground state BSs. The problem is that
they are generally unstable to decay to their ground state. Combining excited states with
the ground state is one way around this.

Although bosons in the same state are indistinguishable, it is possible
to construct non-trivial configurations with bosons in different excited states.
A system of bosons in $P$ different states that only interact with each other gravitationally
can be described by the following Lagrangian density
\begin{equation}\label{Lagrangian}
  {\cal L} = \frac{1}{16 \pi G} R
   - \sum_{n=1}^{P} \frac{1}{2} \left[ g^{ab} \partial_a \bar \phi^{(n)} \partial_b \phi^{(n)}
   + V\left( \left|\phi^{(n)} \right|^2\right) \right] \,,
\end{equation}
where $\phi^{(n)}$ is the particular complex scalar field representing the bosons in the 
$n$-state with $n-1$ nodes. 
\sllnew{Although originally called multi-state, the notation used here would refer to them as $n$-boson stars, a subset of multi-field BSs in different $n$-states.}
The equations of motion are very similar to the standard ones
described in Sect.~\ref{subsection:lagrangian},
with two peculiarities: (i)~there are $n$ independent KG equations (i.e., one for each
state) and (ii)~the stress-energy tensor is now the sum of contributions from each mode. Equilibrium
configurations for this system were found in \cite{2010PhRvD..81d4031B}.

\begin{figure}[htb]
\centerline{
  \includegraphics[width=0.49\textwidth]{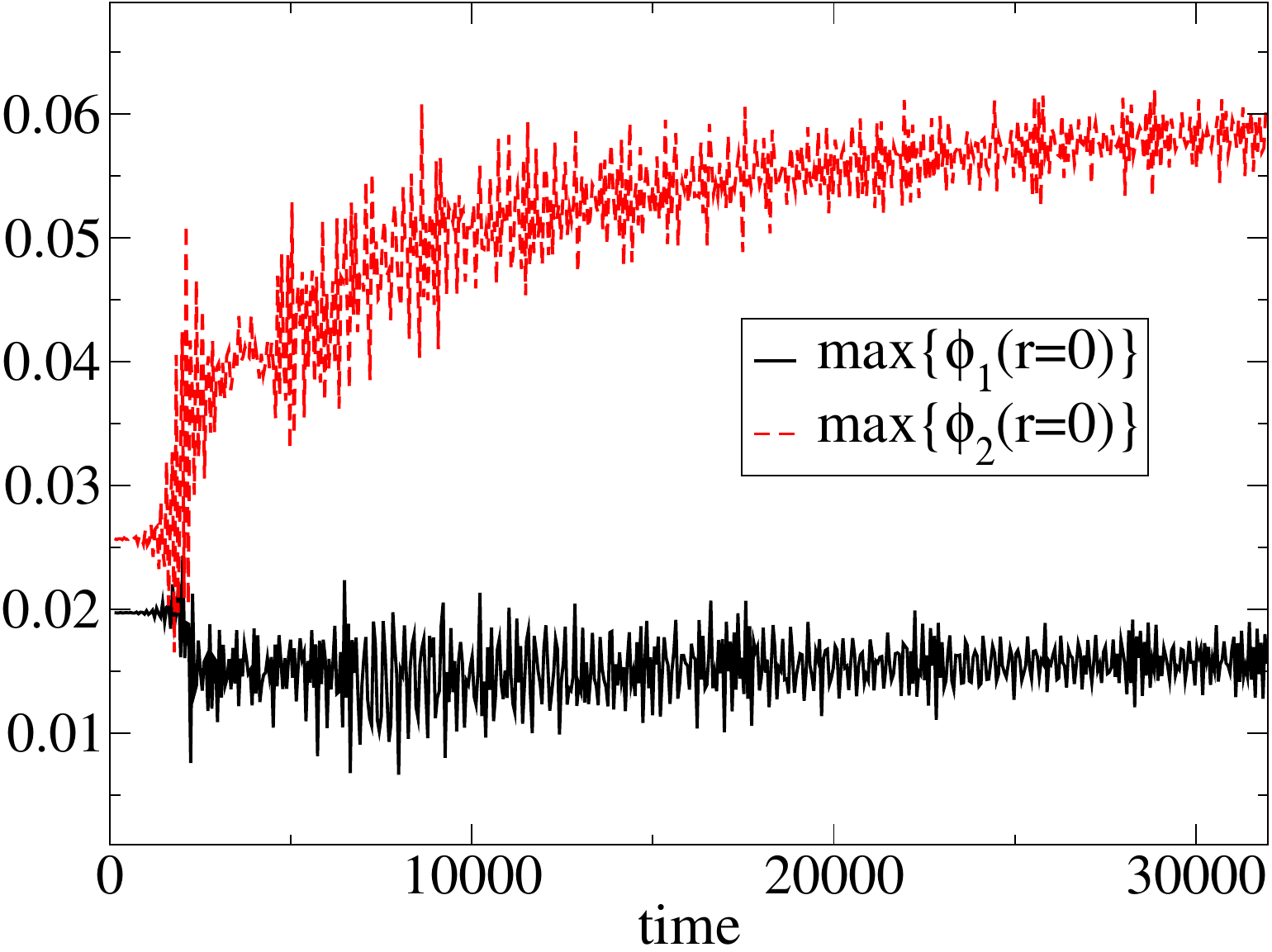}
  \includegraphics[width=0.49\textwidth]{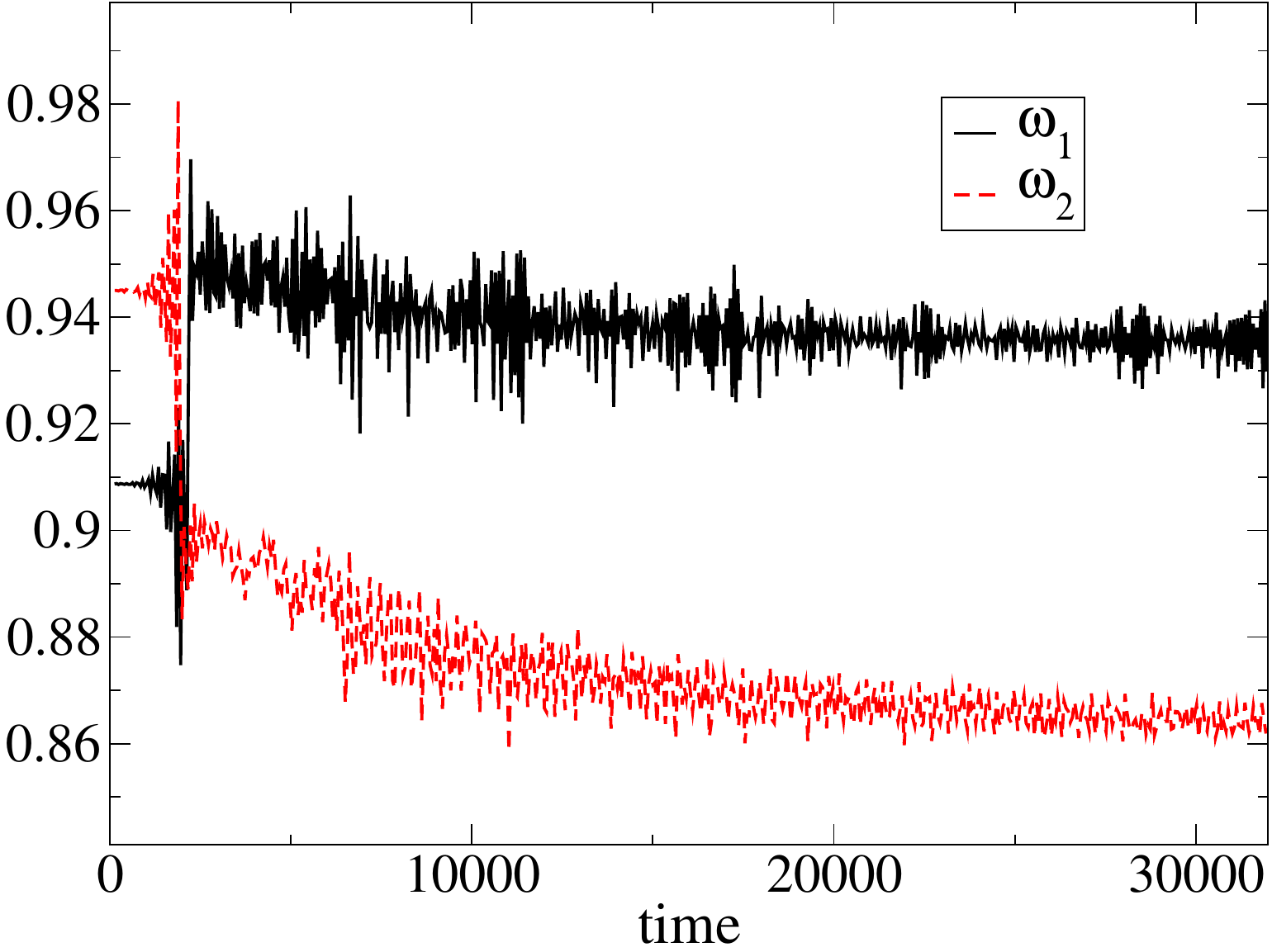}
}
\caption{\emph{Left:} The maximum of the central value of each of the
  two scalar fields constituting the multi-state BS for the fraction
  $\eta=3$, where $\eta \equiv N^{(2)}/N^{(1)}$ defines the relative
  ``amount'' of each state. \emph{Right:} The frequencies associated
  with each of the two states of the multi-state BS.
  At $t=2000$, there is a flip in which the excited state (black
  solid) decays and the scalar field in the ground state (red dashed)
  becomes excited. Discussed in Sect.~\ref{sec:multi}. Reproduced
  with permission from \cite{2010PhRvD..81d4031B}, copyright by APS.}
\label{fig:msbs_evolution}
\end{figure}

In the simplest case of
a multi-field boson, one has the ground state and the first excited state. Such configurations are
stable if the number of particles in the
ground state is larger than the number of particles in the excited
state \citep{2010PhRvD..81d4031B, alicthesis}
\begin{equation}\label{condition_stability_MSBS}
             N^{(1)} \ge N^{(2)} \,.
\end{equation}
This result can be understood as the ground state deepening the gravitational potential of
the excited state, thereby stabilizing it.
Unstable configurations migrate to a stable one via a flip-flop of
the modes; the excited state decays, while the ground state jumps to the first exited state, so
that the condition~(\ref{condition_stability_MSBS}) is satisfied. An example
of this behavior can be observed in Fig.~\ref{fig:msbs_evolution}.

Similar results were found in the Newtonian limit \citep{2010PhRvD..82l3535U},
however, with a slightly higher stability limit $N^{(1)} \ge 1.13\, N^{(2)}$. This work 
stresses that combining several excited states makes it possible to obtain flatter
rotation curves than only with ground state, producing better models for 
galactic dark matter halos (see also discussion of boson stars as an explanation of
dark matter in Sect.~\ref{sec:darkmatter}).

\cite{Hawley:2002zn} \sllnew{first introduced \emph{multi-field} boson stars by considering} two scalar fields each describing boson stars which are phase
shifted in time with respect to each other, studying the dynamics numerically.
In particular, one
can consider multiple scalar fields \emph{with} an explicit interaction (beyond just
gravity) between
them, say $V\left( |\phi^{(1)}|\, |\phi^{(2)}| \right)$.
\cite{2009PhRvD..80f4014B,2009PhRvD..79f4013B} construct such solutions, considering
the individual particle-like configurations for each complex field
as \emph{interacting} with each other.
\sllnew{
\cite{Li:2020ffy} construct rotating boson stars from two, self-interacting, scalar fields.}
\sllnew{The merger of two boson stars described by different scalar fields (so that their interaction was solely via the gravitational field) was found numerically by \cite{Bezares:2018qwa} (binaries are further discussed in Sect.~\ref{dynamics_binary_bs}).
}

\begin{figure}[htb]
		\centerline{
		\includegraphics[width=0.49\textwidth]{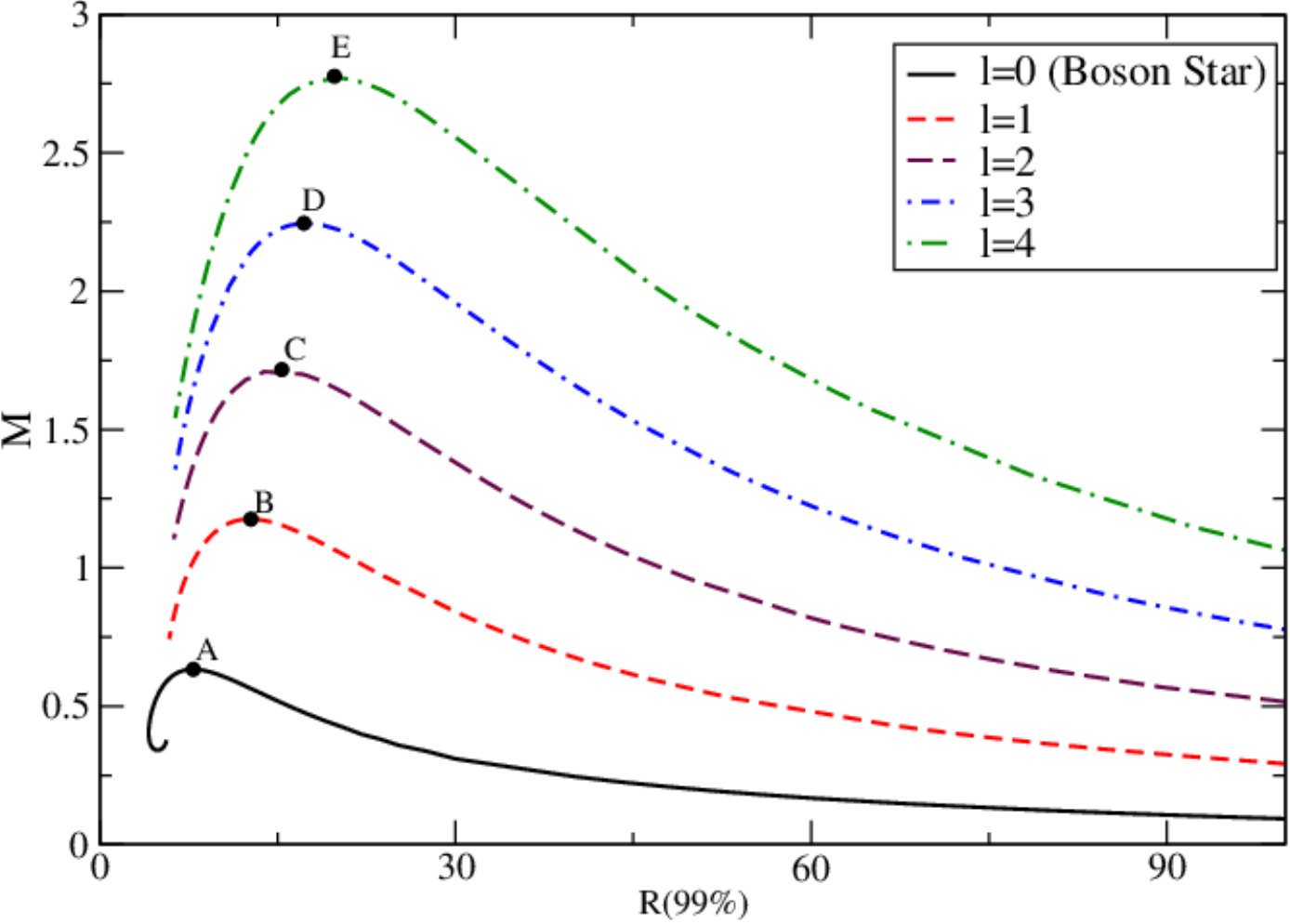}
		\includegraphics[width=0.49\textwidth]{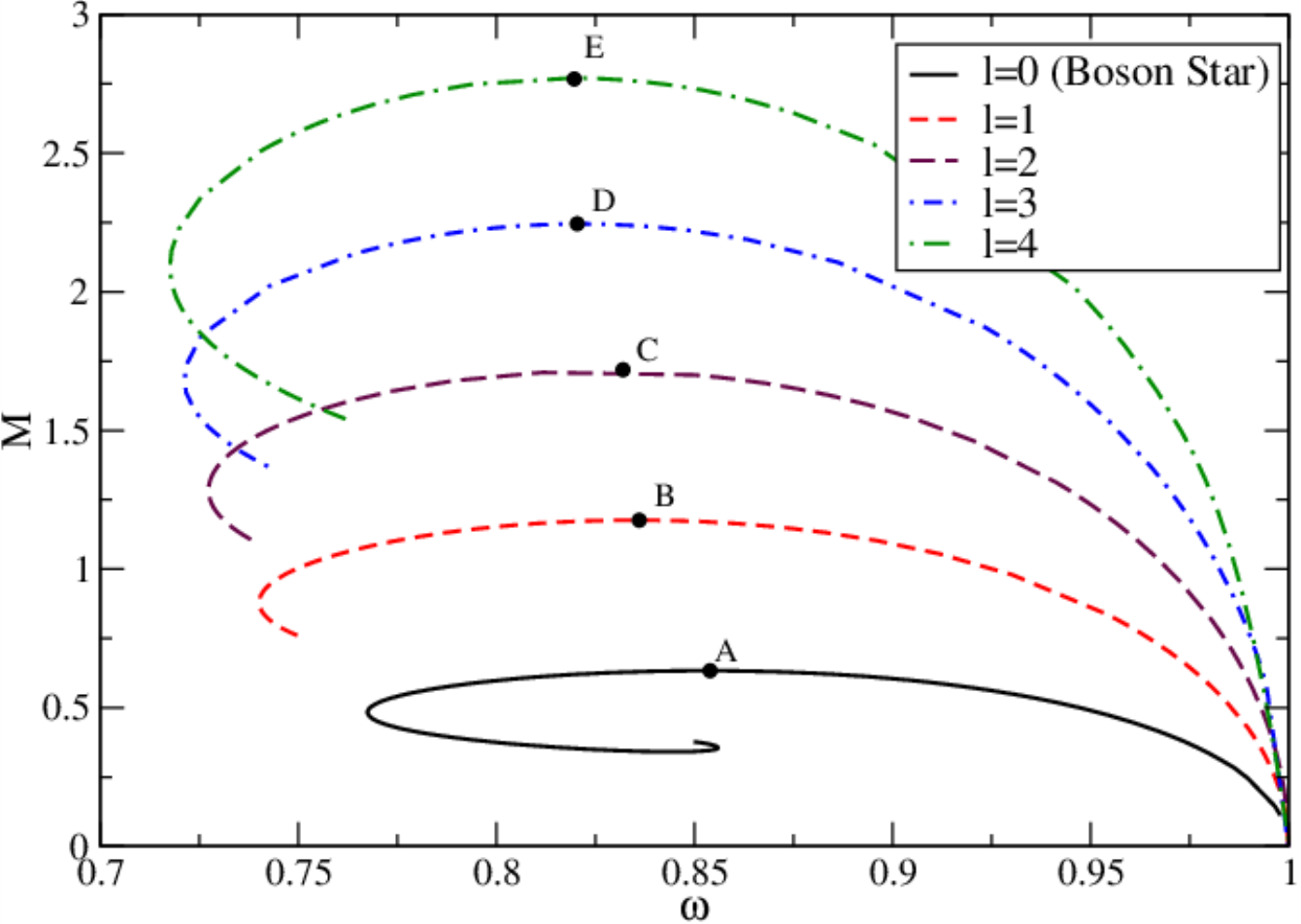}
		}
		\caption{$\ell$-boson stars: (Left) Total mass as a function of the effective radius for equilibrium configurations of different angular momentum number, denoted here as $l$.
                Note that the mass increases more quickly with $k$ than the radius, and hence the compactness increases.
		(Right) Total mass as a function of the frequency of oscillation for the same configurations as in the left panel. Reproduced with permission from \cite{Alcubierre:2018ahf}, copyright by IOP.}
		\label{fig:lboson_massomega}
\end{figure}

\sllnew{
\cite{Alcubierre:2018ahf} extends such solutions, following \cite{Olabarrieta:2007di}
which constructed spherically symmetric configurations of scalar field from specific superpositions of states with angular momentum for their studies of critical collapse~(see Sect.~\ref{sub:critical}). In particular, for a given value of $l$, their field configuration is a sum of fields representing the appropriate spherical harmonics $(\ell,m)$ 
\begin{equation}
\sum\limits_{m=-\ell}^{m=+\ell} \phi_{\ell m} \left(t,{\bf r} \right) = \Psi_\ell(t,r) Y^{\ell m} \left(\vartheta,\varphi\right)
\end{equation}
resulting in a spherical solution that they call \emph{$\ell$-boson stars}. These solutions can be more compact than corresponding regular boson (i.e. $\ell=0$ solutions).} \CPn{ In Fig.~\ref{fig:lboson_massomega} the equilibrium configurations up to $k=4$ are shown, displaying the total mass as a function of the effective radius (left panel) and frequency (right panel), respectively.}
\sllnew{
Indeed, follow-up work finds that for large $\ell$ the compactness of stable solutions approaches roughly half
the Buchdahl limit\footnote{The Buchdahl limit constrains spherically symmetric ``stars'' of ordinary matter to a compactness \CPi{ $M/R \leq 4c^2/\left(9G\right)$, where the maximum value} corresponds to stars of constant density \citep{Buchdahl:1959zz}.} as the solutions become increasingly anisotropic \citep{Alcubierre:2021psa}.
}

\begin{figure}[htb]
		\centerline{\includegraphics[height=8.0cm]{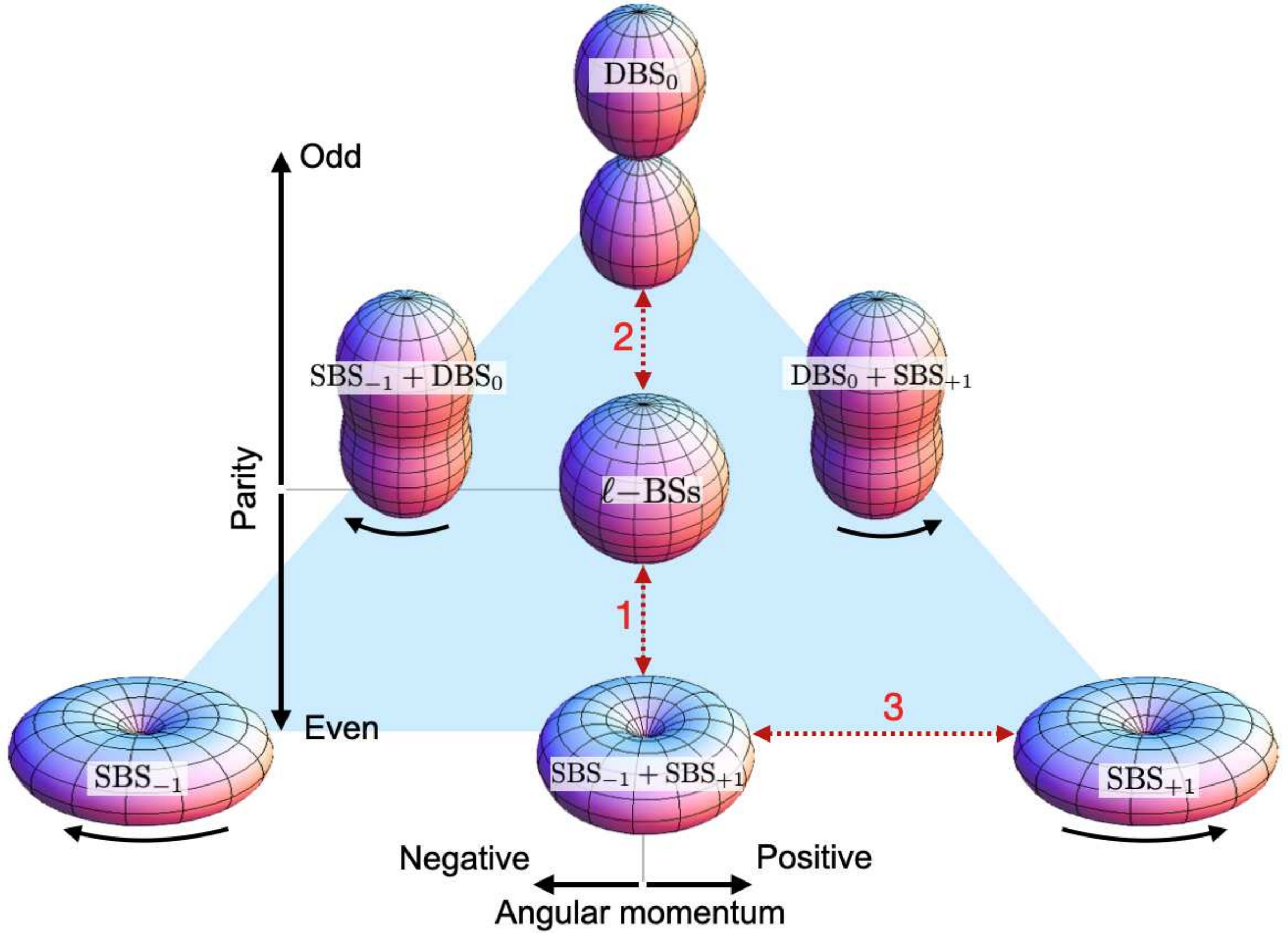}}
		\caption{The energy density distribution of the different $k=1$ hybrid-$\ell$-boson stars.  
                   In this case, we see: dipole BSs ($\mathrm{DBS}_0$),  spinning BSs, ($\mathrm{SBS}_{\pm 1}$), spinning dipolar BSs ($\mathrm{DBS}_0+\mathrm{SBS}_{\pm 1}$), toroidal static BSs ($\mathrm{SBS}_{-1}+\mathrm{SBS}_{+1}$), 
		 and finally $\ell$-BSs ($\mathrm{SBS}_{-1}+\mathrm{DBS}_0+\mathrm{SBS}_{+1}$).
	    Reproduced with permission from \cite{Sanchis-Gual:2021edp}, copyright by APS.}
		\label{fig:hybrid_lboson}
\end{figure}

\sllnew{
\cite{Sanchis-Gual:2021edp} extends $\ell$-boson stars to what they call \emph{hybrid-$\ell$-boson stars} as the combination of  non-spherically symmetric BSs composed of multiple fields with multiple states (some with rotation). They further evolve these states to understand which are stable and which unstable.
}
\CPn{The different solutions of the $k=1$ family are displayed in Fig.~\ref{fig:hybrid_lboson}}.

\subsection{Proca stars}
\label{sec:proca}

\begin{figure}[htb]
\centerline{\includegraphics[height=8.0cm,width=9.0cm]{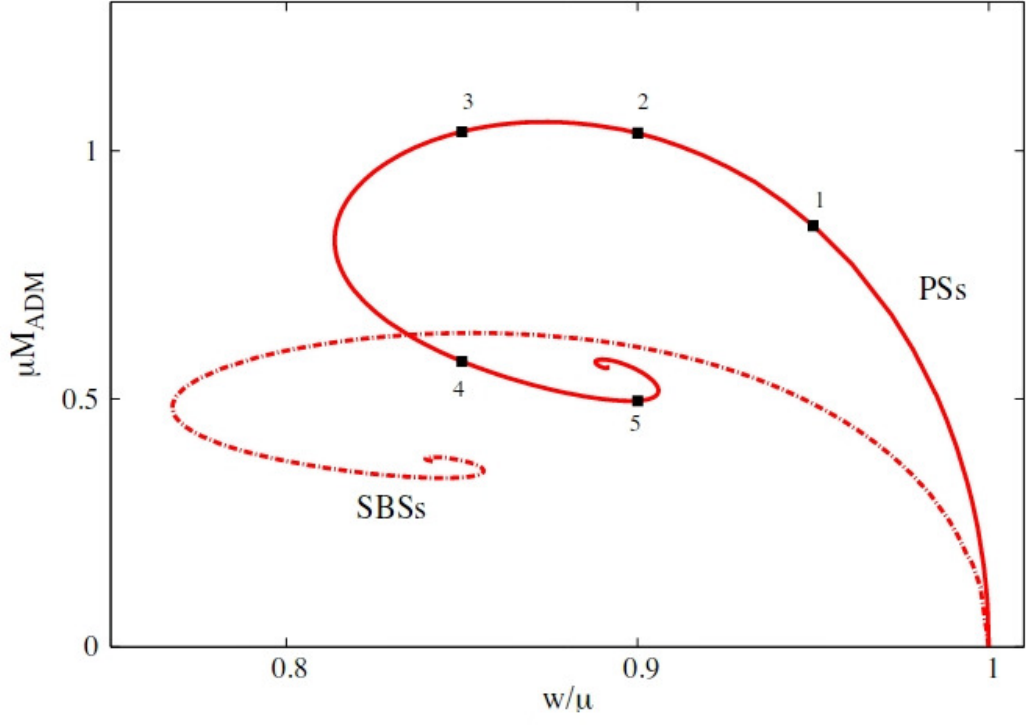}}
\caption{
Comparison of Proca solutions with boson stars. The ADM mass of spherical Proca solutions (solid) and scalar BS solutions (dashed) 
are shown versus oscillation frequency. Here, the mass is expressed in terms of the field mass, $\mu$.
Although the profiles are qualitatively similar, notice that the maximum mass of the Proca solutions is almost twice that of BSs. Reproduced with permission from \cite{2017PhRvD..95j4028S}, copyright by APS.}
\label{fig:proca}
\end{figure}

%
%
Boson stars can be understood as condensates of massive spin 0 bosonic particles modeled by a scalar field. Recently, analogous self-gravitating solutions, made of massive spin 1 particles, have been found in the novel work of \cite{Brito:2015pxa}. These configurations, modeled by a massive complex vector field $A_a$, are described by the Proca action for the matter sector
\begin{eqnarray}\label{Lagrangian_Proca}
  {\cal L_M} = 
   - \frac{1}{4} F_{ab} {\bar F}^{ab} - \frac{1}{2} m^2 A_a {\bar A}^a \,,
\end{eqnarray}
where $m$ is the mass of the Proca field and $F_{ab}$ the field strength satisfying $F_{ab} = \nabla_a A_b - \nabla_b A_a$. The system of equations obtained by performing the variations on the action forms
the Einstein--Proca system. The evolution equations for the Proca field are
\begin{eqnarray}\label{Proca1}
  \nabla_a F^{ab} = m^2 A^b \,, 
\end{eqnarray}
which implies that the Lorentz condition $\nabla_a A^a =0$ is not a gauge choice like in Maxwell equations, but instead a dynamical requirement. The Einstein equations include now the stress-energy tensor 
\begin{eqnarray}\label{Tmunu_Proca}
  T_{ab} = - F_{c(a} {{\bar F}_{b)}}^{~c}
  - \frac{1}{4} g_{ab} F_{cd} {\bar F}^{cd} 
  + m^2 \left[ A_{(a} {\bar A}_{b)} - \frac{1}{2} g_{ab} A_{c} {\bar A}^c \right].
\end{eqnarray}
Like in the scalar case, there is a global $U(1)$ invariance of the action under transformations $A_a \rightarrow e^{i\theta} A_a$, implying
the existence of a conserved 4-current due to Noether's theorem
\begin{eqnarray}\label{Proca2}
  J^a = \frac{i}{2} \left[ {\bar F}^{ab} A_b - F^{ab} {\bar A}_b \right].
\end{eqnarray}
In addition to carrying a conserved Noether charge, Proca stars share many other features with boson stars. Both have a harmonic time dependence but
solutions exist only for a limited range of frequencies. Both achieve  a maximum ADM mass, which for Proca stars is $M_{\max}=1.058 M^2_{\mathrm{Planck}}/m$, larger, but of the same order, than those for (mini-)boson stars.
Fig.~\ref{fig:proca} displays the masses of both BSs and Proca stars versus
their (internal) oscillation frequencies.
The maximum mass solution separates stable from unstable configurations.
Different types of Proca stars are also possible, such as those  with rotation \citep{Brito:2015pxa}, charge \citep{Garcia:2016ldc}, or in anti-de~Sitter spacetime \citep{Duarte:2016lig}.
Numerical evolutions of these configurations have been performed for instance in \cite{Sanchis-Gual:2017bhw}.

\sllnew{
Rotating Proca stars have different stability properties than rotating boson stars which
were discussed earlier in Sect.~\ref{subsection:varieties_rotating}.
Binaries composed of Proca stars are discussed in Sect.~\ref{dynamics_binary_bs}.
}
\sllnewer{\cite{Gorghetto:2022sue} study Proca stars with a dark, \emph{real}, vector potential, $A_a$, arising from vacuum fluctuations in the early universe and serving as a cosmological source of dark matter, discussed in Sect.~\ref{sec:darkmatter}.
}

\subsection{Kerr black holes with scalar hair \& superradiance}
\label{scalarclouds}

Closely related to a BS, one can instead construct stable configurations of
a complex scalar field around a rotating black hole \citep{Hod:2012px}.  Such solutions are akin to
a BS with a black hole embedded at its center. As such, the scalar field
serves as \emph{scalar hair} (see the review about no-scalar-hair theorems by \citealt{Herdeiro:2015waa}).

To find such solutions, one  proceeds in much the same fashion as the construction of rotating solutions (Sect.~\ref{subsection:varieties_rotating}).
In particular, because rotation is required to achieve a stable configuration,
one works in axisymmetry and assumes a harmonic ansatz for both the
internal and azimuthal rotations
\begin{equation}
  \phi( {\bf r},t) = e^{-i \omega t} e^{i  m \phi} \psi(r,\theta).
  \label{eq:superradiant}
\end{equation}
Here $\omega$ is the (complex) angular frequency and $m$ must be an integer ($m=\pm 1, \pm 2, \dots$) for continuity in the azimuthal direction.

Instead of solving the full system of equations, a first approximation
can be obtained by solving the linearized scalar field equations on a fixed
spacetime \citep{Herdeiro:2015gia}. 
Within such a linear approximation, one finds that non-rotating (Schwarzschild) BHs do not allow
for bound states with strictly real $\omega$ \citep{2014PhRvL.112v1101H}.
However, \emph{quasi-bound} states can exist with $\Im(\omega) <0$
in which the scalar field decays, infalling into the BH.

For a Kerr black hole with angular momentum $J$, mass $M$, and horizon radius in the equatorial plane $r_H$, one can identify
the angular velocity of the horizon as $\Omega_H\equiv J/(2 M^2 r_H)$.  For such rotating BHs, there is a critical frequency 
$\omega_c \equiv m \Omega_H$ separating disparate behavior. For $\omega=\omega_c$, the frequency is strictly real allowing for regular bound states known as \emph{scalar clouds}.

As $\omega$ increases above $\omega_c$, its imaginary part becomes negative, allowing again only for quasi-bound states with
a time-decaying scalar field. In contrast,
as $\omega$ decreases below $\omega_c$, its imaginary part becomes positive, indicating growth of the scalar field
in Eq.~(\ref{eq:superradiant}). This growth of the massive field is called the \emph{superradiant instability} (for a recent review of superradiance see \citealt{Brito:2015oca}) and results in
the extraction of energy, charge, and angular momentum from the black hole.
For a rigorous treatment of this instability and a proof of boundedness
see the work of \cite{Dafermos:2014jwa}.

In \cite{2014PhRvD..90j3526K}, an analog of a boson star (see 
Sect.~\ref{sec:analog} for physical analogs of BSs) is used to study superradiance.
BSs have also been found as the zero
radius limit of hairy black holes in AdS${}_4$ (see Sect.~\ref{sec:other} for BSs in AdS), and these hairy BHs are proposed as the end state of
the superradiant instability \citep{Dias:2016pma}.

\begin{figure}[htb]
\centerline{\includegraphics[height=8.0cm,width=10.0cm]{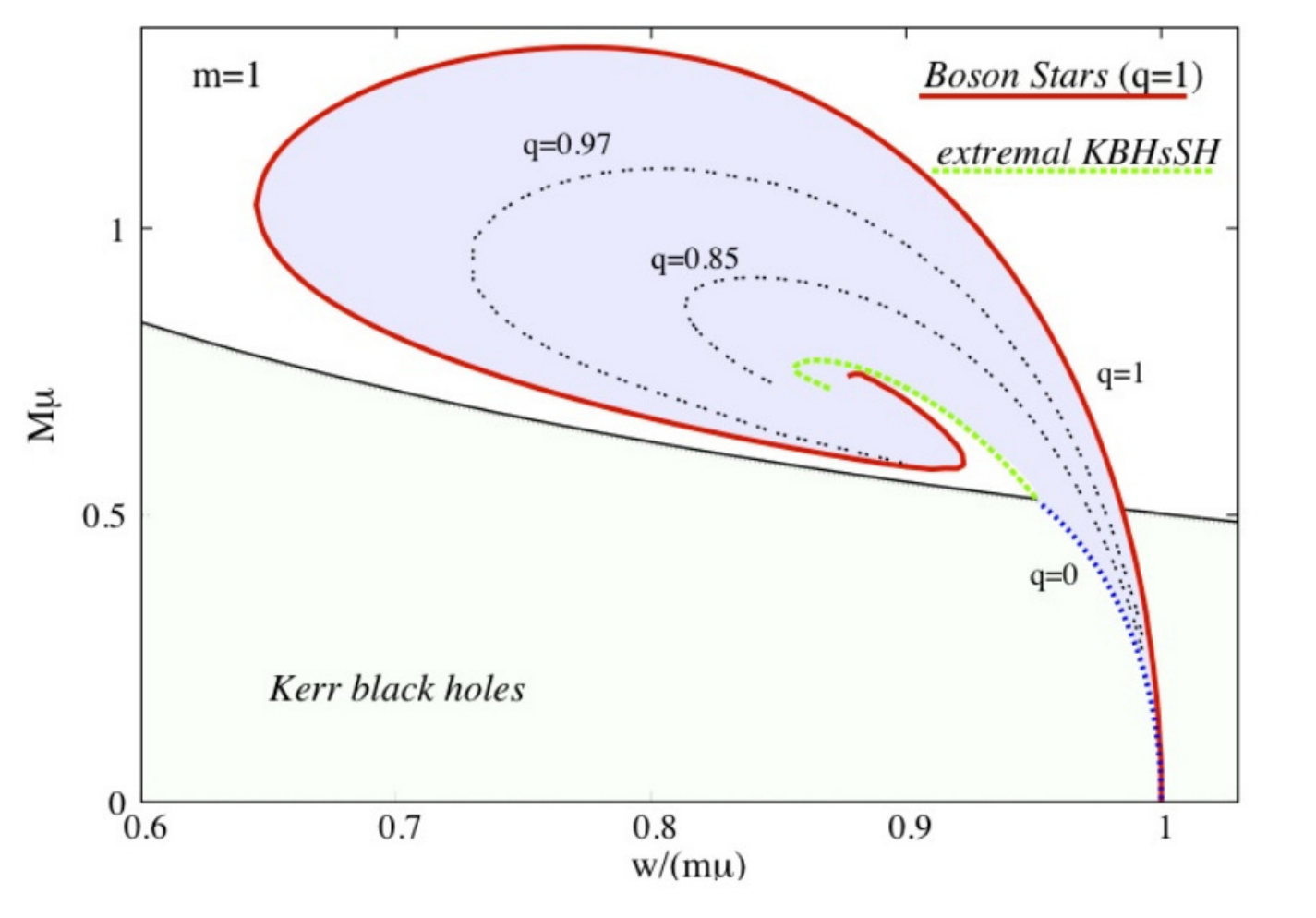}}
\caption{Domain of existence for hairy black holes.
The ADM mass of the solutions versus the oscillation frequency of the scalar field frequency.  Solutions for a range of values of $q$ 
interpolating between Kerr ($q=0$) and BSs ($q=1$) all with azimuthal quantum number $m=1$.
For $0<q<1$, solutions describe rotating BHs surrounded by a scalar cloud, constituting scalar hair for the BH.
Reproduced with permission from \cite{Herdeiro:2015gia}, copyright by IOP.}
\label{fig:hairyBH}
\end{figure}

These solutions persist when solving the fully nonlinear system in which  the harmonic ansatz of Eq.~(\ref{eq:superradiant})
 implies that the stress-energy tensor is independent of $\{t,\phi\}$,
and are generically known as \emph{Kerr BHs with scalar hair} \citep{2014PhRvL.112v1101H}.
As reviewed by \cite{Herdeiro:2015gia}, solutions can be parametrized in such a way that connects pure Kerr BHs
with pure BSs. In particular, defining $q \equiv kN/J$ where $N$ is the number of bosonic particles as in Eq.~(\ref{Noether_charge}) and where $k$ is the integer ``quantum'' number associated with the angular momentum as in Eq.~(\ref{quantization_J}),
Kerr BHs are described by the vanishing of the scalar field, $q=0$, and BSs are described by the vanishing of the horizon, $q=1$.
Fig.~\ref{fig:hairyBH} shows the space of solutions interpolating between these two limits.

More recent work has extended these solutions. For example,
a self-interacting potential with a quartic term was considered in \cite{Herdeiro:2015tia,Herdeiro:2016gxs}, producing a larger amplitude scalar field but not a more massive black hole than with the non-self-interacting potential.
Coupling the scalar field to the electromagnetic field allows for charged clouds \citep{Delgado:2016jxq}.
Kerr black holes with Proca hair (see Sec.~\ref{sec:proca} for a description of Proca stars) were constructed in \cite{Herdeiro:2016tmi}.
\sllnew{
Evolutions by \cite{East:2017ovw} of black holes with a Proca field in axisymmetry find that superradiance saturates as expected when the frequency of the field matches the horizon frequency and find the resulting stationary states ``plausibly'' the same as those
constructed in \cite{Herdeiro:2016tmi}.
}

Superradiant instabilities are likely to be weaker for hairy black holes than for Kerr black holes with the same global charge \citep{2014arXiv1406.1225H}. %
A recent review on the physical properties of Kerr black holes with scalar hair can be found in \cite{Herdeiro:2015gia}, and prospects for testing whether BHs have hair
is reviewed in \cite{Cardoso:2016ryw}.

\cite{Chodosh:2015oma} study the scalar cloud solutions analytically and demonstrate existence.
They also consider certain uniqueness and stability properties of solutions close to Kerr and review past analytic work in this area.

\subsection{Alternative theories of gravity}
\label{sec:alt}

Instead of modifying the scalar field potential, one can consider
alternative theories of gravity. Constraints on such theories  are already significant given the great success of general relativity \citep{will},
and more strict bounds might be set with present and future astrophysical observations \citep{Berti:2015itd}. However, 
the fast advance of electromagnetic observations and the rise of gravitational
wave astronomy promise much more in this area, in particular in the context of compact objects that probe strong-field gravity.

An ambitious effort is begun
in \cite{Pani:2011xm}, which studies a very general gravitational Lagrangian
(``extended scalar-tensor theories'')
with both fluid stars and boson stars. The goal is for observations of compact stars to constrain such theories of gravity. General theoretical bounds on the mass to radius ratio of stable compact objects (i.e., both neutron and boson stars) can be set for extended gravity theories, in particular for  scalar tensor theories \citep{Burikham:2016cwz}.

Scalar tensor theories allow for
\emph{spontaneous scalarization} in which the 
scalar component of the gravity theory transitions to a non-trivial configuration
analogously to ferromagnetism with neutron stars \citep{Damour:1996ke}.
\sllnew{
Studies have found the existence of scalarized boson stars \citep{Brihaye:2019puo} and hairy black holes \citep{Kleihaus:2015iea}. Spontaneous scalarization has also been found in the evolution of single boson stars including only the massive term in the potential \citep{Alcubierre:2010ea}.
}

\CP{A special class of tensor-multi-scalar theories has been considered, which admits a new type of compact object solution; the tensor-multi-scalar solitons formed by a condensate of the gravitational scalars, which can be understood as a generalization of the standard boson star \citep{Yazadjiev:2019oul}. Soon after, the same authors also found mixed configurations of tensor-multi-scalar solitons and relativistic neutron stars \citep{Doneva:2019krb}. Boson and neutron stars have also been studied in a scalar-tensor theory with an explicitly time-dependent real scalar field \citep{Brihaye:2020klz}.
}

One motivation for alternative theories is to explain the apparent existence of
dark matter without resorting to some unknown dark matter component. Perhaps
the most well known of these is MOND (modified Newtonian dynamics) in which
gravity is modified only at large distances \citep{milgromorig,milgromnew} (for a review see \citealt{Famaey:2011kh}).
A nonminimal coupling of the scalar field to the Ricci curvature scalar results in configurations that better resemble dark energy stars than ordinary boson stars \citep{2013CQGra..30n5006H,Marunovic:2015obj}. Boson stars are studied within TeVeS (Tensor-Vector-Scalar), a
relativistic generalization of MOND \citep{2008PhRvD..78d4034C}. In particular,
their evolutions of boson stars develop caustic singularities, and the authors propose
modifications of the theory to avoid such problems.

Bosons star solutions also exist in bi-scalar extensions of Horndeski gravity \citep{Brihaye:2016lin},
\sllnew{
and the properties and stability of fermion-boson stars in Horndeski theories were studied in \cite{Roque:2021lvr}.
}
In addition, solutions have been found in the framework
of teleparallel gravity \citep{2015CQGra..32c5023H} and within conformal gravity and its scalar-tensor extensions \citep{Brihaye:2009ef,Brihaye:2009hf}.
Charged boson stars with a torsion-coupled field have been considered in \cite{Horvat:2015qva} \sllnew{followed with a further study of solutions in an $f({\cal T})$ theory of gravity \citep{Ilijic:2020vzu}.}
\sllnew{
Families of boson stars were constructed in quadratic Palatini $f({\cal R})$ gravity finding significant degeneracy with respect to those in standard GR \citep{Maso-Ferrando:2021ngp}.
}

Recently there has been renewed interest in Einstein--Gauss--Bonnet theory, which appears naturally in the low energy effective action of quantum gravity models. This theory only differs from General Relativity for dimensions $D>4$, and so the easiest non-trivial case is to consider $D=5$. Boson star have been found in (4+1)-dimensional Gauss--Bonnet gravity \citep{2013PhLB..726..906H}. Rotating configurations were constructed in \cite{2014PhRvD..89j4060B}, and its classical instability and existence of ergoregions studied in \cite{Brihaye:2015jja}.
Rotating boson stars in odd-dimensional asymptotically anti-de~Sitter spacetimes in
Einstein--Gauss--Bonnet gravity are studied in \cite{2015PhRvD..91b4009H}.
A non-minimal
coupling between a complex scalar field and the Gauss--Bonnet term was studied in \cite{Baibhav:2016fot}. Coupling Einstein gravity to a complex self-interacting boson field as well as a phantom field allows for new type of configurations, namely boson stars harboring a wormhole at their core \citep{2014PhRvD..90l4038D}.

\subsection{Gauged boson stars}
\label{subsection:varieties_gauged}
%

In 1988, Bartnik and McKinnon published quite unexpected results
showing the existence of particle-like
solutions within $SU(2)$ Yang--Mills coupled to gravity \citep{Bartnik:1988am}.
These solutions, although unstable,
were unexpected because no particle-like solutions are found
in either the Yang--Mills or gravity sectors in isolation.
Recall also that no particle-like solutions were found with gravity coupled
to electromagnetism in early efforts to find Wheeler's geon (however, see 
Sect.~\ref{sec:other} for discussion of \cite{Dias:2011ss}, which finds geons within AdS).

Bartnik and McKinnon generalize from the Abelian $U(1)$ gauge group to the non-Abelian $SU(2)$ group and thereby find these unexpected particle-like solutions.
One can consider, as does \cite{Schunck:2003kk} (see Sect.~IIp), these globally regular solutions (and their generalizations to $SU(n)$ for $n>2$) as \emph{gauged
boson stars} even though these contain no scalar field.
One can instead explicitly include a scalar field doublet coupled to the Yang--Mills gauge field \citep{Brihaye:2004nd} as perhaps a more direct generalization of the ($U(1)$) charged boson stars
discussed in Sect.~\ref{subsection:varieties_charged}.

\cite{2007MPLA...22..273D}
studies BSs formed from a gauge condensate of an $SU(3)$ gauge field,
and \cite{Brihaye:2009hf}
extends the Bartnik--McKinnon solutions to conformal gravity with a Higgs field.


\section{Dynamics of boson stars}
\label{section:dynamics}

In this section, the formation, stability and dynamical evolution
of boson stars are discussed. One approach to the question of stability considers small perturbations
around an equilibrium configuration, so that the system remains in the linearized
regime. Growing modes indicate instability. However, a solution can be linearly stable
and yet have a nonlinear instability. One example is Minkowski space, which, under small
perturbations, relaxes back to flat, but, for sufficiently large perturbations, leads to 
black-hole formation, decidedly not Minkowski. To study nonlinear stability, other methods
are needed. In particular, full numerical evolutions of the Einstein--Klein--Gordon~(EKG)
equations are quite useful for understanding the dynamics of boson stars.

\subsection{Gravitational stability}

A linear stability analysis consists of studying the time evolution of infinitesimal
perturbations about an equilibrium configuration, usually with the additional
constraint that the total number of particles must be conserved. In the case
of spherically symmetric, fermionic stars described by a perfect fluid, it is possible
to find an eigenvalue equation for the perturbations that determines the normal
modes and frequencies of the radial oscillations (see, e.g., \citealt{Font:2001ew}). Stability theorems
also allow for a direct characterization of the stability branches of the 
equilibrium solutions \citep{sorkin,cst94}.
Analogously, one can write a similar eigenvalue equation for boson stars and show
the validity of similar stability theorems.
In addition to these methods, the stability of boson stars has also been studied using mainly two other, independent methods: by applying catastrophe theory
and by solving numerically the time dependent Einstein--Klein--Gordon equations.
Recently, a method utilizing information theory shows promise in analyzing the stability of equilibrium configurations.
All
these methods agree with the results obtained in the linear stability analysis.

\subsubsection{Linear stability analysis}

Assume that a spherically symmetric boson star in an equilibrium configuration
is perturbed only in the radial direction. The equations governing these
small radial perturbations are obtained by linearizing the system of equations in the
standard way; expand the metric and the scalar field functions to first order in
the perturbation and neglect higher order terms in
the equations \citep{1988PhRvD..38.2376G, 1989NuPhB.316..411J}. Considering the collection
of fields for the system $f_i$, one expands them in terms of the background solution ${}^0 f_i$
and perturbation as
\begin{equation}
      f_i(r,t) = {}^0 f_i(r) + {}^1 f_i(r) e^{i \sigma t} \,,
\end{equation}
which assumes harmonic time dependence for the perturbation.
Substitution of this expansion into the system of equations then provides
a linearized system, which reduces to a set of coupled equations that determines the spectrum of
modes ${}^1 f_i$ and eigenvalues $\sigma^2$ 
\begin{equation}
      L_{ij} {}^1 f_i = \sigma^2\, M_{ij} \, {}^1 f_i \,,
\end{equation}
where $L_{ij}$ is a differential operator containing partial derivatives and $M_{ij}$ is 
a matrix depending on the background equilibrium fields ${}^0 f_i$.
Solving this system, known as the pulsation equation, produces the spectrum of eigenmodes and their
eigenvalues $\sigma$. Recently, several powerful techniques have been introduced to compute the quasinormal modes of compact objects in complicated configurations, such as in the presence of interacting fields \citep{Macedo:2016wgh}.

The stability of the star depends crucially
on the sign of the smallest eigenvalue. Because of time reversal symmetry, only $\sigma^2$ enters the
equations \citep{1989NuPhB.315..477L}, and we label the smallest eigenvalue $\sigma_0^2$. If it is negative, the eigenmode grows exponentially
with time and the star
is unstable. On the other hand, for positive eigenvalues the configuration has no unstable modes
and is therefore stable. The critical point at which the stability transitions from stable to unstable therefore
occurs when the smallest eigenvalue vanishes, $\sigma_0 = 0$.

Equilibrium solutions of nonrotating BSs can be parametrized with a single variable,
such as the central value of the scalar field $\phi_c$. We can therefore write the
mass and particle number as
$M=M(\phi_c)$ and $N=N(\phi_c)$, and stability theorems 
indicate that transitions between stable and unstable
configurations occur only at the critical points in the parameter space such that 
\begin{eqnarray}\label{stability_bosonSL}
   \frac{dM}{d\phi_c} =
   \frac{dN}{d\phi_c} = 0 .
\end{eqnarray}
These transitions in stability are completely analogous to those for neutron stars \citep{cst94,sorkin,1965gtgc.book.....H,1984grra.book.....S}.

\sllnew{
A linear analysis of $f$-mode frequencies of massive boson stars in the context of dark matter is undertaken with a goal of establishing a connection between  boson star oscillations and the underlying scalar potential \citep{VasquezFlores:2019eht}.
Other work analyzed perturbations of boson stars constructed with an ultralight, repulsive dark matter field \citep{Lopes:2019eue}.
}

One can generalize this result for fermion-boson stars which contain a number of fermions, $N_F$, in addition
to some number of  bosons, $N$ (see Sect.~\ref{subsection:varieties_fermionicbosonic} for
a discussion of fermion-boson stars). In particular, one looks for critical points in a higher
dimensional parameter space by considering a vector of perturbations, $\mathbf{n}$ in a space spanned
by the total mass at infinity, $M$, and the two particle numbers, $N$ and $N_F$.
Following \cite{1990PhLB..251..511H},  the critical points are such that the directional derivatives vanish
\begin{eqnarray}\label{stability_fermionbosonSL}
   \left. \frac{d M}{d \mathbf{n}}\right \vert_{b} =
   \left. \frac{d N}{d \mathbf{n}} \right \vert_{b} = 
      \left. \frac{d N_F}{d \mathbf{n}} \right \vert_{b} = 0     
\end{eqnarray}
where the subscript $b$ means the value of the quantities at
the critical point. The direction $\mathbf{n}$ at the stability boundary is tangential to the level curves of constant $M$ and $N$; formally speaking, the direction $\mathbf{n}$ is orthogonal to the gradient of the functions at the boundary, $\mathbf{n} \perp \nabla(M , N, N_F )|_b$. 

The condition expressed by Eq.~\ref{stability_fermionbosonSL} reduces to the stability condition of Eq.~\ref{stability_bosonSL} when applied to single parameter solutions,  but it allows for multi-parameter critical curves.
Following the analysis of the fermion-boson star, the condition~\ref{stability_fermionbosonSL} implies that the equilibrium critical configurations manifest themselves at the extreme values of the number of particles when surveyed along a level curve of constant total mass \citep{2013PhRvD..87h4040V,Brito:2015yfh}, namely
\begin{eqnarray}
   \left. \frac{\partial N}{\partial \rho_c}\right \vert_{M=\mathrm{constant}} =
   \left. \frac{\partial N_F}{\partial \rho_c} \right \vert_{M=\mathrm{constant}} & = & 0 \nonumber \\
   \left. \frac{\partial N}{\partial \phi_c}\right \vert_{M=\mathrm{constant}} =
   \left. \frac{\partial N_F}{\partial \phi_c} \right \vert_{M=\mathrm{constant}} & = & 0,
\end{eqnarray}
where $\rho_c$ is the central density of the fermionic component.

Linear perturbation analysis provides a more detailed picture such as the growth rates and the eigenmodes of the perturbations.
For instance, \cite{2013PhRvD..88f4046M} studies the free oscillation spectra of different types of boson stars via perturbation theory.

\cite{1989NuPhB.319..733G} carries out such an analysis for
perturbations that conserve mass and charge. They find the first three perturbative modes and their growth
rates, and they identify
at which precise values of $\phi_c$ these modes become unstable. Starting from small values, they
find that ground state BSs are stable up to the critical point of maximum mass. Further increases in
the central value subsequently encounter additional unstable modes.
This same type of analysis applied to excited state BSs showed
that the same stability criterion applies for perturbations that
conserve the total particle number \citep{1989PhLB..222..447J}. For more general perturbations
that do not conserve particle number, excited states are generally unstable to decaying
to the ground state.

A more involved analysis by \cite{1989NuPhB.315..477L} uses a Hamiltonian formalism
to study BS stability. Considering first order perturbations that conserve mass and
charge ($\delta N = 0$),
their results agree with those of \cite{1989NuPhB.319..733G,1989PhLB..222..447J}.
However, they extend their approach to consider more general perturbations
which do \emph{not} conserve the total number of particles (i.e., $\delta N \neq 0$). To do so,
they must work with the second order quantities. 
They found complex eigenvalues
for the excited states 
that indicate that \emph{excited state boson stars are unstable}.
More detail and discussion on the different stability
analysis can be found in \cite{Jetzer:1991jr}.

\emph{Catastrophe theory} is part of the study of dynamical systems that began in the 1960s and
studies large changes in systems resulting from small changes to certain important parameters (for a physics-oriented
review see \citealt{0034-4885-45-2-002}).
Its use in the context of boson stars is
to evaluate stability, and to do so one constructs
a series of solutions in terms of a limited and appropriate set of parameters. Under certain
conditions, such a series generates a curve smooth everywhere except for certain points.
Within a given smooth expanse between such singular points, the solutions share the same stability properties.
In other words, bifurcations occur at the singular points so that solutions after the singularity gain an
additional, unstable mode.
Much of the recent work in this area confirms the previous conclusions from linear perturbation
analysis \citep{2010PhRvD..81l4041T,2011PhRvD..83h4046T,2011PhRvD..83d4027T,2011PhRvD..84d4054T} and from earlier work with catastrophe theory \citep{Kusmartsev:2008py}.
Another recent work using catastrophe theory finds that rotating stars share
a similar stability picture as nonrotating solutions \citep{Kleihaus:2011sx}.
However, only fast spinning stars are subject to an ergoregion
instability \citep{2008PhRvD..77l4044C}.

A recent and promising alternative method  to determine the  stability bounds of 
self-gravitating astrophysical objects, and in particular of boson stars, makes use of a new measure of shape complexity known as configurational entropy \citep{Gleiser:2015rwa}. 
Their results for the critical stability region agree with those of
traditional perturbation methods with an accuracy of a few percent or better.

\subsubsection{Non-linear stability of  boson stars without angular momentum}

The dynamical evolution of spherically symmetric perturbations of boson stars
has also been studied by solving numerically
the Einstein--Klein--Gordon equations (Sect.~\ref{subsection:3+1_decomposition}),
or its Newtonian limit (Sect.~\ref{subsection:varieties_newtonian}),
the \schrodinger--Poisson system.
The first such work was \cite{1990PhRvD..42..384S} in which the stability
of the ground
state was studied by considering finite perturbations, which
may change the total mass and the particle number (i.e., $\delta N \neq 0$ 
and $\delta M \neq 0$). The results corroborated the linear stability analysis in the
sense that they found a stable and an unstable branch with a transition 
between them at a critical value, $\phi_{\mathrm{crit}}$, of the central scalar field
 corresponding to the maximal BS mass $M_{\max}=0.633\,M^2_{\mathrm{Planck}}/m$.

\begin{figure}[htbp]
\centerline{\includegraphics[width=\textwidth]{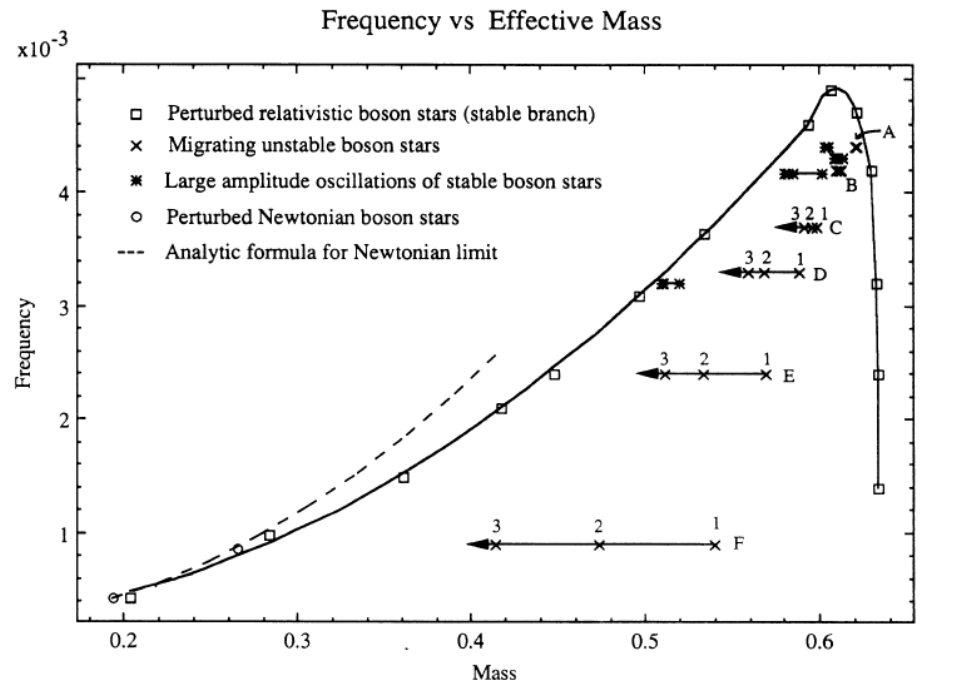}}
\caption{Oscillation frequencies of various boson stars are plotted
  against their mass. Also shown are the oscillation frequencies of
  unstable BSs obtained from the fully nonlinear evolution of the
  dynamical system. Unstable BSs are observed maintaining a constant
  frequency as they approach a stable star configuration. Reproduced
  with permission from \cite{1990PhRvD..42..384S}, copyright by APS.}
\label{fig:freq_mass}
\end{figure}

The perturbed configurations of the stable branch may oscillate and emit scalar 
radiation maintaining a characteristic frequency $\nu$, eventually settling into some other
stable state with less mass than the original. This characteristic frequency can be
approximated in the non-relativistic limit as \citep{1990PhRvD..42..384S}
\begin{equation}
    \nu = \frac{\pi}{4 m R^2} - \frac{m G M}{2 \pi R} \,,
\end{equation}
where $R$ is the effective radius of the star and $M$ its total mass. Scalar
radiation is the only damping mechanism available because spherical symmetry does
not allow for gravitational radiation and because the Klein--Gordon equation has no viscous
or dissipative terms. This process was named
\emph{gravitational cooling}, and it is extremely important in the context of
formation of compact bosonic objects \citealt{1994PhRvL..72.2516S} (see below).
The behavior of perturbed solutions can be represented on
a plot of frequency versus effective mass as in Fig.~\ref{fig:freq_mass}.
Perturbed stars will oscillate with a frequency below its corresponding solid line and
they radiate scalar field to infinity. As they do so, they lose mass by oscillating at constant
frequency, moving leftward on the plot until they settle on the stable branch of
(unperturbed) solutions. 

The perturbed unstable configurations will either collapse to a black hole or migrate to
a stable configuration, depending on the nature of the initial perturbation. If the density
of the star is increased, it will collapse to a black hole. On the other hand, if it is
decreased, the star explodes, expanding quickly as it approaches the stable 
branch.
Along with the expansion, energy in the form of scalar field is radiated away, leaving
a very perturbed stable star, less massive than the original unstable one.

\begin{figure}[htb]
\centerline{\includegraphics[width=\textwidth]{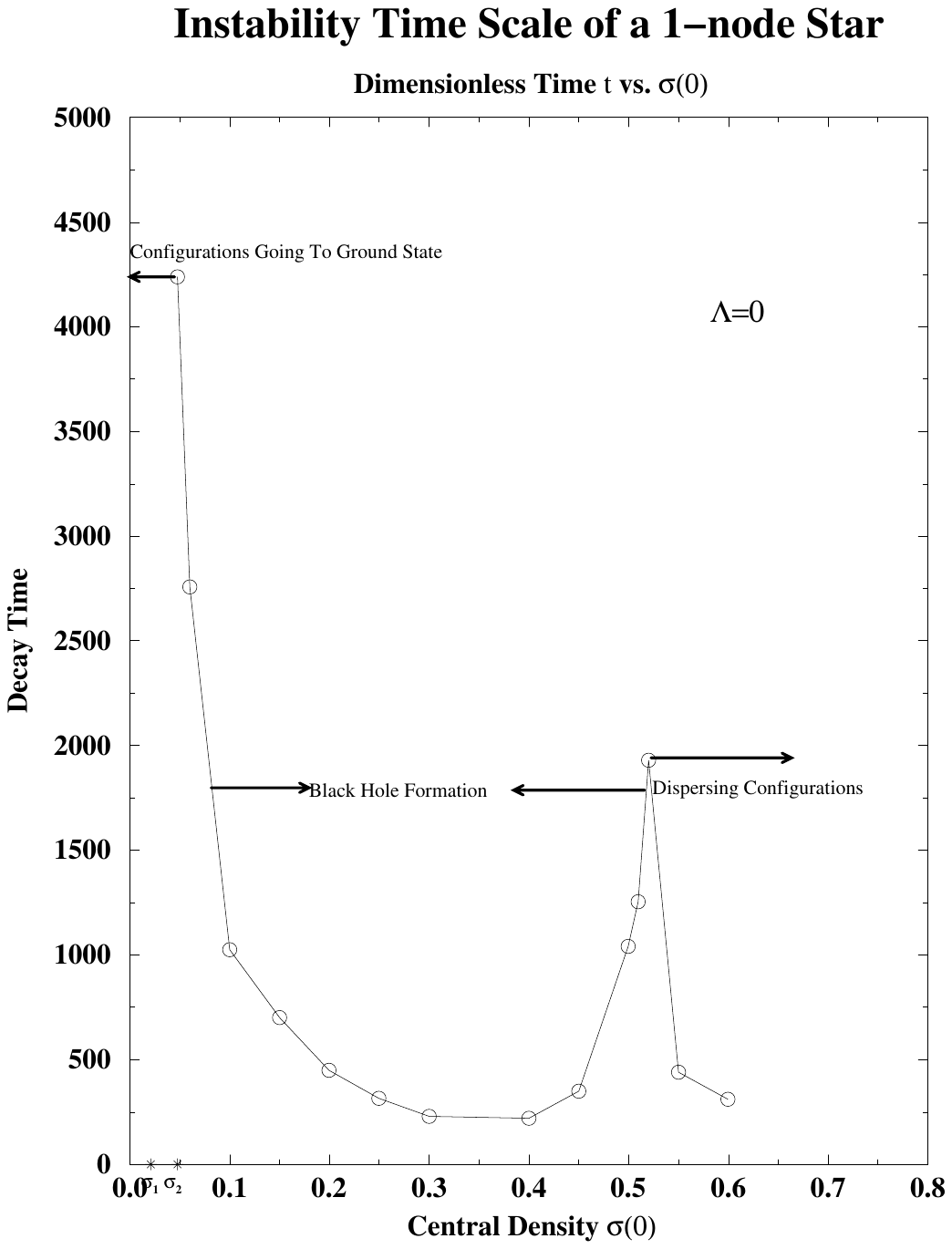}}
\caption{The instability time scale of an excited boson star (the
  first excitation) to one of three end states: (i)~decay to the
  ground state, (ii)~collapse to a black hole, or
  (iii)~dispersal. Reproduced with permission
  from \cite{1998PhRvD..58j4004B}, copyright by APS.}
\label{fig:timescale_decay}
\end{figure}

This analysis was extended to boson stars with self-interaction and to
excited BSs in \cite{1998PhRvD..58j4004B}, showing that both branches of
the excited states were intrinsically unstable under generic perturbations that
do not preserve $M$ and $N$.
The low density excited stars, with masses close to the ground state configurations,
will evolve to ground state boson stars when perturbed. The more massive configurations
form a black hole if the binding energy $E_B = M - N m$ is negative, through a cascade
of intermediate states. The kinetic energy of the stars increases as the configuration
gets closer to $E_B=0$, so that for positive binding energies there is an excess of 
kinetic energy that tends to disperse the bosons to infinity. These results
are summarized in Fig.~\ref{fig:timescale_decay}, which shows the time scale
of the excited star to decay to one of these states.

More recently, the stability of the ground state was revisited with 3D simulations using
a Cartesian grid \citep{2004PhRvD..70d4033G}. The Einstein equations were written in terms
of the BSSN formulation \citep{PhysRevD.52.5428,1999PhRvD..59b4007B}, which is one of the most
commonly used formulations in numerical relativity. Intrinsic numerical error from discretization
served to perturb the ground state for both stable and unstable stars.
It was found that unstable stars with negative binding energy would
collapse and form a black hole, while ones with positive binding energy
would suffer an excess of kinetic energy and disperse to infinity.

That these unstable stars would disperse, instead of simply expanding into some less compact
stable solution, disagrees with
the previous results of \cite{1990PhRvD..42..384S}, and was subsequently further
analyzed in \cite{guzman2009} in spherical symmetry with an explicit
perturbation (i.e., a Gaussian shell of particles, which increases the mass
of the star around 0.1\%). The spherically symmetric results corroborated the previous
3D calculations, suggesting that the slightly perturbed configurations of the unstable branch
have three possible endstates: (i) collapse to BH, (ii) migration to a less dense stable
solution, or (iii) dispersal to infinity, dependent on the sign of the binding energy.

\sllnew{
Recently, \cite{Kain:2021rmk} numerically evolves boson stars and finds the results
agree with a linear perturbation results.
Although excited boson stars are generally unstable,
evolutions in \cite{Sanchis-Gual:2021phr} show that
self-interactions can stabilize excited boson stars. The stabilizing effect of
interactions was previously found for certain rotating solutions \citep{Siemonsen:2020hcg}.
}

\begin{figure}[htb]
\centerline{\includegraphics[width=8.0cm]{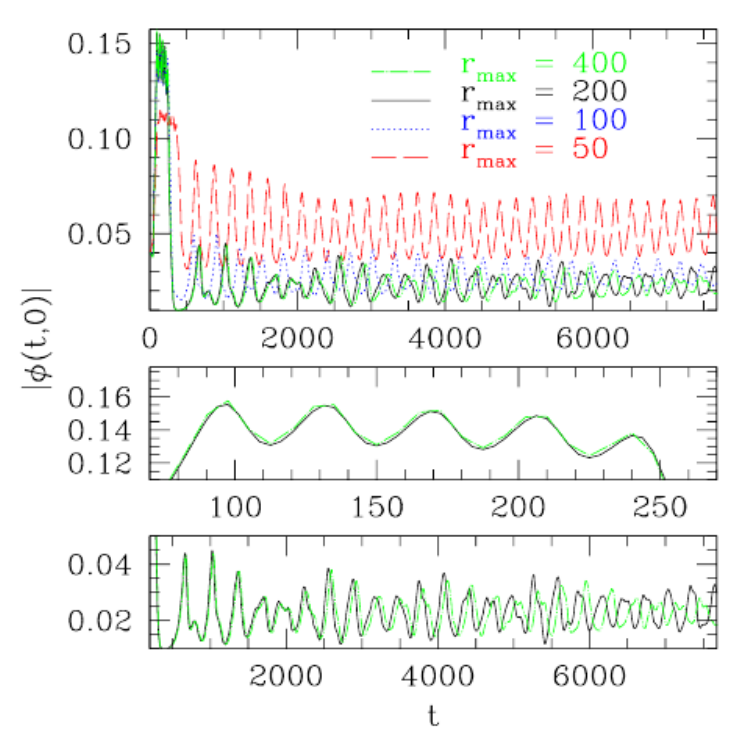}}
\caption{Very long evolutions of a perturbed, slightly sub-critical,
  boson star with differing outer boundaries. The central magnitude of
  the scalar field is shown. 
At early times ($t<250$ and the middle frame), the boson star
demonstrates near-critical behavior with small-amplitude oscillations
about an unstable solution. For late times ($t>250$), the solution
appears converged for the largest two outer boundaries and suggests
that sub-critical boson stars are \emph{not} dispersing. Instead, they
execute large amplitude oscillations about low-density boson
stars. Reproduced with permission from \cite{Lai:2007tj}.}
\label{fig:lai}
\end{figure}

Closely related is the work of \cite{Lai:2007tj} studying BS
critical behavior (discussed in Sect.~\ref{sub:critical}). They tune perturbations of
boson stars so that dynamically the solution approaches some particular unstable 
solution for some finite time. They then study evolutions that ultimately do not collapse to BH, so-called
sub-critical solutions, and find that they do not disperse to infinity, instead oscillating about some
less compact, stable star.
They show results with increasingly distant outer boundary that suggest that this behavior is not a finite-boundary-related
effect (reproduced in Fig.~\ref{fig:lai}).
They use a different form
of perturbation than \cite{guzman2009}, and, being only slightly subcritical, may be working in a regime with non-positive binding
energy. However, it is interesting to consider that if indeed there are three distinct end-states, then
one might expect critical behavior in the transition among the different pairings.
Non-spherical perturbations of boson stars have been studied numerically in \cite{2006gr.qc.....2078B}
with a 3D code to analyze the emitted gravitational waves.

The dynamics of non-standard boson stars have also been studied through numerical simulations in different scenarios. Boson stars in scalar-tensor theories of gravity were considered in \cite{2012PhRvD..86j4044R}, focusing on the study of spontaneous and induced scalarization. Evolutions of fermion-boson stars have confirmed their stability properties and have found the normal modes of oscillations of neutron stars with a dark matter component \citep{2013PhRvD..87h4040V}.
\CP{These studies have been extended by including more realistic equations of state to model the neutron star matter \citep{Nyhan:2022pda}, or more complicated self-interaction potentials for the scalar field \citep{Valdez-Alvarado:2020vqa}.
}

\CPi{The stability of excited fermion-boson stars, characterized by the presence of at least one node in the radial profile of the scalar field, is studied numerically in \cite{DiGiovanni:2021vlu}. Similar examples of this stabilization mechanism have been found in multi-field boson stars (Sect.~\ref{sec:multi}), suggesting that the  mechanism is a purely gravitational effect and does not depend on the type of matter of the companion star.}
\CP{
In \cite{DiGiovanni:2022mkn} the authors also explored the effect of bosonic fields (i.e., either scalar or vector) on unstable, differentially rotating neutron stars subject to the bar-mode instability. They found for a region of the parameter space that the presence of  dark-matter accretion in neutron stars could change the frequency of the expected gravitational wave emission associated with the bar-mode instability.
}

More recently, spherical Proca stars (see Sect.~\ref{sec:proca} for a discussion of such stars) have also been studied numerically \citep{Sanchis-Gual:2017bhw},
confirming that the evolutions of unstable solutions lead to outcomes analogous to those of boson stars (i.e., migration to the stable branch, total dispersion of the scalar field, or  collapse to a black hole).

\begin{figure}[htb]
\centerline{
  \includegraphics[width=6.0cm]{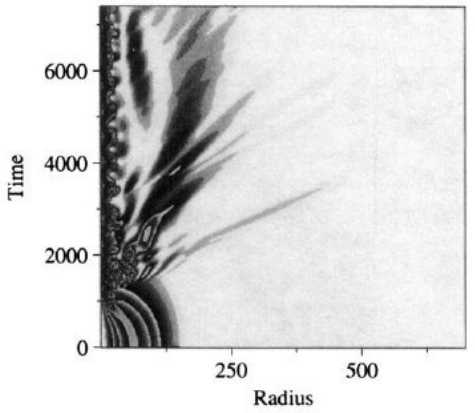}\qquad
  \includegraphics[width=6.0cm]{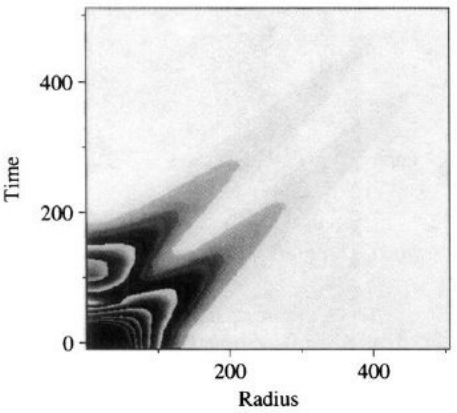}
}
\caption{The evolution of $r^2 \rho$ (where $\rho$ is the energy
  density of the complex scalar field) with massive field~(left) and
  massless~(right). In the massive case, much of the scalar field
  collapses and a perturbed boson star is formed at the center,
  settling down by gravitational cooling. In the massless case, the
  scalar field bounces through the origin and then disperses without
  forming any compact object. Reproduced with permission
  from \cite{1994PhRvL..72.2516S}, copyright by APS.}
\label{fig:formation}
\end{figure}

The issue of \textbf{formation} of boson stars has been addressed in \cite{1994PhRvL..72.2516S}
by performing numerical evolutions of the EKG system with different initial Gaussian
distributions describing unbound states (i.e., the kinetic energy is larger than the potential
energy). Quite independent of the initial condition, the scalar field
collapses and settles down to a bound state by ejecting some of the
scalar energy during each bounce.
The ejected scalar field carries away excess and ever-decreasing amounts of kinetic energy, 
as the system becomes bounded.
After a few free-fall times of the initial configuration, the scalar
field has settled into a perturbed boson star on the stable branch. This process is
the already mentioned \emph{gravitational cooling}, and allows for the formation of compact soliton stars 
(boson stars for complex scalar fields and oscillatons for real scalar fields).
Although these evolutions assumed spherical symmetry, which does not include important processes
such as fragmentation or the formation of pancakes, they demonstrate the feasibility of the
formation mechanism; clouds of scalar field will collapse under their own self-gravity while
shedding excess kinetic energy.
The results also confirm the importance of the mass term in the potential.
By removing the massive term in the simulations, the field collapses, rebounds and completely
disperses to infinity, and no compact object forms. 
The evolution of the scalar field with and without the massive term
is displayed in Fig.~\ref{fig:formation}.

\CP{An analogous gravitational cooling process has been found in simulations modeling the dynamical formation of other types of boson stars, such as Proca-stars \citep{DiGiovanni:2018bvo}, or   even in mixed configurations, as fermion-boson stars \citep{DiGiovanni:2020frc}, where the scalar field condenses around an already existing neutron star.}

\sllnewer{
We also discuss a few recent 
papers \citep{Gorghetto:2022sue,Arvanitaki:2019rax,Levkov:2018kau}
on star formation in various
dark matter models at the end of Sect.~\ref{sec:darkmatter}.
}

\sllnewer{
As we discuss in the next section the stability of rotating boson stars, it is important to note that,
even at the linear level, studies of slowly rotating neutron stars
suggest that rotation induces a coupling of axial and polar modes,
making a linear stability analysis extremely complicated \citep{2005IJMPD..14..543S}.
}

\subsubsection{Non-linear stability of  boson stars with angular momentum}
\label{sec:stability_w_angmom}

\CPi{Much less is known about rotating BSs, which are more difficult to construct and to evolve because they are usually not spherically symmetric (except in some specific cases, like the $\ell$-boson stars). The stability of rotating boson stars was previously thought to be similar to nonrotating stars by using catastrophe theory \citep{Kleihaus:2011sx}.  However, a number of dynamical studies of the merger of two boson stars failed to produce a rotating boson star as the remnant of the merger despite the initial data having angular momentum greater than what would be required for the remnant	to spin at the first level ($k=1$). These results suggested that perhaps rotating boson stars were unstable (or else that the formation of a rotating star dynamically may be difficult).
}
\sllnewer{
See, for instance, \cite{Palenzuela:2007dm,Mundim:2010hi,Bezares:2017mzk,Palenzuela:2017kcg,Bezares:2022obu} as discussed in 
Sect.~\ref{dynamics_binary_bs}.
}

\CPi{Even at the linear level, there is not expected to be a clean decoupling of the scalar and the gravitational modes,  making a linear stability analysis extremely complicated. Due to these difficulties, it seems more plausible to perform dynamical simulations of (possibly perturbed) rotating boson stars to get a better understanding of their stability properties. }

\CPi{The first of these evolved
spinning scalar and vector (i.e., Proca) stars finding that the boson stars are unstable to a non-axisymmetric instability~(NAI) whereas some Proca stars are stable \citep{Sanchis-Gual:2019ljs}. They also studied the formation of such stars via gravitational cooling in the presence of angular momentum.
Further work by many of the same authors
continued the study of stability of rotating stars, finding
that doubly wound $k=2$ Proca stars always decay to $k=1$ stars and that a quartic self-interaction delays the instability of rotating (scalar)  boson stars \citep{DiGiovanni:2020ror}.
}

\begin{figure}[htbp]
		\centerline{\includegraphics[width=10.0cm]{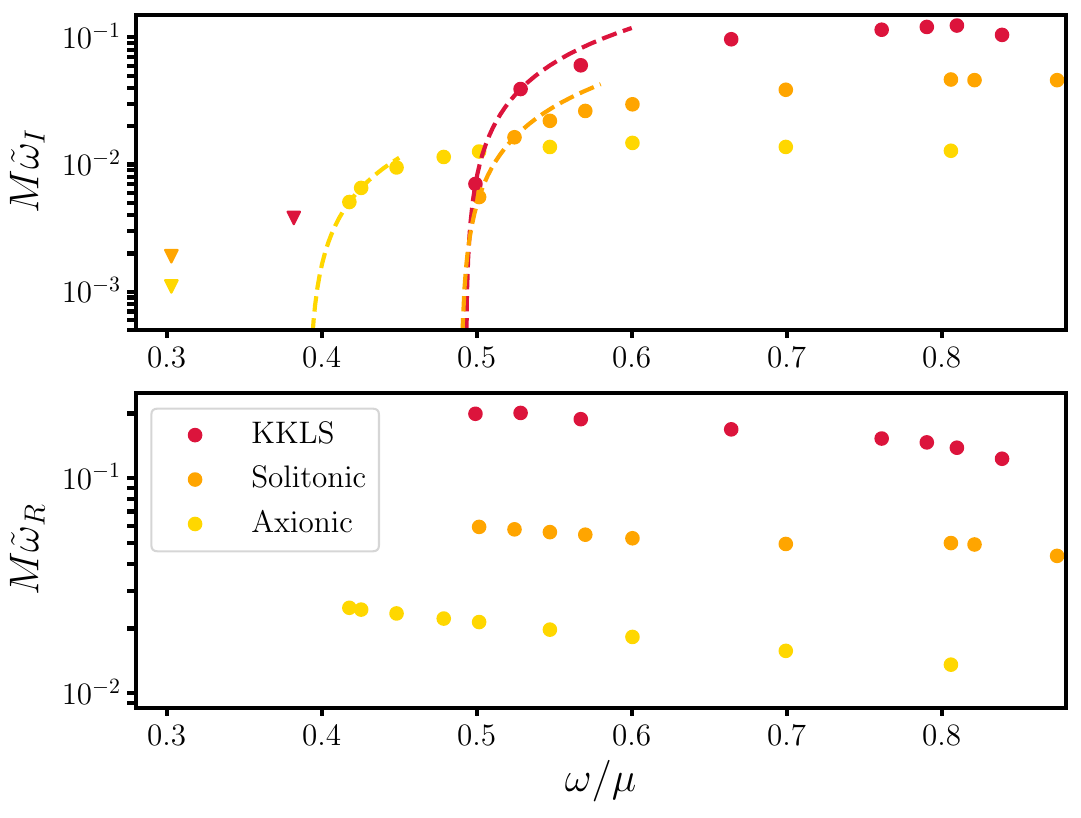}}
		\caption{Mode analysis of rotating BSs. The imaginary (top) and real (bottom) frequency components of the $m=2$ mode as a function of the BS frequency $\omega/\mu$ for families of rotating BS solutions with $k=1$ and three different potentials. The NAI is present for all the potential (i.e., $M \tilde{\omega}_I > 0$) above a critical frequency. 
			Reproduced with permission from \cite{Siemonsen:2020hcg}, copyright by APS.}
		\label{fig:rotatingBS_NAI}
\end{figure}

\CPi{
\cite{Siemonsen:2020hcg} clarified these issues by constructing and evolving a number of cases, finding that the non-axisymmetric instability of $k=1$ is always present for mini-boson stars (see Sect.~\ref{subsection:spherical}). Since all nonlinear scalar self-interactions reduce to the mass term for sufficiently small field values, this holds also for all scalar BS in the Newtonian limit. Interestingly, nonlinear interactions added to the scalar potential can quench the non-axisymmetric instability for compact solutions (relative to each specific potential)}.
%
%
\sllnew{The real and imaginary components of the frequencies of this $m=2$ mode,  displayed in Fig.~\ref{fig:rotatingBS_NAI} for a few representative self-interaction
potentials, demonstrate that there exists a
critical frequency at which the imaginary frequency changes sign, indicating a change in stability (i.e., the NAI shuts off for frequencies below the critical).}
        
\sllnewer{
Also relevant is the work of \cite{Dmitriev:2021utv}, which studies analytically the
stability of rotating boson stars in
the Newtonian limit. In particular, they find instability if self-interactions 
are either attractive or negligibly small, but they find stability of the $k=1$ solution 
with sufficiently strong repulsive self-interaction. These results are consistent
with the work of \cite{Siemonsen:2020hcg}.
}

\begin{figure}[htbp]
		\centerline{\includegraphics[width=10.0cm]{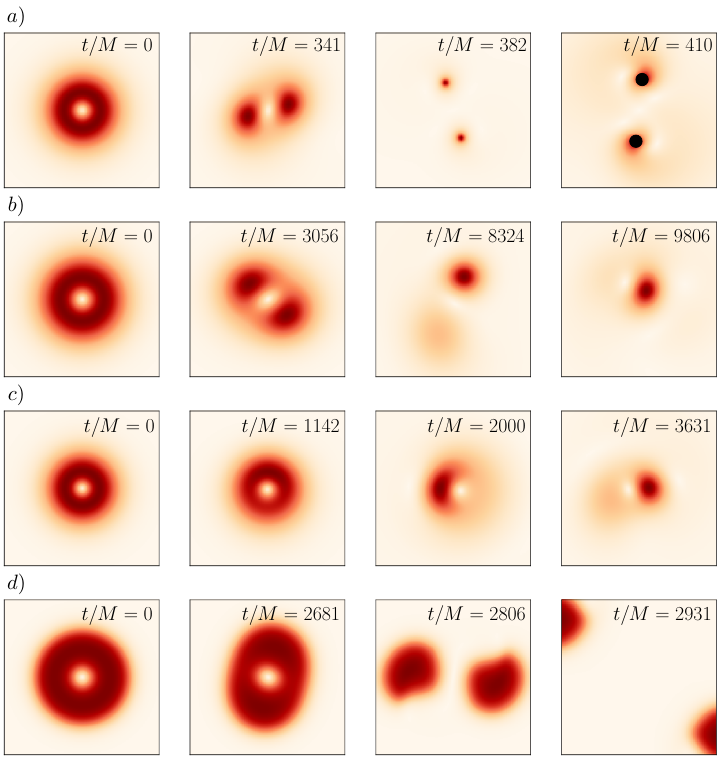}}
		\caption{Series of snapshots showing the evolution of $|\phi|^2$ in four scenarios where an unstable rotating BS with $k=1$  undergoes the NAI reaching different end states. From top to bottom: a)~a mini-BS 
		that collapses to a binary BH (the regions inside the apparent horizons are indicated in black), b)~a rotating mini-BS resulting in a non-rotating with non-negligible linear momentum, c)~a BS with strong self-interaction yielding a non-rotating BS with large linear momentum, and
		d)~an axion BS where the NAI results in the fragmentation of the star
		into two equal-mass non-rotating BSs. 
		Reproduced with permission from \cite{Siemonsen:2020hcg}, copyright by APS.}
		\label{fig:rotatingBS_NAI_evolution}
\end{figure}

\CPi{Finally, the NAI in rotating boson stars leads to a diverse range of dynamics, represented in Fig.~\ref{fig:rotatingBS_NAI_evolution}, including fragmentation into multiple unbound non-rotating stars, and formation of binary black holes.
}

\CPi{
	The $\ell$-boson stars have been evolved in spherical symmetry to study their stability \citep{Alcubierre:2019qnh}. They find results similar to standard $\ell=0$ boson stars with a change in stability occurring for solutions with maximum mass.
	A subsequent linear perturbation analysis of radial modes confirms the existence of both a stable and unstable branch \citep{Alcubierre:2021mvs}.
	Evolving $\ell$-boson stars beyond spherical symmetry with a full 3D code confirms
	this stability picture, but evidence for zero modes suggests that these solutions are
	part of a wider class of less symmetric solutions \citep{Jaramillo:2020rsv}.
}

\CPi{The formation, stability and final state of scalar clouds around black holes have been investigated through nonlinear numerical simulations in \cite{Okawa:2015fsa} }.

\subsection{Dynamics of binary boson stars}
\label{dynamics_binary_bs}
%

The dynamics of binary boson stars is sufficiently complicated that it generally
requires numerical solutions. The necessary lack of symmetry and the resolution
requirement dictated by the harmonic time dependence of the scalar field combine
so that significant computational resources must be expended for such a study.
However, boson stars serve as simple proxies for compact objects without the difficulties
(shocks and surfaces) associated with perfect fluid stars, and, as such, binary BS systems
have been studied in the two-body problem of general relativity. When sufficiently distant
from each other, the precise structure of the star should be irrelevant
as suggested by Damour's ``effacement theorem'' \citep{Hawking:1987en}.
\CP{According to this theorem, one could construct approximate initial data for a binary boson star system as a superposition of boosted single boson star solutions \citep{Bezares:2017mzk}. This simple recipe can be further improved by following the procedure described in \cite{Helfer:2021brt}. }

First attempts at binary boson-star simulations assumed the Newtonian limit,
since the SP system is simpler than the EKG one. Numerical evolutions of Newtonian
binaries showed that in head-on collisions with small velocities, the stars merge forming
a perturbed star \citep{1998PhDT........16C}. With larger velocities,
they demonstrate solitonic behavior by passing through each other, producing an
interference pattern during the interaction but roughly retaining 
their original shapes afterwards \citep{PhysRevA.66.063609}.
\cite{1998PhDT........16C} simulated
coalescing binaries, although the lack of resolution in 
these 3D simulations did not allow for strong conclusions. 

\begin{figure}[htb]
\centerline{\includegraphics[width=0.5\textwidth]{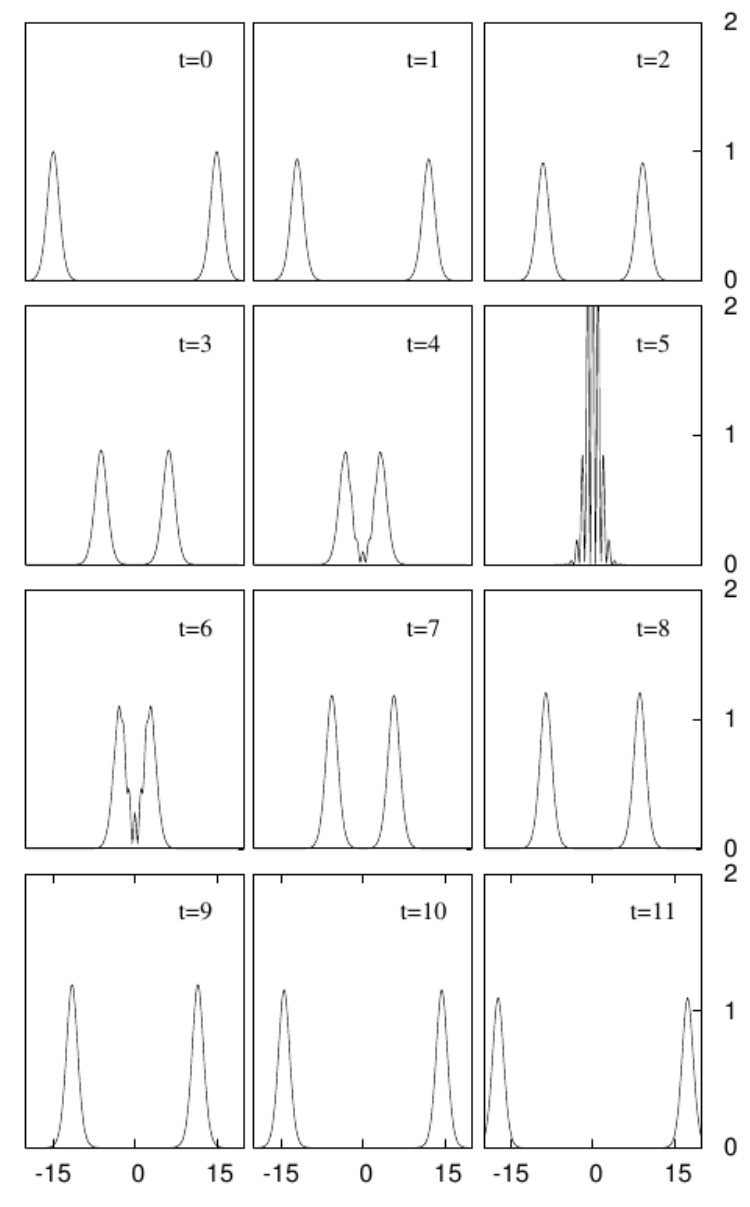}}

\caption{Collision of identical boson stars with large kinetic energy
  in the Newtonian limit. The total energy (i.e., the sum of
  kinetic, gravitational and self-interaction) is positive and the
  collision displays solitonic behavior. Contrast this with the
  gravity-dominated collision displayed in
  Fig.~\ref{fig:bin_new_eqmass_merge}. Reproduced with permission
  from \cite{2006PhRvD..74j3002B}, copyright by APS.}
\label{fig:bin_new_eqmass_soliton}
\end{figure}

\begin{figure}[htb]
\centerline{\includegraphics[width=0.5\textwidth]{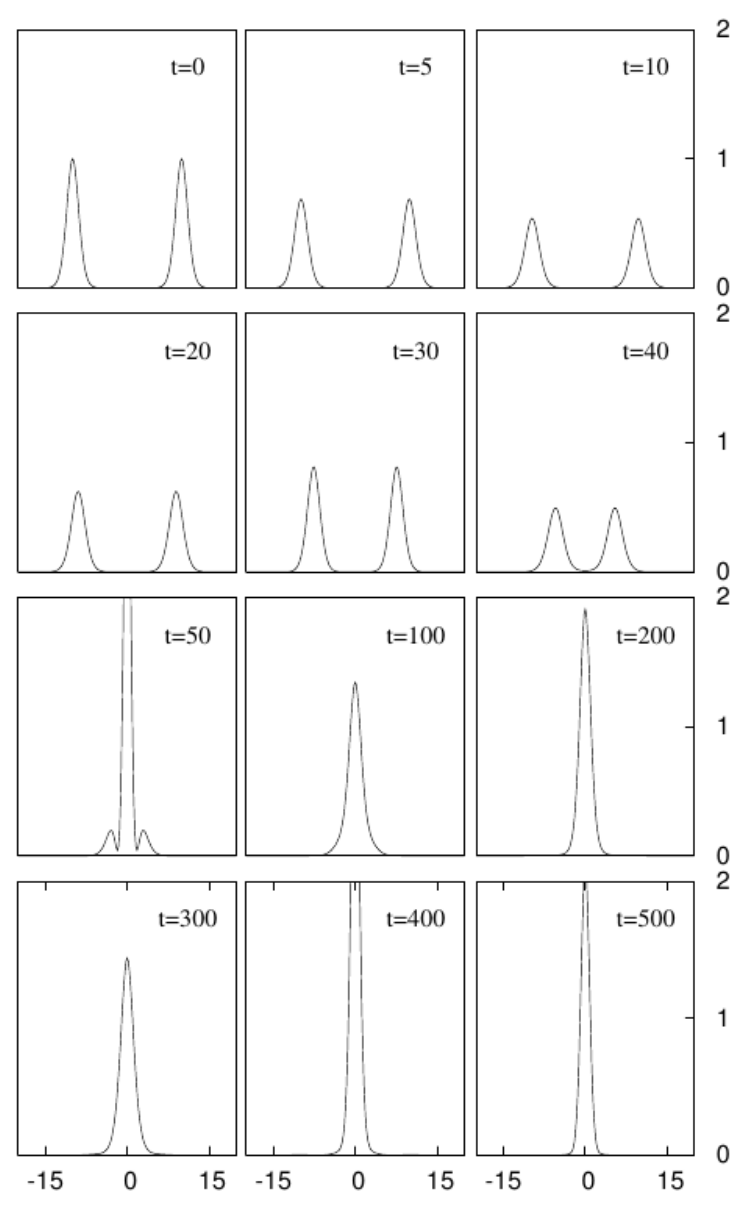}}
\caption{Collision of identical boson stars with small kinetic energy
  in the Newtonian limit. The total energy is dominated by the
  gravitational energy and is therefore negative. The collision leads
  to the formation of a single, gravitationally bound object,
  oscillating with large perturbations. This contrasts with the large
  kinetic energy case (and therefore positive total energy) displayed
  in Fig.~\ref{fig:bin_new_eqmass_soliton}. Reproduced with
  permission from \cite{2006PhRvD..74j3002B}, copyright by APS.}
\label{fig:bin_new_eqmass_merge}
\end{figure}

The head-on 
case was revisited in \cite{2006PhRvD..74j3002B} with a 2D axisymmetric code.
In particular, these evolutions show that the final state will depend on the total energy
of the system (e.g. the sum of kinetic, gravitational and self-interaction
energies). If the total energy is positive, the stars exhibit solitonic behavior
both for identical stars (see Fig.~\ref{fig:bin_new_eqmass_soliton}) and non-identical stars.
When the total energy is negative, the gravitational force is the main driver
of the dynamics of the system. This case produces a true collision, forming a single
object with large perturbations, which slowly decays by gravitational cooling, as 
displayed in Fig.~\ref{fig:bin_new_eqmass_merge}.

The first simulations of boson stars with full general relativity were reported
in \cite{1999PhDT........44B}, where the gravitational waves were computed
for a head-on collision. The general behavior is similar to the one
displayed for the Newtonian limit; the stars attract each other through
their gravitational interaction and then merge to produce
a largely perturbed boson star. However, in this case the merger of the
binary was promptly followed by collapse to a black hole, an outcome not possible 
when working within Newtonian gravity instead of general relativity.
Unfortunately, very little detail was given on the dynamics.

Much more elucidating was work in axisymmetry \citep{2005PhDT.........2L}, in which head-on collisions of identical boson
stars were studied in the context of critical collapse
(discussed in Sect.~\ref{sub:critical}) with general relativity. Stars with identical masses
of $M = 0.47 \approx 0.75\,M_{\max}$ were chosen, and so it is not surprising that for small
initial momenta the stars merged together to form an unstable single star
(i.e., its mass was larger than the maximum allowed mass, $M_{\max}$). The unstable
\emph{hypermassive} star subsequently collapsed
to a black hole. However, for large initial momentum the stars passed 
through each other, displaying a form of solitonic behavior since the individual
identities were recovered after the interaction. The stars showed
a particular interference pattern during the overlap, much like that
displayed in Figs.~\ref{fig:soliton} and \ref{fig:bin_new_eqmass_soliton}.

Another study considered the very high speed, head-on collision of BSs \citep{Choptuik:2009ww}.
Beginning with two identical boson stars boosted with Lorentz factors ranging as high
as 4, the stars generally demonstrate solitonic behavior upon collision,
as shown in the insets of Fig.~\ref{fig:hoop}. This work is further discussed in Sect.~\ref{hoop}.

The interaction of non-identical boson stars was studied in \cite{Palenzuela:2006wp}
using a 3D Cartesian code to simulate head-on collisions of stars
initially at rest. It was found that, for a given separation, the merger of
two stars would produce an unstable star that collapses to a black hole if the
initial individual mass were $M \ge 0.26 \approx 0.4\,M_{\max}$. For smaller masses,
the resulting star would avoid gravitational collapse and its features 
would strongly depend on the initial configuration.
The parameterization of the initial data
was written as a superposition of the single boson-star
solution $\phi_0(\mathbf{r})$, located at different positions $\mathbf{r_1}$ and $\mathbf{r_2}$
\begin{equation}
   \phi = {}^{(1)}\phi_0(\mathbf{r_1}) e^{i \omega t} 
        + {}^{(2)}\phi_0(\mathbf{r_2}) e^{i (\epsilon \omega t + \theta)}  . 
\label{eq:bsid}
\end{equation}
Many different initial configurations are possible with this parameterization.
The precise solution $\phi_0$ is unaffected by changing the direction of rotation (within 
the complex plane) via $\epsilon=\pm 1$ or by a phase shift $\theta$.

\begin{figure}[htbp]
\centerline{\includegraphics[width=8.0cm]{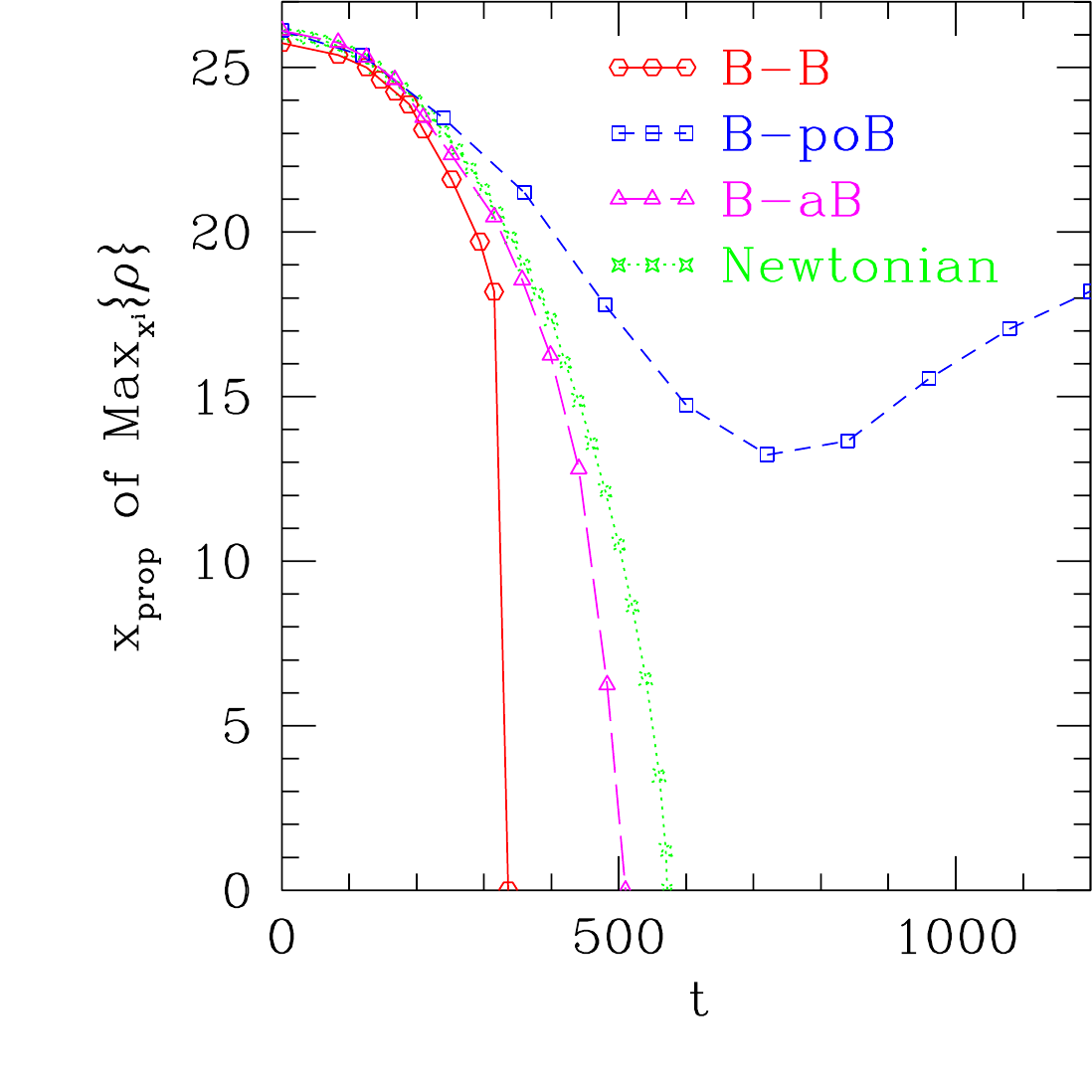}}
\caption{The position of the center of one BS in a head-on binary as a function of time
for (i) [B-B] identical BSs, (ii) [B-poB] opposite phase pair, and (iii) [B-aB] a boson--anti-boson pair. A simple argument is made which qualitatively matches these numerical results, as discussed in
Sect.~\ref{dynamics_binary_bs}. Also shown is the expected trajectory from a simple Newtonian two-body estimate.
Reproduced with permission from \cite{Palenzuela:2006wp}, copyright by APS.}
\label{fig:binary_headon}
\end{figure}

When $\epsilon=-1$, the Noether charge changes sign and the compact
object is then known as an \emph{anti-boson star}. Three particular binary cases were studied in detail:
  (i)~identical boson stars ($\epsilon=1$, $\theta=0$),
 (ii)~the pair in phase opposition ($\epsilon=1$, $\theta=\pi$), and
(iii)~a boson--anti-boson pair ($\epsilon=-1$, $\theta=0$).
The trajectories of the centers of the stars are
displayed in Fig.~\ref{fig:binary_headon}, together with a simple estimate of the expected
trajectory assuming Newtonian gravity. The figure makes  clear that the
merger depends strongly on the kind of
pair considered, that is, on the interaction between the scalar fields.

A simple energy argument is made in \cite{Palenzuela:2006wp} to understand the differing behavior.
In the weak gravity limit when the stars are well separated, one can consider the local energy
density between the two stars. In addition to the contribution due to each star separately, a
remaining term $\Delta$ results from the interaction of the two stars and it is precisely this term that
will depend on the parameters $\epsilon$ and $\theta$. This term takes the simple form
\begin{eqnarray}
   \Delta = \Delta_0\, \cos [ (1-\epsilon) \omega\, t - \theta ] \,,
\end{eqnarray}
where $\Delta_0$ is a positive definite quantity.
One then observes that the identical pair will have an increased energy density $\Delta=+\Delta_0$
resulting in a deeper (and more attractive) gravitational well between the stars.
In contrast, the pair with opposite phases has a decreased energy density $\Delta=-\Delta_0$
between them, resulting in a gravitational well less attractive than the area surrounding it.
This less attractive well results in an effective repulsion relative to the identical pairing.
The boson--anti-boson pair has an interaction that
is harmonic in time $\Delta=\Delta_0 \cos \left(2\omega t\right)$ and therefore sometimes positive and sometimes negative. However, if the time scale of interaction is not particularly fast, then
the interaction averages to zero. Note that the boson--anti-boson pair trajectory is the closest
to the simple Newtonian estimate.
The qualitative behavior
agrees very well with the numerical results.

The orbital case was later studied in \cite{Palenzuela:2007dm}. This case
is much more involved both from the computational point of view (i.e., there is less
symmetry in the problem) and from the theoretical point of view, since for the
final object to settle into a stationary, rotating boson star it must satisfy
the additional quantization condition 
for the angular momentum of Eq.~(\ref{quantization_J}).

One simulation consisted
of an identical pair each with individual mass $M=0.5$, with small orbital 
angular momentum such that $J \le N$. In this case, the binary merges forming a rotating bar that 
oscillates for some time before ultimately splitting apart. This can be considered
as a scattered interaction, which could not settle down to a stable boson star
unless all the angular momentum was radiated.

\begin{figure}[htbp]
\centerline{\includegraphics[width=8.0cm]{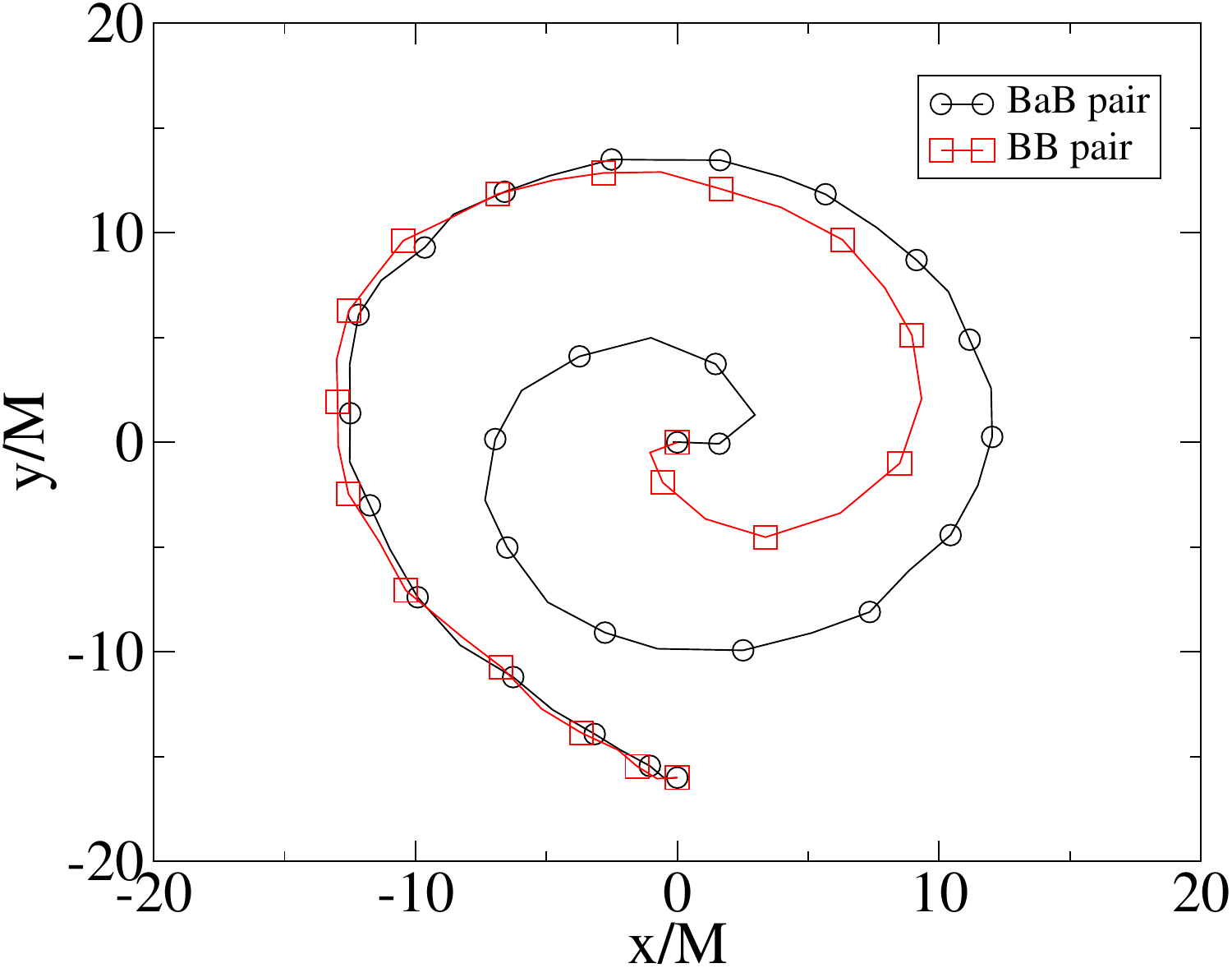}}
\caption{The position of the center of one BS within an orbiting
  binary as a function of time for the two cases: (i)~[B-B] identical
  BSs and (ii)~[B-poB] opposite phase pair. Notice that the orbits are
  essentially identical at early times (and large separations), but
  that they start to deviate from each other on closer approach. This
  is consistent with the internal structure of each member of the
  binary being irrelevant at large separations. Reproduced with
  permission from \cite{Palenzuela:2007dm}, copyright by APS.}
\label{fig:binary_inpiral}
\end{figure}

In the case of boson--anti-boson
pair, the total Noether charge is already trivial, and the final object
resembles the structure of a rotating dipole. The pair in opposition of phase
was not considered because of the repulsive effect from the interaction.
The cases with very small angular momentum $J \ll N$ or with $J \le N$ 
collapsed to a black hole soon after the merger. The trajectories for this latter
case are displayed in Fig.~\ref{fig:binary_inpiral}, indicating that the internal structure 
of the star is irrelevant (as per the effacement theorem \citealt{Hawking:1987en})
until the scalar fields overlap.

Other simulations of orbiting, identical binaries have been performed
within the conformally flat approximation instead of full GR, which neglects 
gravitational waves~(GW) \citep{Mundim:2010hi}.
Three different qualitative behaviours were found. For high
angular momentum, the stars orbit for comparatively long times around each other. For intermediate
values, the stars merged and formed a pulsating and rotating boson star. For low angular momentum,
the merger produces a black hole. No evidence was found of the stars splitting apart after the merger.

Three dimensional simulations of solitonic core mergers colliding two or more
 boson stars in the Newtonian limit (\schrodinger-Poisson) are studied in the context of
dark matter with
 different mass ratios, phases and orbital angular momentum \citep{Schwabe:2016rze}. The final core mass does not
depend strongly on the phase difference nor on the angular momentum.
\cite{Cotner:2016aaq} also studies
collisions within the \schrodinger--Poisson system and discusses implications for dark matter.
However, this work focuses on the head-on case and includes
effects of different mass ratios, relative phases, self-couplings, and separation distances.
Interestingly, analytic estimates are compared to the numerical simulations \citep{Cotner:2016aaq}.

The dynamics of particularly compact boson stars %
are interesting to contrast with the dynamics of black holes because, at least in part,
we now have observations of the gravitational waves from binary BH mergers (discussed more in Sect.~\ref{sec:gravitationalwaves}).
 To this end, the study
of the head-on collision of solitonic boson  stars (which can be quite compact) \citep{Cardoso:2016oxy}
found
the dynamics to be qualitatively similar to those observed previously with mini-boson stars \citep{Palenzuela:2006wp}.
However, the gravitational waves emitted displayed significant differences and, in some cases, closely resembled the signal
from a binary black hole merger.

\begin{figure}[htbp]
\centerline{\includegraphics[width=\textwidth]{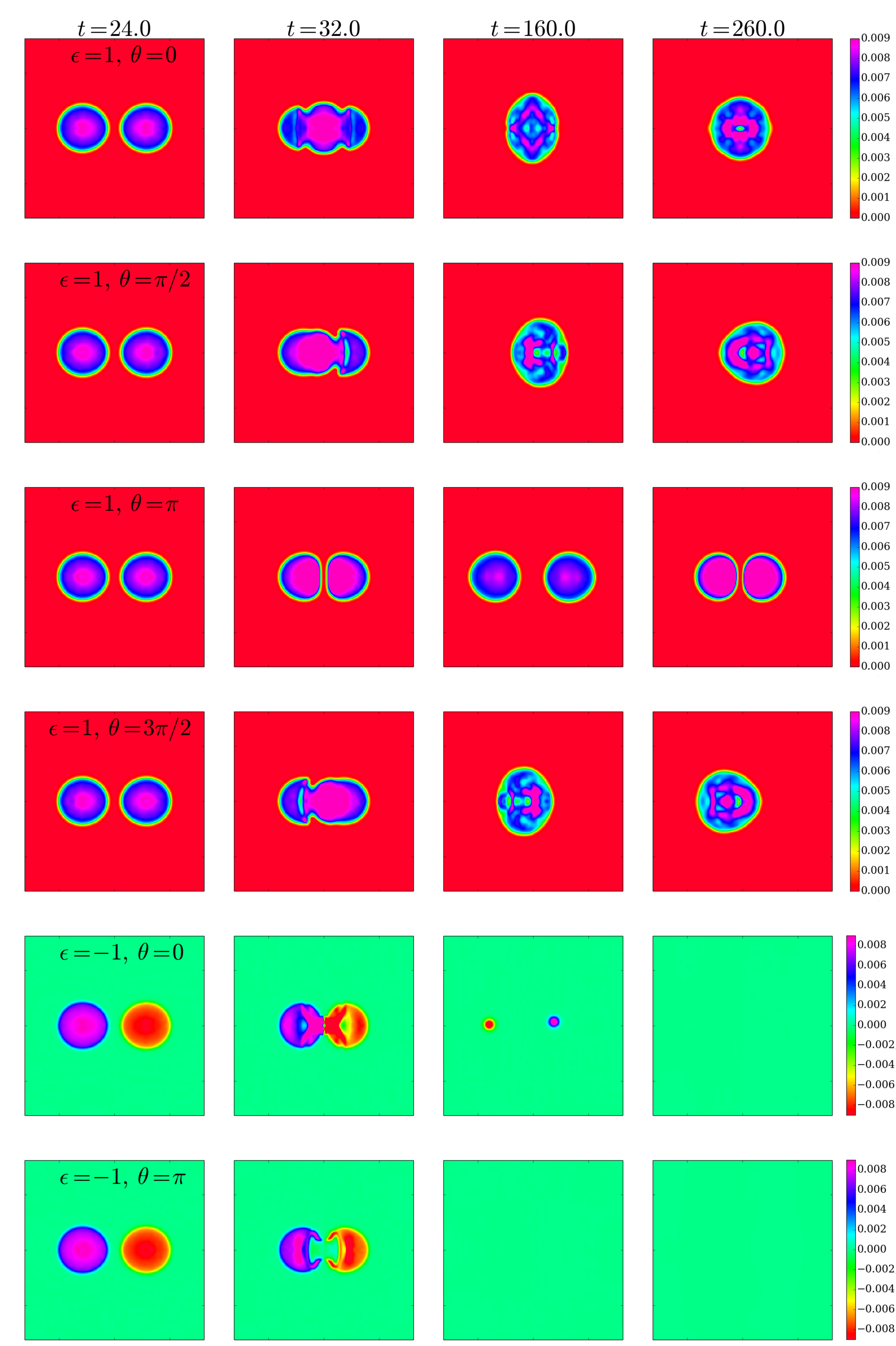}}
\caption{Snapshots in time of the Noether charge density in the $z=0$ plane for 
head-on binary collisions of compact solitonic boson stars. Each row corresponds 
to a  different boson-boson and boson-anti-boson case studied with a phase shift $\theta$
as described by Eq.~(\ref{eq:bsid}).
The collision of the stars occurs approximately at $t=28$. The result of the  
boson-boson merger is a single boson star except in the case with $\theta=\pi$. 
The stars in the boson-anti-boson case annihilate each other during the merger. 
Reproduced with permission from \cite{Bezares:2017mzk}, copyright by APS.}
\label{fig:bbs_headon}
\end{figure}

\begin{figure}[htb]
\centerline{\includegraphics[height=6.0cm,width=8.0cm]{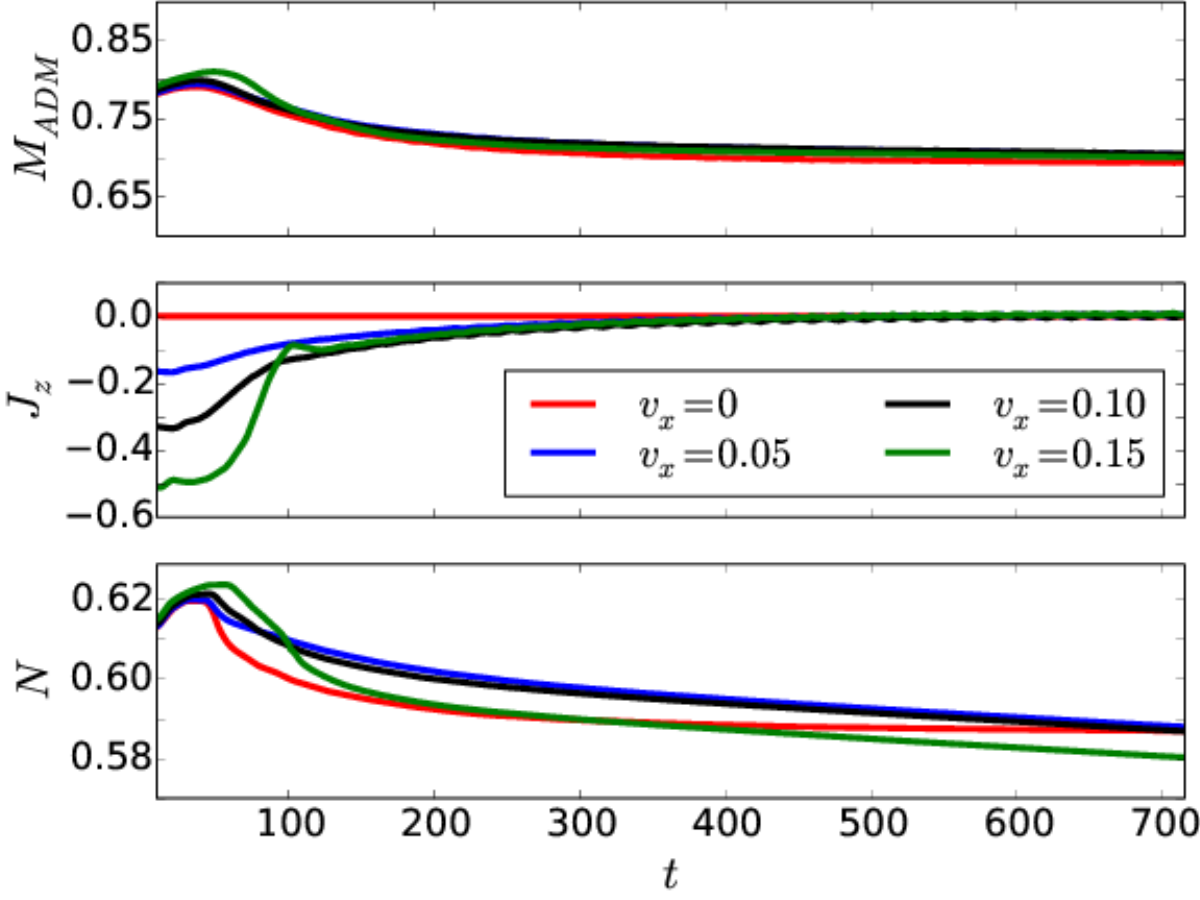}}
\caption{ADM mass (top panel), angular momentum $J_{z}$ (middle panel), and Noether charge (bottom panel) as functions of time for the orbital binary collisions of compact solitonic boson stars with different tangential boost velocities. During the coalescence, approximately 5\% of the mass and Noether charge is radiated, as well as most of the angular momentum.
Reproduced with permission from \cite{Bezares:2017mzk}, copyright by APS.}
\label{fig:bbs_orbital:MJN}
\end{figure}

These studies have been extended to the orbital case in \cite{Bezares:2017mzk}. Surprisingly, for stars not so massive as to collapse promptly, the merger %
does not lead to a rotating boson star but instead to a non-rotating perturbed BS (snapshots of some of these simulations are shown in Fig.~\ref{fig:bbs_headon}).
As apparent in  Fig.~\ref{fig:bbs_orbital:MJN}, the system radiates most of its angular momentum via scalar radiation and gravitational waves soon after the merger.

\begin{figure}[htbp]
\centerline{\includegraphics[width=\textwidth]{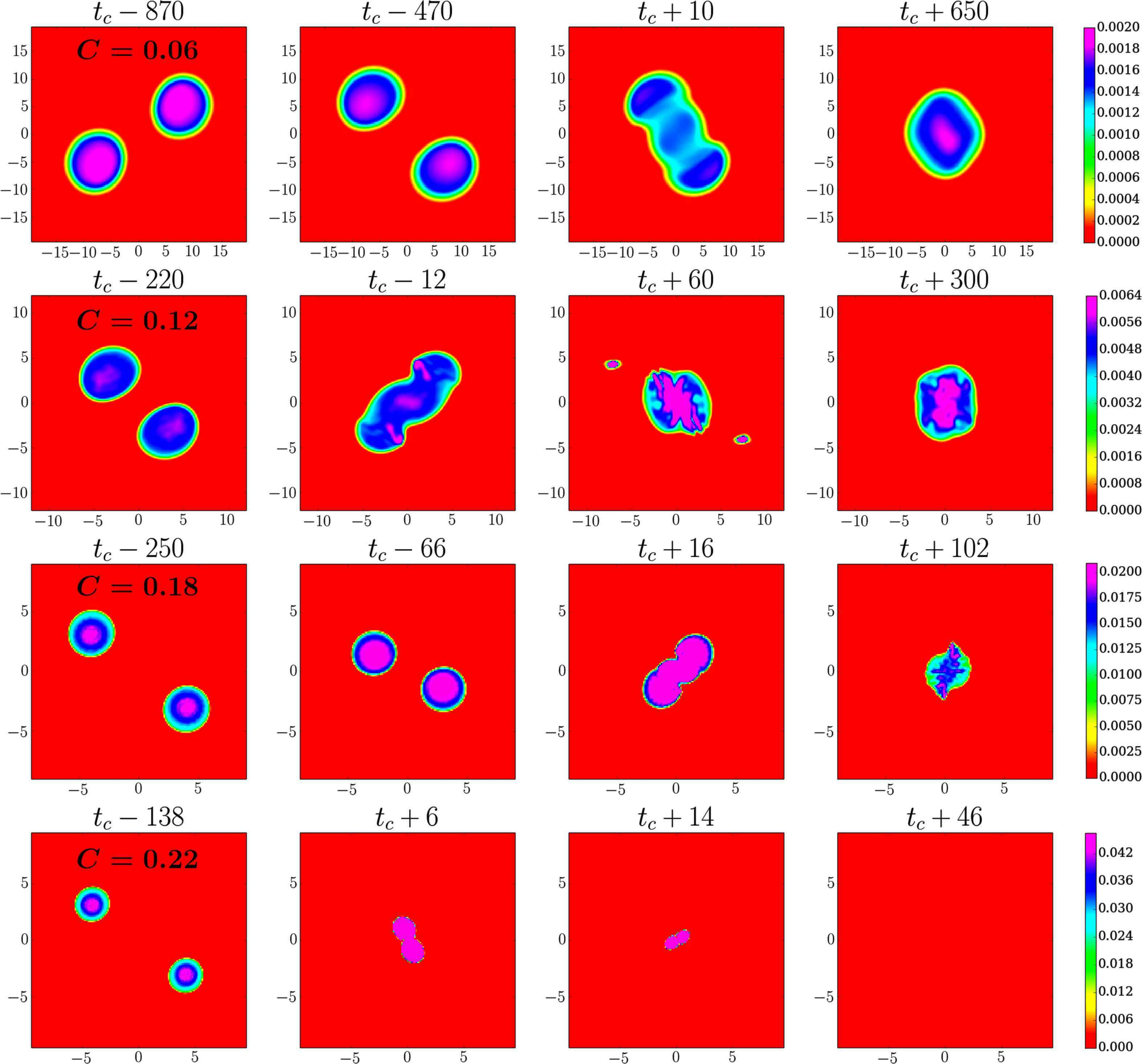}}
\caption{Snapshots from the mergers of four different, equal mass, solitonic BS binaries with increasing compactnesses (from top row to bottom, each BS of the binary has compactness 0.06, 0.12, 0.18, and 0.22). Shown is the Noether charge density on the orbital plane before and after the time at which the stars first make contact, $t_c$. The most compact case collapses to a BH while the $C=0.12$ case ejects two blobs of scalar field.
Reproduced with permission from \cite{Palenzuela:2017kcg}, copyright by the authors.}
\label{fig:q1bbs}
\end{figure}

\begin{figure}[htbp]
\centerline{\includegraphics[width=\textwidth]{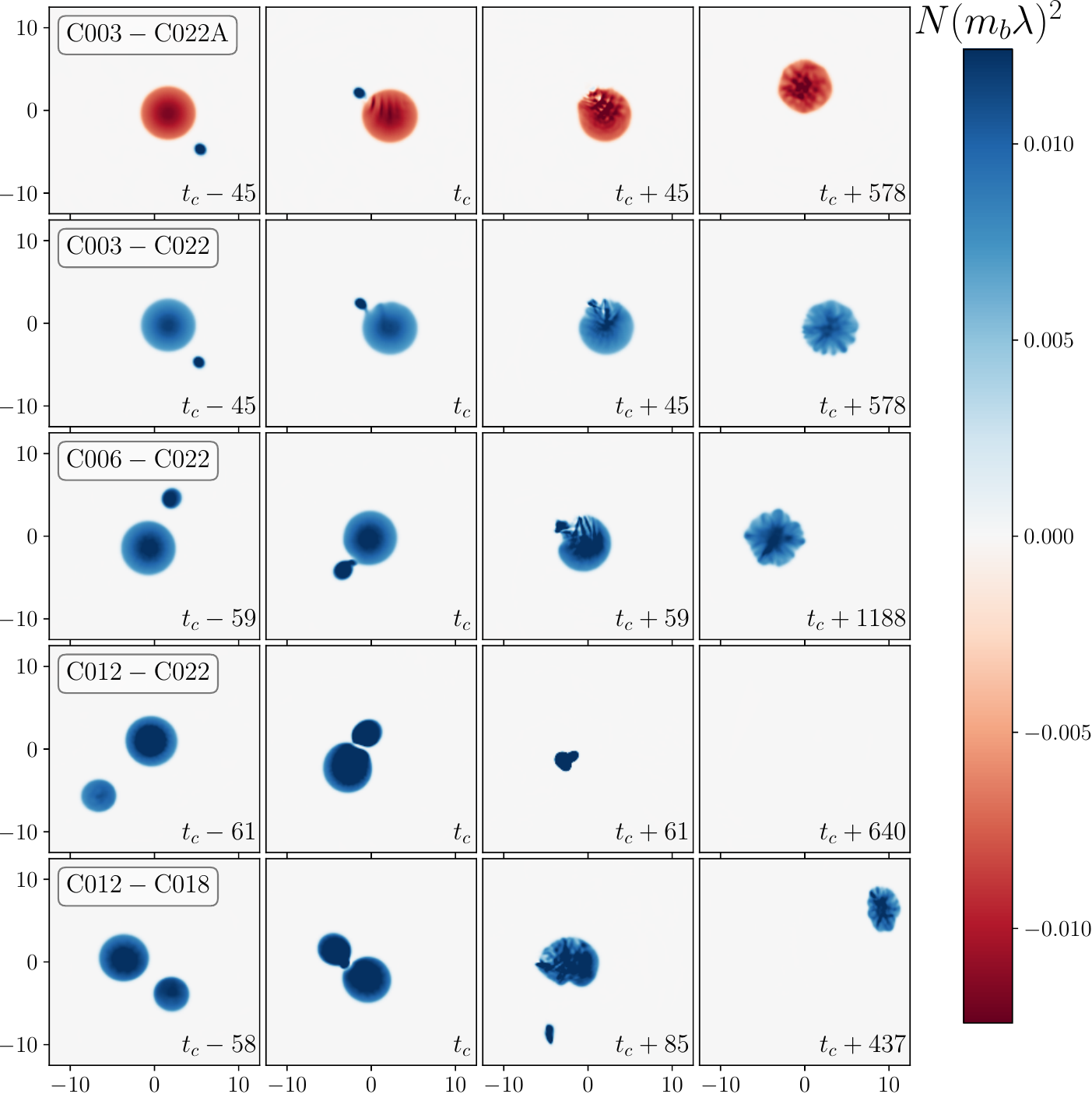}}
\caption{Snapshots from the mergers of four different, unequal mass, solitonic BS binaries (similar to Fig.~\ref{fig:q1bbs} which shows only equal mass binaries).
Displayed is the Noether charge density on the orbital plane before and after the contact time, $t_c$.
The top two rows consist of a star with compactness 0.03 paired with one with compactness 0.22, but the top row begins with the compact star having the opposite charge from its stellar pair. For this case, the small star is annihilated, leaving a less massive remnant. Continuing downward, the next row has stars with compactnesses 0.06 and 0.12, followed by 0.12 paired with 0.22 and with 0.18. Note that this final case ejects a single scalar blob in contrast to the pair of blobs ejected in the second row of Fig.~\ref{fig:q1bbs}.
Reproduced with permission from \cite{Bezares:2022obu}, copyright by APS.}
\label{fig:qne1bbs}
\end{figure}

\CP{Similar results were found in \cite{Palenzuela:2017kcg}, revealing that the remnant settles down to a non-rotating boson star, emitting significant gravitational radiation during this post-merger state} \CPn{ (see Fig.~\ref{fig:q1bbs}).}
\CP{The unequal mass case, with mass ratios up to $q=M_1/M_2=23$, was considered in \cite{Bezares:2022obu},}
\CPn{ and snapshots of these evolutions are shown Fig.~\ref{fig:qne1bbs}.}
\CP{Similar to the equal-mass case, the merger produces either a non-spinning boson star or a spinning black hole, depending on the initial masses and on the binary angular momentum. Interestingly, in contrast to the equal-mass case, one of the mechanisms to dissipate angular momentum is now asymmetric (i.e., the ejection of a scalar field \emph{blob}), and leads to large kick velocities which could produce wandering remnant boson stars \citep{Bezares:2022obu}.}

\begin{figure}[htbp]
		\centerline{\includegraphics[width=\textwidth]{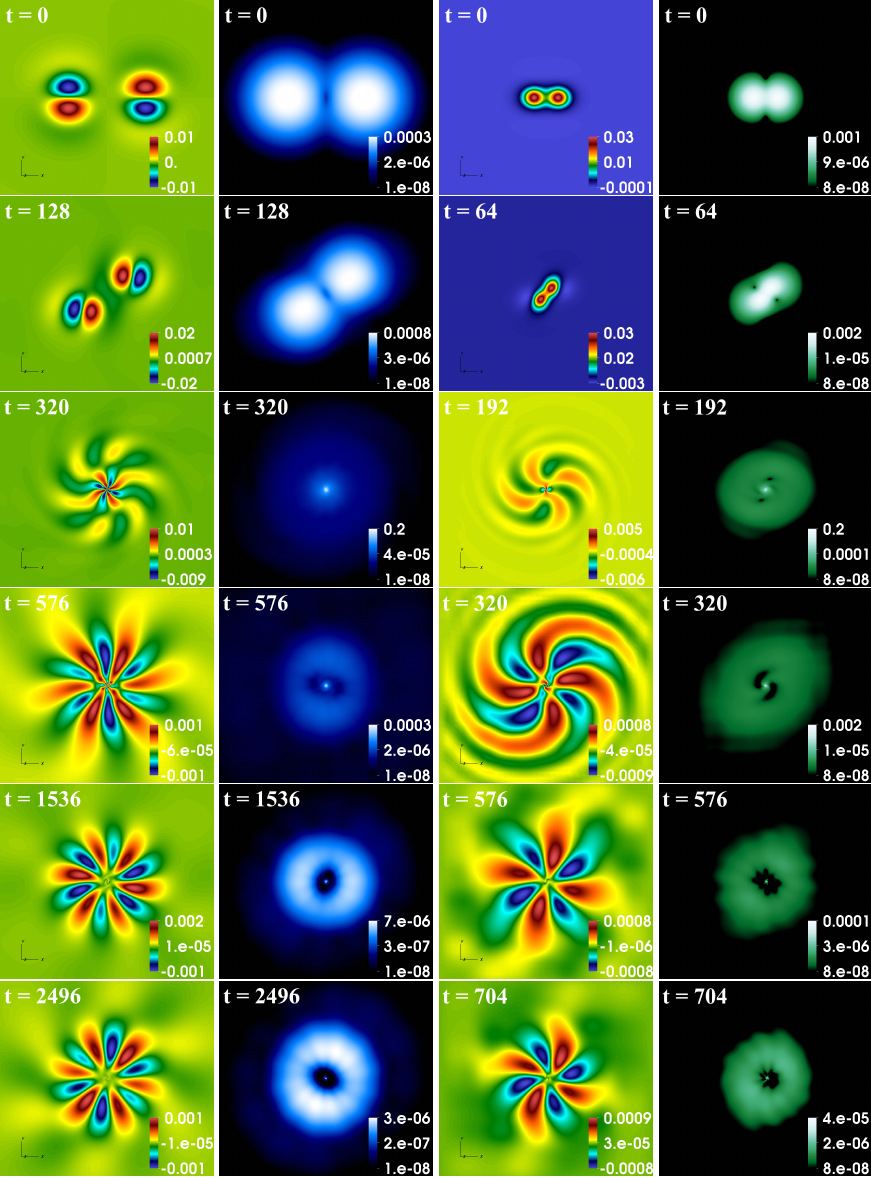}}
		\caption{Mergers of orbiting, non-spinning: Proca stars (left two columns) and scalar boson stars (right two columns).
			Snapshots along the equator show: (leftmost) the  real part of the scalar Proca potential, $\mathcal{X}_{\phi}$, (middle left) Proca energy density, (middle right) the real part of the scalar field, $\phi$, (rightmost) the scalar energy density.
			These evolutions suggest that mergers result in a BH synchronized with an $m=6$ Proca cloud or an $m=4$ scalar cloud, respectively.
			Reproduced with permission from \cite{Sanchis-Gual:2020mzb}, copyright by APS.}
		\label{fig:atoms}
\end{figure}

\CP{Similar head-on and orbital simulations have been performed for equal mass Proca-stars, showing behavior analogous to boson stars \citep{PhysRevD.99.024017}. In the head-on case, the collisions of Proca stars with sufficiently small compactness form a stable Proca star remnant, whereas the merger of more compact Proca stars eventually forms a black hole. For binaries with orbital angular momentum, the merger of highly compact stars forms a Kerr black hole surrounded by a transient Proca field remnant. For low compactness, the binary forms a massive remnant with angular momentum which decays to a non-rotating Proca star. Interestingly, some of the mergers of orbiting boson stars lead to the formation of so-called \emph{synchronized gravitational atoms} \citep{Sanchis-Gual:2020mzb}, rotating black holes surrounded by stationary bosonic clouds which were not found in previous works. After the formation of a horizon, the BH spins up by accreting the bosonic field until the remnant reaches a stationary state. However, in order to spin up to synchronization, fine tuning of
the initial data is required: the synchronized gravitational atom will not be formed if either there is not enough
angular momentum available for the black hole to spin as fast as the scalar field cloud, or there is too much, overshooting the angular velocity of the cloud \citep{Sanchis-Gual:2020mzb}.} 
\CPn{Snapshots of two particular mergers, one with Proca stars and the other with scalar BSs, are shown in
Fig.~\ref{fig:atoms} in which the azimuthal structure of the cloud is apparent.}

\CP{All the previous simulations assumed the scalar field to be in a coherent state, meaning that both stars are represented with the same scalar field. The incoherent case, where a different scalar field constitutes each star such that they interact with each other only through gravity, was considered in \cite{Bezares:2018qwa}.}
	
\CP{Besides collisions of boson and Proca-stars, other types of boson star mergers have been considered. For instance, the head-on collision of $\ell$-boson stars was
studied in \cite{Jaramillo:2022zwg}. Despite being spherically symmetric, $\ell$-boson stars have a (hidden) frame of reference, used in defining their individual multipolar fields. In addition to explorations with different angles between the axes of the two colliding stars, the authors also considered the coherent and incoherent cases, when both stars are made of either the same or different scalar fields. The simulations reproduce the generic features of boson star mergers: (i) the collision of two sufficiently massive stars leads to black hole formation, and (ii) below a certain mass threshold the remnant is a quasi-stationary bound state. However, this remnant generically deviates from spherical symmetry, and it seems to belong to the previously reported larger family of equilibrium multi-field boson stars of which $\ell$-boson stars are a symmetry enhanced point \citep{Sanchis-Gual:2021edp}.}

\CP{In addition to mergers of two boson stars, there have been attempts to study mixed binaries consisting of a boson star with another compact object. In \cite{Dietrich:2018bvi} they studied the merger of boson stars with a neutron star. They found that, depending on the
mass of the boson star, the merger remnant can either be a black hole or a neutron star
surrounded by a bosonic cloud.}

\CPn{Closely related are the investigations of the internal structure of neutron stars through the gravitational waves produced during their coalescence. The LIGO-Virgo-KAGRA facilities could  potentially observe the effects of dark matter particles trapped within their interior (i.e., dark matter cores) that, on long timescales, can condense to a boson star. Simulations of such binaries, modeled as two fermion-boson stars, might help constrain the amount of dark matter trapped in the interior of neutron stars \citep{Bezares:2019jcb}}.

\sllnew{
If dark matter is composed of a scalar field (see Sect.~\ref{sec:darkmatter}), then the interaction of a black hole
with a large boson star is interesting for its astrophysical implications.
}
\CP{The piercing of a large and heavy boson star by a black hole with one-tenth its mass was considered in \cite{Cardoso:2022vpj}. The simulations show that the black hole is slowed down by accretion and dynamical friction with the boson star, with a large fraction (i.e., usually more than 95\%) of the boson star material accreting onto it instead of dispersing to infinity, even when the black hole collides with a large velocity.
These studies have been extended by the same authors in \cite{Cardoso:2022nzc}, considering the accretion of a boson star onto a central black hole, representing a dark matter halo which hosts a parasitic supermassive black hole. Numerical simulations allowed them to provide a general expression for the lifetime of the boson star. Such lifetimes can be large enough to allow the dark matter halos to survive until the present time.
}

\clearpage
\section{Boson stars in astronomy}
\label{section:astro}

Scalar fields are often employed by astronomers and cosmologists in their efforts to
model the Universe. Most models of inflation adopt a scalar field as the \emph{inflaton} field,
the vacuum energy of which drives the exponential inflation of the Universe. Dark energy also 
motivates many scalar field models, such as \emph{k-essence} and \emph{phantom} energy models.
It is therefore not surprising that boson stars, as compact configurations of scalar field,
are called upon to provide consequences similar to those observed.

\subsection{As astrophysical stellar objects}
\label{section:astrostellar}

We have already discussed a number of similarities between boson stars and
models of neutron stars. Just as one can parameterize models of neutron stars
by their central densities, one can consider a 1-parameter family of boson stars
according to the central magnitude of the scalar field. 
The mass is then a function of this parameter, and one finds the existence of a local maximum across which
solutions transition from stable to unstable, just as is the case for neutron stars. Similarly, models of neutron
stars can be constructed with different equations of state, whereas boson stars
are constructed with differing scalar field potentials. 

One difference of consequence concerns the stellar surface. Neutron stars of course
have a surface at which the fluid density is discontinuous, as discussed for example
in \cite{Gundlach:2010gy,Gundlach:2009ft}. In contrast, the scalar field that constitutes
the boson star is smooth everywhere and lacks a particular surface. In its place, one
generally defines a radius that encompasses some percentage (e.g.\ 99\%) of the stellar mass.
Such a difference could have observational consequences when matter accretes onto either
type of star.

It is still an open question whether some of the stars already observed and interpreted as neutron
stars could instead be astrophysical boson stars. In a similar fashion, it is not known whether
many, if not all, of the stars we observe already have a bosonic component that has
settled into the gravitational well of the star (see
Sect.~\ref{subsection:varieties_fermionicbosonic} for a discussion of fermion-boson stars).
The bosonic contribution may arise from exotic matter which could appear at high densities
inside the neutron star or from some sort of dark matter accretion \citep{2014JCAP...05..013G}. This possibility has gained popularity recently
and there have been several attempts to constrain the properties of weakly
interacting dark matter particles~(WIMPs) by examining signatures related to their accretion and/or
annihilation inside stars (for instance, see \citealt{PhysRevD.82.063531} and works cited in the introduction).

\sllnew{In addition to the number of bosons, fermion-boson stars are also characterized by the number of fermions present. \cite{DiGiovanni:2021ejn} argue that this additional freedom over standard boson stars 
can mitigate disagreement between expected and observed masses and radii of neutron stars reported in recent multi-messenger observations and nuclear-physics experiments. For instance,  the LVK merger event
GW190814 reported a secondary mass of $M\approx 2.50$--$2.67\,M_{\odot}$, which would 
either be the most massive neutron star or lightest black hole yet observed.
Such a high mass could be explained if the neutron star contains a Bose--Einstein condensate in addition to the regular neutron star material \citep{DiGiovanni:2021ejn}.}

Recently, it was suggested that, due to the stronger gravitational field of neutron stars compared
to other stars such as white dwarfs and main sequence stars, WIMPs will
accrete more efficiently, leading to two different possibilities. If the dark matter is its
own antiparticle, it will self-annihilate and heat the neutron star. This temperature increase 
could be observable in old stars, especially if they are close to the galactic
center \citep{PhysRevD.82.063531,PhysRevD.81.123521}. If WIMPs do not self-annihilate, they will
settle in the center of the star forming a fermion-boson star 
(as discussed in Sect.~\ref{subsection:varieties_fermionicbosonic}). The accretion 
of dark matter would then increase the star's compactness 
until the star collapses \citep{PhysRevD.81.123521}
 (see discussion of BSs as a source of dark matter in Sect.~\ref{sec:darkmatter}).
\cite{2011PhRvD..84b4043N} follow such work by considering the result of a collision between
a BH and a boson star. In particular, they consider the problem as a perturbation of a black hole via
scalar accretion and analyze the resulting gravitational-wave output.

Because of the similarities between boson stars and neutron stars, one finds that boson
stars are often used in place of the other. This is especially so within numerical work
because boson stars are easier to evolve than neutron star models. One can, for example compare 
the gravitational-wave signature of a boson-star merger with that of more
conventional compact object binaries consisting of BHs and/or NSs.
Differentiating BSs from other compact objects with gravitational-wave observations
is discussed further in Sect.~\ref{sec:gravitationalwaves}.

With the continued advancement in observation, both in the
electromagnetic and gravitational spectra, perhaps soon we will have
evidence for these questions. At the same time, further study of boson
stars can help identify possible distinguishing observational effects
in these bands. One example where knowledge is lacking is the
interaction between boson stars with a magnetic field. Whereas a
neutron star can source its own magnetic field and a neutral star can
obtain an induced charge when moving with respect to a magnetic field,
we are aware of no studies of the interaction of boson stars with a
magnetic field.

\subsection{Compact alternatives to black holes}
\label{sec:mimickers}

As a localized scalar field configuration, a boson star can be constructed as a non-interacting 
compact object, as long as one does not include any explicit coupling to 
electromagnetic or other fields.
In that respect, it resembles a BH, although it lacks a horizon. Can observations of purported BHs be
fully explained by massive boson stars?
See \cite{lrr-2008-9} for a review of such observations.

Neutron stars also lack horizons, but, in contrast to a boson star, have a hard surface. 
A hard surface is important because one would expect accretion onto such a surface to have
observable consequences. Can a boson star avoid such consequences?
\cite{2004ApJ...606.1112Y} consider the
viability of $10\,M_{\odot}$ boson stars as BH candidates in X-ray binaries.
They find that accreting gas collects not at the surface (which the star lacks), but
instead at the center, which ultimately should lead to Type~I X-ray bursts.
Because these bursts are not observed, the case against boson stars as black hole
mimickers is weakened (at least for BH candidates in X-ray binaries).

\cite{2009PhRvD..80h4023G} consider a simplified model of accretion and searches for
boson-star configurations that would mimic an accreting black hole. Although they find
matches, they argue that light deflection about a boson star will differ from the BH they
mimic because of the lack of a photon sphere. Further work studies the scalar field tails
about boson stars and compares them to those of BHs \citep{2010PhRvD..82b3005L}.
If indeed a boson star collapses to a BH, then one could hope to observe the QNM of the massive scalar field, as described in \cite{Hod:2011zz}.
Differences between accretion structures surrounding boson
stars and black holes are analyzed in \cite{Meliani:2015zta},
showing that the accretion tori around boson stars have different
characteristics than in the vicinity of a black hole. \CP{Similar differences have been reported regarding tidal disruption clouds orbiting either a spherically symmetric compact boson star or a Schwarzschild black hole \citep{Teodoro:2020gps}. The simulations showed the formation of a ring-like structure around the boson star which is not present in the black hole scenario.}
Further studies on the subject include  disk \citep{Meliani:2016rfe} and supersonic winds \citep{Gracia-Linares:2016gvy} accreting onto boson stars.

Some of the strongest evidence for the existence of BHs is found at the center of most
galaxies. Observational evidence strongly suggests supermassive objects (of the order of millions of solar
mass) occupying a small region (of order an astronomical unit) which is easily explained by a supermassive 
BH \citep{2012AAS...21925201B}.
\sllnew{
However, some argue for the viability of supermassive
}
boson stars at galactic centers \citep{2000PhRvD..62j4012T}.
There could potentially be differences in the (electromagnetic)
spectrum between a black hole and a boson star, but there is considerable freedom
in adjusting the boson star potential to tweak the expected spectrum \citep{Guzman:2007zz}.
However, there are stringent constraints on BH alternatives to Sgr~A* by the low luminosity in
the near infrared \citep{Broderick:2005xa}. In particular, the low luminosity implies
a bound on the accretion rate assuming a hard surface radiating
thermally and, therefore, the observational evidence favors a black hole because it lacks such a surface. 
In particular, although a BS lacks a surface, any material it accretes would
accumulate and that material would have a surface that would radiate thermally.

\sllnew{We discuss here three methods to test the nature of astrophysical black hole candidates: X-ray observations, gravitational wave observations, and very long baseline interferometry~(VLBI).}

The analysis of X-ray reflection spectroscopy with data provided by the current X-ray missions can only provide weak constraints on boson stars \citep{Cao:2016zbh}, Proca stars \citep{Shen:2016acv}, and hairy Kerr BHs \citep{Ni:2016rhz}. The quasi-periodic oscillations (QPOs) observed in the X-ray flux emitted by accreting compact objects also provide a powerful tool both to constrain deviations from Kerr and to search for exotic compact objects.
Therefore, a future eXTP mission or LOFT-like mission could set very stringent constraints on black holes with bosonic hair and on (scalar or Proca) boson 
stars \citep{Franchini:2016yvq}.

\CP{Exotic objects, such as boson stars and gravastars \citep{Mazur:2001fv},}  
\CP{can be massive and compact enough to be easily confused with black holes. Nevertheless, these objects differ from black holes in having nonzero tidal deformabilities, which can allow one to distinguish binaries containing such objects from binary black holes using GW observations. It was found that such constraints can be used to rule out some simple models of boson stars \citep{Johnson-Mcdaniel:2018cdu}. Gravitational waves	
produced by extreme-mass-ratio inspirals into the supermassive compact object at the center of a galaxy, could also clarify whether it is  a black hole or a rotating boson star that lacks a horizon \citep{Zhang:2021ojz}.}
\CPn{Also, a stellar-mass object inspiralling around a supermassive boson star generically  excites resonant-modes \citep{2013PhRvD..88f4046M}, producing a characteristic imprint on the gravitational-wave emission which may discriminate between black holes and other horizonless compact objects.}

VLBI, on the other hand, may be able to resolve Sgr~A*,
our closest supermassive black hole, located at the center of our galaxy. 
\sllnew{
The Event Horizon Telescope~(EHT) uses a large collection of telescopes to create an Earth-scale interferometer to resolve supermassive black holes. So far, they have produced images both for Sgr~A* \citep{EventHorizonTelescope:2022xnr} and for M87 \citep{EventHorizonTelescope:2019dse}, which is much further away but also much larger.
}
These images allow the study of so-called \emph{BH shadows},
that is, the gravitational lensing and redshift effect due to the BH on the radiation from background sources~\sllnew{\cite{Gralla:2019xty}}.

\begin{figure}[htb]
\centerline{  
     \includegraphics[height=6.0cm]{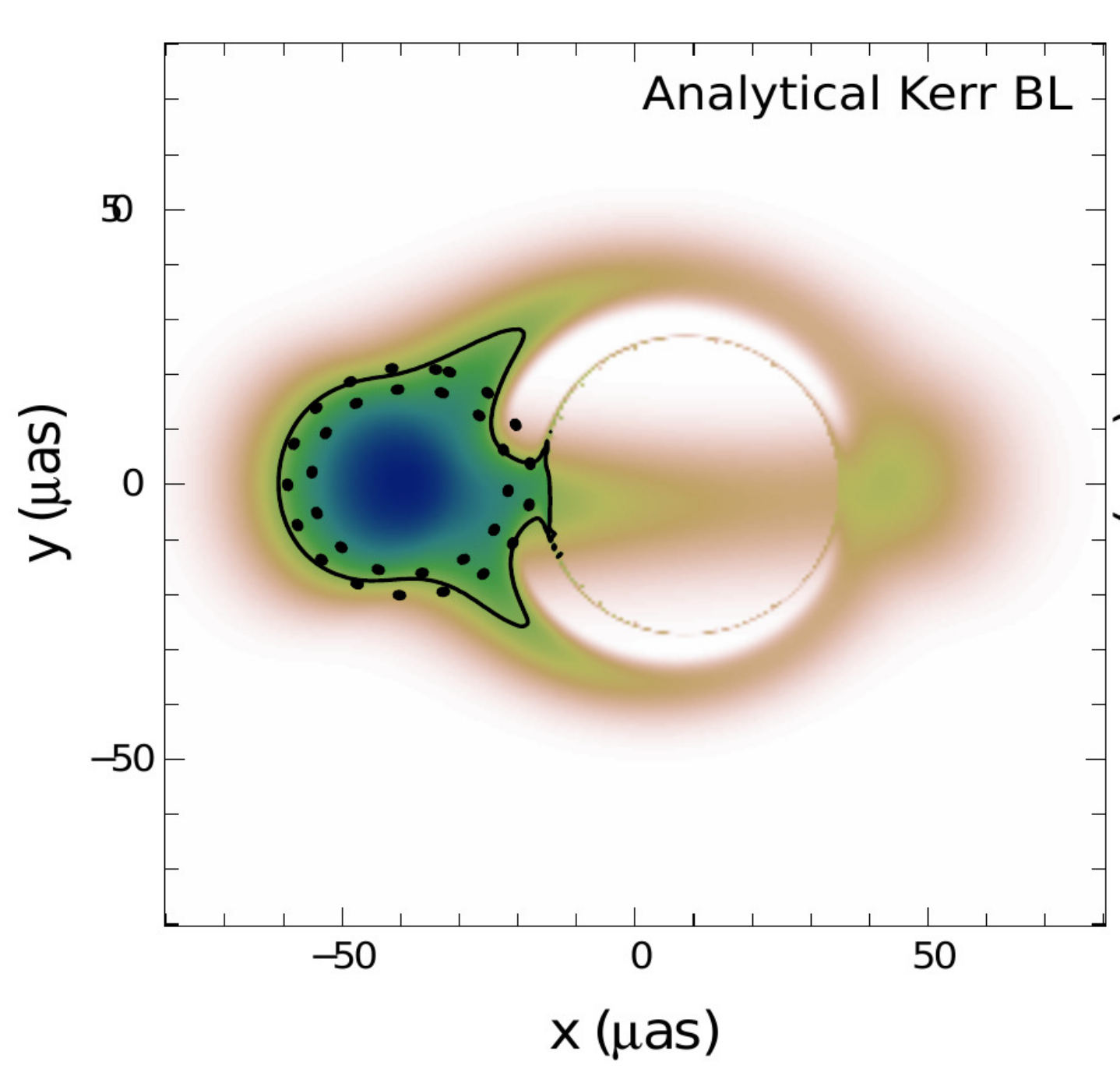}
     \includegraphics[height=5.8cm]{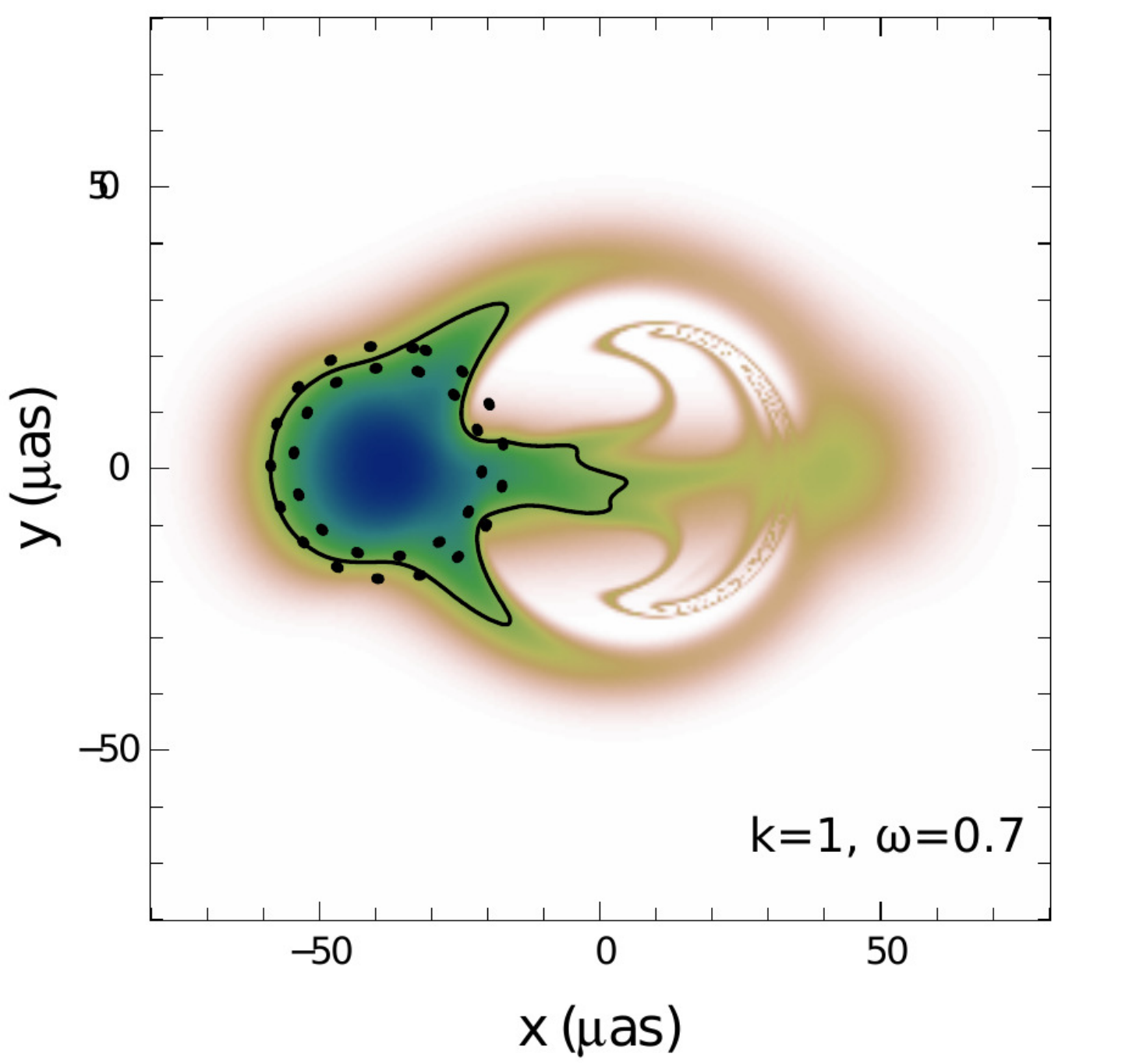}
}
\caption{Computed images as might be expected from the EHT for: (left) a Kerr black hole and (right) a fast spinning boson star with accretion according to  certain assumptions.
The similarity in images indicates that ruling out a BS candidate in images of Sgr~A* 
may prove difficult.
Reproduced with permission from \cite{Vincent:2015xta}, copyright by IOP.}
\label{fig:eht}
\end{figure}

\begin{figure}[htbp]
\centerline{\includegraphics[width=\textwidth]{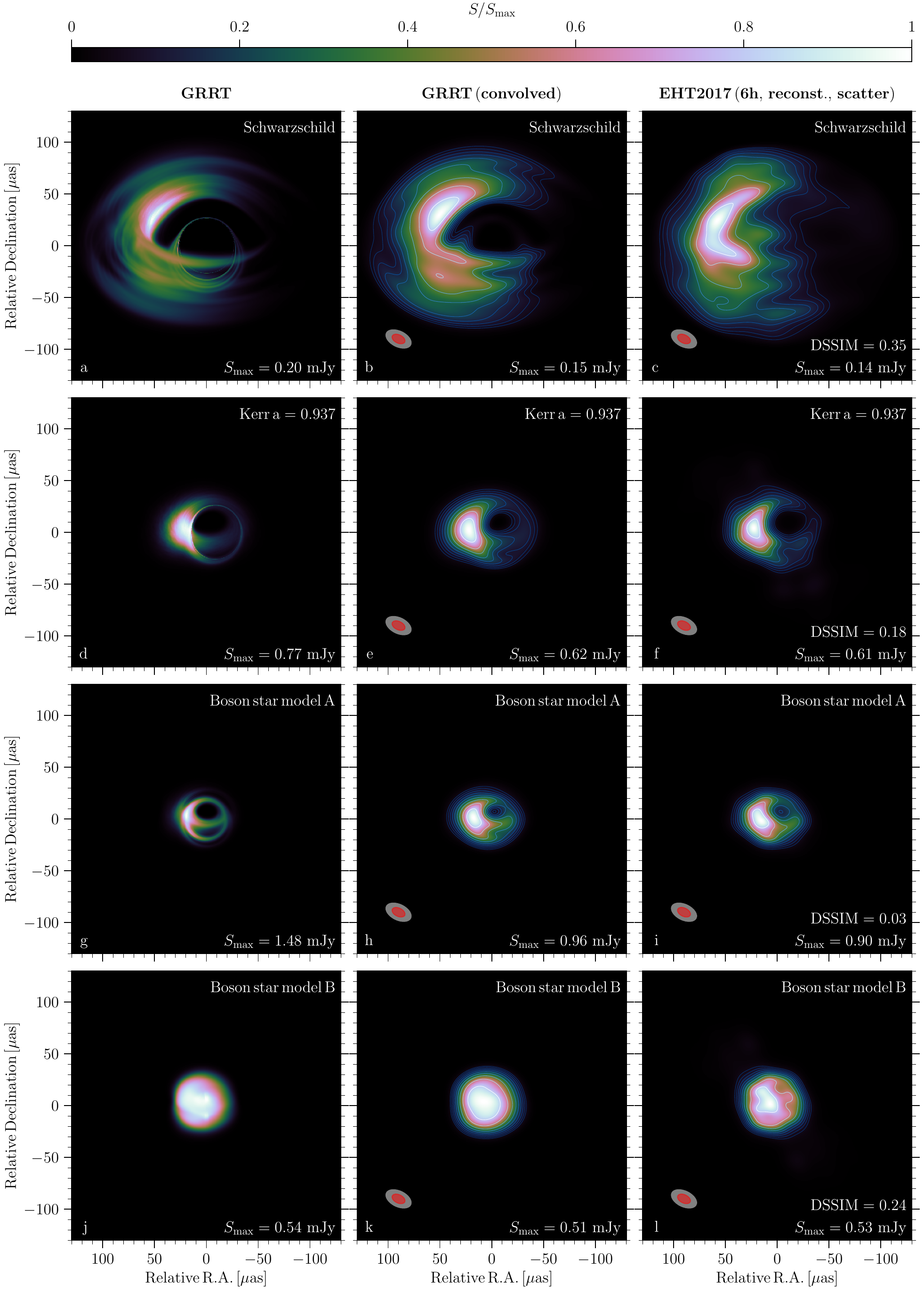}}
\caption{Synthetic images of various compact objects (Schwarzschild and Kerr black holes and two different boson stars).
     Note that the central dark area of the BSs is smaller than the black hole cases and that the BS images are more symmetric.
Reproduced with permission from \cite{Olivares:2018abq}, copyright by the authors.}
\label{fig:sgraoptions}
\end{figure}

Images of an accretion torus around Sgr~A*, assuming this compact object is a boson star,
are computed in \cite{Vincent:2015xta}. However, their results demonstrate that
very relativistic rotating boson stars produce images extremely similar to Kerr black holes, making them difficult to distinguish from a black hole. %
Figure~\ref{fig:eht} displays images predicted from this work for both a BH and a BS which
appear quite similar. The conclusion of \cite{Vincent:2015xta} expresses a number
of interesting caveats, and this study is also discussed as part of a more wide ranging paper
about efforts to firmly establish Sgr~A* as a BH \citep{Eckart:2017bhq}.
\CP{More recent and accurate simulations of the accretion flow onto Sgr~A*, assuming it is either a black hole or a non-rotating boson star, found that under realistic astronomical observing conditions the differences in the appearance are large enough to be detectable \citep{Olivares:2018abq}. These differences arise from dynamical effects directly related to the absence of an event horizon: the accumulation of matter in the form of either a small torus or a spheroidal cloud in the interior of the boson star, and the absence of an evacuated high-magnetization funnel in the polar regions. The mechanism behind these effects is general enough to apply to other horizonless and surfaceless black hole mimickers, strengthening confidence in the ability of the EHT to identify such objects.}
\sllnew{Examples of the resulting images are displayed in Fig.~\ref{fig:sgraoptions}.}

\CP{However, some of these differences might disappear, or at least diminish, by considering a Proca star instead of a boson star.  Even without a light ring, the Proca star can potentially mimic the shadow of a near-equilibrium Schwarzschild BH with the same mass under at least some observational conditions \citep{Herdeiro:2021lwl}. Similar results are obtained when considering the shadows of boson and Proca stars with thin accretion disks \citep{Rosa:2022tfv}.
}

It has also been shown in \cite{Cunha:2015yba} that hairy Kerr BHs can exhibit very distinct shadows from those of their vacuum counterparts when the light source is sufficiently far away from the BH. 
These differences remain, albeit less dramatically, when the BH is surrounded by an emitting torus of matter \citep{Vincent:2016sjq}.

Other studies have also studied the difference in appearance of a BS with that of the presumed BH in the center of our galaxy. 
\cite{2013arXiv1301.1396B} argue that, because BSs have an extended mass 
distribution that is transparent to 
electromagnetic radiation, the resulting strong gravitational lensing images of the S stars in the galactic center would yield much brighter images than a BH of similar mass.  \cite{2013CQGra..30i5014H} study BSs with a nonminimally coupled
scalar field and makes a similar argument about bright images.

One can also consider differences between the motion of celestial bodies about BSs versus BHs.
In particular, finding general geodesic motion of test particles in the space-time of boson stars generally 
requires numerical integration.
Geodesics around non-compact boson star were studied in \cite{2013PhRvD..88d4025D}, finding additional bound orbits of massive test particles close to the center of the star that are not present in the Schwarzschild case and that could be used to make predictions about extreme-mass-ratio inspirals (EMRIs), such as the stars orbiting Sagittarius~A*.
One can also compute the mass parameters of compact objects from redshifts and blueshifts emitted by geodesic particles around them \citep{Becerril:2016qxf}. The motion of charged, massive test particles in the spacetime of charged boson stars was considered in \cite{2014PhRvD..89h4048B}, \CP{and the trajectories of a spinning test particles in rotating boson star in \cite{Zhang:2021xhp,Zhang:2022qzw}}.

There are other possible BH mimickers, and a popular
recent one is the \emph{gravastar} \citep{Mazur:2001fv}. Common among all these
alternatives %
is the lack of an event horizon.
Both gravastars and BSs undergo an ergoregion instability for high
spin $J/ (G M^2) >0.4$ \citep{2008PhRvD..77l4044C}. As mentioned above for BSs, gravitational
waves may similarly be able to distinguish gravastars from BHs \citep{2009PhRvD..80l4047P}.
In order to reach the high compactnesses needed to mimic a BH, one can
       adopt specialized potentials \citep{Cardoso:2016oxy}, but an alternative
       is to embed the BS within a global monopole as studied in \cite{Reid:2015hiq} and \cite{2014CQGra..31d5010M}.
%

\subsection{As source of gravitational waves}
\label{sec:gravitationalwaves}
%

The era of gravitational-wave astronomy began in 2015, precisely one hundred years after Einstein's development of GR. In particular, during the first observational run O1, LIGO directly detected the gravitational waves from the inspiral, merger, and ringdown of a BH binary \citep{Abbott:2016blz}. \CP{This observation has since been followed by many others during O2, which included VIRGO, and during O3, also with KAGRA, with a total of 90 events detected by mid-2022 \citep{LIGOScientific:2021djp}. All the gravitational-wave observations so far are consistent with merging binaries of black holes and neutron stars. These detections are helping to ensure the development and completion during the next decade of space-based gravitational
wave observatories such as LISA \citep{Armano:2017oco}, as well as the Einstein Telescope, the third generation of ground-based detectors \citep{Evans:2016mbw}.}

Now that we have actual GW observations in hand, it behooves us to extract as much science as
possible from this new window on the Universe. Much work has already appeared
examining the implications of these initial 
detections \citep{Yunes:2016jcc,Yagi:2016jml,TheLIGOScientific:2016src}. In this paper, of course, we are concerned with the implications for BSs:
(i)~could these extent observations actually represent the signal from a pair of boson stars 
instead of BHs?
(ii)~might we observe a signal from boson stars, and, if so, what templates will we need?
or
(iii)~can we place tight bounds excluding the existence of boson stars?

\sllnew{In terms of boson stars, a binary}
is the most natural GW source.
However, at
early times, the precise structure
of the stars is irrelevant and the signatures are largely the same whether the binary
is composed of NSs, BHs, or BSs \citep{Bezares:2018qwa}. However, during the late inspiral and
merger, internal structure becomes important. In particular for boson stars, 
the relative phase determines the GW signature \citep{Palenzuela:2006wp,Palenzuela:2007dm,Cardoso:2016oxy}.

\CP{Gravitational-waves produced during the coalescence of 
boson stars can be used to constrain the fundamental coupling constants of a scalar field theory \CPi{(i.e., the self-interaction potential)}, 
in much the same way that GWs from binary 
neutron star coalescences help constrain the microscopic interaction of matter at ultrahigh density. Waveform models for the inspiral of boson stars with quartic interactions, including  spin-induced quadrupolar and tidal-deformability contributions, were constructed in \cite{Pacilio:2020jza}. Further analysis showed that future instruments such as the Einstein Telescope and the Laser Interferometer Space Antenna can provide strong complementary bounds on bosonic self-interactions. 
The gravitational radiation background generated from boson star binaries formed in locally dense clusters, with a formation rate tracked by the regular star formation rate, has been estimated in \cite{Croon:2018ftb}, as well as the dependence of the frequency window on the parameters of the model (i.e., the boson field mass and repulsive self-coupling). With these estimates of the GW background, future observations from  detectors such as LISA and the International Pulsar Timing Array \citep{2010CQGra..27h4013H} may be able to set constraints on scalar field theories.}

\sllnew{GW observations of binaries involving either two neutron stars (GW170817 and GW190425) or one neutron star and a black hole (GW190814, GW200105, and GW200115)
offer the potential of testing certain models of dark matter admixed neutron stars.
A study of the mixed binary (BH-NS) observations
found that the dark matter particle mass is mostly unconstrained by these observations \citep{Wystub:2021qrn}. 
Other work \citep{Lee:2021yyn} considers the ensuing constraints if one assumes that the $2.6\,M_{\odot}$ secondary 
in the GW190814 observation is a compact object other than a black hole. In particular, if the object is a QCD axion admixed neutron
star, then they constrain the axion mass to an already excluded range.
\cite{Karkevandi:2021ygv} study the impact of self-interacting bosonic asymmetric dark matter on various observable properties of fermion-boson stars with either a dense dark matter core or an extended dark halo.
Their combined analysis of the mass-radius relation and the tidal deformability constraints set by the LIGO/Virgo Collaboration, sets a stringent constraint on the dark matter fraction below 5\% \citep{Karkevandi:2021ygv}. 
}

\begin{figure}[htbp]
\centerline{\includegraphics[height=6.0cm,width=8.0cm]{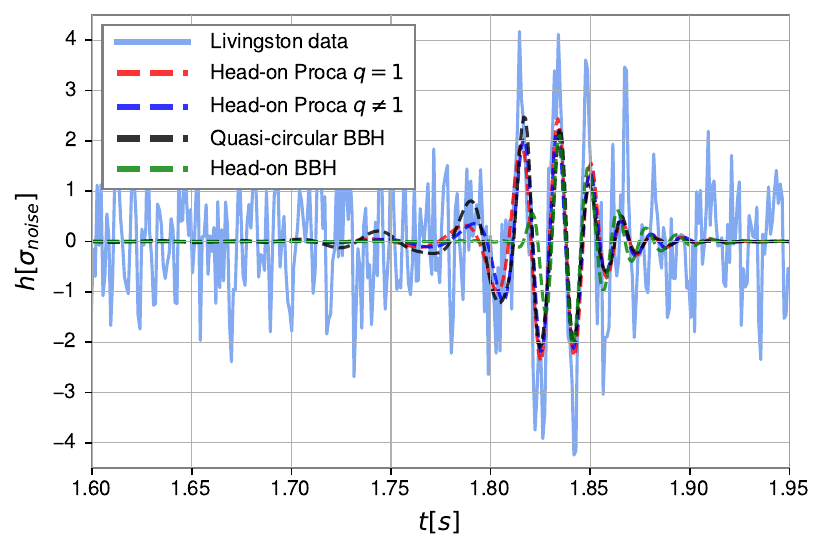}}
\centerline{\includegraphics[height=6.0cm,width=8.0cm]{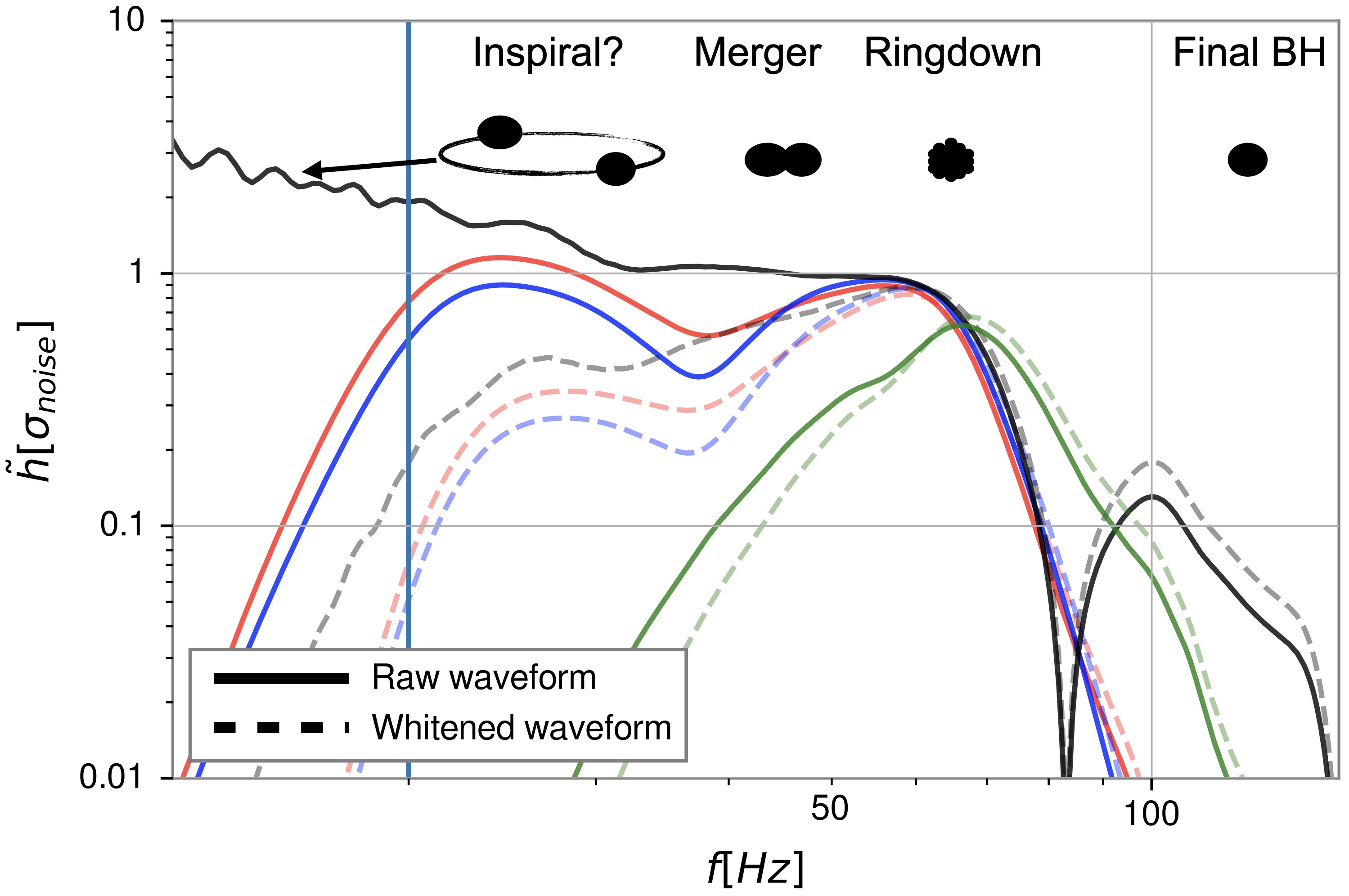}}
\caption{Comparison of the predicted \CP{strains} from a variety of compact object mergers \CPn{(i.e, spinning Proca stars and black holes)} with the GW190521 signal \CPn{observed} by LIGO Livingston. \emph{Top:} Time domain comparison and \emph{Bottom:} Frequency domain comparison. 
The equal mass ($q=1$) and unequal mass ($q \ne 1$) mergers \CP{of spinning Proca stars} are remarkably faithful to the observation.
Reproduced with permission from \cite{Bustillo:2020syj}, copyright by APS.}
\label{fig:gw190521}
\end{figure}

Gravitational waves may be an ideal messenger for revealing dark matter (discussed in Sect.~\ref{sec:darkmatter}).
If new dark sector particles can form exotic compact objects (ECOs) of astronomical size, then the first evidence for such objects --and their underlying microphysical description-- may arise in gravitational-wave observations. The relationship between the
macroscopic properties of ECOs, such as their GW signatures, with their microscopic properties, and hence new particles, was studied in \cite{Giudice:2016zpa}. 
The GW efficiency of compact binaries generally is examined in \cite{Hanna:2016uhs}. \CP{More recently, a systematic search for exotic compact mergers in Advanced LIGO and Virgo events has been performed, focusing on head-on mergers of Proca stars \citep{Bustillo:2020syj,CalderonBustillo:2022cja}. 
}
\sllnew{Although their Proca star merger hypothesis  is statistically rejected in favor of a black hole merger for some events, remarkably for others the mergers are somewhat
better fit by a Proca star merger.
}
Fig.~\ref{fig:gw190521} shows such a comparison for the specific case of the GW190521 observation.

\CPn{The first all-sky search for long-duration, quasimonochromatic gravitational-wave signals emitted by ultralight scalar boson clouds around spinning black holes was performed by the LVK collaboration using O3 data \citep{PhysRevD.105.102001}. This kind of search, which presumably will become routine during upcoming LVK runs and when 3G detectors come online, have the potential not only to unveil the existence of (fuzzy) dark matter haloes but also to constrain the mass of the bosonic particle from which boson stars are made of. In this search no evidence for such signals was found, setting an exclusion region in the boson mass/black hole mass plane and the maximum detectable distance for a given boson mass.}

\begin{figure}[htb]
\centerline{\includegraphics[height=6.0cm,width=8.0cm]{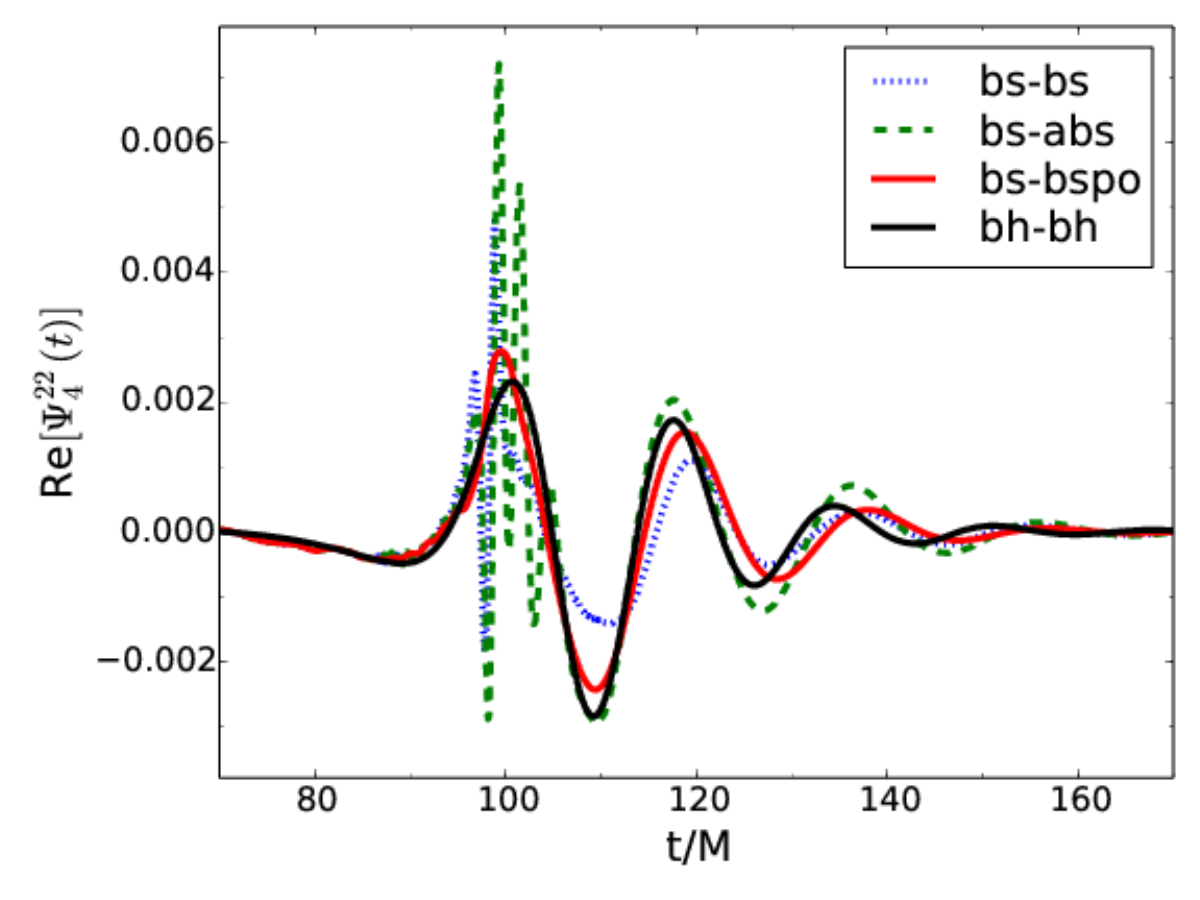}}
\caption{Gravitational waves, represented by the $l=m=2$ mode of the Newman-Penrose scalar, $\Psi_4$, emitted during the head-on collision of two solitonic BSs. 
For all configurations, the final object is massive enough to promptly collapse to a BH. However, for the boson--boson and boson--anti-boson configurations
the late inspiral signatures differ significantly from the corresponding binary
black-hole signal.
Reproduced with permission from \cite{Cardoso:2016oxy}, copyright by the authors.}
\label{fig:bbs_headon_psi4}
\end{figure}

Along the same lines, the tidal Love numbers for different ECOs, including different families of boson stars, are calculated in \cite{Cardoso:2017cfl}. The tidal Love number,
which encodes the deformability of a self-gravitating object within an external tidal field,
depends significantly both on the object's internal structure and on the dynamics of the gravitational field. Present and future gravitational-wave detectors can potentially measure this quantity in a binary inspiral of compact objects and impose constraints on boson stars.
Direct numerical simulations in head-on collision already have shown similarities 
in the gravitational waves emitted by black holes and boson stars in some cases \citep{Cardoso:2016oxy}. Figure~\ref{fig:bbs_headon_psi4} compares the expected GW signal of a BH binary with various
BS binaries.

One can also examine supermassive BHs and ask whether they could instead be some form
of BS. In particular,
the observation of gravitational waves from such objects may  be able to 
distinguish BHs from BSs \citep{2006IJMPD..15.2209B}. Such a test would occur in the
bandwidth for a space-based observatory such as the %
LISA mission \citep{newlisa}.
Because BSs allow for orbits within what would otherwise be a black hole event horizon,
geodesics will exhibit extreme pericenter precession resulting in potentially
distinguishable gravitational radiation \citep{Kesden:2004qx}.
In any case, observations of supermassive objects at the centers of galaxies can be used
to constrain the scalar field parameters of possible mimickers \citep{barranco-2011}.
In \cite{2013PhRvD..88f4046M} the authors construct mini-boson, boson and solitonic boson stars and analyze the gravitational and scalar response of boson star spacetimes to an inspiralling stellar-mass object.

\subsection{As origin of dark matter}
\label{sec:darkmatter}

Studies of stellar orbits within various galaxies produce \emph{rotation curves}
which indicate galactic mass within the radius of the particular orbit. The
discovery that these curves remain flat at large radius suggests the existence of
a large \emph{halo} of massive, yet dark, matter that holds the galaxy together
despite its large rotation (see \cite{Feng:2010gw} for a review). However, the precise form of matter that could fulfill
the observational constraints is still very much unclear. Scalar fields are an often
used tool in the cosmologist's toolkit, but one 
cannot have a regular, static configuration of scalar field to serve as the halo \citep{Pena:1997cy}
(see \citealt{Dias:2011at} as discussed in Sect.~\ref{sec:other} for a discussion of \emph{rotating}
boson stars with embedded, rotating BH solutions).
Instead, some form of boson star represents a possible candidate for providing the necessary dark mass.

Compact binaries
are the primary target of LIGO, but instead of neutron stars or black holes, \cite{Soni:2016yes} study the expected signal from binaries consisting of SU(N) glueball objects, 
one of the simplest models of dark matter.
More discussion of the merger of two BSs and the production of GW can
be found
in Sect.~\ref{dynamics_binary_bs}.
At the lower frequencies targeted by LISA,
if galaxies generally possess some extended, supermassive configuration, then the inspiral
of small compact body into this field will result in both dynamical friction and dark matter accretion,
in addition to radiation-reaction \citep{2013ApJ...774...48M}. These dynamical effects may potentially
be encoded on observable gravitational waves from the inspiral.

Boson stars can be matched onto
the observational constraints for galactic dark matter halos \citep{Lee:2008ab,2008arXiv0812.3470S}.
\CP{For instance, astrometric and spectroscopic observations of the orbital motion of S2 around 
Sgr~A* might narrow the allowed range for the mass of an ultralight boson forming a solitonic dark matter core in the innermost part of the halo \citep{DellaMonica:2022kow}.}
However, multi-\CP{field} boson stars that superpose
various boson-star solutions (e.g., an unexcited solution with an excited solution)
can perhaps find better fits to the constraints \citep{2010PhRvD..82l3535U}.
Boson stars at the galactic scale may not exhibit general relativistic effects and
can be effectively considered as 
Bose--Einstein condensates (BEC) with angular momentum \citep{RindlerDaller:2011kx}.

\CP{Boson stars can be a good descriptions of dark matter haloes if the fields are very light. In \cite{Annulli:2020ilw} the authors studied the dynamical response of Newtonian boson stars when excited by external matter (stars, planets or black holes) in their vicinities, including the first self-consistent calculation of dynamical friction acting on moving bodies in these backgrounds.}

\sllnew{
\cite{Laha:2018zav} proposes using fast radio bursts to look for gravitational lensing by boson stars. 
\cite{Choi:2019mva} argue that by 
combining such lensing with gravitational wave observations, one can probe the dynamics of 
boson stars.
}

Representing dark matter as BSs also offers certain computational benefits, avoiding some of the costs of modeling
the particles themselves with an $N$-body scheme. For example, \cite{Davidson:2016uok} study structure formation of
an axion dark matter model with ground state solutions of the appropriate \schrodinger-Poisson system along with quantum pressure term (see Eq.~\ref{eq:SP}).
Even if dark matter consists of clumps of weakly interacting massive particles~(WIMPs) instead of BSs, Mendes and Yang map
clumps of such particles to perturbed boson stars and study their tidal deformability, bypassing the
large computational cost of studying the dynamics of these WIMPs with an $N$-body code \citep{Mendes:2016vdr}.
Tidal deformability of BSs was also studied recently in the context of testing strong-field general relativity \citep{Cardoso:2017cfl}.

\CP{One can also consider a more general framework in which the dark matter halo is produced by an $N$-body system of boson stars. For instance, \cite{Amin:2022pzv} 
investigate the differences in the small-scale structure resulting from either vector dark matter (VDM) or scalar dark matter (SDM) using 3+1 dimensional simulations of the \schrodinger--Poisson system. Starting with a collection of idealized halos (self-gravitating solitons) as an initial condition, the system dynamically evolves to an approximately spherically symmetric configuration that has a core surrounded by a halo of interference patterns in the mass density. Their results point towards the possibility of distinguishing VDM from SDM using astrophysical and terrestrial observations. 
}

Instead of galactic scale BSs, one could instead argue for the accumulation of bosonic field in neutron
stars. Such solutions contain the ``normal'' fermionic matter as well as a bosonic component (discussed above
in Sect.~\ref{subsection:varieties_fermionicbosonic}). However, the accumulation of additional mass
in a neutron star, already the expected last stage before complete collapse to black hole, might conceivably lead
to the star's collapse. \CP{The observations of black holes with masses $\approx 1\,M_{\odot}$, which cannot be produced via stellar evolution, could be explained by the accumulation of dark matter triggering gravitational collapse in the star centers \citep{Garani:2021gvc}. } If indeed collapse can be expected, then the existence of old neutron stars would
place constraints on such a form of dark matter \citep{2012arXiv1204.2564F, 2013PhRvD..88c5004J,2013PhRvD..87e5012B}.
In the face of such arguments, \cite{2013PhRvD..87l3537K, 2013PhRvD..87l3507B} instead argue that a broad range of realistic models survive such
constraints. Most recently, \cite{Brito:2015yga} argue with perturbation and numerical methods that old stars are
in fact stable to the accretion of light bosons by an efficient gravitational cooling mechanism (see also \citealt{2016arXiv160705146B}).
Another dark matter model arising from a scalar field is 
\emph{wave dark matter} \citep{Bray:2014dca,
                                2013arXiv1301.0255B,
                                Goetz:2015qoa,
                                Goetz:2015lwa}.
In particular, they examine Tully--Fisher relationships predicted by this
wave dark matter model \citep{Bray:2014dca,Goetz:2015lwa}.
High-resolution simulations of a non-relativistic Bose--Einstein condensate within this model reproduce  the large scale structure of standard cold dark matter
while differing inside galaxies \citep{Schive:2014dra}.

Other studies solve the Gross--Pitaevskii equation for a Bose--Einstein condensate as a model of dark matter stars and study its stability properties \citep{2012JCAP...06..001L,2015PhRvD..91d4041M,Marsh:2015wka}.

The solitonic nature of boson stars (see Fig.~\ref{fig:soliton}) lends itself naturally to the wonderful
observation of dark matter in the Bullet Cluster \citep{Lee:2008mq}.
\cite{1475-7516-2010-01-007} attempt to determine a minimum galactic mass from such a match.

Interestingly, \cite{Barranco:2011pv} foregoe boson stars
and instead look for quasi-stationary scalar solutions about a Schwarzschild
black hole that could conceivably survive for cosmological times.
Another approach is to use scalar fields for both the dark matter halo and the supermassive, central object.
\cite{2010JCAP...11..002A} look for such a match, but find no suitable solutions.
Quite a number of more exotic models viably fit within current constraints, including those
using Q-balls \citep{Doddato:2011hx}.

\sllnewer{
Instead of beginning with boson stars as dark matter, recent work considers the \textbf{formation} of soliton stars from the existence of a bosonic field.
\cite{Gorghetto:2022sue} consider a dark vector boson and argue that such a field will be produced during inflation in the early universe from vacuum fluctuations.
Such fluctuations would then condense into gravitationally bound solitons, Proca stars,
and survive as a significant component of dark matter.
\cite{Arvanitaki:2019rax} posit a cosmological \emph{large-misalignment mechanism} that 
can lead to the formation of axion stars and study its observational consequences.
\cite{Levkov:2018kau} provide another well motivated formation channel in which
virialized dark matter bosons (dark QCD axions or Fuzzy Dark Matter) condense into
stars in the kinetic regime.
}

Section~\ref{dynamics_binary_bs} discusses the dynamics of boson stars including
some references commenting on the implications of the dynamics for dark matter.


\section{Boson stars in mathematical relativity}
\label{section:math}

Although the experimental foundation for the existence of boson stars
is completely lacking, on the theoretical and mathematical front,
boson stars are well studied. Recent work includes a mathematical
approach in terms of large and small data \citep{2009arXiv0910.2721F},
followed up by studying singularity
formation \citep{2011arXiv1103.3140L} and
uniqueness \citep{2009arXiv0905.3105F,lenzmann2009} for a certain boson
star equation. In \cite{ozawa2009}, they study radial solutions
of the semi-relativistic Hartree type equations in terms of global
well-posedness. \cite{Bicak:2010tt} demonstrate stationarity of
time periodic scalar field solutions.

Already discussed in Sect.~\ref{scalarclouds} has been the \emph{no hair conjecture} in the context of
BSs holding a central BH within.
Beyond just existence, however,
boson stars are often employed mathematically to study dynamics. Here, we concentrate on a few of these topics that
have attracted recent interest.

\subsection{Black-hole critical behavior}
\label{sub:critical}

If one considers some initial distribution of energy and watches it evolve,
generally one arrives at one of three states. If the energy is
sufficiently weak in terms of its gravity, the energy might end
up \emph{dispersing} to larger and large distances. However, if the energy
is instead quite large, then perhaps it will concentrate until
a \emph{black hole} is formed. Or, if the form of the energy supports it,
some of the energy will condense into a \emph{stationary state}.

In his seminal work, \cite{choptuik} considers a real, massless
scalar field and numerically evolves various initial configurations, finding
either dispersion or black-hole formation. By parameterizing these initial
configurations, say by the amplitude of an initial pulse $p$, and by tuning
this parameter, he was able to study the \emph{threshold for black-hole
formation} at which he found fascinating black-hole critical behavior. 
In particular, his numerical work suggested that continued tuning could produce
as small a black hole as one wished. This behavior is analogous to a phase
transition in which the black-hole mass serves as an order parameter.
Similar to phase transitions, one can categorize two types of transition that
distinguish between whether the black-hole mass varies continuously (Type~II)
or discontinuously (Type~I). For Choptuik's work with a massless field, the
transition is therefore of Type~II because the black-hole mass varies from zero
continuously to infinitesimal values.

Subsequent work has since established that this critical behavior can
be considered as occurring in the neighborhood of a \emph{separatrix} between
the \emph{basins of attraction} of the two end states. 
For $p=p^*$, the system is precisely critical and
remains on the (unstable) separatrix.
Similarly
other models find such threshold behavior occurring between a stationary
state and black-hole formation. Critical behavior about stationary solutions
necessarily involve black-hole formation ``turning-on'' at finite mass, and is
therefore categorized as Type~I critical behavior.

The critical surface, therefore, appears as a \emph{co-dimension 1} surface,
which evolutions increasingly approach as one tunes the parameter $p$.
The distance from criticality $|p-p^*|$ serves as a measure of the extent
to which a particular initial configuration has excited the unstable mode that drives solutions
away from this surface. For Type~II critical behavior, the mass of
the resulting black-hole mass scales as a power law in this distance, whereas
for Type~I critical behavior, it is the survival time of the critical solution
that scales as a power law.
See \cite{Gundlach:2007gc} for a recent review.

We have seen that boson stars represent stationary solutions of Einstein's
equations and, thus, one would correctly guess that they may occur within
Type~I black-hole critical behavior. To look for such behavior, \cite{Hawley:2000dt} begin their evolutions with 
boson-star solutions and then perturb them both dynamically and
gravitationally. They, therefore, included in their evolutionary system
a distinct, free, massless, real scalar field which couples to the boson
star purely through its gravity. 

\begin{figure}[htbp]
\centerline{\includegraphics[width=8.0cm]{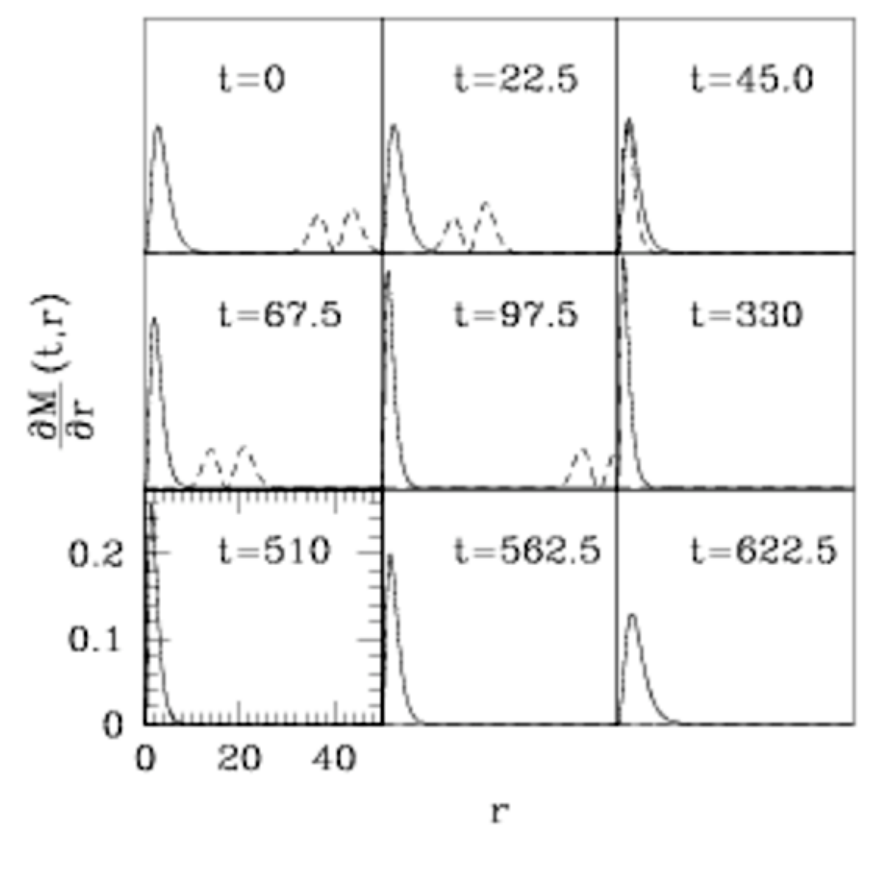}}
\caption{Evolution of a boson star (solid line) perturbed by a shell
  of scalar field (dashed line). Shown is the mass density $\partial M
  / \partial r$ for each contribution. By $t \approx 100$ the real
  scalar field pulse has departed the central region and perturbed the
  boson star into an unstable, compact configuration. Contrast the
  $t=0$ frame with that of $t=97.5$ and note the increase in
  compaction. This unstable BS survives until $t\approx 500$ only
  because the initial perturbation has been tuned to one part in
  $10^{15}$ and indicates Type~I critical behavior. Reproduced with
  permission from \cite{Lai:2007tj}.}
\label{fig:lai_evolve}
\end{figure}

The initial data, therefore, consisted of a boson star surrounded by
a distant, surrounding shell of real scalar field parametrized by the amplitude
of the shell. For small perturbations, the boson star oscillated about
an unstable boson star before settling into a low mass, stable solution
(see Fig.~\ref{fig:lai_evolve}).
For large perturbations, the real scalar field serves to compress the initial
star and, after a period of oscillation about an unstable boson star, the
complex field collapses to a black hole. By tuning the initial perturbation,
they find a longer and longer lived unstable boson star, which serves as
the critical solution (see Fig.~\ref{fig:lai}).
The survival time $\tau$ obeys a power law in terms of the distance from
criticality $|p-p^*|$
\begin{equation}
\tau \propto \gamma \ln |p - p^*| \,,
\end{equation}
where $\gamma$ is a real constant that depends on the characteristic
instability rate of the particular unstable boson star approached in
the critical regime.

\sllnew{
\cite{Jimenez-Vazquez:2022fix} subsequently studied the same system with different initial data. In particular, they adopted a Gaussian pulse as initial data, finding, as expected, that increasing the initial width of the pulse sent the system
from Type~II to Type~I critical behavior (for another system in which the initial pulse width determines the fate of the system see \cite{2013arXiv1304.4166B} mentioned in Sect.~\ref{sec:other}). In the large width, Type~I regime, they found
unstable boson stars acting as the critical solutions in agreement with 
\cite{Hawley:2000dt}.
}

One can also consider these BSs in axisymmetry in which non-spherically symmetric modes could potentially
become important. A first step in this direction studied spherically symmetric BSs within
conformally flat gravity (which does not allow for gravitational waves) in axisymmetry \citep{rousseau}.
Later, better resolution using adaptive mesh refinement within full
general relativity was achieved
by \cite{2005PhDT.........2L,Lai:2007tj}, which upheld the results
found within spherical symmetry. This work thus suggests that there are either no
additional, unstable, axisymmetric modes or that such unstable modes are extremely slowly growing.

A very different type of critical behavior was also investigated by
\cite{2005PhDT.........2L}. By boosting identical boson stars
toward each other and adjusting their initial momenta, he was able to
tune to the threshold for black-hole formation. At the threshold, he
found that the time till black-hole formation scaled consistent with
Type~I critical behavior and conjectured that the critical solution
was itself an unstable boson star. This is one of the few fully
nonlinear critical searches in less symmetry than spherical symmetry,
and the first of Type~I behavior in less symmetry. A related study
colliding neutron stars instead of boson stars similarly finds Type~I
critical behavior \citep{Jin:2006gm} and subsequently confirmed
by \cite{Kellermann:2010rt}.

The gauged stars discussed in Sect.~\ref{subsection:varieties_gauged} 
also serve as critical solutions in spherical
symmetry \citep{Choptuik:1996yg, Choptuik:1999gh, Millward:2002pk}.

\subsection{Hoop conjecture}
\label{hoop}

An interesting use of boson stars was made by \cite{Choptuik:2009ww}.
They sought to answer classically whether the ultra-relativistic collision
of two particles results in black-hole formation. Such a question clearly
has relevance to hopes of producing black holes at the
LHC (see, e.g., \cite{Landsberg:2006mm,Park:2012fe,Sirunyan:2017anm}).
Guidance on
this question is provided by Thorne's \emph{Hoop Conjecture} \citep{thorne_hoop}
which suggests that, if one squeezes energy into some spherical space of
dimension less than the Schwarzschild radius for that energy, then a black hole is formed.

\begin{figure}[htbp]
\centerline{\includegraphics[width=\textwidth]{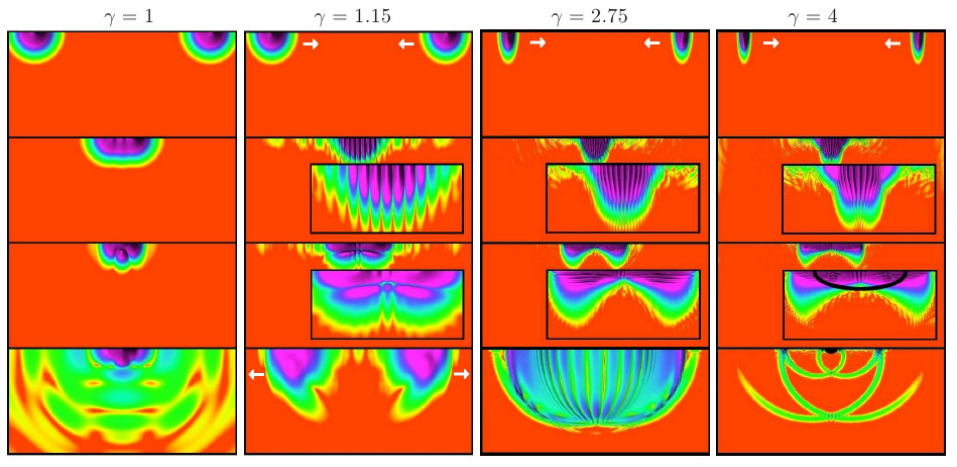}}
\caption{Evolutions of the head-on collisions of identical boson stars
  boosted toward each other with initial Lorentz factors $\gamma$ as
  indicated. Time flows downward within each column and the top edge
  displays the axis of symmetry. The color-scale indicates the value
  of $|\phi|$. In the middle frames one sees the interference pattern
  characteristic of high kinetic energy BS collisions (as mentioned in
  Fig.~\ref{fig:soliton}). In the last column on the right, the
  collision produces a BH with apparent horizon indicated by the black
  oval in the third frame. Reproduced with permission
  from \cite{Choptuik:2009ww}, copyright by APS.}
\label{fig:hoop}
\end{figure}

They, therefore, numerically collide boson stars head-on at relativistic energies to study 
black-hole formation from just such dynamical ``squeezing''.
Here, the nature of boson stars is largely irrelevant as they serve as
simple bundles of energy that can be accelerated (see Fig.~\ref{fig:hoop}).
However, unlike using
boosted black-hole solutions, the choice of boson stars avoids any type of
bias or predisposition to formation of a black hole. In addition, a number
of previous studies of boson star head-on collisions showed interesting
interference effects at energies below the threshold for black-hole
formation \citep{bosonunpub,PhysRevA.66.063609,2005PhDT.........2L,Mundim:2010hi}.
Indeed, it has been proposed that such an interference pattern could
be evidence for the bosonic nature of dark matter because of evidence that
an ideal fluid fails to produce such a pattern \citep{2011PhRvD..83j3513G}.

\cite{Choptuik:2009ww}
find that indeed black-hole formation occurs at energies \emph{below} that
estimated by the Hoop Conjecture. This result is
only a classical result consistent with the conjecture, but if it had not held,
then there would have been no reason to expect a quantum theory to 
be consistent with it.

\subsection{Other dimensions and anti-de~Sitter spacetime}
\label{sec:other}

Much work has been invested recently in considering physics in other
dimensions. Motivation comes from various ideas including string theory (more dimensions)
such as the AdS/CFT correspondence
and holography (one fewer dimensions) \citep{Maldacena:1997re,McGreevy:2009xe,Polchinski:2010hw}.
Another source of motivation comes from the
fact that higher dimensional black holes can have different properties than those
in three spatial dimensions \citep{lrr-2008-6}. 
Perhaps BSs will similarly display novel properties in other dimensions.

In lower dimensional AdS (2+1) spacetimes, early work in 1998 studied exact solutions 
of boson stars \citep{Sakamoto:1998hq, Degura:1998hw, Sakamoto:1998aj}. Higher dimensional 
scenarios were apparently first considered qualitatively a few years
later in the context of brane world models \citep{Stojkovic:2001qi}. This discussion
was followed with a detailed analysis of the 3, 4, and 5 dimensional AdS
solutions \citep{Astefanesei:2003qy}.

\cite{Fodor:2009kg} consider oscillatons in higher dimensions
and measures the scalar mass loss rate for dimensions 3, 4, and 5.
They extend this work considering inflationary spacetimes \citep{2010PhRvD..82d4043F}.
\cite{2014PhLB..739....1B} and \cite{Herdeiro:2015kha} construct higher dimensional black hole solutions
(Myers--Perry BHs) with scalar hair, and, in so doing, they find higher dimensional, rotating BS
solutions.
\sllnew{
More recently, \cite{Blazquez-Salcedo:2019qrz} construct boson and Dirac stars in various dimensions and finds a mass gap in higher dimensions in which the family of solutions does not connect to Minkowski space.
}

The axisymmetric rotating BSs discussed in Sect.~\ref{subsection:varieties_rotating} satisfy a coupled
set of nonlinear, elliptic PDEs in two dimensions. One might therefore suspect that adding other dimensions will
only make things more difficult. As it turns out, however, moving to four spatial dimensions provides for
another angular momentum, \emph{independent} of the one along the \textit{z}-direction (for example). Each of these
angular momenta are associated with their own orthogonal plane of rotation. And so if one chooses solutions
with equal magnitudes for each of these momenta, the solutions depend on only a \emph{single} radial coordinate.
This choice results in the remarkable simplification that one need only solve ODEs to find rotating
solutions \citep{Kunz:2006eh}. 

In \cite{Hartmann:2010pm}, they extend this idea by assuming an ansatz for two complex scalar fields with equal magnitudes
of angular momentum in the two independent directions. Letting the complex doublet be denoted by $\Phi$, the ansatz
takes the form
\begin{equation}
\Phi = \phi(r) e^{i\omega t} \left(
                                     \begin{array}{c}
                                     \sin \theta e^{i\varphi_1} \\
                                     \cos \theta e^{i\varphi_2}
                                     \end{array} 
                             \right)
\end{equation}
in terms of the two angular coordinates $\varphi_1$ and $\varphi_2$. One observes that the BS
 (i)~retains a profile $\phi(r)$,
(ii)~possesses harmonic time dependence,
and (iii)~maintains single-valuedness in the two angles (the ansatz assumes a rotational quantum number of one).
They find solutions that are both globally regular and asymptotically flat but these solutions appear
only stable with weak gravitational coupling \citep{Hartmann:2010pm}.
Solutions have since been constructed
in AdS${}_3$ \citep{2012JPhA...45K4025S,2014PhRvD..89d4018S},
in higher odd-dimensional AdS spacetimes \citep{2014PhRvD..89d4017S},
and
in Gauss--Bonnet gravity \citep{2015PhRvD..91b4009H} (see Sect.~\ref{sec:alt} for BS in alternative theories of gravity).

The work of \cite{Dias:2011at} makes ingenious use of this 5D ansatz to construct rotating black holes with only a single Killing vector. They set the
potential of \cite{Hartmann:2010pm} to zero so that the scalar fields are massless and they add a (negative) cosmological constant to work in anti-de~Sitter (AdS). \sllnew{Some of their solutions represent a black hole embedded inside a rotating BS.} They
find solutions for rotating black holes in 5D AdS that correspond to
a \emph{bar mode} for rotating neutron stars in 3D (see also \citealt{Shibata:2010wz} for a numerical evolution of a black hole in higher dimensions which demonstrates such bar formation; see \citealt{lrr-2008-6} for a review of black holes in higher dimensions).

One might expect such a non-symmetric black hole to settle into a more symmetric state via the emission of gravitational waves. However, AdS provides for an essentially reflecting boundary in which the black hole can be in equilibrium.
The distortion of the
higher dimensional black hole also has a correspondence with the discrete values of the angular momentum of the corresponding boson star. For higher values of the rotational quantum number, the black hole develops
multiple ``lobes'' about its center.
Very compact BSs constructed with this single Killing vector posses an ergoregion \citep{2015PhRvD..92d4049B}.

This construction can be
extended to arbitrary odd-dimensional AdS spacetimes \citep{Stotyn:2011ns}.
Finding the solutions perturbatively, they explicitly show that these solutions approach 
(i)~the boson star and (ii)~the Myers--Perry black-hole solutions in AdS \citep{Myers:1986un}
in different limits.
Boson stars, along with neutron stars and black holes, in five dimensions
are discussed in \cite{Brihaye:2016ibz}, and 
see \cite{lrr-2008-6} for a review of black holes in higher dimensions.

In AdS${}_4$ this ansatz cannot be used, and the construction of spinning boson stars requires
the solution of the appropriate multidimensional PDEs as is done in \cite{2012PhLB..717..450R}.

Interest in the dynamics of AdS spacetimes increased significantly with the work of \cite{Bizon:2011gg} who studied the collapse of a scalar field in spherically
symmetric, global AdS${}_4$. They argued that a non-zero initial amplitude for the scalar
field would generically result in gravitational collapse to black hole via turbulent
instability. In particular, fully nonlinear numerical evolutions of small amplitude configurations of scalar field generically resulted in a continued sharpening of the initial pulse as it reflected off the AdS boundary. This instability in the bulk is considered the mechanism that
achieves thermal equilibration in the conformal theory on the boundary.

Many studies followed trying to answer the many questions arising from this work.
Did this instability extend to any initial amplitude? Was the instability tied to
the precise structure of AdS or instead simply to the fact that the spacetime was bounded?

One question in particular concerned the implications of this instability for localized
solutions which might naturally be expected to extend their stability in asymptotically
flat spacetimes. To that end, \cite{2013arXiv1304.4166B} studied
boson stars in AdS, and found that indeed they are stable. In the course of
understanding how the boson stars were stable, this work found a whole class of initial
data that appear immune to the instability.
Later work added to this class, namely breather solutions in AdS \citep{Fodor:2015eia}.
Linear perturbation analysis of spherically symmetric Proca stars in AdS suggests that
these too will be stable \citep{Duarte:2016lig}.

The authors of \cite{Dias:2011at} also report on the existence of geons in 3+1 AdS ``which can be viewed as gravitational analogs of boson stars'' \citep{Dias:2011ss} (recall that boson stars themselves arose from Wheeler's desire to construct local electrovacuum solutions). These bundles of gravitational energy are stable to first order due to the confining boundary condition adopted with AdS.
The instability of these geons, black holes, and boson stars were studied
in \cite{Dias:2011at} in the context of the turbulent instability reported in \cite{Bizon:2011gg}, but later these authors argued for their nonlinear stability \citep{2012CQGra..29w5019D}.

\cite{2010JHEP...10..045B} also study black-hole solutions in 5D AdS. They find solutions for black holes with scalar hair that resemble a boson star with a BH in its center.
The stability of charged boson stars with a massive scalar field in five-dimensional AdS was
studied in \cite{2013CQGra..30k5009B}.
Also in AdS${}_5$, Buchel studies boson stars in a type~IIB supergravity approximation to string
theory in which the $U(1)$ symmetry of the complex field is gauged instead of global \citep{Buchel:2015rwa,Buchel:2015sma}.
A range of solutions, including Q-balls and shell solutions, for different values of the cosmological
constant have similarly been constructed \citep{2013PhRvD..88l4033H,2012PhRvD..86j4008H,2013PhRvD..87d4003H}.

\cite{2010JHEP...10..045B} also study black hole solutions in 5D AdS. They find solutions for black holes with scalar hair that resemble a boson star with a BH in its center.

Earlier work with BSs in lower dimensional AdS was reported in \cite{Astefanesei:2003qy}.

Boson stars in AdS with charge are constructed in \cite{2012arXiv1209.2378H}
\sllnew{and in the large charge limit by \cite{Guo:2020bqz}.}
They are also used as the background for a study of \emph{entanglement
entropy} \citep{2013PhRvD..87j6006N} (for a review of holographic entanglement entropy see \citealt{Rangamani:2016dms}).
Charged boson stars with spin in AdS have also been studied \citep{2014PhLB..728..328K}.
See \cite{Gentle:2011kv} for a review of charged scalar solitons in AdS.

\subsection{Analog gravity and physical systems}
\label{sec:analog}

The study of the correspondence between gravitating systems
and analogous physical systems goes by the name of \emph{analog gravity} \citep{Barcelo:2005fc}.
One example of such an analog is the \emph{acoustic} or \emph{dumb} hole,
analogous to a black hole,
that requires information to flow in a particular direction. For such a system the analog
of Hawking radiation is expected, and, remarkably, such radiation may have already  been
measured \citep{Unruh:2014hua}.

Analogs exist for BS as well.
Recent work of \cite{Roger:2016slp} finds an interesting
optical analog of Newtonian
BSs.
So far this analog appears to be mostly associated with corresponding equations
of motion as opposed to some deep physical correspondence that might reveal critical
insight.

A more concrete analog is the formation of a Bose--Einstein condensate 
such as studied in \cite{2014PhRvD..90j3526K} in the context of superradiance
(see Sect.~\ref{scalarclouds}). However, note that
as mentioned in Sect.~\ref{sec:origins}, ground state BSs can be considered as
condensed states of bosons without invoking any analogy \citep{Chavanis:2015zua,Chavanis:2016shp}.
%
%
%

%
%
%
\section{Open software}
\label{section:opensoftware}

%
\sllnew{
A number of codes and data sets are publicly available, and we collect some of them here in the hopes of making their availability more widely known:
}
\CPi{
\begin{itemize}
\item  Mathematica notebook and data file describing multipolar BSs from \cite{Vaglio:2022flq}.\footnote{
  \url{https://bitbucket.org/paolopani_uniroma1/repository_gmunu/src/master/Boson_Star_Multipoles}}  
\item
  Mathematica notebook to integrate the differential equations to solve for scalar, vector, and charged boson stars.\footnote{
    \url{https://centra.tecnico.ulisboa.pt/network/grit/files/boson-stars/}}
\item
  Tidal Love numbers for exotic compact objects (boson stars, gravastars, wormholes, mirrors).\footnote{
    \url{https://centra.tecnico.ulisboa.pt/network/grit/files/tidal-love-numbers/}}
\item
  A general solver in python for boson stars\footnote{
   \url{https://github.com/ThomasHelfer/BosonStar/blob/master/bosonstar/ComplexBosonStar.py}}
\item
  Various files associated with boson stars in AdS.\footnote{
    \url{https://github.com/hansbantilan/bstar}}
\item AMR code evolving the Einstein--Klein--Gordon system in 3D without symmetries (setup to run a single BS).\footnote{
   \url{http://mhduet.liu.edu}}
\item The C++ code used to construct rotating boson stars, as described in \cite{Ontanon:2021hbg}.\footnote{
   \url{https://github.com/sontanon/ROTBOSON}}
\item \textsc{Canuda}, a public library built with the Einstein Toolkit that evolves scalar and proca fields.\footnote{\url{https://doi.org/10.5281/zenodo.3565474}}
\end{itemize}
}

%

\section{Final remarks}
\label{section:final}

Boson stars have a long history as candidates for all manner of phenomena, from fundamental particle,
to galactic dark matter. A huge variety of solutions have been found and their dynamics studied.
Mathematically, BS are fascinating soliton-like solutions. Astrophysically, they represent possible
explanations of black hole candidates and dark matter, with observations constraining BS properties.

\sllnew{
Remarkably, in the five years between the first version (2012) of this review and its first revision (2017), two incredibly
significant experimental results have appeared. The Higgs particle
has been found by the LHC, the first scalar particle, although its instability makes it less than promising as
the fundamental constituent of boson stars.
}
\sllnew{
Far from the quantum particle regime of the LHC, the LIGO-Virgo collaboration directly detected gravitational waves in 2015, which were completely consistent with the merger of a binary black hole system as predicted by general relativity.
Not only does this put an end to the nagging questions about whether LIGO-Virgo can really detect such 
extremely
weak signals, but, as said often in the wake of these detections, 
it opens a new window into some of the most energetic events in the
Universe. 
}

\sllnew{
At the time of this second revision (2022), the LIGO-Virgo-KAGRA collaboration has
directly detected almost a hundred mergers, consistent with the merger of classical compact binaries (i.e., black holes and neutron stars).
Although it is impossible to predict what new phenomena 
will be observed, one can hope that
gravitational waves will further illuminate the nature of compact objects.
}

\sllnew{
In the electromagnetic spectrum, the EHT has produced images of Sgr~A*~(2022) and M87~(2019), both consistent with their being supermassive black holes surrounded by an accretion disk. 
Despite the tremendous achievement and beautiful work by the EHT collaboration, the higher resolution needed to firmly establish the existence of a horizon and definitively exclude a boson star awaits future work.}

With all of this experimental and observational data,
physicists need to provide unambiguous tests and explicit predictions.
Much work on that front is ongoing, trying to tease out observational
differences from alternative models of gravity or alternatives to the standard compact objects 
(BHs and NSs) \citep{Berti:2016rij,Berti:2015itd,Choptuik:2015mma}. Black holes were once
exotic and disbelieved, but now %
BHs are the commonly accepted standard while BSs are proposed as just one of many
exotic compact objects.

Perhaps future work on boson stars will be experimental, if fundamental scalar fields are observed, or
if evidence arises indicating the boson stars uniquely fit galactic dark matter. But regardless of any experimental results found by these remarkable experiments, there will always
be regimes unexplored by experiments where boson stars will find a natural home.


\begin{acknowledgements}
It is our pleasure to thank Juan Barranco, Miguel Bezares, Juan Calder\'on, Francisco Guzm\'{a}n, Carlos Herdeiro, Luis Lehner, Nicolas Sanchis-Gual, and Will East for their helpful comments on the manuscript.
We especially appreciate the careful and critical reading by Bruno Mundim.
We also thank Gyula Fodor and P\'eter Forg\'acs for their kind assistance with the section on
oscillatons and oscillons.

SLL also thanks ICERM at Brown University where work on the second revision occurred and the Perimeter Institute for their hospitality where the first version of this work was completed.
Research at Perimeter Institute is supported by the Government of Canada through Industry
Canada and by the Province of Ontario through the Ministry of Research and Innovation.

This work was also supported by NSF grants PHY-2011383, PHY-1912769, PHY-1607291, PHY-1308621, PHY-0969827, and PHY-0803624 to Long Island University. CP acknowledges support by the grants FPA2013-41042-P, AYA2016-80289-P (AEI/FEDER, UE), 
and PID2019-110301GB-I00 funded by MCIN/AEI/10.13039/501100011033 and by ``ERDF A way of making Europe''.
\end{acknowledgements}

\section*{Declarations}
The authors have no competing interests to declare that are relevant to the content of this article.


\phantomsection
\addcontentsline{toc}{section}{References}
\bibliographystyle{spbasic-FS}      
\bibliography{refs}   

\begin{thebibliography}{484}
\expandafter\ifx\csname url\endcsname\relax
 \def\url#1{\burl{#1}}\fi
\expandafter\ifx\csname urlprefix\endcsname\relax\def\urlprefix{URL }\fi
\providecommand{\bibinfo}[2]{#2}
\providecommand{\eprint}[2][]{\url{#2}}
\providecommand{\doi}[1]{\urlstyle{rm}\url{https://doi.org/#1}}

\bibitem[{Aad et~al.(2012)}]{Aad:2012tfa}
Aad G, et~al. (2012) Observation of a new particle in the search for the
  standard model {H}iggs boson with the {ATLAS} detector at the {LHC}. Phys
  Lett B 716:1--29. \doi{10.1016/j.physletb.2012.08.020}.
  {\href{https://arxiv.org/abs/1207.7214}{{arXiv:1207.7214}}} {[hep-ex]}

\bibitem[{Abbott et~al.(2016{\natexlab{a}})}]{Abbott:2016blz}
Abbott BP, et~al. (2016{\natexlab{a}}) Observation of gravitational waves from
  a binary black hole merger. Phys Rev Lett 116:061102.
  \doi{10.1103/PhysRevLett.116.061102}.
  {\href{https://arxiv.org/abs/1602.03837}{{arXiv:1602.03837}}} {[gr-qc]}

\bibitem[{Abbott et~al.(2016{\natexlab{b}})}]{TheLIGOScientific:2016src}
Abbott BP, et~al. (2016{\natexlab{b}}) Tests of general relativity with
  {GW150914}. Phys Rev Lett 116:221101. \doi{10.1103/PhysRevLett.116.221101}.
  {\href{https://arxiv.org/abs/1602.03841}{{arXiv:1602.03841}}} {[gr-qc]}

\bibitem[{Abbott et~al.(2017)}]{Evans:2016mbw}
Abbott BP, et~al. (2017) Exploring the sensitivity of next generation
  gravitational wave detectors. Class Quantum Grav 34:044001.
  \doi{10.1088/1361-6382/aa51f4}.
  {\href{https://arxiv.org/abs/1607.08697}{{arXiv:1607.08697}}} {[astro-ph.IM]}

\bibitem[{Abbott and et~al.(2022)}]{PhysRevD.105.102001}
Abbott R, et~al (2022) All-sky search for gravitational wave emission from
  scalar boson clouds around spinning black holes in {LIGO} {O3} data. Phys Rev
  D 105:102001. \doi{10.1103/PhysRevD.105.102001}

\bibitem[{Abbott et~al.(2021)}]{LIGOScientific:2021djp}
Abbott R, et~al. (2021) {GWTC-3: Compact Binary Coalescences Observed by LIGO
  and Virgo During the Second Part of the Third Observing Run}. arXiv e-prints
  {\href{https://arxiv.org/abs/2111.03606}{{arXiv:2111.03606}}} {[gr-qc]}

\bibitem[{Adam et~al.(2012)Adam, Kitchen, and Wiseman}]{Adam:2011dn}
Adam A, Kitchen S, Wiseman T (2012) {A numerical approach to finding general
  stationary vacuum black holes}. Class Quantum Grav 29:165002.
  \doi{10.1088/0264-9381/29/16/165002}.
  {\href{https://arxiv.org/abs/1105.6347}{{arXiv:1105.6347}}} {[gr-qc]}

\bibitem[{Adam et~al.(2010)Adam, Grandi, Klimas, S{\'{a}}nchez-Guill{\'{e}}n,
  and Wereszczy{\'{n}}ski}]{2010GReGr..42.2663A}
Adam C, Grandi N, Klimas P, S{\'{a}}nchez-Guill{\'{e}}n J, Wereszczy{\'{n}}ski
  A (2010) Compact boson stars in k field theories. Gen Relativ Gravit
  42:2663--2701. \doi{10.1007/s10714-010-1006-4}.
  {\href{https://arxiv.org/abs/0908.0218}{{arXiv:0908.0218}}} {[hep-th]}

\bibitem[{Adam et~al.(2022)Adam, Castelo, Mart\'\i{}n-Caro, Huidobro,
  V\'azquez, and Wereszczynski}]{Adam:2022nlq}
Adam C, Castelo J, Mart\'\i{}n-Caro AG, Huidobro M, V\'azquez R, Wereszczynski
  A (2022) {Universal relations for rotating Boson Stars}. arXiv e-prints
  {\href{https://arxiv.org/abs/2203.16558}{{arXiv:2203.16558}}} {[gr-qc]}

\bibitem[{Agnihotri et~al.(2009)Agnihotri, Schaffner-Bielich, and
  Mishustin}]{2009PhRvD..79h4033A}
Agnihotri P, Schaffner-Bielich J, Mishustin IN (2009) Boson stars with
  repulsive self-interactions. Phys Rev D 79:084033.
  \doi{10.1103/PhysRevD.79.084033}.
  {\href{https://arxiv.org/abs/0812.2770}{{arXiv:0812.2770}}}

\bibitem[{Akhoury and Gauthier(2008)}]{2008arXiv0804.3437A}
Akhoury R, Gauthier CS (2008) Galactic halos and black holes in non-canonical
  scalar field theories. ArXiv e-prints
  {\href{https://arxiv.org/abs/0804.3437}{{arXiv:0804.3437}}} {[hep-th]}

\bibitem[{Akiyama et~al.(2019)}]{EventHorizonTelescope:2019dse}
Akiyama K, et~al. (2019) {First M87 Event Horizon Telescope Results. I. The
  Shadow of the Supermassive Black Hole}. Astrophys J Lett 875:L1.
  \doi{10.3847/2041-8213/ab0ec7}.
  {\href{https://arxiv.org/abs/1906.11238}{{arXiv:1906.11238}}} {[astro-ph.GA]}

\bibitem[{Akiyama et~al.(2022)}]{EventHorizonTelescope:2022xnr}
Akiyama K, et~al. (2022) {First Sagittarius A* Event Horizon Telescope Results.
  I. The Shadow of the Supermassive Black Hole in the Center of the Milky Way}.
  Astrophys J Lett 930:L12. \doi{10.3847/2041-8213/ac6674}

\bibitem[{Alcubierre(2008)}]{alcubierre2008introduction}
Alcubierre M (2008) Introduction to 3+1 Numerical Relativity, International
  Series of Monographs on Physics, vol 140. Oxford University Press, Oxford;
  New York

\bibitem[{Alcubierre et~al.(2003)Alcubierre, Becerril, Guzm{\'{a}}n, Matos,
  N{\'{u}}{\~{n}}ez, and Ure{\~{n}}a-L{\'{o}}pez}]{2003CQGra..20.2883A}
Alcubierre M, Becerril R, Guzm{\'{a}}n FS, Matos T, N{\'{u}}{\~{n}}ez D,
  Ure{\~{n}}a-L{\'{o}}pez LA (2003) Numerical studies of
  $\phi^{2}$-oscillatons. Class Quantum Grav 20:2883--2903.
  \doi{10.1088/0264-9381/20/13/332}.
  {\href{https://arxiv.org/abs/gr-qc/0301105}{{arXiv:gr-qc/0301105}}}

\bibitem[{Alcubierre et~al.(2010)Alcubierre, Degollado, N{\'{u}}{\~{n}}ez,
  Ruiz, and Salgado}]{Alcubierre:2010ea}
Alcubierre M, Degollado JC, N{\'{u}}{\~{n}}ez D, Ruiz M, Salgado M (2010)
  Dynamic transition to spontaneous scalarization in boson stars. Phys Rev D
  81:124018. \doi{10.1103/PhysRevD.81.124018}.
  {\href{https://arxiv.org/abs/1003.4767}{{arXiv:1003.4767}}} {[gr-qc]}

\bibitem[{Alcubierre et~al.(2018)Alcubierre, Barranco, Bernal, Degollado,
  Diez-Tejedor, Megevand, Nunez, and Sarbach}]{Alcubierre:2018ahf}
Alcubierre M, Barranco J, Bernal A, Degollado JC, Diez-Tejedor A, Megevand M,
  Nunez D, Sarbach O (2018) {$\ell$-Boson stars}. Class Quantum Grav 35:19LT01.
  \doi{10.1088/1361-6382/aadcb6}.
  {\href{https://arxiv.org/abs/1805.11488}{{arXiv:1805.11488}}} {[gr-qc]}

\bibitem[{Alcubierre et~al.(2019)Alcubierre, Barranco, Bernal, Degollado,
  Diez-Tejedor, Megevand, N\'u\~nez, and Sarbach}]{Alcubierre:2019qnh}
Alcubierre M, Barranco J, Bernal A, Degollado JC, Diez-Tejedor A, Megevand M,
  N\'u\~nez D, Sarbach O (2019) {Dynamical evolutions of $\ell$-boson stars in
  spherical symmetry}. Class Quantum Grav 36:215013.
  \doi{10.1088/1361-6382/ab4726}.
  {\href{https://arxiv.org/abs/1906.08959}{{arXiv:1906.08959}}} {[gr-qc]}

\bibitem[{Alcubierre et~al.(2021)Alcubierre, Barranco, Bernal, Degollado,
  Diez-Tejedor, Megevand, N\'u\~nez, and Sarbach}]{Alcubierre:2021mvs}
Alcubierre M, Barranco J, Bernal A, Degollado JC, Diez-Tejedor A, Megevand M,
  N\'u\~nez D, Sarbach O (2021) {On the linear stability of
  \ensuremath{\ell}-boson stars with respect to radial perturbations}. Class
  Quantum Grav 38:174001. \doi{10.1088/1361-6382/ac0160}.
  {\href{https://arxiv.org/abs/2103.15012}{{arXiv:2103.15012}}} {[gr-qc]}

\bibitem[{Alcubierre et~al.(2022)Alcubierre, Barranco, Bernal, Degollado,
  Diez-Tejedor, Jaramillo, Megevand, N\'u\~nez, and
  Sarbach}]{Alcubierre:2021psa}
Alcubierre M, Barranco J, Bernal A, Degollado JC, Diez-Tejedor A, Jaramillo V,
  Megevand M, N\'u\~nez D, Sarbach O (2022) {Extreme \ensuremath{\ell}-boson
  stars}. Class Quantum Grav 39:094001. \doi{10.1088/1361-6382/ac5fc2}.
  {\href{https://arxiv.org/abs/2112.04529}{{arXiv:2112.04529}}} {[gr-qc]}

\bibitem[{Alic(2009)}]{alicthesis}
Alic D (2009) Theoretical issues in numerical relativity simulations. PhD
  thesis, Universitat de les Illes Balears, Palma.
  \urlprefix\url{http://hdl.handle.net/10803/9438}

\bibitem[{Amaro-Seoane et~al.(2010)Amaro-Seoane, Barranco, Bernal, and
  Rezzolla}]{2010JCAP...11..002A}
Amaro-Seoane P, Barranco J, Bernal A, Rezzolla L (2010) Constraining scalar
  fields with stellar kinematics and collisional dark matter. J Cosmol
  Astropart Phys 2010(11):002. \doi{10.1088/1475-7516/2010/11/002}.
  {\href{https://arxiv.org/abs/1009.0019}{{arXiv:1009.0019}}} {[astro-ph.CO]}

\bibitem[{Amin et~al.(2022)Amin, Jain, Karur, and Mocz}]{Amin:2022pzv}
Amin MA, Jain M, Karur R, Mocz P (2022) {Small-scale structure in vector dark
  matter}. JCAP 08(08):014. \doi{10.1088/1475-7516/2022/08/014}.
  {\href{https://arxiv.org/abs/2203.11935}{{arXiv:2203.11935}}} {[astro-ph.CO]}

\bibitem[{Annulli et~al.(2020)Annulli, Cardoso, and Vicente}]{Annulli:2020ilw}
Annulli L, Cardoso V, Vicente R (2020) {Stirred and shaken: Dynamical behavior
  of boson stars and dark matter cores}. Phys Lett B 811:135944.
  \doi{10.1016/j.physletb.2020.135944}.
  {\href{https://arxiv.org/abs/2007.03700}{{arXiv:2007.03700}}} {[astro-ph.HE]}

\bibitem[{Antusch and Orani(2016)}]{Antusch:2015ziz}
Antusch S, Orani S (2016) Impact of other scalar fields on oscillons after
  hilltop inflation. J Cosmol Astropart Phys 2016(03):026.
  \doi{10.1088/1475-7516/2016/03/026}.
  {\href{https://arxiv.org/abs/1511.02336}{{arXiv:1511.02336}}} {[hep-ph]}

\bibitem[{Antusch et~al.(2017)Antusch, Cefala, and Orani}]{Antusch:2016con}
Antusch S, Cefala F, Orani S (2017) Gravitational waves from oscillons after
  inflation. Phys Rev Lett 118:011303. \doi{10.1103/PhysRevLett.118.011303}.
  {\href{https://arxiv.org/abs/1607.01314}{{arXiv:1607.01314}}} {[astro-ph.CO]}

\bibitem[{Armano et~al.(2017)}]{Armano:2017oco}
Armano M, et~al. (2017) Charge-induced force-noise on free-falling test masses:
  results from {LISA} pathfinder. Phys Rev Lett 118:171101.
  \doi{10.1103/PhysRevLett.118.171101}.
  {\href{https://arxiv.org/abs/1702.04633}{{arXiv:1702.04633}}} {[astro-ph.IM]}

\bibitem[{Arnowitt et~al.(1962)Arnowitt, Deser, and Misner}]{adm1962}
Arnowitt R, Deser S, Misner CW (1962) The dynamics of general relativity. In:
  Witten L (ed) Gravitation: An Introduction to Current Research. Wiley, New
  York; London, pp 227--265. \doi{10.1007/s10714-008-0661-1}.
  {\href{https://arxiv.org/abs/gr-qc/0405109}{{arXiv:gr-qc/0405109}}}

\bibitem[{Arod{\'{z}} et~al.(2009)Arod{\'{z}}, Karkowski, and
  {\'{S}}wierczy{\'{n}}ski}]{2009PhRvD..80f7702A}
Arod{\'{z}} H, Karkowski J, {\'{S}}wierczy{\'{n}}ski Z (2009) Spinning
  {Q}-balls in the complex signum-{G}ordon model. Phys Rev D 80:067702.
  \doi{10.1103/PhysRevD.80.067702}.
  {\href{https://arxiv.org/abs/0907.2801}{{arXiv:0907.2801}}} {[hep-th]}

\bibitem[{Arvanitaki et~al.(2020)Arvanitaki, Dimopoulos, Galanis, Lehner,
  Thompson, and Van~Tilburg}]{Arvanitaki:2019rax}
Arvanitaki A, Dimopoulos S, Galanis M, Lehner L, Thompson JO, Van~Tilburg K
  (2020) {Large-misalignment mechanism for the formation of compact axion
  structures: Signatures from the {QCD} axion to fuzzy dark matter}. Phys Rev D
  101:083014. \doi{10.1103/PhysRevD.101.083014}.
  {\href{https://arxiv.org/abs/1909.11665}{{arXiv:1909.11665}}} {[astro-ph.CO]}

\bibitem[{Astefanesei and Radu(2003)}]{Astefanesei:2003qy}
Astefanesei D, Radu E (2003) Boson stars with negative cosmological constant.
  Nucl Phys B 665:594--622. \doi{10.1016/S0550-3213(03)00482-6}.
  {\href{https://arxiv.org/abs/gr-qc/0309131}{{arXiv:gr-qc/0309131}}}

\bibitem[{Baibhav and Maity(2017)}]{Baibhav:2016fot}
Baibhav V, Maity D (2017) Boson stars in higher-derivative gravity. Phys Rev D
  95:024027. \doi{10.1103/PhysRevD.95.024027}.
  {\href{https://arxiv.org/abs/1609.07225}{{arXiv:1609.07225}}} {[gr-qc]}

\bibitem[{Balakrishna(1999)}]{1999PhDT........44B}
Balakrishna J (1999) A numerical study of boson stars: {E}instein equations
  with a matter source. PhD thesis, Washington University, St. Louis.
  {\href{https://arxiv.org/abs/gr-qc/9906110}{{arXiv:gr-qc/9906110}}}

\bibitem[{Balakrishna et~al.(1998)Balakrishna, Seidel, and
  Suen}]{1998PhRvD..58j4004B}
Balakrishna J, Seidel E, Suen WM (1998) Dynamical evolution of boson stars.
  {II}. excited states and self-interacting fields. Phys Rev D 58:104004.
  \doi{10.1103/PhysRevD.58.104004}.
  {\href{https://arxiv.org/abs/gr-qc/9712064}{{arXiv:gr-qc/9712064}}}

\bibitem[{Balakrishna et~al.(2006)Balakrishna, Bondarescu, Daues, Guzm{\'{a}}n,
  and Seidel}]{2006gr.qc.....2078B}
Balakrishna J, Bondarescu R, Daues G, Guzm{\'{a}}n FS, Seidel E (2006)
  Evolution of 3d boson stars with waveform extraction. Class Quantum Grav
  23:2631--2652. \doi{10.1088/0264-9381/23/7/024}.
  {\href{https://arxiv.org/abs/gr-qc/602078}{{arXiv:gr-qc/602078}}}

\bibitem[{Balakrishna et~al.(2008)Balakrishna, Bondarescu, Daues, and
  Bondarescu}]{2008PhRvD..77b4028B}
Balakrishna J, Bondarescu R, Daues G, Bondarescu M (2008) Numerical simulations
  of oscillating soliton stars: Excited states in spherical symmetry and ground
  state evolutions in 3d. Phys Rev D 77:024028.
  \doi{10.1103/PhysRevD.77.024028}.
  {\href{https://arxiv.org/abs/0710.4131}{{arXiv:0710.4131}}} {[gr-qc]}

\bibitem[{Bao and Dong(2011)}]{2011JCoPh.230.5449B}
Bao W, Dong X (2011) Numerical methods for computing ground states and dynamics
  of nonlinear relativistic hartree equation for boson stars. J Comput Phys
  230:5449--5469. \doi{10.1016/j.jcp.2011.03.051}

\bibitem[{{Barcel{\'o}} et~al.(2011){Barcel{\'o}}, {Liberati}, and
  {Visser}}]{Barcelo:2005fc}
{Barcel{\'o}} C, {Liberati} S, {Visser} M (2011) {Analogue Gravity}. Living Rev
  Relativ 14:3. \doi{10.12942/lrr-2011-3}

\bibitem[{Barranco and Bernal(2011{\natexlab{a}})}]{barranco-2011}
Barranco J, Bernal A (2011{\natexlab{a}}) Constraining scalar field properties
  with boson stars as black hole mimickers. In: Ure{\~{n}}a-L{\'{o}}pez LA,
  Morales-T{\'{e}}cotl HA, Linares-Romero R, Santos-Rodr{\'{\i}}guez E,
  Estrada-Jim{\'{e}}nez S (eds) VIII Workshop of the Gravitation and
  Mathematical Physics Division of the Mexican Physical Society. AIP Conference
  Proceedings, vol 1396. American Institute of Physics, Melville, NY, pp
  171--175. \doi{10.1063/1.3647542}.
  {\href{https://arxiv.org/abs/1108.1208}{{arXiv:1108.1208}}} {[astro-ph.CO]}

\bibitem[{Barranco and Bernal(2011{\natexlab{b}})}]{2011PhRvD..83d3525B}
Barranco J, Bernal A (2011{\natexlab{b}}) Self-gravitating system made of
  axions. Phys Rev D 83:043525. \doi{10.1103/PhysRevD.83.043525}.
  {\href{https://arxiv.org/abs/1001.1769}{{arXiv:1001.1769}}} {[astro-ph.CO]}

\bibitem[{Barranco et~al.(2011)Barranco, Bernal, Degollado, Diez-Tejedor,
  Megevand, Alcubierre, N{\'{u}}{\~{n}}ez, and Sarbach}]{Barranco:2011pv}
Barranco J, Bernal A, Degollado JC, Diez-Tejedor A, Megevand M, Alcubierre M,
  N{\'{u}}{\~{n}}ez D, Sarbach O (2011) Are black holes a serious threat to
  scalar field dark matter models? Phys Rev D 84:083008.
  \doi{10.1103/PhysRevD.84.083008}.
  {\href{https://arxiv.org/abs/1108.0931}{{arXiv:1108.0931}}} {[gr-qc]}

\bibitem[{Bartnik and McKinnon(1988)}]{Bartnik:1988am}
Bartnik R, McKinnon J (1988) Particlelike solutions of the
  {E}instein--{Y}ang--{M}ills equations. Phys Rev Lett 61:141--144.
  \doi{10.1103/PhysRevLett.61.141}

\bibitem[{Basu et~al.(2010)Basu, Bhattacharya, Bhattacharyya, Loganayagam,
  Minwalla, and Umesh}]{2010JHEP...10..045B}
Basu P, Bhattacharya J, Bhattacharyya S, Loganayagam R, Minwalla S, Umesh V
  (2010) Small hairy black holes in global {AdS} spacetime. J High Energy Phys
  2010(10):045. \doi{10.1007/JHEP10(2010)045}.
  {\href{https://arxiv.org/abs/1003.3232}{{arXiv:1003.3232}}} {[hep-th]}

\bibitem[{Battye and Sutcliffe(2000)}]{2000NuPhB.590..329B}
Battye RA, Sutcliffe PM (2000) {Q}-ball dynamics. Nucl Phys B 590:329--363.
  \doi{10.1016/S0550-3213(00)00506-X}.
  {\href{https://arxiv.org/abs/hep-th/0003252}{{arXiv:hep-th/0003252}}}

\bibitem[{Baumgarte and Shapiro(1999)}]{1999PhRvD..59b4007B}
Baumgarte TW, Shapiro SL (1999) Numerical integration of {E}instein's field
  equations. Phys Rev D 59:024007. \doi{10.1103/PhysRevD.59.024007}.
  {\href{https://arxiv.org/abs/gr-qc/9810065}{{arXiv:gr-qc/9810065}}} {[gr-qc]}

\bibitem[{Baumgarte and Shapiro(2010)}]{baumgarte2010numerical}
Baumgarte TW, Shapiro SL (2010) Numerical Relativity: Solving {E}instein's
  Equations on the Computer. Cambridge University Press, Cambridge; New York

\bibitem[{Becerril et~al.(2016)Becerril, Valdez-Alvarado, and
  Nucamendi}]{Becerril:2016qxf}
Becerril R, Valdez-Alvarado S, Nucamendi U (2016) Obtaining mass parameters of
  compact objects from redshifts and blueshifts emitted by geodesic particles
  around them. Phys Rev D 94:124024. \doi{10.1103/PhysRevD.94.124024}.
  {\href{https://arxiv.org/abs/1610.01718}{{arXiv:1610.01718}}} {[gr-qc]}

\bibitem[{{Bell} et~al.(2013){Bell}, {Melatos}, and
  {Petraki}}]{2013PhRvD..87l3507B}
{Bell} NF, {Melatos} A, {Petraki} K (2013) Realistic neutron star constraints
  on bosonic asymmetric dark matter. Phys Rev D 87:123507.
  \doi{10.1103/PhysRevD.87.123507}.
  {\href{https://arxiv.org/abs/1301.6811}{{arXiv:1301.6811}}} {[hep-ph]}

\bibitem[{Bernal and Guzm{\'{a}}n(2006{\natexlab{a}})}]{2006PhRvD..74j3002B}
Bernal A, Guzm{\'{a}}n FS (2006{\natexlab{a}}) Scalar field dark matter:
  Head-on interaction between two structures. Phys Rev D 74:103002.
  \doi{10.1103/PhysRevD.74.103002}.
  {\href{https://arxiv.org/abs/astro-ph/0610682}{{arXiv:astro-ph/0610682}}}

\bibitem[{Bernal and Guzm{\'{a}}n(2006{\natexlab{b}})}]{2006PhRvD..74f3504B}
Bernal A, Guzm{\'{a}}n FS (2006{\natexlab{b}}) Scalar field dark matter:
  Nonspherical collapse and late-time behavior. Phys Rev D 74:063504.
  \doi{10.1103/PhysRevD.74.063504}.
  {\href{https://arxiv.org/abs/astro-ph/0608523}{{arXiv:astro-ph/0608523}}}

\bibitem[{Bernal et~al.(2010)Bernal, Barranco, Alic, and
  Palenzuela}]{2010PhRvD..81d4031B}
Bernal A, Barranco J, Alic D, Palenzuela C (2010) Multistate boson stars. Phys
  Rev D 81:044031. \doi{10.1103/PhysRevD.81.044031}.
  {\href{https://arxiv.org/abs/0908.2435}{{arXiv:0908.2435}}} {[gr-qc]}

\bibitem[{Berti and Cardoso(2006)}]{2006IJMPD..15.2209B}
Berti E, Cardoso V (2006) Supermassive black holes or boson stars? hair
  counting with gravitational wave detectors. Int J Mod Phys D 15:2209--2216.
  \doi{10.1142/S0218271806009637}.
  {\href{https://arxiv.org/abs/gr-qc/0605101}{{arXiv:gr-qc/0605101}}}

\bibitem[{Berti et~al.(2016)Berti, Cardoso, Crispino, Gualtieri, Herdeiro, and
  Sperhake}]{Berti:2016rij}
Berti E, Cardoso V, Crispino LCB, Gualtieri L, Herdeiro C, Sperhake U (2016)
  Numerical relativity and high energy physics: Recent developments. Int J Mod
  Phys D 25:1641022. \doi{10.1142/S0218271816410224}, proceedings, 3rd
  Amazonian Symposium on Physics and 5th NRHEP Network Meeting is approaching:
  Celebrating 100 Years of General Relativity: Belem, Brazil.
  {\href{https://arxiv.org/abs/1603.06146}{{arXiv:1603.06146}}} {[gr-qc]}

\bibitem[{Berti et~al.(2015)}]{Berti:2015itd}
Berti E, et~al. (2015) Testing general relativity with present and future
  astrophysical observations. Class Quantum Grav 32:243001.
  \doi{10.1088/0264-9381/32/24/243001}.
  {\href{https://arxiv.org/abs/1501.07274}{{arXiv:1501.07274}}} {[gr-qc]}

\bibitem[{Bezares and Palenzuela(2018)}]{Bezares:2018qwa}
Bezares M, Palenzuela C (2018) {Gravitational Waves from Dark Boson Star binary
  mergers}. Class Quantum Grav 35(23):234002. \doi{10.1088/1361-6382/aae87c}.
  {\href{https://arxiv.org/abs/1808.10732}{{arXiv:1808.10732}}} {[gr-qc]}

\bibitem[{{Bezares} et~al.(2017){Bezares}, {Palenzuela}, and
  {Bona}}]{Bezares:2017mzk}
{Bezares} M, {Palenzuela} C, {Bona} C (2017) Final fate of compact boson star
  mergers. Phys Rev D 95:124005. \doi{10.1103/PhysRevD.95.124005}.
  {\href{https://arxiv.org/abs/1705.01071}{{arXiv:1705.01071}}} {[gr-qc]}

\bibitem[{Bezares et~al.(2019)Bezares, Vigan\`o, and
  Palenzuela}]{Bezares:2019jcb}
Bezares M, Vigan\`o D, Palenzuela C (2019) {Gravitational wave signatures of
  dark matter cores in binary neutron star mergers by using numerical
  simulations}. Phys Rev D 100(4):044049. \doi{10.1103/PhysRevD.100.044049}.
  {\href{https://arxiv.org/abs/1905.08551}{{arXiv:1905.08551}}} {[gr-qc]}

\bibitem[{Bezares et~al.(2022)Bezares, Bo\v{s}kovi\'c, Liebling, Palenzuela,
  Pani, and Barausse}]{Bezares:2022obu}
Bezares M, Bo\v{s}kovi\'c M, Liebling S, Palenzuela C, Pani P, Barausse E
  (2022) {Gravitational waves and kicks from the merger of unequal mass, highly
  compact boson stars}. Phys Rev D 105(6):064067.
  \doi{10.1103/PhysRevD.105.064067}.
  {\href{https://arxiv.org/abs/2201.06113}{{arXiv:2201.06113}}} {[gr-qc]}

\bibitem[{Bhatt and Sreekanth(2009)}]{2009arXiv0910.1972B}
Bhatt JR, Sreekanth V (2009) Boson stars: Chemical potential and quark
  condensates. ArXiv e-prints
  {\href{https://arxiv.org/abs/0910.1972}{{arXiv:0910.1972}}} {[hep-ph]}

\bibitem[{Bi{\v{c}}{\'{a}}k et~al.(2010)Bi{\v{c}}{\'{a}}k, Scholtz, and
  Tod}]{Bicak:2010tt}
Bi{\v{c}}{\'{a}}k J, Scholtz M, Tod P (2010) On asymptotically flat solutions
  of {E}instein's equations periodic in time {II}. spacetimes with scalar-field
  sources. Class Quantum Grav 27:175011. \doi{10.1088/0264-9381/27/17/175011}.
  {\href{https://arxiv.org/abs/1008.0248}{{arXiv:1008.0248}}} {[gr-qc]}

\bibitem[{{Bin-Nun}(2013)}]{2013arXiv1301.1396B}
{Bin-Nun} AY (2013) Method for detecting a boson star at {Sgr A*} through
  gravitational lensing. ArXiv e-prints
  {\href{https://arxiv.org/abs/1301.1396}{{arXiv:1301.1396}}} {[gr-qc]}

\bibitem[{Bizo{\'{n}} and Rostworowski(2011)}]{Bizon:2011gg}
Bizo{\'{n}} P, Rostworowski A (2011) On weakly turbulent instability of anti-de
  {S}itter space. Phys Rev Lett 107:031102.
  \doi{10.1103/PhysRevLett.107.031102}.
  {\href{https://arxiv.org/abs/1104.3702}{{arXiv:1104.3702}}} {[gr-qc]}

\bibitem[{Bl\'azquez-Salcedo et~al.(2019)Bl\'azquez-Salcedo, Knoll, and
  Radu}]{Blazquez-Salcedo:2019qrz}
Bl\'azquez-Salcedo JL, Knoll C, Radu E (2019) {Boson and Dirac stars in $D\geq
  4$ dimensions}. Phys Lett B 793:161--168.
  \doi{10.1016/j.physletb.2019.04.035}.
  {\href{https://arxiv.org/abs/1902.05851}{{arXiv:1902.05851}}} {[gr-qc]}

\bibitem[{Boehle et~al.(2012)Boehle, Ghez, Schoedel, Yelda, and
  Meyer}]{2012AAS...21925201B}
Boehle A, Ghez A, Schoedel R, Yelda S, Meyer L (2012) New orbital analysis of
  stars at the {G}alactic center using speckle holography. In: AAS 219th
  Meeting. Bull. Am. Astron. Soc., vol~44. American Astronomical Society,
  Washington, DC

\bibitem[{Bogolyubski{\u{\i}} and Makhan'kov(1977)}]{1977ZhPmR..25..120B}
Bogolyubski{\u{\i}} IL, Makhan'kov VG (1977) Dynamics of spherically
  symmetrical pulsons of large amplitude. JETP Lett 25:107--110

\bibitem[{Bona et~al.(2009)Bona, Palenzuela-Luque, and
  Bona-Casas}]{bona2009elements}
Bona C, Palenzuela-Luque C, Bona-Casas C (2009) Elements of Numerical
  Relativity and Relativistic Hydrodynamics: From {E}instein's Equations to
  Astrophysical Simulations, Lecture Notes in Physics, vol 783, 2nd edn.
  Springer, Berlin; New York. \doi{10.1007/978-3-642-01164-1}

\bibitem[{Bosch et~al.(2016)Bosch, Green, and Lehner}]{PhysRevLett.116.141102}
Bosch P, Green SR, Lehner L (2016) Nonlinear evolution and final fate of
  charged anti--{de~Sitter} black hole superradiant instability. Phys Rev Lett
  116:141102. \doi{10.1103/PhysRevLett.116.141102}

\bibitem[{Bo\v{s}kovi\'c and Barausse(2022)}]{Boskovic:2021nfs}
Bo\v{s}kovi\'c M, Barausse E (2022) {Soliton boson stars, Q-balls and the
  causal Buchdahl bound}. JCAP 02:032. \doi{10.1088/1475-7516/2022/02/032}.
  {\href{https://arxiv.org/abs/2111.03870}{{arXiv:2111.03870}}} {[gr-qc]}

\bibitem[{Braaten and Zhang(2019)}]{Braaten:2019knj}
Braaten E, Zhang H (2019) {Colloquium : The physics of axion stars}. Rev Mod
  Phys 91:041002. \doi{10.1103/RevModPhys.91.041002}

\bibitem[{Brady et~al.(2002)Brady, Choptuik, Gundlach, and
  Neilsen}]{Brady:2002iz}
Brady PR, Choptuik MW, Gundlach C, Neilsen DW (2002) Black-hole threshold
  solutions in stiff fluid collapse. Class Quantum Grav 19:6359--6376.
  \doi{10.1088/0264-9381/19/24/306}.
  {\href{https://arxiv.org/abs/gr-qc/0207096}{{arXiv:gr-qc/0207096}}}

\bibitem[{{Bramante} et~al.(2013){Bramante}, {Fukushima}, and
  {Kumar}}]{2013PhRvD..87e5012B}
{Bramante} J, {Fukushima} K, {Kumar} J (2013) Constraints on bosonic dark
  matter from observation of old neutron stars. Phys Rev D 87:055012.
  \doi{10.1103/PhysRevD.87.055012}.
  {\href{https://arxiv.org/abs/1301.0036}{{arXiv:1301.0036}}} {[hep-ph]}

\bibitem[{Bray and Goetz(2014)}]{Bray:2014dca}
Bray HL, Goetz AS (2014) Wave dark matter and the {T}ully-{F}isher relation.
  ArXiv e-prints {\href{https://arxiv.org/abs/1409.7347}{{arXiv:1409.7347}}}
  {[astro-ph.GA]}

\bibitem[{{Bray} and {Parry}(2013)}]{2013arXiv1301.0255B}
{Bray} HL, {Parry} AR (2013) Modeling wave dark matter in dwarf spheroidal
  galaxies. ArXiv e-prints
  {\href{https://arxiv.org/abs/1301.0255}{{arXiv:1301.0255}}} {[astro-ph.GA]}

\bibitem[{Brihaye and Delsate(2016)}]{Brihaye:2016ibz}
Brihaye Y, Delsate T (2016) Boson stars, neutron stars and black holes in five
  dimensions. ArXiv e-prints
  {\href{https://arxiv.org/abs/1607.07488}{{arXiv:1607.07488}}} {[gr-qc]}

\bibitem[{Brihaye and Hartmann(2009)}]{2009PhRvD..79f4013B}
Brihaye Y, Hartmann B (2009) Angularly excited and interacting boson stars and
  $q$ balls. Phys Rev D 79:064013. \doi{10.1103/PhysRevD.79.064013}.
  {\href{https://arxiv.org/abs/0812.3968}{{arXiv:0812.3968}}} {[hep-ph]}

\bibitem[{Brihaye and Hartmann(2016)}]{Brihaye:2015jja}
Brihaye Y, Hartmann B (2016) Minimal boson stars in 5 dimensions: classical
  instability and existence of ergoregions. Class Quantum Grav 33:065002.
  \doi{10.1088/0264-9381/33/6/065002}.
  {\href{https://arxiv.org/abs/1509.04534}{{arXiv:1509.04534}}} {[hep-th]}

\bibitem[{Brihaye and Hartmann(2019)}]{Brihaye:2019puo}
Brihaye Y, Hartmann B (2019) {Spontaneous scalarization of boson stars}. JHEP
  09:049. \doi{10.1007/JHEP09(2019)049}.
  {\href{https://arxiv.org/abs/1903.10471}{{arXiv:1903.10471}}} {[gr-qc]}

\bibitem[{Brihaye and Hartmann(2022)}]{Brihaye:2021mqk}
Brihaye Y, Hartmann B (2022) {Boson stars and black holes with wavy scalar
  hair}. Phys Rev D 105:104063. \doi{10.1103/PhysRevD.105.104063}.
  {\href{https://arxiv.org/abs/2112.12830}{{arXiv:2112.12830}}} {[gr-qc]}

\bibitem[{{Brihaye} and {Riedel}(2014)}]{2014PhRvD..89j4060B}
{Brihaye} Y, {Riedel} J (2014) Rotating boson stars in five-dimensional
  {E}instein-{G}auss-{B}onnet gravity. Phys Rev D 89:104060.
  \doi{10.1103/PhysRevD.89.104060}.
  {\href{https://arxiv.org/abs/1310.7223}{{arXiv:1310.7223}}} {[gr-qc]}

\bibitem[{Brihaye and Verbin(2009)}]{Brihaye:2009ef}
Brihaye Y, Verbin Y (2009) Spherical structures in conformal gravity and its
  scalar-tensor extension. Phys Rev D 80:124048.
  \doi{10.1103/PhysRevD.80.124048}.
  {\href{https://arxiv.org/abs/0907.1951}{{arXiv:0907.1951}}} {[gr-qc]}

\bibitem[{Brihaye and Verbin(2010)}]{Brihaye:2009hf}
Brihaye Y, Verbin Y (2010) Spherical non-{A}belian solutions in conformal
  gravity. Phys Rev D 81:044041. \doi{10.1103/PhysRevD.81.044041}.
  {\href{https://arxiv.org/abs/0910.0973}{{arXiv:0910.0973}}} {[gr-qc]}

\bibitem[{Brihaye et~al.(2005)Brihaye, Hartmann, and Radu}]{Brihaye:2004nd}
Brihaye Y, Hartmann B, Radu E (2005) Boson stars in {SU(2)}
  {Y}ang-{M}ills-scalar field theories. Phys Lett B 607:17--26.
  \doi{10.1016/j.physletb.2004.12.020}.
  {\href{https://arxiv.org/abs/hep-th/0411207}{{arXiv:hep-th/0411207}}}
  {[hep-th]}

\bibitem[{Brihaye et~al.(2009{\natexlab{a}})Brihaye, Caebergs, and
  Delsate}]{Brihaye:2009dx}
Brihaye Y, Caebergs T, Delsate T (2009{\natexlab{a}})
  Charged-spinning-gravitating {Q}-balls. ArXiv e-prints
  {\href{https://arxiv.org/abs/0907.0913}{{arXiv:0907.0913}}} {[gr-qc]}

\bibitem[{Brihaye et~al.(2009{\natexlab{b}})Brihaye, Caebergs, Hartmann, and
  Minkov}]{2009PhRvD..80f4014B}
Brihaye Y, Caebergs T, Hartmann B, Minkov M (2009{\natexlab{b}}) Symmetry
  breaking in (gravitating) scalar field models describing interacting boson
  stars and {Q}-balls. Phys Rev D 80:064014. \doi{10.1103/PhysRevD.80.064014}.
  {\href{https://arxiv.org/abs/0903.5419}{{arXiv:0903.5419}}} {[gr-qc]}

\bibitem[{{Brihaye} et~al.(2013){Brihaye}, {Hartmann}, and
  {Tojiev}}]{2013CQGra..30k5009B}
{Brihaye} Y, {Hartmann} B, {Tojiev} S (2013) Stability of charged solitons and
  formation of boson stars in five-dimensional anti-de {S}itter spacetime.
  Class Quantum Grav 30:115009. \doi{10.1088/0264-9381/30/11/115009}.
  {\href{https://arxiv.org/abs/1301.2452}{{arXiv:1301.2452}}} {[hep-th]}

\bibitem[{{Brihaye} et~al.(2014{\natexlab{a}}){Brihaye}, {Diemer}, and
  {Hartmann}}]{2014PhRvD..89h4048B}
{Brihaye} Y, {Diemer} V, {Hartmann} B (2014{\natexlab{a}}) Charged {Q}-balls
  and boson stars and dynamics of charged test particles. Phys Rev D 89:084048.
  \doi{10.1103/PhysRevD.89.084048}.
  {\href{https://arxiv.org/abs/1402.1055}{{arXiv:1402.1055}}} {[gr-qc]}

\bibitem[{{Brihaye} et~al.(2014{\natexlab{b}}){Brihaye}, {Herdeiro}, and
  {Radu}}]{2014PhLB..739....1B}
{Brihaye} Y, {Herdeiro} C, {Radu} E (2014{\natexlab{b}}) Myers-perry black
  holes with scalar hair and a mass gap. Phys Lett B 739:1--7.
  \doi{10.1016/j.physletb.2014.10.019}.
  {\href{https://arxiv.org/abs/1408.5581}{{arXiv:1408.5581}}} {[gr-qc]}

\bibitem[{{Brihaye} et~al.(2015){Brihaye}, {Hartmann}, and
  {Riedel}}]{2015PhRvD..92d4049B}
{Brihaye} Y, {Hartmann} B, {Riedel} J (2015) Self-interacting boson stars with
  a single {K}illing vector field in anti-de {S}itter space-time. Phys Rev D
  92:044049. \doi{10.1103/PhysRevD.92.044049}.
  {\href{https://arxiv.org/abs/1404.1874}{{arXiv:1404.1874}}} {[gr-qc]}

\bibitem[{Brihaye et~al.(2016)Brihaye, Cisterna, and Erices}]{Brihaye:2016lin}
Brihaye Y, Cisterna A, Erices C (2016) Boson stars in biscalar extensions of
  {H}orndeski gravity. Phys Rev D 93:124057. \doi{10.1103/PhysRevD.93.124057}.
  {\href{https://arxiv.org/abs/1604.02121}{{arXiv:1604.02121}}} {[hep-th]}

\bibitem[{Brihaye et~al.(2020)Brihaye, Ducobu, and Hartmann}]{Brihaye:2020klz}
Brihaye Y, Ducobu L, Hartmann B (2020) {Boson and neutron stars with increased
  density}. Phys Lett B 811:135906. \doi{10.1016/j.physletb.2020.135906}.
  {\href{https://arxiv.org/abs/2004.08292}{{arXiv:2004.08292}}} {[gr-qc]}

\bibitem[{{Brito}(2016)}]{2016arXiv160705146B}
{Brito} R (2016) Fundamental fields around compact objects: Massive spin-2
  fields, superradiant instabilities and stars with dark matter cores. PhD
  thesis, Lisboa University.
  {\href{https://arxiv.org/abs/1607.05146}{{arXiv:1607.05146}}} {[gr-qc]}

\bibitem[{Brito et~al.(2015{\natexlab{a}})Brito, Cardoso, and
  Okawa}]{Brito:2015yga}
Brito R, Cardoso V, Okawa H (2015{\natexlab{a}}) Accretion of dark matter by
  stars. Phys Rev Lett 115:111301. \doi{10.1103/PhysRevLett.115.111301}.
  {\href{https://arxiv.org/abs/1508.04773}{{arXiv:1508.04773}}} {[gr-qc]}

\bibitem[{Brito et~al.(2015{\natexlab{b}})Brito, Cardoso, and
  Pani}]{Brito:2015oca}
Brito R, Cardoso V, Pani P (2015{\natexlab{b}}) Superradiance, Lecture Notes in
  Physics, vol 906. Springer, Cham. \doi{10.1007/978-3-319-19000-6}.
  {\href{https://arxiv.org/abs/1501.06570}{{arXiv:1501.06570}}} {[gr-qc]}

\bibitem[{Brito et~al.(2016{\natexlab{a}})Brito, Cardoso, Herdeiro, and
  Radu}]{Brito:2015pxa}
Brito R, Cardoso V, Herdeiro CAR, Radu E (2016{\natexlab{a}}) {P}roca stars:
  Gravitating {B}ose--{E}instein condensates of massive spin 1 particles. Phys
  Lett B 752:291--295. \doi{10.1016/j.physletb.2015.11.051}.
  {\href{https://arxiv.org/abs/1508.05395}{{arXiv:1508.05395}}} {[gr-qc]}

\bibitem[{Brito et~al.(2016{\natexlab{b}})Brito, Cardoso, Macedo, Okawa, and
  Palenzuela}]{Brito:2015yfh}
Brito R, Cardoso V, Macedo CFB, Okawa H, Palenzuela C (2016{\natexlab{b}})
  Interaction between bosonic dark matter and stars. Phys Rev D 93:044045.
  \doi{10.1103/PhysRevD.93.044045}.
  {\href{https://arxiv.org/abs/1512.00466}{{arXiv:1512.00466}}} {[astro-ph.SR]}

\bibitem[{Broderick and Narayan(2006)}]{Broderick:2005xa}
Broderick AE, Narayan R (2006) On the nature of the compact dark mass at the
  {G}alactic center. Astrophys J Lett 638:L21--L24. \doi{10.1086/500930}.
  {\href{https://arxiv.org/abs/astro-ph/0512211}{{arXiv:astro-ph/0512211}}}

\bibitem[{Buchdahl(1959)}]{Buchdahl:1959zz}
Buchdahl HA (1959) {General Relativistic Fluid Spheres}. Phys Rev 116:1027.
  \doi{10.1103/PhysRev.116.1027}

\bibitem[{Buchel(2015)}]{Buchel:2015rwa}
Buchel A (2015) {AdS} boson stars in string theory. ArXiv e-prints
  {\href{https://arxiv.org/abs/1510.08415}{{arXiv:1510.08415}}} {[hep-th]}

\bibitem[{Buchel and Buchel(2015)}]{Buchel:2015sma}
Buchel A, Buchel M (2015) On stability of nonthermal states in strongly coupled
  gauge theories. ArXiv e-prints
  {\href{https://arxiv.org/abs/1509.00774}{{arXiv:1509.00774}}} {[hep-th]}

\bibitem[{Buchel et~al.(2013)Buchel, Liebling, and
  Lehner}]{2013arXiv1304.4166B}
Buchel A, Liebling SL, Lehner L (2013) Boson stars in {AdS} spacetime. Phys Rev
  D 87:123006. \doi{10.1103/PhysRevD.87.123006}.
  {\href{https://arxiv.org/abs/1304.4166}{{arXiv:1304.4166}}} {[gr-qc]}

\bibitem[{Burikham et~al.(2016)Burikham, Harko, and Lake}]{Burikham:2016cwz}
Burikham P, Harko T, Lake MJ (2016) Mass bounds for compact spherically
  symmetric objects in generalized gravity theories. Phys Rev D 94:064070.
  \doi{10.1103/PhysRevD.94.064070}.
  {\href{https://arxiv.org/abs/1606.05515}{{arXiv:1606.05515}}} {[gr-qc]}

\bibitem[{Bustillo et~al.(2021)Bustillo, Sanchis-Gual, Torres-Forn\'e, Font,
  Vajpeyi, Smith, Herdeiro, Radu, and Leong}]{Bustillo:2020syj}
Bustillo JC, Sanchis-Gual N, Torres-Forn\'e A, Font JA, Vajpeyi A, Smith R,
  Herdeiro C, Radu E, Leong SHW (2021) {GW190521 as a Merger of Proca Stars: A
  Potential New Vector Boson of $8.7\times 10^{-13}$ eV}. Phys Rev Lett
  126:081101. \doi{10.1103/PhysRevLett.126.081101}.
  {\href{https://arxiv.org/abs/2009.05376}{{arXiv:2009.05376}}} {[gr-qc]}

\bibitem[{Calderon~Bustillo et~al.(2022)Calderon~Bustillo, Sanchis-Gual, Leong,
  Chandra, Torres-Forne, Font, Herdeiro, Radu, Wong, and
  Li}]{CalderonBustillo:2022cja}
Calderon~Bustillo J, Sanchis-Gual N, Leong SHW, Chandra K, Torres-Forne A, Font
  JA, Herdeiro C, Radu E, Wong ICF, Li TGF (2022) {Searching for vector
  boson-star mergers within LIGO-Virgo intermediate-mass black-hole merger
  candidates}. arXiv e-prints
  {\href{https://arxiv.org/abs/2206.02551}{{arXiv:2206.02551}}} {[gr-qc]}

\bibitem[{Cao et~al.(2016)Cao, Cardenas-Avendano, Zhou, Bambi, Herdeiro, and
  Radu}]{Cao:2016zbh}
Cao Z, Cardenas-Avendano A, Zhou M, Bambi C, Herdeiro CAR, Radu E (2016) Iron
  k$\alpha$ line of boson stars. J Cosmol Astropart Phys 20156(10):003.
  \doi{10.1088/1475-7516/2016/10/003}.
  {\href{https://arxiv.org/abs/1609.00901}{{arXiv:1609.00901}}} {[gr-qc]}

\bibitem[{Cardoso and Gualtieri(2016)}]{Cardoso:2016ryw}
Cardoso V, Gualtieri L (2016) Testing the black hole `no-hair' hypothesis.
  Class Quantum Grav 33:174001. \doi{10.1088/0264-9381/33/17/174001}.
  {\href{https://arxiv.org/abs/1607.03133}{{arXiv:1607.03133}}} {[gr-qc]}

\bibitem[{Cardoso et~al.(2008)Cardoso, Pani, Cadoni, and
  Cavagli{\`{a}}}]{2008PhRvD..77l4044C}
Cardoso V, Pani P, Cadoni M, Cavagli{\`{a}} M (2008) Ergoregion instability of
  ultracompact astrophysical objects. Phys Rev D 77:124044.
  \doi{10.1103/PhysRevD.77.124044}.
  {\href{https://arxiv.org/abs/0709.0532}{{arXiv:0709.0532}}} {[gr-qc]}

\bibitem[{Cardoso et~al.(2016)Cardoso, Hopper, Macedo, Palenzuela, and
  Pani}]{Cardoso:2016oxy}
Cardoso V, Hopper S, Macedo CFB, Palenzuela C, Pani P (2016) Gravitational-wave
  signatures of exotic compact objects and of quantum corrections at the
  horizon scale. Phys Rev D 94:084031. \doi{10.1103/PhysRevD.94.084031}.
  {\href{https://arxiv.org/abs/1608.08637}{{arXiv:1608.08637}}} {[gr-qc]}

\bibitem[{Cardoso et~al.(2017)Cardoso, Franzin, Maselli, Pani, and
  Raposo}]{Cardoso:2017cfl}
Cardoso V, Franzin E, Maselli A, Pani P, Raposo G (2017) Testing strong-field
  gravity with tidal love numbers. Phys Rev D 95:084014.
  \doi{10.1103/PhysRevD.95.084014}.
  {\href{https://arxiv.org/abs/1701.01116}{{arXiv:1701.01116}}} {[gr-qc]}

\bibitem[{Cardoso et~al.(2022{\natexlab{a}})Cardoso, Ikeda, Vicente, and
  Zilh\~ao}]{Cardoso:2022nzc}
Cardoso V, Ikeda T, Vicente R, Zilh\~ao M (2022{\natexlab{a}}) {Parasitic black
  holes: the swallowing of a fuzzy dark matter soliton}. arXiv e-prints
  {\href{https://arxiv.org/abs/2207.09469}{{arXiv:2207.09469}}} {[gr-qc]}

\bibitem[{Cardoso et~al.(2022{\natexlab{b}})Cardoso, Ikeda, Zhong, and
  Zilh\~ao}]{Cardoso:2022vpj}
Cardoso V, Ikeda T, Zhong Z, Zilh\~ao M (2022{\natexlab{b}}) {Piercing of a
  boson star by a black hole}. Phys Rev D 106(4):044030.
  \doi{10.1103/PhysRevD.106.044030}.
  {\href{https://arxiv.org/abs/2206.00021}{{arXiv:2206.00021}}} {[gr-qc]}

\bibitem[{Cardoso et~al.(2022{\natexlab{c}})Cardoso, Macedo, Maeda, and
  Okawa}]{Cardoso:2021ehg}
Cardoso V, Macedo CFB, Maeda Ki, Okawa H (2022{\natexlab{c}}) {ECO-spotting:
  looking for extremely compact objects with bosonic fields}. Class Quantum
  Grav 39:034001. \doi{10.1088/1361-6382/ac41e7}.
  {\href{https://arxiv.org/abs/2112.05750}{{arXiv:2112.05750}}} {[gr-qc]}

\bibitem[{Carloni and Rosa(2019)}]{Carloni:2019cyo}
Carloni S, Rosa JaL (2019) {Derrick\textquoteright{}s theorem in curved
  spacetime}. Phys Rev D 100:025014. \doi{10.1103/PhysRevD.100.025014}.
  {\href{https://arxiv.org/abs/1906.00702}{{arXiv:1906.00702}}} {[gr-qc]}

\bibitem[{Chan et~al.(2022)Chan, Sibiryakov, and Xue}]{Chan:2022bkz}
Chan JHH, Sibiryakov S, Xue W (2022) {Condensation and evaporation of boson
  stars}. arXiv e-prints
  {\href{https://arxiv.org/abs/2207.04057}{{arXiv:2207.04057}}} {[astro-ph.CO]}

\bibitem[{Chatrchyan et~al.(2012)}]{Chatrchyan:2012xdj}
Chatrchyan S, et~al. (2012) Observation of a new boson at a mass of 125 {GeV}
  with the {CMS} experiment at the {LHC}. Phys Lett B 716:30--61.
  \doi{10.1016/j.physletb.2012.08.021}.
  {\href{https://arxiv.org/abs/1207.7235}{{arXiv:1207.7235}}} {[hep-ex]}

\bibitem[{Chavanis(2011)}]{2011PhRvD..84d3531C}
Chavanis PH (2011) Mass-radius relation of {N}ewtonian self-gravitating
  {B}ose-{E}instein condensates with short-range interactions. {I}. analytical
  results. Phys Rev D 84:043531. \doi{10.1103/PhysRevD.84.043531}.
  {\href{https://arxiv.org/abs/1103.2050}{{arXiv:1103.2050}}} {[astro-ph.CO]}

\bibitem[{Chavanis(2012)}]{2012AA...537A.127C}
Chavanis PH (2012) Growth of perturbations in an expanding universe with
  {B}ose-{E}instein condensate dark matter. Astron Astrophys 537:A127.
  \doi{10.1051/0004-6361/201116905}.
  {\href{https://arxiv.org/abs/1103.2698}{{arXiv:1103.2698}}} {[astro-ph.CO]}

\bibitem[{Chavanis(2015)}]{Chavanis:2015zua}
Chavanis PH (2015) Self-gravitating {B}ose-{E}instein condensates. In: Calmet X
  (ed) Quantum Aspects of Black Holes. Fundamental Theories of Physics, vol
  178. Springer, Cham, pp 151--194. \doi{10.1007/978-3-319-10852-0_6}

\bibitem[{{Chavanis} and {Harko}(2012)}]{2011arXiv1108.3986C}
{Chavanis} PH, {Harko} T (2012) {B}ose-{E}instein condensate general
  relativistic stars. Phys Rev D 86:064011. \doi{10.1103/PhysRevD.86.064011}.
  {\href{https://arxiv.org/abs/1108.3986}{{arXiv:1108.3986}}} {[astro-ph.SR]}

\bibitem[{Chavanis and Matos(2017)}]{Chavanis:2016shp}
Chavanis PH, Matos T (2017) Covariant theory of {B}ose-{E}instein condensates
  in curved spacetimes with electromagnetic interactions: the hydrodynamic
  approach. Eur Phys J Plus 132:30. \doi{10.1140/epjp/i2017-11292-4}.
  {\href{https://arxiv.org/abs/1606.07041}{{arXiv:1606.07041}}} {[gr-qc]}

\bibitem[{Cho et~al.(2009)Cho, Ozawa, Sasaki, and Shim}]{ozawa2009}
Cho Y, Ozawa T, Sasaki H, Shim Y (2009) Remarks on the semirelativistic
  {H}artree equations. Discrete Contin Dyn Syst A 23:1277--1294.
  \doi{10.3934/dcds.2009.23.1277}

\bibitem[{Chodosh and Shlapentokh-Rothman(2015)}]{Chodosh:2015oma}
Chodosh O, Shlapentokh-Rothman Y (2015) Time-periodic
  {E}instein-{K}lein-{G}ordon bifurcations of {K}err. ArXiv e-prints
  {\href{https://arxiv.org/abs/1510.08025}{{arXiv:1510.08025}}} {[gr-qc]}

\bibitem[{Choi et~al.(2009)Choi, Lai, Choptuik, Hirschmann, Liebling, and
  Pretorius}]{bosonunpub}
Choi D, Lai CW, Choptuik MW, Hirschmann EW, Liebling SL, Pretorius F (2009)
  Dynamics of axisymmetric (head-on) boson star collisions,
  \urlprefix\url{http://laplace.physics.ubc.ca/Group/Papers/choi-etal-prd-05/choi-etal-prd-05.pdf},
  unpublished

\bibitem[{Choi(1998)}]{1998PhDT........16C}
Choi DI (1998) Numerical studies of nonlinear {S}chr\"odinger and
  {K}lein-{G}ordon systems: Techniques and applications. PhD thesis, The
  University of Texas, Austin.
  \urlprefix\url{http://laplace.physics.ubc.ca/Members/matt/Doc/Theses/}

\bibitem[{Choi(2002)}]{PhysRevA.66.063609}
Choi DI (2002) Collision of gravitationally bound {B}ose-{E}instein
  condensates. Phys Rev A 66:063609. \doi{10.1103/PhysRevA.66.063609}

\bibitem[{Choi et~al.(2019)Choi, He, and Schiappacasse}]{Choi:2019mva}
Choi G, He HJ, Schiappacasse ED (2019) {Probing Dynamics of Boson Stars by Fast
  Radio Bursts and Gravitational Wave Detection}. JCAP 10:043.
  \doi{10.1088/1475-7516/2019/10/043}.
  {\href{https://arxiv.org/abs/1906.02094}{{arXiv:1906.02094}}} {[astro-ph.CO]}

\bibitem[{Choptuik et~al.(2019)Choptuik, Masachs, and Way}]{Choptuik:2019zji}
Choptuik M, Masachs R, Way B (2019) {Multioscillating Boson Stars}. Phys Rev
  Lett 123:131101. \doi{10.1103/PhysRevLett.123.131101}.
  {\href{https://arxiv.org/abs/1904.02168}{{arXiv:1904.02168}}} {[gr-qc]}

\bibitem[{Choptuik(1993)}]{choptuik}
Choptuik MW (1993) Universality and scaling in gravitational collapse of a
  massless scalar field. Phys Rev Lett 70:9--12. \doi{10.1103/PhysRevLett.70.9}

\bibitem[{Choptuik and Pretorius(2010)}]{Choptuik:2009ww}
Choptuik MW, Pretorius F (2010) Ultrarelativistic particle collisions. Phys Rev
  Lett 104:111101. \doi{10.1103/PhysRevLett.104.111101}.
  {\href{https://arxiv.org/abs/0908.1780}{{arXiv:0908.1780}}} {[gr-qc]}

\bibitem[{Choptuik et~al.(1996)Choptuik, Chmaj, and
  Bizo{\'{n}}}]{Choptuik:1996yg}
Choptuik MW, Chmaj T, Bizo{\'{n}} P (1996) Critical behavior in gravitational
  collapse of a {Y}ang-{M}ills field. Phys Rev Lett 77:424--427.
  \doi{10.1103/PhysRevLett.77.424}.
  {\href{https://arxiv.org/abs/gr-qc/9603051}{{arXiv:gr-qc/9603051}}}

\bibitem[{Choptuik et~al.(1999)Choptuik, Hirschmann, and
  Marsa}]{Choptuik:1999gh}
Choptuik MW, Hirschmann EW, Marsa RL (1999) New critical behavior in
  {E}instein-{Y}ang-{M}ills collapse. Phys Rev D 60:124011.
  \doi{10.1103/PhysRevD.60.124011}.
  {\href{https://arxiv.org/abs/gr-qc/9903081}{{arXiv:gr-qc/9903081}}}

\bibitem[{Choptuik et~al.(2015)Choptuik, Lehner, and
  Pretorius}]{Choptuik:2015mma}
Choptuik MW, Lehner L, Pretorius F (2015) Probing strong-field gravity through
  numerical simulations. In: Ashtekar A, Berger BK, Isenberg J, MacCallum M
  (eds) General Relativity and Gravitation: A Centennial Perspective. Cambridge
  University Press, Cambridge, pp 361--411. \doi{10.1017/CBO9781139583961.011}.
  {\href{https://arxiv.org/abs/1502.06853}{{arXiv:1502.06853}}} {[gr-qc]}

\bibitem[{{Chru{\'s}ciel} et~al.(2012){Chru{\'s}ciel}, {Costa}, and
  {Heusler}}]{lrr-2012-7}
{Chru{\'s}ciel} PT, {Costa} JL, {Heusler} M (2012) Stationary black holes:
  Uniqueness and beyond. Living Rev Relativity 15:7. \doi{10.12942/lrr-2012-7}.
  {\href{https://arxiv.org/abs/1205.6112}{{arXiv:1205.6112}}}

\bibitem[{Coleman(1985)}]{Coleman:1985ki}
Coleman SR (1985) {Q}-balls. Nucl Phys B 262:263--283.
  \doi{10.1016/0550-3213(85)90286-X}

\bibitem[{Collodel and Doneva(2022)}]{Collodel:2022jly}
Collodel LG, Doneva DD (2022) {Solitonic Boson Stars: Numerical solutions
  beyond the thin-wall approximation}. arXiv e-prints
  {\href{https://arxiv.org/abs/2203.08203}{{arXiv:2203.08203}}} {[gr-qc]}

\bibitem[{Collodel et~al.(2019)Collodel, Kleihaus, and Kunz}]{Collodel:2019ohy}
Collodel LG, Kleihaus B, Kunz J (2019) {Structure of rotating charged boson
  stars}. Phys Rev D 99:104076. \doi{10.1103/PhysRevD.99.104076}.
  {\href{https://arxiv.org/abs/1901.11522}{{arXiv:1901.11522}}} {[gr-qc]}

\bibitem[{Collodel et~al.(2020)Collodel, Doneva, and
  Yazadjiev}]{Collodel:2019uns}
Collodel LG, Doneva DD, Yazadjiev SS (2020) {Rotating tensor-multiscalar
  solitons}. Phys Rev D 101:044021. \doi{10.1103/PhysRevD.101.044021}.
  {\href{https://arxiv.org/abs/1912.02498}{{arXiv:1912.02498}}} {[gr-qc]}

\bibitem[{Colpi et~al.(1986)Colpi, Shapiro, and
  Wasserman}]{1986PhRvL..57.2485C}
Colpi M, Shapiro SL, Wasserman I (1986) Boson stars: Gravitational equilibria
  of self-interacting scalar fields. Phys Rev Lett 57:2485--2488.
  \doi{10.1103/PhysRevLett.57.2485}

\bibitem[{Contaldi et~al.(2008)Contaldi, Wiseman, and
  Withers}]{2008PhRvD..78d4034C}
Contaldi CR, Wiseman T, Withers B (2008) Teves gets caught on caustics. Phys
  Rev D 78:044034. \doi{10.1103/PhysRevD.78.044034}.
  {\href{https://arxiv.org/abs/0802.1215}{{arXiv:0802.1215}}} {[gr-qc]}

\bibitem[{Cook(2000)}]{cook_living_review}
Cook GB (2000) Initial data for numerical relativity. Living Rev Relativity
  3:5. \doi{10.12942/lrr-2000-5}.
  {\href{https://arxiv.org/abs/gr-qc/0007085}{{arXiv:gr-qc/0007085}}} {[gr-qc]}

\bibitem[{Cook et~al.(1994)Cook, Shapiro, and Teukolsky}]{cst94}
Cook GB, Shapiro SL, Teukolsky SA (1994) Rapidly rotating neutron stars in
  general relativity: Realistic equations of state. Astrophys J 424:823--845.
  \doi{10.1086/173934}

\bibitem[{Cotner(2016)}]{Cotner:2016aaq}
Cotner E (2016) Collisional interactions between self-interacting
  nonrelativistic boson stars: Effective potential analysis and numerical
  simulations. Phys Rev D 94:063503. \doi{10.1103/PhysRevD.94.063503}.
  {\href{https://arxiv.org/abs/1608.00547}{{arXiv:1608.00547}}} {[astro-ph.CO]}

\bibitem[{Croon et~al.(2018)Croon, Gleiser, Mohapatra, and Sun}]{Croon:2018ftb}
Croon D, Gleiser M, Mohapatra S, Sun C (2018) {Gravitational Radiation
  Background from Boson Star Binaries}. Phys Lett B 783:158--162.
  \doi{10.1016/j.physletb.2018.03.055}.
  {\href{https://arxiv.org/abs/1802.08259}{{arXiv:1802.08259}}} {[hep-ph]}

\bibitem[{Cunha et~al.(2015)Cunha, Herdeiro, Radu, and
  Runarsson}]{Cunha:2015yba}
Cunha PVP, Herdeiro CAR, Radu E, Runarsson HF (2015) Shadows of {K}err black
  holes with scalar hair. Phys Rev Lett 115:211102.
  \doi{10.1103/PhysRevLett.115.211102}.
  {\href{https://arxiv.org/abs/1509.00021}{{arXiv:1509.00021}}} {[gr-qc]}

\bibitem[{Cunha et~al.(2022)Cunha, Herdeiro, Radu, and
  Sanchis-Gual}]{Cunha:2022gde}
Cunha PVP, Herdeiro C, Radu E, Sanchis-Gual N (2022) {The fate of the
  light-ring instability}. arXiv e-prints
  {\href{https://arxiv.org/abs/2207.13713}{{arXiv:2207.13713}}} {[gr-qc]}

\bibitem[{Dafermos et~al.(2014)Dafermos, Rodnianski, and
  Shlapentokh-Rothman}]{Dafermos:2014jwa}
Dafermos M, Rodnianski I, Shlapentokh-Rothman Y (2014) A scattering theory for
  the wave equation on {K}err black hole exteriors. ArXiv e-prints
  {\href{https://arxiv.org/abs/1412.8379}{{arXiv:1412.8379}}} {[gr-qc]}

\bibitem[{Damour(1987)}]{Hawking:1987en}
Damour T (1987) The problem of motion in {N}ewtonian and {E}insteinian gravity.
  In: Hawking SW, Israel W (eds) Three Hundred Years of Gravitation. Cambridge
  University Press, Cambridge; New York, pp 128--198

\bibitem[{Damour and Esposito-Far{\`{e}}se(1996)}]{Damour:1996ke}
Damour T, Esposito-Far{\`{e}}se G (1996) Tensor-scalar gravity and
  binary-pulsar experiments. Phys Rev D 54:1474--1491.
  \doi{10.1103/PhysRevD.54.1474}.
  {\href{https://arxiv.org/abs/gr-qc/9602056}{{gr-qc/9602056}}}

\bibitem[{Danzmann(2017)}]{newlisa}
Danzmann Kea (2017) {LISA}: {L}aser {I}nterferometer {S}pace {A}ntenna. a
  proposal in response to the {ESA} call for {L3} mission concepts. Tech. rep.,
  Max Planck Institute for Gravitational Physics (Albert Einstein Institute),
  Potsdam.
  \urlprefix\url{https://www.elisascience.org/files/publications/LISA_L3_20170120.pdf}

\bibitem[{Dariescu and Dariescu(2010)}]{2010ChPhL..27a1101C}
Dariescu C, Dariescu MA (2010) Boson nebulae charge. Chinese Phys Lett
  27:011101. \doi{10.1088/0256-307X/27/1/011101}

\bibitem[{Davidson and Schwetz(2016)}]{Davidson:2016uok}
Davidson S, Schwetz T (2016) Rotating drops of axion dark matter. Phys Rev D
  93:123509. \doi{10.1103/PhysRevD.93.123509}.
  {\href{https://arxiv.org/abs/1603.04249}{{arXiv:1603.04249}}} {[astro-ph.CO]}

\bibitem[{Degura et~al.(2001)Degura, Sakamoto, and Shiraishi}]{Degura:1998hw}
Degura Y, Sakamoto K, Shiraishi K (2001) Black holes with scalar hair in
  (2+1)-dimensions. Grav Cosmol 7:153--158.
  {\href{https://arxiv.org/abs/gr-qc/9805011}{{arXiv:gr-qc/9805011}}} {[gr-qc]}

\bibitem[{Delgado et~al.(2016)Delgado, Herdeiro, Radu, and
  Runarsson}]{Delgado:2016jxq}
Delgado JFM, Herdeiro CAR, Radu E, Runarsson H (2016) {K}err--{N}ewman black
  holes with scalar hair. Phys Lett B 761:234--241.
  \doi{10.1016/j.physletb.2016.08.032}.
  {\href{https://arxiv.org/abs/1608.00631}{{arXiv:1608.00631}}} {[gr-qc]}

\bibitem[{Delgado et~al.(2020)Delgado, Herdeiro, and Radu}]{Delgado:2020udb}
Delgado JFM, Herdeiro CAR, Radu E (2020) {Rotating Axion Boson Stars}. JCAP
  06:037. \doi{10.1088/1475-7516/2020/06/037}.
  {\href{https://arxiv.org/abs/2005.05982}{{arXiv:2005.05982}}} {[gr-qc]}

\bibitem[{Della~Monica and de~Martino(2022)}]{DellaMonica:2022kow}
Della~Monica R, de~Martino I (2022) {Shutting the allowed mass range of the
  ultralight bosons with S2 star}. arXiv e-prints
  {\href{https://arxiv.org/abs/2206.03980}{{arXiv:2206.03980}}} {[gr-qc]}

\bibitem[{Derrick(1964)}]{Derrick:1964ww}
Derrick GH (1964) Comments on nonlinear wave equations as models for elementary
  particles. J Math Phys 5:1252--1254. \doi{10.1063/1.1704233}

\bibitem[{Di~Giovanni et~al.(2018)Di~Giovanni, Sanchis-Gual, Herdeiro, and
  Font}]{DiGiovanni:2018bvo}
Di~Giovanni F, Sanchis-Gual N, Herdeiro CAR, Font JA (2018) {Dynamical
  formation of Proca stars and quasistationary solitonic objects}. Phys Rev D
  98(6):064044. \doi{10.1103/PhysRevD.98.064044}.
  {\href{https://arxiv.org/abs/1803.04802}{{arXiv:1803.04802}}} {[gr-qc]}

\bibitem[{Di~Giovanni et~al.(2020)Di~Giovanni, Fakhry, Sanchis-Gual, Degollado,
  and Font}]{DiGiovanni:2020frc}
Di~Giovanni F, Fakhry S, Sanchis-Gual N, Degollado JC, Font JA (2020)
  {Dynamical formation and stability of fermion-boson stars}. Phys Rev D
  102(8):084063. \doi{10.1103/PhysRevD.102.084063}.
  {\href{https://arxiv.org/abs/2006.08583}{{arXiv:2006.08583}}} {[gr-qc]}

\bibitem[{{Di Giovanni} et~al.(2020){Di Giovanni}, Sanchis-Gual,
  Cerd\'a-Dur\'an, Zilh\~ao, Herdeiro, Font, and Radu}]{DiGiovanni:2020ror}
{Di Giovanni} F, Sanchis-Gual N, Cerd\'a-Dur\'an P, Zilh\~ao M, Herdeiro C,
  Font JA, Radu E (2020) {Dynamical bar-mode instability in spinning bosonic
  stars}. Phys Rev D 102:124009. \doi{10.1103/PhysRevD.102.124009}.
  {\href{https://arxiv.org/abs/2010.05845}{{arXiv:2010.05845}}} {[gr-qc]}

\bibitem[{Di~Giovanni et~al.(2021)Di~Giovanni, Fakhry, Sanchis-Gual, Degollado,
  and Font}]{DiGiovanni:2021vlu}
Di~Giovanni F, Fakhry S, Sanchis-Gual N, Degollado JC, Font JA (2021) {A
  stabilization mechanism for excited fermion\textendash{}boson stars}. Class
  Quantum Grav 38(19):194001. \doi{10.1088/1361-6382/ac1b45}.
  {\href{https://arxiv.org/abs/2105.00530}{{arXiv:2105.00530}}} {[gr-qc]}

\bibitem[{Di~Giovanni et~al.(2022{\natexlab{a}})Di~Giovanni, Sanchis-Gual,
  Cerd\'a-Dur\'an, and Font}]{DiGiovanni:2021ejn}
Di~Giovanni F, Sanchis-Gual N, Cerd\'a-Dur\'an P, Font JA (2022{\natexlab{a}})
  {Can fermion-boson stars reconcile multimessenger observations of compact
  stars?} Phys Rev D 105(6):063005. \doi{10.1103/PhysRevD.105.063005}.
  {\href{https://arxiv.org/abs/2110.11997}{{arXiv:2110.11997}}} {[gr-qc]}

\bibitem[{Di~Giovanni et~al.(2022{\natexlab{b}})Di~Giovanni, Sanchis-Gual,
  Guerra, Miravet-Ten\'es, Cerd\'a-Dur\'an, and Font}]{DiGiovanni:2022mkn}
Di~Giovanni F, Sanchis-Gual N, Guerra D, Miravet-Ten\'es M, Cerd\'a-Dur\'an P,
  Font JA (2022{\natexlab{b}}) {Impact of ultralight bosonic dark matter on the
  dynamical bar-mode instability of rotating neutron stars}. Phys Rev D
  106(4):044008. \doi{10.1103/PhysRevD.106.044008}.
  {\href{https://arxiv.org/abs/2206.00977}{{arXiv:2206.00977}}} {[gr-qc]}

\bibitem[{Dias and Masachs(2017)}]{Dias:2016pma}
Dias {\'{O}}JC, Masachs R (2017) Hairy black holes and the endpoint of
  {AdS}$_4$ charged superradiance. J High Energy Phys 2017(02):128.
  \doi{10.1007/JHEP02(2017)128}.
  {\href{https://arxiv.org/abs/1610.03496}{{arXiv:1610.03496}}} {[hep-th]}

\bibitem[{Dias et~al.(2011)Dias, Horowitz, and Santos}]{Dias:2011at}
Dias {\'{O}}JC, Horowitz GT, Santos JE (2011) Black holes with only one
  {K}illing field. J High Energy Phys 2011(07):115.
  \doi{10.1007/JHEP07(2011)115}.
  {\href{https://arxiv.org/abs/1105.4167}{{arXiv:1105.4167}}} {[hep-th]}

\bibitem[{{Dias} et~al.(2012){Dias}, {Horowitz}, {Marolf}, and
  {Santos}}]{2012CQGra..29w5019D}
{Dias} {\'O}JC, {Horowitz} GT, {Marolf} D, {Santos} JE (2012) On the nonlinear
  stability of asymptotically anti-de {S}itter solutions. Class Quantum Grav
  29:235019. \doi{10.1088/0264-9381/29/23/235019}.
  {\href{https://arxiv.org/abs/1208.5772}{{arXiv:1208.5772}}} {[gr-qc]}

\bibitem[{Dias et~al.(2012)Dias, Horowitz, and Santos}]{Dias:2011ss}
Dias {\'{O}}JC, Horowitz GT, Santos JE (2012) Gravitational turbulent
  instability of anti-de {S}itter space. Class Quantum Grav 29:194002.
  \doi{10.1088/0264-9381/29/19/194002}.
  {\href{https://arxiv.org/abs/1109.1825}{{arXiv:1109.1825}}} {[hep-th]}

\bibitem[{Dias et~al.(2016)Dias, Santos, and Way}]{Dias:2015nua}
Dias {\'{O}}JC, Santos JE, Way B (2016) Numerical methods for finding
  stationary gravitational solutions. Class Quantum Grav 33:133001.
  \doi{10.1088/0264-9381/33/13/133001}.
  {\href{https://arxiv.org/abs/1510.02804}{{arXiv:1510.02804}}} {[hep-th]}

\bibitem[{Dias et~al.(2021)Dias, Masachs, and Rodgers}]{Dias:2021acy}
Dias OJC, Masachs R, Rodgers P (2021) {Boson stars and solitons confined in a
  Minkowski box}. JHEP 04:236. \doi{10.1007/JHEP04(2021)236}.
  {\href{https://arxiv.org/abs/2101.01203}{{arXiv:2101.01203}}} {[gr-qc]}

\bibitem[{{Diemer} et~al.(2013){Diemer}, {Eilers}, {Hartmann}, {Schaffer}, and
  {Toma}}]{2013PhRvD..88d4025D}
{Diemer} V, {Eilers} K, {Hartmann} B, {Schaffer} I, {Toma} C (2013) Geodesic
  motion in the space-time of a noncompact boson star. Phys Rev D 88:044025.
  \doi{10.1103/PhysRevD.88.044025}.
  {\href{https://arxiv.org/abs/1304.5646}{{arXiv:1304.5646}}} {[gr-qc]}

\bibitem[{Dietrich et~al.(2019)Dietrich, Ossokine, and
  Clough}]{Dietrich:2018bvi}
Dietrich T, Ossokine S, Clough K (2019) {Full 3D numerical relativity
  simulations of neutron star\textendash{}boson star collisions with BAM}.
  Class Quantum Grav 36(2):025002. \doi{10.1088/1361-6382/aaf43e}.
  {\href{https://arxiv.org/abs/1807.06959}{{arXiv:1807.06959}}} {[gr-qc]}

\bibitem[{{Diez-Tejedor} and {Gonzalez-Morales}(2013)}]{2013PhRvD..88f7302D}
{Diez-Tejedor} A, {Gonzalez-Morales} AX (2013) No-go theorem for static scalar
  field dark matter halos with no noether charges. Phys Rev D 88:067302.
  \doi{10.1103/PhysRevD.88.067302}.
  {\href{https://arxiv.org/abs/1306.4400}{{arXiv:1306.4400}}} {[gr-qc]}

\bibitem[{Dmitriev et~al.(2021)Dmitriev, Levkov, Panin, Pushnaya, and
  Tkachev}]{Dmitriev:2021utv}
Dmitriev AS, Levkov DG, Panin AG, Pushnaya EK, Tkachev II (2021) {Instability
  of rotating Bose stars}. Phys Rev D 104:023504.
  \doi{10.1103/PhysRevD.104.023504}.
  {\href{https://arxiv.org/abs/2104.00962}{{arXiv:2104.00962}}} {[gr-qc]}

\bibitem[{Doddato and McDonald(2012)}]{Doddato:2011hx}
Doddato F, McDonald J (2012) New {Q}-ball solutions in gauge-mediation,
  {A}ffleck-{D}ine baryogenesis and gravitino dark matter. J Cosmol Astropart
  Phys 2012(06):031. \doi{10.1088/1475-7516/2012/06/031}.
  {\href{https://arxiv.org/abs/1111.2305}{{arXiv:1111.2305}}} {[hep-ph]}

\bibitem[{Doneva and Yazadjiev(2020)}]{Doneva:2019krb}
Doneva DD, Yazadjiev SS (2020) {Mixed configurations of tensor-multiscalar
  solitons and neutron stars}. Phys Rev D 101(2):024009.
  \doi{10.1103/PhysRevD.101.024009}.
  {\href{https://arxiv.org/abs/1909.00473}{{arXiv:1909.00473}}} {[gr-qc]}

\bibitem[{Duarte and Brito(2016)}]{Duarte:2016lig}
Duarte M, Brito R (2016) Asymptotically anti-de {S}itter {P}roca stars. Phys
  Rev D 94:064055. \doi{10.1103/PhysRevD.94.064055}.
  {\href{https://arxiv.org/abs/1609.01735}{{arXiv:1609.01735}}} {[gr-qc]}

\bibitem[{Dymnikova et~al.(2000)Dymnikova, Koziel, Khlopov, and
  Rubin}]{Dymnikova:2000dy}
Dymnikova I, Koziel L, Khlopov M, Rubin S (2000) Quasilumps from first order
  phase transitions. Grav Cosmol 6:311--318.
  {\href{https://arxiv.org/abs/hep-th/0010120}{{arXiv:hep-th/0010120}}}
  {[hep-th]}

\bibitem[{Dzhunushaliev et~al.(2007)Dzhunushaliev, Myrzakulov, and
  Myrzakulov}]{2007MPLA...22..273D}
Dzhunushaliev V, Myrzakulov K, Myrzakulov R (2007) Boson stars from a gauge
  condensate. Mod Phys Lett A 22:273--281. \doi{10.1142/S0217732307022669}.
  {\href{https://arxiv.org/abs/gr-qc/0604110}{{arXiv:gr-qc/0604110}}}

\bibitem[{Dzhunushaliev et~al.(2008)Dzhunushaliev, Folomeev, Myrzakulov, and
  Singleton}]{2008JHEP...07..094D}
Dzhunushaliev V, Folomeev V, Myrzakulov R, Singleton D (2008) Non-singular
  solutions to {E}instein-{K}lein-{G}ordon equations with a phantom scalar
  field. J High Energy Phys 2008(07):094. \doi{10.1088/1126-6708/2008/07/094}.
  {\href{https://arxiv.org/abs/0805.3211}{{arXiv:0805.3211}}} {[gr-qc]}

\bibitem[{Dzhunushaliev et~al.(2011)Dzhunushaliev, Folomeev, and
  Singleton}]{Dzhunushaliev:2011ma}
Dzhunushaliev V, Folomeev V, Singleton D (2011) Chameleon stars. Phys Rev D
  84:084025. \doi{10.1103/PhysRevD.84.084025}.
  {\href{https://arxiv.org/abs/1106.1267}{{arXiv:1106.1267}}} {[astro-ph.SR]}

\bibitem[{{Dzhunushaliev} et~al.(2014){Dzhunushaliev}, {Folomeev}, {Hoffmann},
  {Kleihaus}, and {Kunz}}]{2014PhRvD..90l4038D}
{Dzhunushaliev} V, {Folomeev} V, {Hoffmann} C, {Kleihaus} B, {Kunz} J (2014)
  Boson stars with nontrivial topology. Phys Rev D 90:124038.
  \doi{10.1103/PhysRevD.90.124038}.
  {\href{https://arxiv.org/abs/1409.6978}{{arXiv:1409.6978}}} {[gr-qc]}

\bibitem[{East and Pretorius(2017)}]{East:2017ovw}
East WE, Pretorius F (2017) {Superradiant Instability and Backreaction of
  Massive Vector Fields around Kerr Black Holes}. Phys Rev Lett 119:041101.
  \doi{10.1103/PhysRevLett.119.041101}.
  {\href{https://arxiv.org/abs/1704.04791}{{arXiv:1704.04791}}} {[gr-qc]}

\bibitem[{Eby et~al.(2016)Eby, Kouvaris, Nielsen, and
  Wijewardhana}]{Eby:2015hsq}
Eby J, Kouvaris C, Nielsen NG, Wijewardhana LCR (2016) Boson stars from
  self-interacting dark matter. J High Energy Phys 2016(02):028.
  \doi{10.1007/JHEP02(2016)028}.
  {\href{https://arxiv.org/abs/1511.04474}{{arXiv:1511.04474}}} {[hep-ph]}

\bibitem[{Eckart et~al.(2017)Eckart, H{\"u}ttemann, Kiefer, Britzen,
  Zaja{\v{c}}ek, L{\"a}mmerzahl, St{\"o}ckler, Valencia-S, Karas, and
  Garc{\'i}a-Mar{\'i}n}]{Eckart:2017bhq}
Eckart A, H{\"u}ttemann A, Kiefer C, Britzen S, Zaja{\v{c}}ek M, L{\"a}mmerzahl
  C, St{\"o}ckler M, Valencia-S M, Karas V, Garc{\'i}a-Mar{\'i}n M (2017) The
  {M}ilky {W}ay's supermassive black hole: How good a case is it? a challenge
  for astrophysics \& philosophy of science. Found Phys 47:553--624.
  \doi{10.1007/s10701-017-0079-2}.
  {\href{https://arxiv.org/abs/1703.09118}{{arXiv:1703.09118}}} {[astro-ph.HE]}

\bibitem[{Emparan and Reall(2008)}]{lrr-2008-6}
Emparan R, Reall HS (2008) Black holes in higher dimensions. Living Rev
  Relativity 11:6. \doi{10.12942/lrr-2008-6}.
  {\href{https://arxiv.org/abs/0801.3471}{{arXiv:0801.3471}}} {[hep-th]}

\bibitem[{Eto et~al.(2011)Eto, Hashimoto, Iida, and Miwa}]{Eto:2010vi}
Eto M, Hashimoto K, Iida H, Miwa A (2011) Chiral magnetic effect from
  {Q}-balls. Phys Rev D 83:125033. \doi{10.1103/PhysRevD.83.125033}.
  {\href{https://arxiv.org/abs/1012.3264}{{arXiv:1012.3264}}} {[hep-ph]}

\bibitem[{Famaey and {McGaugh}(2012)}]{Famaey:2011kh}
Famaey B, {McGaugh} SS (2012) {M}odified {N}ewtonian {D}ynamics (mond):
  Observational phenomenology and relativistic extensions. Living Rev
  Relativity 15:10. \doi{10.12942/lrr-2012-10}.
  {\href{https://arxiv.org/abs/1112.3960}{{arXiv:1112.3960}}} {[astro-ph.CO]}

\bibitem[{{Fan} et~al.(2012){Fan}, {Yang}, and {Chang}}]{2012arXiv1204.2564F}
{Fan} Yz, {Yang} Rz, {Chang} J (2012) Constraining asymmetric bosonic
  non-interacting dark matter with neutron stars. ArXiv e-prints
  {\href{https://arxiv.org/abs/1204.2564}{{arXiv:1204.2564}}} {[astro-ph.HE]}

\bibitem[{Faraoni(2012)}]{Faraoni:2012hn}
Faraoni V (2012) Correspondence between a scalar field and an effective perfect
  fluid. Phys Rev D 85:024040. \doi{10.1103/PhysRevD.85.024040}.
  {\href{https://arxiv.org/abs/1201.1448}{{arXiv:1201.1448}}} {[gr-qc]}

\bibitem[{Feng(2010)}]{Feng:2010gw}
Feng JL (2010) Dark matter candidates from particle physics and methods of
  detection. Annu Rev Astron Astrophys 48:495--545.
  \doi{10.1146/annurev-astro-082708-101659}.
  {\href{https://arxiv.org/abs/1003.0904}{{arXiv:1003.0904}}} {[astro-ph.CO]}

\bibitem[{Fodor et~al.(2008)Fodor, Forg{\'{a}}cs, Horv{\'{a}}th, and
  Lukacs}]{Fodor:2008es}
Fodor G, Forg{\'{a}}cs P, Horv{\'{a}}th Z, Lukacs A (2008) Small amplitude
  quasi-breathers and oscillons. Phys Rev D 78:025003.
  \doi{10.1103/PhysRevD.78.025003}.
  {\href{https://arxiv.org/abs/0802.3525}{{arXiv:0802.3525}}} {[hep-th]}

\bibitem[{Fodor et~al.(2009{\natexlab{a}})Fodor, Forg{\'{a}}cs, Horv{\'{a}}th,
  and Mezei}]{Fodor:2008du}
Fodor G, Forg{\'{a}}cs P, Horv{\'{a}}th Z, Mezei M (2009{\natexlab{a}})
  Computation of the radiation amplitude of oscillons. Phys Rev D 79:065002.
  \doi{10.1103/PhysRevD.79.065002}.
  {\href{https://arxiv.org/abs/0812.1919}{{arXiv:0812.1919}}} {[hep-th]}

\bibitem[{Fodor et~al.(2009{\natexlab{b}})Fodor, Forg{\'{a}}cs, Horv{\'{a}}th,
  and Mezei}]{Fodor:2009xw}
Fodor G, Forg{\'{a}}cs P, Horv{\'{a}}th Z, Mezei M (2009{\natexlab{b}})
  Oscillons in dilaton-scalar theories. J High Energy Phys 2009(08):106.
  \doi{10.1088/1126-6708/2009/08/106}.
  {\href{https://arxiv.org/abs/0906.4160}{{arXiv:0906.4160}}} {[hep-th]}

\bibitem[{Fodor et~al.(2009{\natexlab{c}})Fodor, Forg{\'{a}}cs, Horv{\'{a}}th,
  and Mezei}]{Fodor:2009kf}
Fodor G, Forg{\'{a}}cs P, Horv{\'{a}}th Z, Mezei M (2009{\natexlab{c}})
  Radiation of scalar oscillons in 2 and 3 dimensions. Phys Lett B
  674:319--324. \doi{10.1016/j.physletb.2009.03.054}.
  {\href{https://arxiv.org/abs/0903.0953}{{arXiv:0903.0953}}} {[hep-th]}

\bibitem[{Fodor et~al.(2010{\natexlab{a}})Fodor, Forg{\'{a}}cs, and
  Mezei}]{2010PhRvD..82d4043F}
Fodor G, Forg{\'{a}}cs P, Mezei M (2010{\natexlab{a}}) Boson stars and
  oscillatons in an inflationary universe. Phys Rev D 82:044043.
  \doi{10.1103/PhysRevD.82.044043}.
  {\href{https://arxiv.org/abs/1007.0388}{{arXiv:1007.0388}}} {[gr-qc]}

\bibitem[{Fodor et~al.(2010{\natexlab{b}})Fodor, Forgacs, and
  Mezei}]{fodor-2010-81}
Fodor G, Forgacs P, Mezei M (2010{\natexlab{b}}) Mass loss and longevity of
  gravitationally bound oscillating scalar lumps (oscillatons) in d-dimensions.
  Phys Rev D 81:064029. \urlprefix\url{doi:10.1103/PhysRevD.81.064029}

\bibitem[{Fodor et~al.(2010{\natexlab{c}})Fodor, Forg{\'{a}}cs, and
  Mezei}]{Fodor:2009kg}
Fodor G, Forg{\'{a}}cs P, Mezei M (2010{\natexlab{c}}) Mass loss and longevity
  of gravitationally bound oscillating scalar lumps (oscillatons) in
  {D}-dimensions. Phys Rev D 81:064029. \doi{10.1103/PhysRevD.81.064029}.
  {\href{https://arxiv.org/abs/0912.5351}{{arXiv:0912.5351}}} {[gr-qc]}

\bibitem[{Fodor et~al.(2015)Fodor, Forg{\'a}cs, and
  Grandcl{\'e}ment}]{Fodor:2015eia}
Fodor G, Forg{\'a}cs P, Grandcl{\'e}ment P (2015) Self-gravitating scalar
  breathers with negative cosmological constant. Phys Rev D 92:025036.
  \doi{10.1103/PhysRevD.92.025036}.
  {\href{https://arxiv.org/abs/1503.07746}{{arXiv:1503.07746}}} {[gr-qc]}

\bibitem[{Font(2008)}]{lrr-2008-7}
Font JA (2008) Numerical hydrodynamics and magnetohydrodynamics in general
  relativity. Living Rev Relativity 11:7. \doi{10.12942/lrr-2008-7}

\bibitem[{Font et~al.(2002)Font, Goodale, Iyer, Miller, Rezzolla, Seidel,
  Stergioulas, Suen, and Tobias}]{Font:2001ew}
Font JA, Goodale T, Iyer S, Miller M, Rezzolla L, Seidel E, Stergioulas N, Suen
  WM, Tobias M (2002) Three-dimensional numerical general relativistic
  hydrodynamics. {II}. long-term dynamics of single relativistic stars. Phys
  Rev D 65:084024. \doi{10.1103/PhysRevD.65.084024}.
  {\href{https://arxiv.org/abs/gr-qc/0110047}{{arXiv:gr-qc/0110047}}}

\bibitem[{Franchini et~al.(2017)Franchini, Pani, Maselli, Gualtieri, Herdeiro,
  Radu, and Ferrari}]{Franchini:2016yvq}
Franchini N, Pani P, Maselli A, Gualtieri L, Herdeiro CAR, Radu E, Ferrari V
  (2017) Constraining black holes with light boson hair and boson stars using
  epicyclic frequencies and quasiperiodic oscillations. Phys Rev D 95:124025.
  \doi{10.1103/PhysRevD.95.124025}.
  {\href{https://arxiv.org/abs/1612.00038}{{arXiv:1612.00038}}} {[astro-ph.HE]}

\bibitem[{Frank and Lenzmann(2009{\natexlab{a}})}]{2009arXiv0910.2721F}
Frank RL, Lenzmann E (2009{\natexlab{a}}) On ground states for the
  $l^2$-critical boson star equation. ArXiv e-prints
  {\href{https://arxiv.org/abs/0910.2721}{{arXiv:0910.2721}}} {[math.AP]}

\bibitem[{Frank and Lenzmann(2009{\natexlab{b}})}]{2009arXiv0905.3105F}
Frank RL, Lenzmann E (2009{\natexlab{b}}) Uniqueness of ground states for the
  $l^2$-critical boson star equation. ArXiv e-prints
  {\href{https://arxiv.org/abs/0905.3105}{{arXiv:0905.3105}}} {[math.AP]}

\bibitem[{Friedberg et~al.(1987{\natexlab{a}})Friedberg, Lee, and
  Pang}]{1987PhRvD..35.3640F}
Friedberg R, Lee TD, Pang Y (1987{\natexlab{a}}) Mini-soliton stars. Phys Rev D
  35:3640--3657. \doi{10.1103/PhysRevD.35.3640}

\bibitem[{Friedberg et~al.(1987{\natexlab{b}})Friedberg, Lee, and
  Pang}]{1987PhRvD..35.3658F}
Friedberg R, Lee TD, Pang Y (1987{\natexlab{b}}) Scalar soliton stars and black
  holes. Phys Rev D 35:3658--3677. \doi{10.1103/PhysRevD.35.3658}

\bibitem[{Friedman et~al.(1988)Friedman, Ipser, and Sorkin}]{sorkin}
Friedman JL, Ipser JR, Sorkin RD (1988) Turning-point method for axisymmetric
  stability of rotating relativistic stars. Astrophys J 325:722--724.
  \doi{10.1086/166043}

\bibitem[{Garani et~al.(2022)Garani, Levkov, and Tinyakov}]{Garani:2021gvc}
Garani R, Levkov D, Tinyakov P (2022) {Solar mass black holes from neutron
  stars and bosonic dark matter}. Phys Rev D 105(6):063019.
  \doi{10.1103/PhysRevD.105.063019}.
  {\href{https://arxiv.org/abs/2112.09716}{{arXiv:2112.09716}}} {[hep-ph]}

\bibitem[{Garfinkle and Isenberg(2003)}]{Garfinkle:2003an}
Garfinkle D, Isenberg J (2003) {Critical behavior in Ricci flow}. arXiv
  e-prints {\href{https://arxiv.org/abs/math/0306129}{{arXiv:math/0306129}}}

\bibitem[{Gentle et~al.(2012)Gentle, Rangamani, and Withers}]{Gentle:2011kv}
Gentle SA, Rangamani M, Withers B (2012) A soliton menagerie in {AdS}. J High
  Energy Phys 2012(05):106. \doi{10.1007/JHEP05(2012)106}.
  {\href{https://arxiv.org/abs/1112.3979}{{arXiv:1112.3979}}} {[hep-th]}

\bibitem[{Gervalle(2022)}]{Gervalle:2022fze}
Gervalle R (2022) {Chains of rotating boson stars}. Phys Rev D 105:124052.
  \doi{10.1103/PhysRevD.105.124052}.
  {\href{https://arxiv.org/abs/2206.03982}{{arXiv:2206.03982}}} {[gr-qc]}

\bibitem[{{Giangrandi} et~al.(2022){Giangrandi}, {Sagun}, {Ivanytskyi},
  {Provid{\^e}ncia}, and {Dietrich}}]{2022arXiv220910905G}
{Giangrandi} E, {Sagun} V, {Ivanytskyi} O, {Provid{\^e}ncia} C, {Dietrich} T
  (2022) {The effects of self-interacting bosonic dark matter on neutron star
  properties}. arXiv e-prints
  {\href{https://arxiv.org/abs/2209.10905}{{arXiv:2209.10905}}} {[astro-ph.HE]}

\bibitem[{Giudice et~al.(2016)Giudice, McCullough, and
  Urbano}]{Giudice:2016zpa}
Giudice GF, McCullough M, Urbano A (2016) Hunting for dark particles with
  gravitational waves. J Cosmol Astropart Phys 2016(10):001.
  \doi{10.1088/1475-7516/2016/10/001}.
  {\href{https://arxiv.org/abs/1605.01209}{{arXiv:1605.01209}}} {[hep-ph]}

\bibitem[{Gleiser(1988)}]{1988PhRvD..38.2376G}
Gleiser M (1988) Stability of boson stars. Phys Rev D 38:2376--2385.
  \doi{10.1103/PhysRevD.38.2376}

\bibitem[{Gleiser and Jiang(2015)}]{Gleiser:2015rwa}
Gleiser M, Jiang N (2015) Stability bounds on compact astrophysical objects
  from information-entropic measure. Phys Rev D 92:044046.
  \doi{10.1103/PhysRevD.92.044046}.
  {\href{https://arxiv.org/abs/1506.05722}{{arXiv:1506.05722}}} {[gr-qc]}

\bibitem[{Gleiser and Watkins(1989)}]{1989NuPhB.319..733G}
Gleiser M, Watkins R (1989) Gravitational stability of scalar matter. Nucl Phys
  B 319:733--746. \doi{10.1016/0550-3213(89)90627-5}

\bibitem[{Goetz(2015{\natexlab{a}})}]{Goetz:2015qoa}
Goetz AS (2015{\natexlab{a}}) The {E}instein-{K}lein-{G}ordon equations, wave
  dark matter, and the {T}ully-{F}isher relation. PhD thesis, Duke University.
  {\href{https://arxiv.org/abs/1507.02626}{{arXiv:1507.02626}}} {[gr-qc]}

\bibitem[{Goetz(2015{\natexlab{b}})}]{Goetz:2015lwa}
Goetz AS (2015{\natexlab{b}}) {T}ully-{F}isher scalings and boundary conditions
  for wave dark matter. ArXiv e-prints
  {\href{https://arxiv.org/abs/1502.04976}{{arXiv:1502.04976}}} {[astro-ph.GA]}

\bibitem[{Gonz{\'{a}}lez and Guzm{\'{a}}n(2011)}]{2011PhRvD..83j3513G}
Gonz{\'{a}}lez JA, Guzm{\'{a}}n FS (2011) Interference pattern in the collision
  of structures in the {B}ose-{E}instein condensate dark matter model:
  Comparison with fluids. Phys Rev D 83:103513.
  \doi{10.1103/PhysRevD.83.103513}.
  {\href{https://arxiv.org/abs/1105.2066}{{arXiv:1105.2066}}} {[astro-ph.CO]}

\bibitem[{Gorghetto et~al.(2022)Gorghetto, Hardy, March-Russell, Song, and
  West}]{Gorghetto:2022sue}
Gorghetto M, Hardy E, March-Russell J, Song N, West SM (2022) {Dark photon
  stars: formation and role as dark matter substructure}. JCAP 08:018.
  \doi{10.1088/1475-7516/2022/08/018}.
  {\href{https://arxiv.org/abs/2203.10100}{{arXiv:2203.10100}}} {[hep-ph]}

\bibitem[{Gourgoulhon(2012)}]{Gourgoulhon:2007ue}
Gourgoulhon E (2012) 3+1 Formalism in General Relativity: Bases of Numerical
  Relativity, Lecture Notes in Physics, vol 846. Springer, Berlin; New York.
  \doi{10.1007/978-3-642-24525-1}.
  {\href{https://arxiv.org/abs/gr-qc/0703035}{{arXiv:gr-qc/0703035}}} {[gr-qc]}

\bibitem[{Gracia-Linares and Guzman(2016)}]{Gracia-Linares:2016gvy}
Gracia-Linares M, Guzman FS (2016) Accretion of supersonic winds on boson
  stars. Phys Rev D 94:064077. \doi{10.1103/PhysRevD.94.064077}.
  {\href{https://arxiv.org/abs/1609.06398}{{arXiv:1609.06398}}} {[gr-qc]}

\bibitem[{Gralla et~al.(2019)Gralla, Holz, and Wald}]{Gralla:2019xty}
Gralla SE, Holz DE, Wald RM (2019) {Black Hole Shadows, Photon Rings, and
  Lensing Rings}. Phys Rev D 100:024018. \doi{10.1103/PhysRevD.100.024018}.
  {\href{https://arxiv.org/abs/1906.00873}{{arXiv:1906.00873}}} {[astro-ph.HE]}

\bibitem[{Grandcl{\'e}ment(2016)}]{Grandclement:2016eng}
Grandcl{\'e}ment P (2016) Light rings and light points of boson stars. ArXiv
  e-prints {\href{https://arxiv.org/abs/1612.07507}{{arXiv:1612.07507}}}
  {[gr-qc]}

\bibitem[{Grandcl{\'{e}}ment et~al.(2011)Grandcl{\'{e}}ment, Fodor, and
  Forg{\'{a}}cs}]{PhysRevD.84.065037}
Grandcl{\'{e}}ment P, Fodor G, Forg{\'{a}}cs P (2011) Numerical simulation of
  oscillatons: Extracting the radiating tail. Phys Rev D 84:065037.
  \doi{10.1103/PhysRevD.84.065037}

\bibitem[{{Grandcl{\'e}ment} et~al.(2014){Grandcl{\'e}ment}, {Som{\'e}}, and
  {Gourgoulhon}}]{2014PhRvD..90b4068G}
{Grandcl{\'e}ment} P, {Som{\'e}} C, {Gourgoulhon} E (2014) Models of rotating
  boson stars and geodesics around them: New type of orbits. Phys Rev D
  90:024068. \doi{10.1103/PhysRevD.90.024068}.
  {\href{https://arxiv.org/abs/1405.4837}{{arXiv:1405.4837}}} {[gr-qc]}

\bibitem[{Guenther(1995)}]{1995PhDT........25G}
Guenther RL (1995) A numerical study of the time dependent schr\"{o}dinger
  equation coupled with {N}ewtonian gravity. PhD thesis, The University of
  Texas, Austin.
  \urlprefix\url{http://laplace.physics.ubc.ca/Members/matt/Doc/Theses/}

\bibitem[{Guerra et~al.(2019)Guerra, Macedo, and Pani}]{Guerra:2019srj}
Guerra D, Macedo CFB, Pani P (2019) {Axion boson stars}. JCAP 09:061.
  \doi{10.1088/1475-7516/2019/09/061}, [Erratum: JCAP 06, E01 (2020)].
  {\href{https://arxiv.org/abs/1909.05515}{{arXiv:1909.05515}}} {[gr-qc]}

\bibitem[{Gundlach and Leveque(2011)}]{Gundlach:2010gy}
Gundlach C, Leveque RJ (2011) Universality in the run-up of shock waves to the
  surface of a star. J Fluid Mech 676:237--264. \doi{10.1017/jfm.2011.42}.
  {\href{https://arxiv.org/abs/1008.2834}{{arXiv:1008.2834}}} {[astro-ph.SR]}

\bibitem[{Gundlach and Mart{\'{\i}}n-Garc{\'{\i}}a(2007)}]{Gundlach:2007gc}
Gundlach C, Mart{\'{\i}}n-Garc{\'{\i}}a JM (2007) Critical phenomena in
  gravitational collapse. Living Rev Relativity 10:5.
  \doi{10.12942/lrr-2007-5}.
  {\href{https://arxiv.org/abs/0711.4620}{{arXiv:0711.4620}}} {[gr-qc]}

\bibitem[{Gundlach and Please(2009)}]{Gundlach:2009ft}
Gundlach C, Please C (2009) Generic behaviour of nonlinear sound waves near the
  surface of a star: Smooth solutions. Phys Rev D 79:067501.
  \doi{10.1103/PhysRevD.79.067501}.
  {\href{https://arxiv.org/abs/0901.4928}{{arXiv:0901.4928}}} {[astro-ph.SR]}

\bibitem[{Guo et~al.(2021)Guo, Liu, L\"u, and Pang}]{Guo:2020bqz}
Guo SF, Liu HS, L\"u H, Pang Y (2021) {Large-charge limit of AdS boson stars
  with mixed boundary conditions}. JHEP 04:220. \doi{10.1007/JHEP04(2021)220}.
  {\href{https://arxiv.org/abs/2101.00017}{{arXiv:2101.00017}}} {[hep-th]}

\bibitem[{{G{\"u}ver} et~al.(2014){G{\"u}ver}, {Emre Erkoca}, {Hall Reno}, and
  {Sarcevic}}]{2014JCAP...05..013G}
{G{\"u}ver} T, {Emre Erkoca} A, {Hall Reno} M, {Sarcevic} I (2014) On the
  capture of dark matter by neutron stars. J Cosmol Astropart Phys
  2014(05):013. \doi{10.1088/1475-7516/2014/05/013}.
  {\href{https://arxiv.org/abs/1201.2400}{{arXiv:1201.2400}}} {[hep-ph]}

\bibitem[{Guzm{\'{a}}n(2004)}]{2004PhRvD..70d4033G}
Guzm{\'{a}}n FS (2004) Evolving spherical boson stars on a {3D} {C}artesian
  grid. Phys Rev D 70:044033. \doi{10.1103/PhysRevD.70.044033}.
  {\href{https://arxiv.org/abs/gr-qc/0407054}{{arXiv:gr-qc/0407054}}}

\bibitem[{Guzm{\'{a}}n(2007)}]{Guzman:2007zz}
Guzm{\'{a}}n FS (2007) Scalar fields: At the threshold of astrophysics. J Phys
  Conf Ser 91:012003. \doi{10.1088/1742-6596/91/1/012003}

\bibitem[{Guzm{\'{a}}n(2009)}]{guzman2009}
Guzm{\'{a}}n FS (2009) The three dynamical fates of boson stars. Rev Mex Fis
  55:321--326.
  \urlprefix\url{http://www.scielo.org.mx/scielo.php?pid=S0035-001X2009000400011&nrm=iso&script=sci_arttext}

\bibitem[{Guzm{\'{a}}n and Rueda-Becerril(2009)}]{2009PhRvD..80h4023G}
Guzm{\'{a}}n FS, Rueda-Becerril JM (2009) Spherical boson stars as black hole
  mimickers. Phys Rev D 80:084023. \doi{10.1103/PhysRevD.80.084023}.
  {\href{https://arxiv.org/abs/1009.1250}{{arXiv:1009.1250}}} {[astro-ph.HE]}

\bibitem[{Guzm{\'{a}}n and Ure{\~{n}}a-L{\'{o}}pez(2006)}]{2006ApJ...645..814G}
Guzm{\'{a}}n FS, Ure{\~{n}}a-L{\'{o}}pez LA (2006) Gravitational cooling of
  self-gravitating {B}ose condensates. Astrophys J 645:814--819.
  \doi{10.1086/504508}.
  {\href{https://arxiv.org/abs/astro-ph/0603613}{{arXiv:astro-ph/0603613}}}

\bibitem[{Hanna et~al.(2017)Hanna, Johnson, and Lehner}]{Hanna:2016uhs}
Hanna C, Johnson MC, Lehner L (2017) Estimating gravitational radiation from
  super-emitting compact binary systems. Phys Rev D 124042.
  \doi{10.1103/PhysRevD.95.124042}.
  {\href{https://arxiv.org/abs/1611.03506}{{arXiv:1611.03506}}} {[gr-qc]}

\bibitem[{Harrison et~al.(1965)Harrison, Thorne, Wakano, and
  Wheeler}]{1965gtgc.book.....H}
Harrison BK, Thorne KS, Wakano M, Wheeler JA (1965) Gravitation Theory and
  Gravitational Collapse. University of Chicago Press, Chicago

\bibitem[{{Hartmann} and {Riedel}(2012)}]{2012PhRvD..86j4008H}
{Hartmann} B, {Riedel} J (2012) Glueball condensates as holographic duals of
  supersymmetric {Q}-balls and boson stars. Phys Rev D 86:104008.
  \doi{10.1103/PhysRevD.86.104008}.
  {\href{https://arxiv.org/abs/1204.6239}{{arXiv:1204.6239}}} {[hep-th]}

\bibitem[{{Hartmann} and {Riedel}(2013)}]{2013PhRvD..87d4003H}
{Hartmann} B, {Riedel} J (2013) Supersymmetric {Q}-balls and boson stars in
  (d+1) dimensions. Phys Rev D 87:044003. \doi{10.1103/PhysRevD.87.044003}.
  {\href{https://arxiv.org/abs/1210.0096}{{arXiv:1210.0096}}} {[hep-th]}

\bibitem[{Hartmann et~al.(2010)Hartmann, Kleihaus, Kunz, and
  List}]{Hartmann:2010pm}
Hartmann B, Kleihaus B, Kunz J, List M (2010) Rotating boson stars in five
  dimensions. Phys Rev D 82:084022. \doi{10.1103/PhysRevD.82.084022}.
  {\href{https://arxiv.org/abs/1008.3137}{{arXiv:1008.3137}}} {[gr-qc]}

\bibitem[{{Hartmann} et~al.(2012){Hartmann}, {Kleihaus}, {Kunz}, and
  {Schaffer}}]{2012PhLB..714..120H}
{Hartmann} B, {Kleihaus} B, {Kunz} J, {Schaffer} I (2012) Compact boson stars.
  Phys Lett B 714:120--126. \doi{10.1016/j.physletb.2012.06.067}.
  {\href{https://arxiv.org/abs/1205.0899}{{arXiv:1205.0899}}} {[gr-qc]}

\bibitem[{{Hartmann} et~al.(2013{\natexlab{a}}){Hartmann}, {Kleihaus}, {Kunz},
  and {Schaffer}}]{2013PhRvD..88l4033H}
{Hartmann} B, {Kleihaus} B, {Kunz} J, {Schaffer} I (2013{\natexlab{a}}) Compact
  (a)ds boson stars and shells. Phys Rev D 88:124033.
  \doi{10.1103/PhysRevD.88.124033}.
  {\href{https://arxiv.org/abs/1310.3632}{{arXiv:1310.3632}}} {[gr-qc]}

\bibitem[{{Hartmann} et~al.(2013{\natexlab{b}}){Hartmann}, {Riedel}, and
  {Suciu}}]{2013PhLB..726..906H}
{Hartmann} B, {Riedel} J, {Suciu} R (2013{\natexlab{b}}) {G}auss-{B}onnet boson
  stars. Phys Lett B 726:906--912. \doi{10.1016/j.physletb.2013.09.050}.
  {\href{https://arxiv.org/abs/1308.3391}{{arXiv:1308.3391}}} {[gr-qc]}

\bibitem[{Hawley and Choptuik(2000)}]{Hawley:2000dt}
Hawley SH, Choptuik MW (2000) Boson stars driven to the brink of black hole
  formation. Phys Rev D 62:104024. \doi{10.1103/PhysRevD.62.104024}.
  {\href{https://arxiv.org/abs/gr-qc/0007039}{{arXiv:gr-qc/0007039}}}

\bibitem[{Hawley and Choptuik(2003)}]{Hawley:2002zn}
Hawley SH, Choptuik MW (2003) Numerical evidence for `multiscalar stars'. Phys
  Rev D 67:024010. \doi{10.1103/PhysRevD.67.024010}.
  {\href{https://arxiv.org/abs/gr-qc/0208078}{{arXiv:gr-qc/0208078}}}

\bibitem[{Helfer et~al.(2022)Helfer, Sperhake, Croft, Radia, Ge, and
  Lim}]{Helfer:2021brt}
Helfer T, Sperhake U, Croft R, Radia M, Ge BX, Lim EA (2022) {Malaise and
  remedy of binary boson-star initial data}. Class Quantum Grav 39(7):074001.
  \doi{10.1088/1361-6382/ac53b7}.
  {\href{https://arxiv.org/abs/2108.11995}{{arXiv:2108.11995}}} {[gr-qc]}

\bibitem[{{Henderson} et~al.(2015){Henderson}, {Mann}, and
  {Stotyn}}]{2015PhRvD..91b4009H}
{Henderson} LJ, {Mann} RB, {Stotyn} S (2015) {G}auss-{B}onnet boson stars with
  a single {K}illing vector. Phys Rev D 91:024009.
  \doi{10.1103/PhysRevD.91.024009}.
  {\href{https://arxiv.org/abs/1403.1865}{{arXiv:1403.1865}}} {[gr-qc]}

\bibitem[{Henriques et~al.(1989)Henriques, Liddle, and
  Moorhouse}]{1989PhLB..233...99H}
Henriques AB, Liddle AR, Moorhouse RG (1989) Combined boson-fermion stars. Phys
  Lett B 233:99--106. \doi{10.1016/0370-2693(89)90623-0}

\bibitem[{Henriques et~al.(1990)Henriques, Liddle, and
  Moorhouse}]{1990NuPhB.337..737H}
Henriques AB, Liddle AR, Moorhouse RG (1990) Combined boson-fermion stars:
  Configurations and stability. Nucl Phys B 337:737--761.
  \doi{10.1016/0550-3213(90)90514-E}

\bibitem[{{Henriques} et~al.(1990){Henriques}, {Liddle}, and
  {Moorhouse}}]{1990PhLB..251..511H}
{Henriques} AB, {Liddle} AR, {Moorhouse} RG (1990) Stability of boson-fermion
  stars. Phys Lett B 251:511--516. \doi{10.1016/0370-2693(90)90789-9}

\bibitem[{{Herdeiro} and {Radu}(2014{\natexlab{a}})}]{2014arXiv1406.1225H}
{Herdeiro} C, {Radu} E (2014{\natexlab{a}}) Ergosurfaces for {K}err black holes
  with scalar hair. Phys Rev D 89:124018. \doi{10.1103/PhysRevD.89.124018}.
  {\href{https://arxiv.org/abs/1406.1225}{{arXiv:1406.1225}}} {[gr-qc]}

\bibitem[{Herdeiro and Radu(2015{\natexlab{a}})}]{Herdeiro:2015gia}
Herdeiro C, Radu E (2015{\natexlab{a}}) Construction and physical properties of
  {K}err black holes with scalar hair. Class Quantum Grav 32:144001.
  \doi{10.1088/0264-9381/32/14/144001}.
  {\href{https://arxiv.org/abs/1501.04319}{{arXiv:1501.04319}}} {[gr-qc]}

\bibitem[{Herdeiro et~al.(2015{\natexlab{a}})Herdeiro, Kunz, Radu, and
  Subagyo}]{Herdeiro:2015kha}
Herdeiro C, Kunz J, Radu E, Subagyo B (2015{\natexlab{a}}) Myers-perry black
  holes with scalar hair and a mass gap: Unequal spins. Phys Lett B 748:30--36.
  \doi{10.1016/j.physletb.2015.06.059}.
  {\href{https://arxiv.org/abs/1505.02407}{{arXiv:1505.02407}}} {[gr-qc]}

\bibitem[{Herdeiro et~al.(2016{\natexlab{a}})Herdeiro, Radu, and
  Runarsson}]{Herdeiro:2016tmi}
Herdeiro C, Radu E, Runarsson H (2016{\natexlab{a}}) {K}err black holes with
  {P}roca hair. Class Quantum Grav 33:154001.
  \doi{10.1088/0264-9381/33/15/154001}.
  {\href{https://arxiv.org/abs/1603.02687}{{arXiv:1603.02687}}} {[gr-qc]}

\bibitem[{Herdeiro et~al.(2022)Herdeiro, Perapechka, Radu, and
  Shnir}]{Herdeiro:2021jgc}
Herdeiro C, Perapechka I, Radu E, Shnir Y (2022) {Spinning gauged boson and
  Dirac stars: A comparative study}. Phys Lett B 824:136811.
  \doi{10.1016/j.physletb.2021.136811}.
  {\href{https://arxiv.org/abs/2111.14475}{{arXiv:2111.14475}}} {[gr-qc]}

\bibitem[{{Herdeiro} and {Radu}(2014{\natexlab{b}})}]{2014PhRvL.112v1101H}
{Herdeiro} CAR, {Radu} E (2014{\natexlab{b}}) {K}err black holes with scalar
  hair. Phys Rev Lett 112:221101. \doi{10.1103/PhysRevLett.112.221101}.
  {\href{https://arxiv.org/abs/1403.2757}{{arXiv:1403.2757}}} {[gr-qc]}

\bibitem[{Herdeiro and Radu(2015{\natexlab{b}})}]{Herdeiro:2015waa}
Herdeiro CAR, Radu E (2015{\natexlab{b}}) Asymptotically flat black holes with
  scalar hair: a review. Int J Mod Phys D 24:1542014.
  \doi{10.1142/S0218271815420146}, proceedings, 7th Black Holes Workshop 2014:
  Aveiro, Portugal, December 18--19, 2014.
  {\href{https://arxiv.org/abs/1504.08209}{{arXiv:1504.08209}}} {[gr-qc]}

\bibitem[{Herdeiro and Radu(2018)}]{Herdeiro:2018wvd}
Herdeiro CAR, Radu E (2018) {Spinning boson stars and hairy black holes with
  nonminimal coupling}. Int J Mod Phys D 27:1843009.
  \doi{10.1142/S0218271818430095}.
  {\href{https://arxiv.org/abs/1803.08149}{{arXiv:1803.08149}}} {[gr-qc]}

\bibitem[{Herdeiro and Radu(2022)}]{Herdeiro:2022gzp}
Herdeiro CAR, Radu E (2022) {On the classicality of bosonic stars}. arXiv
  e-prints {\href{https://arxiv.org/abs/2205.05395}{{arXiv:2205.05395}}}
  {[gr-qc]}

\bibitem[{Herdeiro et~al.(2015{\natexlab{b}})Herdeiro, Radu, and
  R{\'u}narsson}]{Herdeiro:2015tia}
Herdeiro CAR, Radu E, R{\'u}narsson H (2015{\natexlab{b}}) {K}err black holes
  with self-interacting scalar hair: hairier but not heavier. Phys Rev D
  92:084059. \doi{10.1103/PhysRevD.92.084059}.
  {\href{https://arxiv.org/abs/1509.02923}{{arXiv:1509.02923}}} {[gr-qc]}

\bibitem[{Herdeiro et~al.(2016{\natexlab{b}})Herdeiro, Radu, and
  R{\'u}narsson}]{Herdeiro:2016gxs}
Herdeiro CAR, Radu E, R{\'u}narsson HF (2016{\natexlab{b}}) Spinning boson
  stars and {K}err black holes with scalar hair: the effect of
  self-interactions. Int J Mod Phys D 25:1641014.
  \doi{10.1142/S0218271816410145}, proceedings, 3rd Amazonian Symposium on
  Physics and 5th NRHEP Network Meeting is approaching: Celebrating 100 Years
  of General Relativity: Belem, Brazil.
  {\href{https://arxiv.org/abs/1604.06202}{{arXiv:1604.06202}}} {[gr-qc]}

\bibitem[{Herdeiro et~al.(2021{\natexlab{a}})Herdeiro, Kunz, Perapechka, Radu,
  and Shnir}]{Herdeiro:2021mol}
Herdeiro CAR, Kunz J, Perapechka I, Radu E, Shnir Y (2021{\natexlab{a}})
  {Chains of Boson Stars}. Phys Rev D 103:065009.
  \doi{10.1103/PhysRevD.103.065009}.
  {\href{https://arxiv.org/abs/2101.06442}{{arXiv:2101.06442}}} {[gr-qc]}

\bibitem[{Herdeiro et~al.(2021{\natexlab{b}})Herdeiro, Kunz, Perapechka, Radu,
  and Shnir}]{Herdeiro:2020kvf}
Herdeiro CAR, Kunz J, Perapechka I, Radu E, Shnir Y (2021{\natexlab{b}})
  {Multipolar boson stars: macroscopic Bose-Einstein condensates akin to
  hydrogen orbitals}. Phys Lett B 812:136027.
  \doi{10.1016/j.physletb.2020.136027}.
  {\href{https://arxiv.org/abs/2008.10608}{{arXiv:2008.10608}}} {[gr-qc]}

\bibitem[{Herdeiro et~al.(2021{\natexlab{c}})Herdeiro, Pombo, Radu, Cunha, and
  Sanchis-Gual}]{Herdeiro:2021lwl}
Herdeiro CAR, Pombo AM, Radu E, Cunha PVP, Sanchis-Gual N (2021{\natexlab{c}})
  {The imitation game: Proca stars that can mimic the Schwarzschild shadow}.
  JCAP 04:051. \doi{10.1088/1475-7516/2021/04/051}.
  {\href{https://arxiv.org/abs/2102.01703}{{arXiv:2102.01703}}} {[gr-qc]}

\bibitem[{{Hobbs} et~al.(2010){Hobbs}, {Archibald}, {Arzoumanian}, {Backer},
  {Bailes}, {Bhat}, {Burgay}, {Burke-Spolaor}, {Champion}, {Cognard}, {Coles},
  {Cordes}, {Demorest}, {Desvignes}, {Ferdman}, {Finn}, {Freire}, {Gonzalez},
  {Hessels}, {Hotan}, {Janssen}, {Jenet}, {Jessner}, {Jordan}, {Kaspi},
  {Kramer}, {Kondratiev}, {Lazio}, {Lazaridis}, {Lee}, {Levin}, {Lommen},
  {Lorimer}, {Lynch}, {Lyne}, {Manchester}, {McLaughlin}, {Nice}, {Oslowski},
  {Pilia}, {Possenti}, {Purver}, {Ransom}, {Reynolds}, {Sanidas}, {Sarkissian},
  {Sesana}, {Shannon}, {Siemens}, {Stairs}, {Stappers}, {Stinebring},
  {Theureau}, {van Haasteren}, {van Straten}, {Verbiest}, {Yardley}, and
  {You}}]{2010CQGra..27h4013H}
{Hobbs} G, {Archibald} A, {Arzoumanian} Z, {Backer} D, {Bailes} M, {Bhat} NDR,
  {Burgay} M, {Burke-Spolaor} S, {Champion} D, {Cognard} I, {Coles} W, {Cordes}
  J, {Demorest} P, {Desvignes} G, {Ferdman} RD, {Finn} L, {Freire} P,
  {Gonzalez} M, {Hessels} J, {Hotan} A, {Janssen} G, {Jenet} F, {Jessner} A,
  {Jordan} C, {Kaspi} V, {Kramer} M, {Kondratiev} V, {Lazio} J, {Lazaridis} K,
  {Lee} KJ, {Levin} Y, {Lommen} A, {Lorimer} D, {Lynch} R, {Lyne} A,
  {Manchester} R, {McLaughlin} M, {Nice} D, {Oslowski} S, {Pilia} M, {Possenti}
  A, {Purver} M, {Ransom} S, {Reynolds} J, {Sanidas} S, {Sarkissian} J,
  {Sesana} A, {Shannon} R, {Siemens} X, {Stairs} I, {Stappers} B, {Stinebring}
  D, {Theureau} G, {van Haasteren} R, {van Straten} W, {Verbiest} JPW,
  {Yardley} DRB, {You} XP (2010) {The International Pulsar Timing Array
  project: using pulsars as a gravitational wave detector}. Class Quantum Grav
  27:084013. \doi{10.1088/0264-9381/27/8/084013}.
  {\href{https://arxiv.org/abs/0911.5206}{{arXiv:0911.5206}}} {[astro-ph.SR]}

\bibitem[{Hod(2011)}]{Hod:2011zz}
Hod S (2011) Quasinormal resonances of a massive scalar field in a
  near-extremal {K}err black hole spacetime. Phys Rev D 84:044046.
  \doi{10.1103/PhysRevD.84.044046}.
  {\href{https://arxiv.org/abs/1109.4080}{{arXiv:1109.4080}}} {[gr-qc]}

\bibitem[{Hod(2012)}]{Hod:2012px}
Hod S (2012) Stationary scalar clouds around rotating black holes. Phys Rev D
  86:104026. \doi{10.1103/PhysRevD.86.104026}, [Erratum: Phys. Rev. D 86 (2012)
  129902]. {\href{https://arxiv.org/abs/1211.3202}{{arXiv:1211.3202}}}
  {[gr-qc]}

\bibitem[{Hod(2018)}]{Hod:2018dij}
Hod S (2018) {No-go theorem for static boson stars}. Phys Lett B 778:239--241.
  \doi{10.1016/j.physletb.2018.01.036}.
  {\href{https://arxiv.org/abs/1902.05230}{{arXiv:1902.05230}}} {[gr-qc]}

\bibitem[{Hod(2019)}]{Hod:2019yfl}
Hod S (2019) {No-go theorem for spatially regular boson stars made of static
  nonminimally coupled massive scalar fields}. Eur Phys J C 79:26.
  \doi{10.1140/epjc/s10052-019-6546-5}.
  {\href{https://arxiv.org/abs/2008.13384}{{arXiv:2008.13384}}} {[gr-qc]}

\bibitem[{Honda(2000)}]{Honda:2000gv}
Honda EP (2000) Resonant dynamics within the nonlinear {K}lein-{G}ordon
  equation: Much ado about oscillons. PhD thesis, The University of Texas,
  Austin.
  \urlprefix\url{http://laplace.physics.ubc.ca/Members/matt/Doc/Theses/},
  {\href{https://arxiv.org/abs/hep-ph/0009104}{{arXiv:hep-ph/0009104}}}
  {[hep-ph]}

\bibitem[{Honda(2010)}]{Honda:2010gb}
Honda EP (2010) Fractal boundary basins in spherically symmetric $\phi^4$
  theory. Phys Rev D 82:024038. \doi{10.1103/PhysRevD.82.024038}.
  {\href{https://arxiv.org/abs/1006.2421}{{arXiv:1006.2421}}} {[gr-qc]}

\bibitem[{Honda and Choptuik(2002)}]{Honda:2001xg}
Honda EP, Choptuik MW (2002) Fine structure of oscillons in the spherically
  symmetric $\phi^4$ {K}lein-{G}ordon model. Phys Rev D 65:084037.
  \doi{10.1103/PhysRevD.65.084037}.
  {\href{https://arxiv.org/abs/hep-ph/0110065}{{arXiv:hep-ph/0110065}}}
  {[hep-ph]}

\bibitem[{{Horvat} and {Marunovi{\'c}}(2013)}]{2013CQGra..30n5006H}
{Horvat} D, {Marunovi{\'c}} A (2013) Dark energy-like stars from nonminimally
  coupled scalar field. Class Quantum Grav 30:145006.
  \doi{10.1088/0264-9381/30/14/145006}.
  {\href{https://arxiv.org/abs/1212.3781}{{arXiv:1212.3781}}} {[gr-qc]}

\bibitem[{{Horvat} et~al.(2013){Horvat}, {Iliji{\'c}}, {Kirin}, and {Naran{\v
  c}i{\'c}}}]{2013CQGra..30i5014H}
{Horvat} D, {Iliji{\'c}} S, {Kirin} A, {Naran{\v c}i{\'c}} Z (2013) Formation
  of photon spheres in boson stars with a nonminimally coupled field. Class
  Quantum Grav 30:095014. \doi{10.1088/0264-9381/30/9/095014}.
  {\href{https://arxiv.org/abs/1302.4369}{{arXiv:1302.4369}}} {[gr-qc]}

\bibitem[{Horvat et~al.(2015)Horvat, Ilijic, Kirin, and
  Narancic}]{Horvat:2015qva}
Horvat D, Ilijic S, Kirin A, Narancic Z (2015) Note on the charged boson stars
  with torsion-coupled field. Phys Rev D 92:024045.
  \doi{10.1103/PhysRevD.92.024045}.
  {\href{https://arxiv.org/abs/1503.02480}{{arXiv:1503.02480}}} {[gr-qc]}

\bibitem[{{Horvat} et~al.(2015){Horvat}, {Iliji{\'c}}, {Kirin}, and {Naran{\v
  c}i{\'c}}}]{2015CQGra..32c5023H}
{Horvat} D, {Iliji{\'c}} S, {Kirin} A, {Naran{\v c}i{\'c}} Z (2015)
  Nonminimally coupled scalar field in teleparallel gravity: boson stars. Class
  Quantum Grav 32:035023. \doi{10.1088/0264-9381/32/3/035023}.
  {\href{https://arxiv.org/abs/1407.2067}{{arXiv:1407.2067}}} {[gr-qc]}

\bibitem[{{Hu} et~al.(2012){Hu}, {Liu}, and {Pando
  Zayas}}]{2012arXiv1209.2378H}
{Hu} S, {Liu} JT, {Pando Zayas} LA (2012) Charged boson stars in {AdS} and a
  zero temperature phase transition. ArXiv e-prints
  {\href{https://arxiv.org/abs/1209.2378}{{arXiv:1209.2378}}} {[hep-th]}

\bibitem[{Iliji\'c and Sossich(2020)}]{Ilijic:2020vzu}
Iliji\'c S, Sossich M (2020) {Boson stars in $f(T)$ extended theory of
  gravity}. Phys Rev D 102:084019. \doi{10.1103/PhysRevD.102.084019}.
  {\href{https://arxiv.org/abs/2007.12451}{{arXiv:2007.12451}}} {[gr-qc]}

\bibitem[{{Jamison}(2013)}]{2013PhRvD..88c5004J}
{Jamison} AO (2013) Effects of gravitational confinement on bosonic asymmetric
  dark matter in stars. Phys Rev D 88:035004. \doi{10.1103/PhysRevD.88.035004}.
  {\href{https://arxiv.org/abs/1304.3773}{{arXiv:1304.3773}}} {[hep-ph]}

\bibitem[{Jaramillo et~al.(2020)Jaramillo, Sanchis-Gual, Barranco, Bernal,
  Degollado, Herdeiro, and N\'u\~nez}]{Jaramillo:2020rsv}
Jaramillo V, Sanchis-Gual N, Barranco J, Bernal A, Degollado JC, Herdeiro C,
  N\'u\~nez D (2020) {Dynamical \ensuremath{\ell} -boson stars: Generic
  stability and evidence for nonspherical solutions}. Phys Rev D 101:124020.
  \doi{10.1103/PhysRevD.101.124020}.
  {\href{https://arxiv.org/abs/2004.08459}{{arXiv:2004.08459}}} {[gr-qc]}

\bibitem[{Jaramillo et~al.(2022)Jaramillo, Sanchis-Gual, Barranco, Bernal,
  Degollado, Herdeiro, Megevand, and N\'u\~nez}]{Jaramillo:2022zwg}
Jaramillo V, Sanchis-Gual N, Barranco J, Bernal A, Degollado JC, Herdeiro C,
  Megevand M, N\'u\~nez D (2022) {Head-on collisions of \ensuremath{\ell}-boson
  stars}. Phys Rev D 105(10):104057. \doi{10.1103/PhysRevD.105.104057}.
  {\href{https://arxiv.org/abs/2202.00696}{{arXiv:2202.00696}}} {[gr-qc]}

\bibitem[{Jetzer(1989{\natexlab{a}})}]{1989NuPhB.316..411J}
Jetzer P (1989{\natexlab{a}}) Dynamical instability of bosonic stellar
  configurations. Nucl Phys B 316:411--428. \doi{10.1016/0550-3213(89)90038-2}

\bibitem[{Jetzer(1989{\natexlab{b}})}]{1989PhLB..231..433J}
Jetzer P (1989{\natexlab{b}}) Stability of charged boson stars. Phys Lett B
  231:433--438. \doi{10.1016/0370-2693(89)90689-8}

\bibitem[{Jetzer(1989{\natexlab{c}})}]{1989PhLB..222..447J}
Jetzer P (1989{\natexlab{c}}) Stability of excited bosonic stellar
  configurations. Phys Lett B 222:447--452. \doi{10.1016/0370-2693(89)90342-0}

\bibitem[{Jetzer(1990)}]{1990PhLB..243...36J}
Jetzer P (1990) Stability of combined boson-fermion stars. Phys Lett B
  243:36--40. \doi{10.1016/0370-2693(90)90952-3}

\bibitem[{Jetzer(1992)}]{Jetzer:1991jr}
Jetzer P (1992) Boson stars. Phys Rep 220:163--227.
  \doi{10.1016/0370-1573(92)90123-H}

\bibitem[{Jetzer and van~der Bij(1989)}]{1989PhLB..227..341J}
Jetzer P, van~der Bij JJ (1989) Charged boson stars. Phys Lett B 227:341--346.
  \doi{10.1016/0370-2693(89)90941-6}

\bibitem[{Jimenez-Vazquez and Alcubierre(2022)}]{Jimenez-Vazquez:2022fix}
Jimenez-Vazquez E, Alcubierre M (2022) {Critical gravitational collapse of a
  massive complex scalar field}. Phys Rev D 106:044071.
  \doi{10.1103/PhysRevD.106.044071}.
  {\href{https://arxiv.org/abs/2206.01389}{{arXiv:2206.01389}}} {[gr-qc]}

\bibitem[{Jin and Suen(2007)}]{Jin:2006gm}
Jin KJ, Suen WM (2007) Critical phenomena in head-on collisions of neutron
  stars. Phys Rev Lett 98:131101. \doi{10.1103/PhysRevLett.98.131101}.
  {\href{https://arxiv.org/abs/gr-qc/0603094}{{gr-qc/0603094}}}

\bibitem[{Johnson-Mcdaniel et~al.(2020)Johnson-Mcdaniel, Mukherjee, Kashyap,
  Ajith, Del~Pozzo, and Vitale}]{Johnson-Mcdaniel:2018cdu}
Johnson-Mcdaniel NK, Mukherjee A, Kashyap R, Ajith P, Del~Pozzo W, Vitale S
  (2020) {Constraining black hole mimickers with gravitational wave
  observations}. Phys Rev D 102:123010. \doi{10.1103/PhysRevD.102.123010}.
  {\href{https://arxiv.org/abs/1804.08026}{{arXiv:1804.08026}}} {[gr-qc]}

\bibitem[{Kain(2021{\natexlab{a}})}]{Kain:2021rmk}
Kain B (2021{\natexlab{a}}) {Boson stars and their radial oscillations}. Phys
  Rev D 103:123003. \doi{10.1103/PhysRevD.103.123003}.
  {\href{https://arxiv.org/abs/2106.01740}{{arXiv:2106.01740}}} {[gr-qc]}

\bibitem[{Kain(2021{\natexlab{b}})}]{Kain:2021bwd}
Kain B (2021{\natexlab{b}}) {Fermion\textendash{}charged-boson stars}. Phys Rev
  D 104(4):043001. \doi{10.1103/PhysRevD.104.043001}.
  {\href{https://arxiv.org/abs/2108.01404}{{arXiv:2108.01404}}} {[gr-qc]}

\bibitem[{Kan and Shiraishi(2016)}]{Kan:2016xkn}
Kan N, Shiraishi K (2016) Analytical approximation for {N}ewtonian boson stars
  in four and five dimensions -- a poor person's approach to rotating boson
  stars. Phys Rev D 94:104042. \doi{10.1103/PhysRevD.94.104042}.
  {\href{https://arxiv.org/abs/1605.02846}{{arXiv:1605.02846}}} {[gr-qc]}

\bibitem[{Karkevandi et~al.(2022)Karkevandi, Shakeri, Sagun, and
  Ivanytskyi}]{Karkevandi:2021ygv}
Karkevandi DR, Shakeri S, Sagun V, Ivanytskyi O (2022) {Bosonic dark matter in
  neutron stars and its effect on gravitational wave signal}. Phys Rev D
  105(2):023001. \doi{10.1103/PhysRevD.105.023001}.
  {\href{https://arxiv.org/abs/2109.03801}{{arXiv:2109.03801}}} {[astro-ph.HE]}

\bibitem[{Kasuya and Kawasaki(2000)}]{2000PhRvD..61d1301K}
Kasuya S, Kawasaki M (2000) {Q}-ball formation through the {A}ffleck-{D}ine
  mechanism. Phys Rev D 61:041301. \doi{10.1103/PhysRevD.61.041301}.
  {\href{https://arxiv.org/abs/hep-ph/9909509}{{arXiv:hep-ph/9909509}}}

\bibitem[{Kaup(1968)}]{Kaup:1968zz}
Kaup DJ (1968) {K}lein-{G}ordon geon. Phys Rev 172:1331--1342.
  \doi{10.1103/PhysRev.172.1331}

\bibitem[{Kellermann et~al.(2010)Kellermann, Rezzolla, and
  Radice}]{Kellermann:2010rt}
Kellermann T, Rezzolla L, Radice D (2010) Critical phenomena in neutron stars:
  {II}. head-on collisions. Class Quantum Grav 27:235016.
  \doi{10.1088/0264-9381/27/23/235016}.
  {\href{https://arxiv.org/abs/1007.2797}{{arXiv:1007.2797}}} {[gr-qc]}

\bibitem[{Kesden et~al.(2005)Kesden, Gair, and Kamionkowski}]{Kesden:2004qx}
Kesden M, Gair JR, Kamionkowski M (2005) Gravitational-wave signature of an
  inspiral into a supermassive horizonless object. Phys Rev D 71:044015.
  \doi{10.1103/PhysRevD.71.044015}.
  {\href{https://arxiv.org/abs/astro-ph/0411478}{{arXiv:astro-ph/0411478}}}

\bibitem[{{Khachatryan} et~al.(2015){Khachatryan}, {Sirunyan}, {Tumasyan},
  {Adam}, {Bergauer}, {Dragicevic}, {Er{\"o}}, {Friedl}, {Fr{\"u}hwirth},
  {Ghete}, and et~al.}]{2015EPJC...75..212K}
{Khachatryan} V, {Sirunyan} AM, {Tumasyan} A, {Adam} W, {Bergauer} T,
  {Dragicevic} M, {Er{\"o}} J, {Friedl} M, {Fr{\"u}hwirth} R, {Ghete} VM, et~al
  (2015) Precise determination of the mass of the higgs boson and tests of
  compatibility of its couplings with the standard model predictions using
  proton collisions at 7 and 8. Eur Phys J C 75:212.
  \doi{10.1140/epjc/s10052-015-3351-7}.
  {\href{https://arxiv.org/abs/1412.8662}{{arXiv:1412.8662}}} {[hep-ex]}

\bibitem[{{Kichakova} et~al.(2014){Kichakova}, {Kunz}, and
  {Radu}}]{2014PhLB..728..328K}
{Kichakova} O, {Kunz} J, {Radu} E (2014) Spinning gauged boson stars in anti-de
  {S}itter spacetime. Phys Lett B 728:328--335.
  \doi{10.1016/j.physletb.2013.11.061}.
  {\href{https://arxiv.org/abs/1310.5434}{{arXiv:1310.5434}}} {[gr-qc]}

\bibitem[{Kichenassamy(2008)}]{2008CQGra..25x5004K}
Kichenassamy S (2008) Soliton stars in the breather limit. Class Quantum Grav
  25:245004. \doi{10.1088/0264-9381/25/24/245004}

\bibitem[{Kiessling(2009)}]{2009JSP...137.1063K}
Kiessling MKH (2009) Monotonicity of quantum ground state energies: Bosonic
  atoms and stars. J Stat Phys 137:1063--1078. \doi{10.1007/s10955-009-9843-9}.
  {\href{https://arxiv.org/abs/1001.4280}{{arXiv:1001.4280}}} {[math-ph]}

\bibitem[{Kleihaus et~al.(2005)Kleihaus, Kunz, and List}]{2005PhRvD..72f4002K}
Kleihaus B, Kunz J, List M (2005) Rotating boson stars and {Q}-balls. Phys Rev
  D 72:064002. \doi{10.1103/PhysRevD.72.064002}.
  {\href{https://arxiv.org/abs/gr-qc/0505143}{{arXiv:gr-qc/0505143}}}

\bibitem[{Kleihaus et~al.(2008)Kleihaus, Kunz, List, and
  Schaffer}]{2008PhRvD..77f4025K}
Kleihaus B, Kunz J, List M, Schaffer I (2008) Rotating boson stars and
  {Q}-balls. {II}. negative parity and ergoregions. Phys Rev D 77:064025.
  \doi{10.1103/PhysRevD.77.064025}.
  {\href{https://arxiv.org/abs/0712.3742}{{arXiv:0712.3742}}} {[gr-qc]}

\bibitem[{Kleihaus et~al.(2009)Kleihaus, Kunz, L{\"{a}}mmerzahl, and
  List}]{2009PhLB..675..102K}
Kleihaus B, Kunz J, L{\"{a}}mmerzahl C, List M (2009) Charged boson stars and
  black holes. Phys Lett B 675:102--109. \doi{10.1016/j.physletb.2009.03.066}.
  {\href{https://arxiv.org/abs/0902.4799}{{arXiv:0902.4799}}} {[gr-qc]}

\bibitem[{Kleihaus et~al.(2010)Kleihaus, Kunz, L{\"{a}}mmerzahl, and
  List}]{2010PhRvD..82j4050K}
Kleihaus B, Kunz J, L{\"{a}}mmerzahl C, List M (2010) Boson shells harboring
  charged black holes. Phys Rev D 82:104050. \doi{10.1103/PhysRevD.82.104050}.
  {\href{https://arxiv.org/abs/1007.1630}{{arXiv:1007.1630}}} {[gr-qc]}

\bibitem[{{Kleihaus} et~al.(2012){Kleihaus}, {Kunz}, and
  {Schneider}}]{Kleihaus:2011sx}
{Kleihaus} B, {Kunz} J, {Schneider} S (2012) Stable phases of boson stars. Phys
  Rev D 85:024045. \doi{10.1103/PhysRevD.85.024045}.
  {\href{https://arxiv.org/abs/1109.5858}{{arXiv:1109.5858}}} {[gr-qc]}

\bibitem[{Kleihaus et~al.(2015)Kleihaus, Kunz, and
  Yazadjiev}]{Kleihaus:2015iea}
Kleihaus B, Kunz J, Yazadjiev S (2015) Scalarized hairy black holes. Phys Lett
  B 744:406--412. \doi{10.1016/j.physletb.2015.04.014}.
  {\href{https://arxiv.org/abs/1503.01672}{{arXiv:1503.01672}}} {[gr-qc]}

\bibitem[{Kling et~al.(2021)Kling, Rajaraman, and Rivera}]{Kling:2020xjj}
Kling F, Rajaraman A, Rivera FL (2021) {New solutions for rotating boson
  stars}. Phys Rev D 103:075020. \doi{10.1103/PhysRevD.103.075020}.
  {\href{https://arxiv.org/abs/2010.09880}{{arXiv:2010.09880}}} {[hep-th]}

\bibitem[{Kobayashi et~al.(1994)Kobayashi, Kasai, and
  Futamase}]{Kobayashi:1994qi}
Kobayashi Y, Kasai M, Futamase T (1994) Does a boson star rotate? Phys Rev D
  50:7721--7724. \doi{10.1103/PhysRevD.50.7721}

\bibitem[{{Kouvaris} and {Tinyakov}(2013)}]{2013PhRvD..87l3537K}
{Kouvaris} C, {Tinyakov} P (2013) (not)-constraining heavy asymmetric bosonic
  dark matter. Phys Rev D 87:123537. \doi{10.1103/PhysRevD.87.123537}.
  {\href{https://arxiv.org/abs/1212.4075}{{arXiv:1212.4075}}} {[astro-ph.HE]}

\bibitem[{Kouvaris and Tinyakov(2010)}]{PhysRevD.82.063531}
Kouvaris C, Tinyakov PG (2010) Can neutron stars constrain dark matter? Phys
  Rev D 82:063531. \doi{10.1103/PhysRevD.82.063531}

\bibitem[{{K{\"u}hnel} and {Rampf}(2014)}]{2014PhRvD..90j3526K}
{K{\"u}hnel} F, {Rampf} C (2014) Astrophysical {B}ose-{E}instein condensates
  and superradiance. Phys Rev D 90:103526. \doi{10.1103/PhysRevD.90.103526}.
  {\href{https://arxiv.org/abs/1408.0790}{{arXiv:1408.0790}}} {[gr-qc]}

\bibitem[{Kumar et~al.(2015)Kumar, Kulshreshtha, and
  Kulshreshtha}]{Kumar:2015sia}
Kumar S, Kulshreshtha U, Kulshreshtha DS (2015) Boson stars in a theory of
  complex scalar field coupled to gravity. Gen Relativ Gravit 47:76.
  \doi{10.1007/s10714-015-1918-0}.
  {\href{https://arxiv.org/abs/1605.07015}{{arXiv:1605.07015}}} {[hep-th]}

\bibitem[{Kumar et~al.(2016)Kumar, Kulshreshtha, and
  Kulshreshtha}]{Kumar:2016sxx}
Kumar S, Kulshreshtha U, Kulshreshtha DS (2016) Charged compact boson stars and
  shells in the presence of a cosmological constant. Phys Rev D 94:125023.
  \doi{10.1103/PhysRevD.94.125023}

\bibitem[{Kunz et~al.(2006)Kunz, Navarro-Lerida, and Viebahn}]{Kunz:2006eh}
Kunz J, Navarro-Lerida F, Viebahn J (2006) Charged rotating black holes in odd
  dimensions. Phys Lett B 639:362--367. \doi{10.1016/j.physletb.2006.06.066}.
  {\href{https://arxiv.org/abs/hep-th/0605075}{{arXiv:hep-th/0605075}}}
  {[hep-th]}

\bibitem[{Kunz et~al.(2022)Kunz, Loiko, and Shnir}]{Kunz:2021mbm}
Kunz J, Loiko V, Shnir Y (2022) {U(1) gauged boson stars in the
  Einstein-Friedberg-Lee-Sirlin model}. Phys Rev D 105:085013.
  \doi{10.1103/PhysRevD.105.085013}.
  {\href{https://arxiv.org/abs/2112.06626}{{arXiv:2112.06626}}} {[gr-qc]}

\bibitem[{Kusenko and Steinhardt(2001)}]{2001PhRvL..87n1301K}
Kusenko A, Steinhardt PJ (2001) $q$-ball candidates for self-interacting dark
  matter. Phys Rev Lett 87:141301. \doi{10.1103/PhysRevLett.87.141301}.
  {\href{https://arxiv.org/abs/astro-ph/0106008}{{arXiv:astro-ph/0106008}}}

\bibitem[{Kusmartsev et~al.(1991)Kusmartsev, Mielke, and
  Schunck}]{Kusmartsev:2008py}
Kusmartsev FV, Mielke EW, Schunck FE (1991) Gravitational stability of boson
  stars. Phys Rev D 43:3895--3901. \doi{10.1103/PhysRevD.43.3895}.
  {\href{https://arxiv.org/abs/0810.0696}{{arXiv:0810.0696}}} {[astro-ph]}

\bibitem[{Laha(2020)}]{Laha:2018zav}
Laha R (2020) {Lensing of fast radio bursts: Future constraints on primordial
  black hole density with an extended mass function and a new probe of exotic
  compact fermion and boson stars}. Phys Rev D 102(2):023016.
  \doi{10.1103/PhysRevD.102.023016}.
  {\href{https://arxiv.org/abs/1812.11810}{{arXiv:1812.11810}}} {[astro-ph.CO]}

\bibitem[{Lai(2004)}]{2005PhDT.........2L}
Lai CW (2004) A numerical study of boson stars. PhD thesis, The University of
  British Columbia, Vancouver.
  \urlprefix\url{http://laplace.physics.ubc.ca/Members/matt/Doc/Theses/},
  {\href{https://arxiv.org/abs/gr-qc/0410040}{{arXiv:gr-qc/0410040}}}

\bibitem[{Lai and Choptuik(2007)}]{Lai:2007tj}
Lai CW, Choptuik MW (2007) Final fate of subcritical evolutions of boson stars.
  ArXiv e-prints {\href{https://arxiv.org/abs/0709.0324}{{arXiv:0709.0324}}}
  {[gr-qc]}

\bibitem[{Landea and Garc{\'{\i}}a(2016)}]{Garcia:2016ldc}
Landea IS, Garc{\'{\i}}a F (2016) Charged {P}roca stars. Phys Rev D 94:104006.
  \doi{10.1103/PhysRevD.94.104006}.
  {\href{https://arxiv.org/abs/1608.00011}{{arXiv:1608.00011}}} {[hep-th]}

\bibitem[{Landsberg(2006)}]{Landsberg:2006mm}
Landsberg GL (2006) Black holes at future colliders and beyond. J Phys G: Nucl
  Part Phys 32:R337--R365. \doi{10.1088/0954-3899/32/9/R02}.
  {\href{https://arxiv.org/abs/hep-ph/0607297}{{arXiv:hep-ph/0607297}}}
  {[hep-ph]}

\bibitem[{{Latifah} et~al.(2014){Latifah}, {Sulaksono}, and
  {Mart}}]{2014PhRvD..90l7501L}
{Latifah} S, {Sulaksono} A, {Mart} T (2014) Bosons star at finite temperature.
  Phys Rev D 90:127501. \doi{10.1103/PhysRevD.90.127501}.
  {\href{https://arxiv.org/abs/1412.1556}{{arXiv:1412.1556}}} {[astro-ph.SR]}

\bibitem[{de~Lavallaz and Fairbairn(2010)}]{PhysRevD.81.123521}
de~Lavallaz A, Fairbairn M (2010) Neutron stars as dark matter probes. Phys Rev
  D 81:123521. \doi{10.1103/PhysRevD.81.123521}

\bibitem[{Lee et~al.(2021)Lee, Chu, and Lin}]{Lee:2021yyn}
Lee BKK, Chu Mc, Lin LM (2021) {Could the GW190814 Secondary Component Be a
  Bosonic Dark Matter Admixed Compact Star?} Astrophys J 922(2):242.
  \doi{10.3847/1538-4357/ac2735}.
  {\href{https://arxiv.org/abs/2110.05538}{{arXiv:2110.05538}}} {[astro-ph.HE]}

\bibitem[{Lee(2010)}]{Lee:2008ab}
Lee JW (2010) Is dark matter a {BEC} or scalar field? J Korean Phys Soc 54.
  {\href{https://arxiv.org/abs/0801.1442}{{arXiv:0801.1442}}} {[astro-ph]}

\bibitem[{Lee and Lim(2010)}]{1475-7516-2010-01-007}
Lee JW, Lim S (2010) Minimum mass of galaxies from {BEC} or scalar field dark
  matter. J Cosmol Astropart Phys 2010(01):007.
  \doi{10.1088/1475-7516/2010/01/007}

\bibitem[{Lee et~al.(2008)Lee, Lim, and Choi}]{Lee:2008mq}
Lee JW, Lim S, Choi D (2008) Bec dark matter can explain collisions of galaxy
  clusters. ArXiv e-prints
  {\href{https://arxiv.org/abs/0805.3827}{{arXiv:0805.3827}}} {[hep-ph]}

\bibitem[{Lee(1987)}]{1987PhRvD..35.3637L}
Lee TD (1987) Soliton stars and the critical masses of black holes. Phys Rev D
  35:3637--3639. \doi{10.1103/PhysRevD.35.3637}

\bibitem[{Lee and Pang(1989)}]{1989NuPhB.315..477L}
Lee TD, Pang Y (1989) Stability of mini-boson stars. Nucl Phys B 315:477--516.
  \doi{10.1016/0550-3213(89)90365-9}

\bibitem[{Lee and Pang(1992)}]{1992PhR...221..251L}
Lee TD, Pang Y (1992) Nontopological solitons. Phys Rep 221:251--350.
  \doi{10.1016/0370-1573(92)90064-7}

\bibitem[{Lenzmann(2009)}]{lenzmann2009}
Lenzmann E (2009) Uniqueness of ground states for pseudorelativistic hartree
  equations. Analysis {\&} PDE 2:1--27. \doi{10.2140/apde.2009.2.1}

\bibitem[{Lenzmann and Lewin(2011)}]{2011arXiv1103.3140L}
Lenzmann E, Lewin M (2011) On singularity formation for the $l^2$-critical
  boson star equation. Nonlinearity 24:3515--3540.
  \doi{10.1088/0951-7715/24/12/009}.
  {\href{https://arxiv.org/abs/1103.3140}{{arXiv:1103.3140}}} {[math.AP]}

\bibitem[{Levkov et~al.(2018)Levkov, Panin, and Tkachev}]{Levkov:2018kau}
Levkov DG, Panin AG, Tkachev II (2018) {Gravitational Bose-Einstein
  condensation in the kinetic regime}. Phys Rev Lett 121:151301.
  \doi{10.1103/PhysRevLett.121.151301}.
  {\href{https://arxiv.org/abs/1804.05857}{{arXiv:1804.05857}}} {[astro-ph.CO]}

\bibitem[{Li et~al.(2020)Li, Sun, Hu, Song, and Wang}]{Li:2019mlk}
Li HB, Sun S, Hu TT, Song Y, Wang YQ (2020) {Rotating multistate boson stars}.
  Phys Rev D 101:044017. \doi{10.1103/PhysRevD.101.044017}.
  {\href{https://arxiv.org/abs/1906.00420}{{arXiv:1906.00420}}} {[gr-qc]}

\bibitem[{Li et~al.(2021)Li, Zeng, Song, and Wang}]{Li:2020ffy}
Li HB, Zeng YB, Song Y, Wang YQ (2021) {Self-interacting multistate boson
  stars}. JHEP 04:042. \doi{10.1007/JHEP04(2021)042}.
  {\href{https://arxiv.org/abs/2006.11281}{{arXiv:2006.11281}}} {[gr-qc]}

\bibitem[{{Li} et~al.(2012){Li}, {Harko}, and {Cheng}}]{2012JCAP...06..001L}
{Li} XY, {Harko} T, {Cheng} KS (2012) Condensate dark matter stars. J Cosmol
  Astropart Phys 2012(06):001. \doi{10.1088/1475-7516/2012/06/001}.
  {\href{https://arxiv.org/abs/1205.2932}{{arXiv:1205.2932}}} {[astro-ph.CO]}

\bibitem[{Liddle and Madsen(1992)}]{1992IJMPD...1..101L}
Liddle AR, Madsen MS (1992) The structure and formation of boson stars. Int J
  Mod Phys D 1:101--143. \doi{10.1142/S0218271892000057}

\bibitem[{Liu and Peng(2022)}]{Liu:2022iyj}
Liu G, Peng Y (2022) {A no-go theorem for scalar fields with couplings from
  Ginzburg\textendash{}Landau models}. Eur Phys J C 82:570.
  \doi{10.1140/epjc/s10052-022-10497-5}.
  {\href{https://arxiv.org/abs/2204.03833}{{arXiv:2204.03833}}} {[gr-qc]}

\bibitem[{Lopes and Panotopoulos(2020)}]{Lopes:2019eue}
Lopes I, Panotopoulos G (2020) {Radial oscillations of boson stars made of
  ultralight repulsive dark matter}. Nucl Phys B 961:115266.
  \doi{10.1016/j.nuclphysb.2020.115266}.
  {\href{https://arxiv.org/abs/1904.07191}{{arXiv:1904.07191}}} {[gr-qc]}

\bibitem[{Lora-Clavijo et~al.(2010)Lora-Clavijo, Cruz-Osorio, and
  Guzm{\'{a}}n}]{2010PhRvD..82b3005L}
Lora-Clavijo FD, Cruz-Osorio A, Guzm{\'{a}}n FS (2010) Evolution of a massless
  test scalar field on boson star space-times. Phys Rev D 82:023005.
  \doi{10.1103/PhysRevD.82.023005}.
  {\href{https://arxiv.org/abs/1007.1162}{{arXiv:1007.1162}}} {[gr-qc]}

\bibitem[{Lue and Weinberg(2000)}]{Lue:2000qr}
Lue A, Weinberg EJ (2000) Gravitational properties of monopole spacetimes near
  the black hole threshold. Phys Rev D 61:124003.
  \doi{10.1103/PhysRevD.61.124003}.
  {\href{https://arxiv.org/abs/hep-th/0001140}{{arXiv:hep-th/0001140}}}
  {[hep-th]}

\bibitem[{Lynn(1989)}]{Lynn:1988rb}
Lynn BW (1989) Q-stars. Nucl Phys 321:465--480.
  \doi{10.1016/0550-3213(89)90352-0}

\bibitem[{{Macedo} et~al.(2013{\natexlab{a}}){Macedo}, {Pani}, {Cardoso}, and
  {Crispino}}]{2013PhRvD..88f4046M}
{Macedo} CFB, {Pani} P, {Cardoso} V, {Crispino} LCB (2013{\natexlab{a}})
  Astrophysical signatures of boson stars: Quasinormal modes and inspiral
  resonances. Phys Rev D 88:064046. \doi{10.1103/PhysRevD.88.064046}.
  {\href{https://arxiv.org/abs/1307.4812}{{arXiv:1307.4812}}} {[gr-qc]}

\bibitem[{{Macedo} et~al.(2013{\natexlab{b}}){Macedo}, {Pani}, {Cardoso}, and
  {Crispino}}]{2013ApJ...774...48M}
{Macedo} CFB, {Pani} P, {Cardoso} V, {Crispino} LCB (2013{\natexlab{b}}) Into
  the lair: Gravitational-wave signatures of dark matter. Astrophys J 774:48.
  \doi{10.1088/0004-637X/774/1/48}.
  {\href{https://arxiv.org/abs/1302.2646}{{arXiv:1302.2646}}} {[gr-qc]}

\bibitem[{Macedo et~al.(2016)Macedo, Cardoso, Crispino, and
  Pani}]{Macedo:2016wgh}
Macedo CFB, Cardoso V, Crispino LCB, Pani P (2016) Quasinormal modes of
  relativistic stars and interacting fields. Phys Rev D 93:064053.
  \doi{10.1103/PhysRevD.93.064053}.
  {\href{https://arxiv.org/abs/1603.02095}{{arXiv:1603.02095}}} {[gr-qc]}

\bibitem[{{Madarassy} and {Toth}(2015)}]{2015PhRvD..91d4041M}
{Madarassy} EJM, {Toth} VT (2015) Evolution and dynamical properties of
  {B}ose-{E}instein condensate dark matter stars. Phys Rev D 91:044041.
  \doi{10.1103/PhysRevD.91.044041}.
  {\href{https://arxiv.org/abs/1412.7152}{{arXiv:1412.7152}}} {[hep-ph]}

\bibitem[{Maldacena(1998)}]{Maldacena:1997re}
Maldacena JM (1998) The large-$n$ limit of superconformal field theories and
  supergravity. Adv Theor Math Phys 2:231--252. \doi{10.1023/A:1026654312961}.
  {\href{https://arxiv.org/abs/hep-th/9711200}{{arXiv:hep-th/9711200}}}

\bibitem[{Marsh and Pop(2015)}]{Marsh:2015wka}
Marsh DJE, Pop AR (2015) Axion dark matter, solitons and the cusp--core
  problem. Mon Not R Astron Soc 451:2479--2492. \doi{10.1093/mnras/stv1050}.
  {\href{https://arxiv.org/abs/1502.03456}{{arXiv:1502.03456}}} {[astro-ph.CO]}

\bibitem[{{Marunovi{\'c}}(2015)}]{Marunovic:2015obj}
{Marunovi{\'c}} A (2015) Boson stars with nonminimal coupling. ArXiv e-prints
  {\href{https://arxiv.org/abs/1512.05718}{{arXiv:1512.05718}}} {[gr-qc]}

\bibitem[{{Marunovi{\'c}} and {Murkovi{\'c}}(2014)}]{2014CQGra..31d5010M}
{Marunovi{\'c}} A, {Murkovi{\'c}} M (2014) A novel black hole mimicker: a boson
  star and a global monopole nonminimally coupled to gravity. Class Quantum
  Grav 31:045010. \doi{10.1088/0264-9381/31/4/045010}.
  {\href{https://arxiv.org/abs/1308.6489}{{arXiv:1308.6489}}} {[gr-qc]}

\bibitem[{Mas\'o-Ferrando et~al.(2021)Mas\'o-Ferrando, Sanchis-Gual, Font, and
  Olmo}]{Maso-Ferrando:2021ngp}
Mas\'o-Ferrando A, Sanchis-Gual N, Font JA, Olmo GJ (2021) {Boson stars in
  Palatini gravity}. Class Quantum Grav 38:194003.
  \doi{10.1088/1361-6382/ac1fd0}.
  {\href{https://arxiv.org/abs/2103.15705}{{arXiv:2103.15705}}} {[gr-qc]}

\bibitem[{Mazur and Mottola(2001)}]{Mazur:2001fv}
Mazur PO, Mottola E (2001) Gravitational condensate stars: An alternative to
  black holes. ArXiv e-prints
  {\href{https://arxiv.org/abs/gr-qc/0109035}{{arXiv:gr-qc/0109035}}}

\bibitem[{McGreevy(2010)}]{McGreevy:2009xe}
McGreevy J (2010) Holographic duality with a view toward many-body physics. Adv
  High Energy Phys 2010:723105. \doi{10.1155/2010/723105}.
  {\href{https://arxiv.org/abs/0909.0518}{{arXiv:0909.0518}}} {[hep-th]}

\bibitem[{Meliani et~al.(2015)Meliani, Vincent, Grandcl{\'e}ment, Gourgoulhon,
  Monceau-Baroux, and Straub}]{Meliani:2015zta}
Meliani Z, Vincent FH, Grandcl{\'e}ment P, Gourgoulhon E, Monceau-Baroux R,
  Straub O (2015) Circular geodesics and thick tori around rotating boson
  stars. Class Quantum Grav 32:235022. \doi{10.1088/0264-9381/32/23/235022}.
  {\href{https://arxiv.org/abs/1510.04191}{{arXiv:1510.04191}}} {[astro-ph.HE]}

\bibitem[{Meliani et~al.(2016)Meliani, Grandcl{\'e}ment, Casse, Vincent,
  Straub, and Dauvergne}]{Meliani:2016rfe}
Meliani Z, Grandcl{\'e}ment P, Casse F, Vincent FH, Straub O, Dauvergne F
  (2016) Gr-amrvac code applications: accretion onto compact objects, boson
  stars versus black holes. Class Quantum Grav 33:155010.
  \doi{10.1088/0264-9381/33/15/155010}

\bibitem[{Mendes and Yang(2016)}]{Mendes:2016vdr}
Mendes RFP, Yang H (2016) Tidal deformability of dark matter clumps. ArXiv
  e-prints {\href{https://arxiv.org/abs/1606.03035}{{arXiv:1606.03035}}}
  {[astro-ph.CO]}

\bibitem[{Michelangeli and Schlein(2012)}]{2012CMaPh.311..645M}
Michelangeli A, Schlein B (2012) Dynamical collapse of boson stars. Commun Math
  Phys 311:645--687. \doi{10.1007/s00220-011-1341-7}.
  {\href{https://arxiv.org/abs/1005.3135}{{arXiv:1005.3135}}} {[math-ph]}

\bibitem[{Mielke(2016)}]{Mielke:2016war}
Mielke EW (2016) Rotating boson stars. Fundamental Theories of Physics
  183:115--131. \doi{10.1007/978-3-319-31299-6_6}

\bibitem[{Mielke and Scherzer(1981)}]{Mielke:1981}
Mielke EW, Scherzer R (1981) Geon-type solutions of the nonlinear
  {H}eisenberg-{K}lein-{G}ordon equation. Phys Rev D 24:2111--2126.
  \doi{10.1103/PhysRevD.24.2111}

\bibitem[{Mielke and Schunck(1999)}]{Mielke:1997re}
Mielke EW, Schunck FE (1999) Boson stars: Early history and recent prospects.
  In: Piran T, Ruffini R (eds) The Eighth Marcel Grossmann Meeting on Recent
  Developments in Theoretical and Experimental General Relativity, Gravitation
  and Relativistic Field Theories. World Scientific, Singapore, pp 1607--1626.
  {\href{https://arxiv.org/abs/gr-qc/9801063}{{arXiv:gr-qc/9801063}}}

\bibitem[{Mielke and Schunck(2002)}]{Mielke:2000im}
Mielke EW, Schunck FE (2002) Boson and axion stars. In: Gurzadyan VG, Jantzen
  RT, Ruffini R (eds) The Ninth Marcel Grossmann Meeting: On recent
  developments in theoretical and experimental general relativity, gravitation,
  and relativistic field theories. World Scientific, Singapore, pp 581--591

\bibitem[{Milgrom(1983)}]{milgromorig}
Milgrom M (1983) A modification of the {N}ewtonian dynamics: Implications for
  galaxies. Astrophys J 270:371--383. \doi{10.1086/161131}

\bibitem[{Milgrom(2011)}]{milgromnew}
Milgrom M (2011) {MOND} -- particularly as modified inertia. Acta Phys Pol B
  42:2175--2184. \doi{10.5506/APhysPolB.42.2175}.
  {\href{https://arxiv.org/abs/1111.1611}{{arXiv:1111.1611}}} {[astro-ph.CO]}

\bibitem[{Millward and Hirschmann(2003)}]{Millward:2002pk}
Millward RS, Hirschmann EW (2003) Critical behavior of gravitating sphalerons.
  Phys Rev D 68:024017. \doi{10.1103/PhysRevD.68.024017}.
  {\href{https://arxiv.org/abs/gr-qc/0212015}{{arXiv:gr-qc/0212015}}}

\bibitem[{Mukherjee et~al.(2015)Mukherjee, Shah, and Bose}]{Mukherjee:2014kqa}
Mukherjee A, Shah S, Bose S (2015) Observational constraints on spinning,
  relativistic {B}ose-{E}instein condensate stars. Phys Rev D 91:084051.
  \doi{10.1103/PhysRevD.91.084051}.
  {\href{https://arxiv.org/abs/1409.6490}{{arXiv:1409.6490}}} {[astro-ph.HE]}

\bibitem[{Mundim(2010)}]{Mundim:2010hi}
Mundim BC (2010) A numerical study of boson star binaries. PhD thesis, The
  University of British Columbia, Vancouver.
  \urlprefix\url{http://laplace.physics.ubc.ca/Members/matt/Doc/Theses/},
  {\href{https://arxiv.org/abs/1003.0239}{{arXiv:1003.0239}}} {[gr-qc]}

\bibitem[{Murariu and Puscasu(2010)}]{Murariu:2010zz}
Murariu G, Puscasu G (2010) Solutions for maxwell-equations' system in a static
  conformal space-time. Rom J Phys 55:47--52

\bibitem[{Murariu et~al.(2008)Murariu, Dariescu, and Dariescu}]{dariescu2008}
Murariu G, Dariescu C, Dariescu MA (2008) Maple routines for bosons on curved
  manifolds. Rom J Phys 53:99--108

\bibitem[{Myers and Perry(1986)}]{Myers:1986un}
Myers RC, Perry MJ (1986) Black holes in higher dimensional space-times. Ann
  Phys (NY) 172:304--347. \doi{10.1016/0003-4916(86)90186-7}

\bibitem[{Ni et~al.(2016)Ni, Zhou, Cardenas-Avendano, Bambi, Herdeiro, and
  Radu}]{Ni:2016rhz}
Ni Y, Zhou M, Cardenas-Avendano A, Bambi C, Herdeiro CAR, Radu E (2016) Iron
  k$\alpha$ line of {K}err black holes with scalar hair. J Cosmol Astropart
  Phys 1607(07):049. \doi{10.1088/1475-7516/2016/07/049}.
  {\href{https://arxiv.org/abs/1606.04654}{{arXiv:1606.04654}}} {[gr-qc]}

\bibitem[{{Nogueira}(2013)}]{2013PhRvD..87j6006N}
{Nogueira} F (2013) Extremal surfaces in asymptotically {AdS} charged boson
  stars backgrounds. Phys Rev D 87:106006. \doi{10.1103/PhysRevD.87.106006}.
  {\href{https://arxiv.org/abs/1301.4316}{{arXiv:1301.4316}}} {[hep-th]}

\bibitem[{N{\'{u}}{\~{n}}ez et~al.(2011)N{\'{u}}{\~{n}}ez, Degollado, and
  Moreno}]{2011PhRvD..84b4043N}
N{\'{u}}{\~{n}}ez D, Degollado JC, Moreno C (2011) Gravitational waves from
  scalar field accretion. Phys Rev D 84:024043.
  \doi{10.1103/PhysRevD.84.024043}.
  {\href{https://arxiv.org/abs/1107.4316}{{arXiv:1107.4316}}} {[gr-qc]}

\bibitem[{Nyhan and Kain(2022)}]{Nyhan:2022pda}
Nyhan JE, Kain B (2022) {Dynamical evolution of fermion-boson stars with
  realistic equations of state}. Phys Rev D 105(12):123016.
  \doi{10.1103/PhysRevD.105.123016}.
  {\href{https://arxiv.org/abs/2206.07715}{{arXiv:2206.07715}}} {[gr-qc]}

\bibitem[{Okawa(2015)}]{Okawa:2015fsa}
Okawa H (2015) Nonlinear evolutions of bosonic clouds around black holes. Class
  Quantum Grav 32:214003. \doi{10.1088/0264-9381/32/21/214003}

\bibitem[{Olabarrieta et~al.(2007)Olabarrieta, Ventrella, Choptuik, and
  Unruh}]{Olabarrieta:2007di}
Olabarrieta I, Ventrella JF, Choptuik MW, Unruh WG (2007) {Critical Behavior in
  the Gravitational Collapse of a Scalar Field with Angular Momentum in
  Spherical Symmetry}. Phys Rev D 76:124014. \doi{10.1103/PhysRevD.76.124014}.
  {\href{https://arxiv.org/abs/0708.0513}{{arXiv:0708.0513}}} {[gr-qc]}

\bibitem[{Olivares et~al.(2020)Olivares, Younsi, Fromm, De~Laurentis, Porth,
  Mizuno, Falcke, Kramer, and Rezzolla}]{Olivares:2018abq}
Olivares H, Younsi Z, Fromm CM, De~Laurentis M, Porth O, Mizuno Y, Falcke H,
  Kramer M, Rezzolla L (2020) {How to tell an accreting boson star from a black
  hole}. Mon Not Roy Astron Soc 497(1):521--535. \doi{10.1093/mnras/staa1878}.
  {\href{https://arxiv.org/abs/1809.08682}{{arXiv:1809.08682}}} {[gr-qc]}

\bibitem[{Ontanon and Alcubierre(2021)}]{Ontanon:2021hbg}
Ontanon S, Alcubierre M (2021) {Rotating boson stars using finite differences
  and global Newton methods}. Class Quantum Grav 38:154003.
  \doi{10.1088/1361-6382/ac0b53}.
  {\href{https://arxiv.org/abs/2103.13993}{{arXiv:2103.13993}}} {[gr-qc]}

\bibitem[{Pacilio et~al.(2020)Pacilio, Vaglio, Maselli, and
  Pani}]{Pacilio:2020jza}
Pacilio C, Vaglio M, Maselli A, Pani P (2020) {Gravitational-wave detectors as
  particle-physics laboratories: Constraining scalar interactions with a
  coherent inspiral model of boson-star binaries}. Phys Rev D 102(8):083002.
  \doi{10.1103/PhysRevD.102.083002}.
  {\href{https://arxiv.org/abs/2007.05264}{{arXiv:2007.05264}}} {[gr-qc]}

\bibitem[{Page(2004)}]{2004PhRvD..70b3002P}
Page DN (2004) Classical and quantum decay of oscillations: Oscillating
  self-gravitating real scalar field solitons. Phys Rev D 70:023002.
  \doi{10.1103/PhysRevD.70.023002}.
  {\href{https://arxiv.org/abs/arXiv:gr-qc/0310006}{{arXiv:gr-qc/0310006}}}

\bibitem[{Palenzuela et~al.(2007)Palenzuela, Olabarrieta, Lehner, and
  Liebling}]{Palenzuela:2006wp}
Palenzuela C, Olabarrieta I, Lehner L, Liebling SL (2007) Head-on collisions of
  boson stars. Phys Rev D 75:064005. \doi{10.1103/PhysRevD.75.064005}.
  {\href{https://arxiv.org/abs/gr-qc/0612067}{{arXiv:gr-qc/0612067}}}

\bibitem[{Palenzuela et~al.(2008)Palenzuela, Lehner, and
  Liebling}]{Palenzuela:2007dm}
Palenzuela C, Lehner L, Liebling SL (2008) Orbital dynamics of binary boson
  star systems. Phys Rev D 77:044036. \doi{10.1103/PhysRevD.77.044036}.
  {\href{https://arxiv.org/abs/0706.2435}{{arXiv:0706.2435}}} {[gr-qc]}

\bibitem[{Palenzuela et~al.(2017)Palenzuela, Pani, Bezares, Cardoso, Lehner,
  and Liebling}]{Palenzuela:2017kcg}
Palenzuela C, Pani P, Bezares M, Cardoso V, Lehner L, Liebling S (2017)
  {Gravitational Wave Signatures of Highly Compact Boson Star Binaries}. Phys
  Rev D 96(10):104058. \doi{10.1103/PhysRevD.96.104058}.
  {\href{https://arxiv.org/abs/1710.09432}{{arXiv:1710.09432}}} {[gr-qc]}

\bibitem[{Pani et~al.(2009)Pani, Berti, Cardoso, Chen, and
  Norte}]{2009PhRvD..80l4047P}
Pani P, Berti E, Cardoso V, Chen Y, Norte R (2009) Gravitational wave
  signatures of the absence of an event horizon: Nonradial oscillations of a
  thin-shell gravastar. Phys Rev D 80:124047. \doi{10.1103/PhysRevD.80.124047}.
  {\href{https://arxiv.org/abs/0909.0287}{{arXiv:0909.0287}}} {[gr-qc]}

\bibitem[{{Pani} et~al.(2011){Pani}, {Berti}, {Cardoso}, and
  {Read}}]{Pani:2011xm}
{Pani} P, {Berti} E, {Cardoso} V, {Read} J (2011) Compact stars in alternative
  theories of gravity: {E}instein-dilaton-{G}auss-{B}onnet gravity. Phys Rev D
  84:104035. \doi{10.1103/PhysRevD.84.104035}.
  {\href{https://arxiv.org/abs/1109.0928}{{arXiv:1109.0928}}} {[gr-qc]}

\bibitem[{Park(2012)}]{Park:2012fe}
Park SC (2012) Black holes and the {LHC}: A review. Prog Part Nucl Phys
  67:617--650. \doi{10.1016/j.ppnp.2012.03.004}.
  {\href{https://arxiv.org/abs/1203.4683}{{arXiv:1203.4683}}} {[hep-ph]}

\bibitem[{Peccei and Quinn(1977)}]{Peccei:1977hh}
Peccei RD, Quinn HR (1977) {CP Conservation in the Presence of Instantons}.
  Phys Rev Lett 38:1440--1443. \doi{10.1103/PhysRevLett.38.1440}

\bibitem[{Pena and Sudarsky(1997)}]{Pena:1997cy}
Pena I, Sudarsky D (1997) Do collapsed boson stars result in new types of black
  holes? Class Quantum Grav 14:3131--3134. \doi{10.1088/0264-9381/14/11/013}

\bibitem[{Peng(2020)}]{Peng:2019uzw}
Peng Y (2020) {No-go theorem for static boson stars with negative cosmological
  constants}. Nucl Phys B 953:114955. \doi{10.1016/j.nuclphysb.2020.114955}.
  {\href{https://arxiv.org/abs/1902.06508}{{arXiv:1902.06508}}} {[gr-qc]}

\bibitem[{Petryk(2006)}]{2006PhDT.........6P}
Petryk RJW (2006) Maxwell-{K}lein-{G}ordon fields in black hole spacetimes. PhD
  thesis, The University of British Columbia, Vancouver.
  \urlprefix\url{http://laplace.physics.ubc.ca/Members/matt/Doc/Theses/}

\bibitem[{Pisano and Tomazelli(1996)}]{1996MPLA...11..647P}
Pisano F, Tomazelli JL (1996) Stars of wimps. Mod Phys Lett A 11:647--651.
  \doi{10.1142/S0217732396000667}.
  {\href{https://arxiv.org/abs/gr-qc/9509022}{{arXiv:gr-qc/9509022}}}

\bibitem[{Polchinski(2010)}]{Polchinski:2010hw}
Polchinski J (2010) Introduction to gauge/gravity duality. In: Proceedings,
  Theoretical Advanced Study Institute in Elementary Particle Physics (TASI
  2010). String Theory and Its Applications: From meV to the Planck Scale:
  Boulder, Colorado, USA, June 1--25, 2010. pp 3--46.
  \doi{10.1142/9789814350525_0001}.
  {\href{https://arxiv.org/abs/1010.6134}{{arXiv:1010.6134}}} {[hep-th]}

\bibitem[{Ponglertsakul et~al.(2016)Ponglertsakul, Winstanley, and
  Dolan}]{Ponglertsakul:2016wae}
Ponglertsakul S, Winstanley E, Dolan SR (2016) Stability of gravitating
  charged-scalar solitons in a cavity. Phys Rev D 94:024031.
  \doi{10.1103/PhysRevD.94.024031}.
  {\href{https://arxiv.org/abs/1604.01132}{{arXiv:1604.01132}}} {[gr-qc]}

\bibitem[{Power and Wheeler(1957)}]{Power:1957zz}
Power EA, Wheeler JA (1957) Thermal geons. Rev Mod Phys 29:480--495.
  \doi{10.1103/RevModPhys.29.480}

\bibitem[{Psaltis(2008)}]{lrr-2008-9}
Psaltis D (2008) Probes and tests of strong-field gravity with observations in
  the electromagnetic spectrum. Living Rev Relativity 11:9.
  \doi{10.12942/lrr-2008-9}.
  {\href{https://arxiv.org/abs/0806.1531}{{arXiv:0806.1531}}}

\bibitem[{{Pugliese} et~al.(2013){Pugliese}, {Quevedo}, {Rueda H.}, and
  {Ruffini}}]{2013PhRvD..88b4053P}
{Pugliese} D, {Quevedo} H, {Rueda H} JA, {Ruffini} R (2013) Charged boson
  stars. Phys Rev D 88:024053. \doi{10.1103/PhysRevD.88.024053}.
  {\href{https://arxiv.org/abs/1305.4241}{{arXiv:1305.4241}}} {[astro-ph.HE]}

\bibitem[{{Radu} and {Subagyo}(2012)}]{2012PhLB..717..450R}
{Radu} E, {Subagyo} B (2012) Spinning scalar solitons in anti-de {S}itter
  spacetime. Phys Lett B 717:450--457. \doi{10.1016/j.physletb.2012.09.050}.
  {\href{https://arxiv.org/abs/1207.3715}{{arXiv:1207.3715}}} {[gr-qc]}

\bibitem[{Rangamani and Takayanagi(2017)}]{Rangamani:2016dms}
Rangamani M, Takayanagi T (2017) Holographic Entanglement Entropy, Lecture
  Notes in Physics, vol 931. Springer, Cham. \doi{10.1007/978-3-319-52573-0}.
  {\href{https://arxiv.org/abs/1609.01287}{{arXiv:1609.01287}}} {[hep-th]}

\bibitem[{Reid and Choptuik(2016)}]{Reid:2015hiq}
Reid GD, Choptuik MW (2016) Nonminimally coupled topological-defect boson
  stars: Static solutions. Phys Rev D 93:044022.
  \doi{10.1103/PhysRevD.93.044022}.
  {\href{https://arxiv.org/abs/1512.02142}{{arXiv:1512.02142}}} {[gr-qc]}

\bibitem[{Rindler-Daller and Shapiro(2012)}]{RindlerDaller:2011kx}
Rindler-Daller T, Shapiro PR (2012) Angular momentum and vortex formation in
  {B}ose-{E}instein-condensed cold dark matter haloes. Mon Not R Astron Soc
  422:135--161. \doi{10.1111/j.1365-2966.2012.20588.x}.
  {\href{https://arxiv.org/abs/1106.1256}{{arXiv:1106.1256}}} {[astro-ph.CO]}

\bibitem[{Roger et~al.(2016)Roger, Maitland, Wilson, Westerberg, Vocke, Wright,
  and Faccio}]{Roger:2016slp}
Roger T, Maitland C, Wilson K, Westerberg N, Vocke D, Wright EM, Faccio D
  (2016) Optical analogues of the {N}ewton-{S}chr\"odinger equation and boson
  star evolution. ArXiv e-prints
  {\href{https://arxiv.org/abs/1611.00924}{{arXiv:1611.00924}}}
  {[physics.optics]}

\bibitem[{Roque and Ure\~na L\'opez(2022)}]{Roque:2021lvr}
Roque AA, Ure\~na L\'opez LA (2022) {Horndeski fermion\textendash{}boson
  stars}. Class Quantum Grav 39:044001. \doi{10.1088/1361-6382/ac4614}.
  {\href{https://arxiv.org/abs/2109.14747}{{arXiv:2109.14747}}} {[gr-qc]}

\bibitem[{Rosa and Rubiera-Garcia(2022)}]{Rosa:2022tfv}
Rosa JaL, Rubiera-Garcia D (2022) {Shadows of boson and Proca stars with thin
  accretion disks}. arXiv e-prints
  {\href{https://arxiv.org/abs/2204.12949}{{arXiv:2204.12949}}} {[gr-qc]}

\bibitem[{Rosen(1966)}]{1966JMP.....7.2066R}
Rosen G (1966) Existence of particlelike solutions to nonlinear field theories.
  J Math Phys 7:2066--2070. \doi{10.1063/1.1704890}

\bibitem[{Rousseau(2003)}]{rousseau}
Rousseau B (2003) Axisymmetric boson stars in the conformally flat
  approximation. Master's thesis, The University of British Columbia,
  Vancouver.
  \urlprefix\url{http://laplace.physics.ubc.ca/Members/matt/Doc/Theses/}

\bibitem[{Ruffini and Bonazzola(1969)}]{PhysRev.187.1767}
Ruffini R, Bonazzola S (1969) Systems of self-gravitating particles in general
  relativity and the concept of an equation of state. Phys Rev 187:1767--1783.
  \doi{10.1103/PhysRev.187.1767}

\bibitem[{{Ruiz} et~al.(2012){Ruiz}, {Degollado}, {Alcubierre},
  {N{\'u}{\~n}ez}, and {Salgado}}]{2012PhRvD..86j4044R}
{Ruiz} M, {Degollado} JC, {Alcubierre} M, {N{\'u}{\~n}ez} D, {Salgado} M (2012)
  Induced scalarization in boson stars and scalar gravitational radiation. Phys
  Rev D 86:104044. \doi{10.1103/PhysRevD.86.104044}.
  {\href{https://arxiv.org/abs/1207.6142}{{arXiv:1207.6142}}} {[gr-qc]}

\bibitem[{Ryder(1996)}]{ryderbook}
Ryder LH (1996) Quantum Field Theory, 2nd edn. Cambridge University Press,
  Cambridge; New York

\bibitem[{Sakai and Tamaki(2012)}]{Sakai:2011wn}
Sakai N, Tamaki T (2012) What happens to {Q}-balls if $q$ is so large? Phys Rev
  D 85:104008. \doi{10.1103/PhysRevD.85.104008}.
  {\href{https://arxiv.org/abs/1112.5559}{{arXiv:1112.5559}}} {[gr-qc]}

\bibitem[{Sakamoto and Shiraishi(1998{\natexlab{a}})}]{Sakamoto:1998hq}
Sakamoto K, Shiraishi K (1998{\natexlab{a}}) Boson stars with large
  selfinteraction in (2+1)-dimensions: An exact solution. J High Energy Phys
  1998(07):015. \doi{10.1088/1126-6708/1998/07/015}.
  {\href{https://arxiv.org/abs/gr-qc/9804067}{{arXiv:gr-qc/9804067}}} {[gr-qc]}

\bibitem[{Sakamoto and Shiraishi(1998{\natexlab{b}})}]{Sakamoto:1998aj}
Sakamoto K, Shiraishi K (1998{\natexlab{b}}) Exact solutions for boson fermion
  stars in (2+1)-dimensions. Phys Rev D 58:124017.
  \doi{10.1103/PhysRevD.58.124017}.
  {\href{https://arxiv.org/abs/gr-qc/9806040}{{arXiv:gr-qc/9806040}}} {[gr-qc]}

\bibitem[{Sanchis-Gual et~al.(2016)Sanchis-Gual, Degollado, Montero, Font, and
  Herdeiro}]{PhysRevLett.116.141101}
Sanchis-Gual N, Degollado JC, Montero PJ, Font JA, Herdeiro C (2016) Explosion
  and final state of an unstable {Reissner}-{Nordstr\"om} black hole. Phys Rev
  Lett 116:141101. \doi{10.1103/PhysRevLett.116.141101}

\bibitem[{{Sanchis-Gual} et~al.(2017{\natexlab{a}}){Sanchis-Gual}, {Herdeiro},
  {Radu}, {Degollado}, and {Font}}]{Sanchis-Gual:2017bhw}
{Sanchis-Gual} N, {Herdeiro} C, {Radu} E, {Degollado} JC, {Font} JA
  (2017{\natexlab{a}}) {Numerical evolutions of spherical {P}roca stars}. \prd
  95:104028. \doi{10.1103/PhysRevD.95.104028}.
  {\href{https://arxiv.org/abs/1702.04532}{{arXiv:1702.04532}}} {[gr-qc]}

\bibitem[{{Sanchis-Gual} et~al.(2017{\natexlab{b}}){Sanchis-Gual}, {Herdeiro},
  {Radu}, {Degollado}, and {Font}}]{2017PhRvD..95j4028S}
{Sanchis-Gual} N, {Herdeiro} C, {Radu} E, {Degollado} JC, {Font} JA
  (2017{\natexlab{b}}) Numerical evolutions of spherical {P}roca stars. Phys
  Rev D 95:104028. \doi{10.1103/PhysRevD.95.104028}.
  {\href{https://arxiv.org/abs/1702.04532}{{arXiv:1702.04532}}} {[gr-qc]}

\bibitem[{Sanchis-Gual et~al.(2019{\natexlab{a}})Sanchis-Gual, Di~Giovanni,
  Zilh\~ao, Herdeiro, Cerd\'a-Dur\'an, Font, and Radu}]{Sanchis-Gual:2019ljs}
Sanchis-Gual N, Di~Giovanni F, Zilh\~ao M, Herdeiro C, Cerd\'a-Dur\'an P, Font
  JA, Radu E (2019{\natexlab{a}}) {Nonlinear Dynamics of Spinning Bosonic
  Stars: Formation and Stability}. Phys Rev Lett 123:221101.
  \doi{10.1103/PhysRevLett.123.221101}.
  {\href{https://arxiv.org/abs/1907.12565}{{arXiv:1907.12565}}} {[gr-qc]}

\bibitem[{Sanchis-Gual et~al.(2019{\natexlab{b}})Sanchis-Gual, Herdeiro, Font,
  Radu, and Di~Giovanni}]{PhysRevD.99.024017}
Sanchis-Gual N, Herdeiro C, Font JA, Radu E, Di~Giovanni F (2019{\natexlab{b}})
  Head-on collisions and orbital mergers of proca stars. Phys Rev D 99:024017.
  \doi{10.1103/PhysRevD.99.024017}

\bibitem[{Sanchis-Gual et~al.(2020)Sanchis-Gual, Zilh\~ao, Herdeiro,
  Di~Giovanni, Font, and Radu}]{Sanchis-Gual:2020mzb}
Sanchis-Gual N, Zilh\~ao M, Herdeiro C, Di~Giovanni F, Font JA, Radu E (2020)
  {Synchronized gravitational atoms from mergers of bosonic stars}. Phys Rev D
  102(10):101504. \doi{10.1103/PhysRevD.102.101504}.
  {\href{https://arxiv.org/abs/2007.11584}{{arXiv:2007.11584}}} {[gr-qc]}

\bibitem[{Sanchis-Gual et~al.(2021)Sanchis-Gual, Di~Giovanni, Herdeiro, Radu,
  and Font}]{Sanchis-Gual:2021edp}
Sanchis-Gual N, Di~Giovanni F, Herdeiro C, Radu E, Font JA (2021) {Multifield,
  Multifrequency Bosonic Stars and a Stabilization Mechanism}. Phys Rev Lett
  126:241105. \doi{10.1103/PhysRevLett.126.241105}.
  {\href{https://arxiv.org/abs/2103.12136}{{arXiv:2103.12136}}} {[gr-qc]}

\bibitem[{Sanchis-Gual et~al.(2022)Sanchis-Gual, Herdeiro, and
  Radu}]{Sanchis-Gual:2021phr}
Sanchis-Gual N, Herdeiro C, Radu E (2022) {Self-interactions can stabilize
  excited boson stars}. Class Quantum Grav 39:064001.
  \doi{10.1088/1361-6382/ac4b9b}.
  {\href{https://arxiv.org/abs/2110.03000}{{arXiv:2110.03000}}} {[gr-qc]}

\bibitem[{Schive et~al.(2014)Schive, Chiueh, and Broadhurst}]{Schive:2014dra}
Schive HY, Chiueh T, Broadhurst T (2014) Cosmic structure as the quantum
  interference of a coherent dark wave. Nature Phys 10:496--499.
  \doi{10.1038/nphys2996}.
  {\href{https://arxiv.org/abs/1406.6586}{{arXiv:1406.6586}}} {[astro-ph.GA]}

\bibitem[{Schunck and Mielke(1996)}]{1996rscc.conf..138S}
Schunck FE, Mielke EW (1996) Rotating boson stars. In: Hehl FW, Puntigam RA,
  Ruder H (eds) Relativity and Scientific Computing: Computer Algebra,
  Numerics, Visualization. Springer, Berlin; New York, pp 138--151.
  \doi{10.1007/978-3-642-95732-1_7}

\bibitem[{Schunck and Mielke(2003)}]{Schunck:2003kk}
Schunck FE, Mielke EW (2003) General relativistic boson stars. Class Quantum
  Grav 20:R301--R356. \doi{10.1088/0264-9381/20/20/201}.
  {\href{https://arxiv.org/abs/0801.0307}{{arXiv:0801.0307}}} {[astro-ph]}

\bibitem[{Schunck and Torres(2000)}]{2000IJMPD...9..601S}
Schunck FE, Torres DF (2000) Boson stars with generic self-interactions. Int J
  Mod Phys D 9:601--618. \doi{10.1142/S0218271800000608}.
  {\href{https://arxiv.org/abs/gr-qc/9911038}{{arXiv:gr-qc/9911038}}}

\bibitem[{Schupp and van~der Bij(1996)}]{Schupp:1995dy}
Schupp B, van~der Bij JJ (1996) {An axially symmetric Newtonian boson star}.
  Phys Lett B 366:85--88. \doi{10.1016/0370-2693(95)01327-X}.
  {\href{https://arxiv.org/abs/astro-ph/9508017}{{arXiv:astro-ph/9508017}}}

\bibitem[{Schwabe et~al.(2016)Schwabe, Niemeyer, and Engels}]{Schwabe:2016rze}
Schwabe B, Niemeyer JC, Engels JF (2016) Simulations of solitonic core mergers
  in ultralight axion dark matter cosmologies. Phys Rev D 94:043513.
  \doi{10.1103/PhysRevD.94.043513}.
  {\href{https://arxiv.org/abs/1606.05151}{{arXiv:1606.05151}}} {[astro-ph.CO]}

\bibitem[{Seidel and Suen(1990)}]{1990PhRvD..42..384S}
Seidel E, Suen WM (1990) Dynamical evolution of boson stars: Perturbing the
  ground state. Phys Rev D 42:384--403. \doi{10.1103/PhysRevD.42.384}

\bibitem[{Seidel and Suen(1991)}]{1991PhRvL..66.1659S}
Seidel E, Suen WM (1991) Oscillating soliton stars. Phys Rev Lett
  66:1659--1662. \doi{10.1103/PhysRevLett.66.1659}

\bibitem[{Seidel and Suen(1994)}]{1994PhRvL..72.2516S}
Seidel E, Suen WM (1994) Formation of solitonic stars through gravitational
  cooling. Phys Rev Lett 72:2516--2519. \doi{10.1103/PhysRevLett.72.2516}.
  {\href{https://arxiv.org/abs/gr-qc/9309015}{{arXiv:gr-qc/9309015}}}

\bibitem[{Sharma et~al.(2008)Sharma, Karmakar, and
  Mukherjee}]{2008arXiv0812.3470S}
Sharma R, Karmakar S, Mukherjee S (2008) Boson star and dark matter. ArXiv
  e-prints {\href{https://arxiv.org/abs/0812.3470}{{arXiv:0812.3470}}}
  {[gr-qc]}

\bibitem[{Shen et~al.(2017)Shen, Zhou, Bambi, Herdeiro, and
  Radu}]{Shen:2016acv}
Shen T, Zhou M, Bambi C, Herdeiro CAR, Radu E (2017) Iron {K}$\alpha$ line of
  {P}roca stars. J Cosmol Astropart Phys 2017(08):014.
  \doi{10.1088/1475-7516/2017/08/014}.
  {\href{https://arxiv.org/abs/1701.00192}{{arXiv:1701.00192}}} {[gr-qc]}

\bibitem[{Shibata and Nakamura(1995)}]{PhysRevD.52.5428}
Shibata M, Nakamura T (1995) Evolution of three-dimensional gravitational
  waves: Harmonic slicing case. Phys Rev D 52:5428--5444.
  \doi{10.1103/PhysRevD.52.5428}

\bibitem[{Shibata and Yoshino(2010)}]{Shibata:2010wz}
Shibata M, Yoshino H (2010) Bar-mode instability of rapidly spinning black hole
  in higher dimensions: Numerical simulation in general relativity. Phys Rev D
  81:104035. \doi{10.1103/PhysRevD.81.104035}.
  {\href{https://arxiv.org/abs/1004.4970}{{arXiv:1004.4970}}} {[gr-qc]}

\bibitem[{Shnir(2022)}]{Shnir:2022lba}
Shnir Y (2022) {Boson Stars}. arXiv e-prints
  {\href{https://arxiv.org/abs/2204.06374}{{arXiv:2204.06374}}} {[gr-qc]}

\bibitem[{Siemonsen and East(2021)}]{Siemonsen:2020hcg}
Siemonsen N, East WE (2021) {Stability of rotating scalar boson stars with
  nonlinear interactions}. Phys Rev D 103:044022.
  \doi{10.1103/PhysRevD.103.044022}.
  {\href{https://arxiv.org/abs/2011.08247}{{arXiv:2011.08247}}} {[gr-qc]}

\bibitem[{Silveira and de~Sousa(1995)}]{1995PhRvD..52.5724S}
Silveira V, de~Sousa CMG (1995) Boson star rotation: A {N}ewtonian
  approximation. Phys Rev D 52:5724--5728. \doi{10.1103/PhysRevD.52.5724}.
  {\href{https://arxiv.org/abs/astro-ph/9508034}{{arXiv:astro-ph/9508034}}}

\bibitem[{Sirunyan et~al.(2017)}]{Sirunyan:2017anm}
Sirunyan AM, et~al. (2017) Search for black holes in high-multiplicity final
  states in proton-proton collisions at sqrt(s) = 13 {TeV}. ArXiv e-prints
  {\href{https://arxiv.org/abs/1705.01403}{{arXiv:1705.01403}}} {[hep-ex]}

\bibitem[{Smoli{\'c}(2015)}]{Smolic:2015txa}
Smoli{\'c} I (2015) Symmetry inheritance of scalar fields. Class Quantum Grav
  32:145010. \doi{10.1088/0264-9381/32/14/145010}.
  {\href{https://arxiv.org/abs/1501.04967}{{arXiv:1501.04967}}} {[gr-qc]}

\bibitem[{Soni and Zhang(2017)}]{Soni:2016yes}
Soni A, Zhang Y (2017) Gravitational waves from {SU($N$)} glueball dark matter.
  Phys Lett B 771:379--384. \doi{10.1016/j.physletb.2017.05.077}.
  {\href{https://arxiv.org/abs/1610.06931}{{arXiv:1610.06931}}} {[hep-ph]}

\bibitem[{de~Sousa et~al.(1998)de~Sousa, Tomazelli, and
  Silveira}]{1998PhRvD..58l3003D}
de~Sousa CMG, Tomazelli JL, Silveira V (1998) Model for stars of interacting
  bosons and fermions. Phys Rev D 58:123003. \doi{10.1103/PhysRevD.58.123003}.
  {\href{https://arxiv.org/abs/gr-qc/9507043}{{arXiv:gr-qc/9507043}}}

\bibitem[{de~Sousa et~al.(2001)de~Sousa, Silveira, and
  Fang}]{2001IJMPD..10..881D}
de~Sousa CMG, Silveira V, Fang LZ (2001) Slowly rotating boson-fermion star.
  Int J Mod Phys D 10:881--892. \doi{10.1142/S0218271801001360}.
  {\href{https://arxiv.org/abs/gr-qc/0012020}{{arXiv:gr-qc/0012020}}}

\bibitem[{{Stavridis} and {Kokkotas}(2005)}]{2005IJMPD..14..543S}
{Stavridis} A, {Kokkotas} KD (2005) {Evolution Equations for Slowly Rotating
  Stars}. Int J Mod Phys D 14:543--571. \doi{10.1142/S021827180500592X}.
  {\href{https://arxiv.org/abs/gr-qc/0411019}{{arXiv:gr-qc/0411019}}} {[gr-qc]}

\bibitem[{Stewart(1982)}]{0034-4885-45-2-002}
Stewart I (1982) Catastrophe theory in physics. Rep Prog Phys 45:185--221.
  \doi{10.1088/0034-4885/45/2/002}

\bibitem[{Stojkovic(2003)}]{Stojkovic:2001qi}
Stojkovic D (2003) Nontopological solitons in brane world models. Phys Rev D
  67:045012. \doi{10.1103/PhysRevD.67.045012}.
  {\href{https://arxiv.org/abs/hep-ph/0111061}{{arXiv:hep-ph/0111061}}}
  {[hep-ph]}

\bibitem[{{Stotyn} and {Mann}(2012)}]{2012JPhA...45K4025S}
{Stotyn} S, {Mann} RB (2012) Another mass gap in the btz geometry? J Phys A
  45:374025. \doi{10.1088/1751-8113/45/37/374025}.
  {\href{https://arxiv.org/abs/1203.0214}{{arXiv:1203.0214}}} {[gr-qc]}

\bibitem[{Stotyn et~al.(2012)Stotyn, Park, McGrath, and Mann}]{Stotyn:2011ns}
Stotyn S, Park M, McGrath P, Mann RB (2012) Black holes and boson stars with
  one {K}illing field in arbitrary odd dimensions. Phys Rev D 85:044036.
  \doi{10.1103/PhysRevD.85.044036}.
  {\href{https://arxiv.org/abs/1110.2223}{{arXiv:1110.2223}}} {[hep-th]}

\bibitem[{{Stotyn} et~al.(2014{\natexlab{a}}){Stotyn}, {Chanona}, and
  {Mann}}]{2014PhRvD..89d4018S}
{Stotyn} S, {Chanona} M, {Mann} RB (2014{\natexlab{a}}) Numerical boson stars
  with a single {K}illing vector. {II}. the d=3 case. Phys Rev D 89:044018.
  \doi{10.1103/PhysRevD.89.044018}.
  {\href{https://arxiv.org/abs/1309.2911}{{arXiv:1309.2911}}} {[hep-th]}

\bibitem[{{Stotyn} et~al.(2014{\natexlab{b}}){Stotyn}, {Leonard}, {Oltean},
  {Henderson}, and {Mann}}]{2014PhRvD..89d4017S}
{Stotyn} S, {Leonard} CD, {Oltean} M, {Henderson} LJ, {Mann} RB
  (2014{\natexlab{b}}) Numerical boson stars with a single {K}illing vector
  {I}. the $d\ge5$ case. Phys Rev D 89:044017.
  \doi{10.1103/PhysRevD.89.044017}.
  {\href{https://arxiv.org/abs/1307.8159}{{arXiv:1307.8159}}} {[hep-th]}

\bibitem[{Straumann(1984)}]{1984grra.book.....S}
Straumann N (1984) General Relativity and Relativistic Astrophysics. Springer,
  Berlin; New York. \doi{10.1007/978-3-642-84439-3}

\bibitem[{Straumann(1992)}]{Straumann:1991pt}
Straumann N (1992) Fermion and boson stars. In: Ehlers J, Sch{\"{a}}fer G (eds)
  Relativistic gravity research with emphasis on experiments and observations.
  Lecture Notes in Physics, vol 410. Springer, Berlin; New York, pp 267--293.
  \doi{10.1007/3-540-56180-3_12}

\bibitem[{Tamaki and Sakai(2010)}]{2010PhRvD..81l4041T}
Tamaki T, Sakai N (2010) Unified picture of {Q}-balls and boson stars via
  catastrophe theory. Phys Rev D 81:124041. \doi{10.1103/PhysRevD.81.124041}.
  {\href{https://arxiv.org/abs/1105.1498}{{arXiv:1105.1498}}} {[gr-qc]}

\bibitem[{Tamaki and Sakai(2011{\natexlab{a}})}]{2011PhRvD..83h4046T}
Tamaki T, Sakai N (2011{\natexlab{a}}) Gravitating {Q}-balls in the
  {A}ffleck-{D}ine mechanism. Phys Rev D 83:084046.
  \doi{10.1103/PhysRevD.83.084046}.
  {\href{https://arxiv.org/abs/1105.3810}{{arXiv:1105.3810}}} {[gr-qc]}

\bibitem[{Tamaki and Sakai(2011{\natexlab{b}})}]{2011PhRvD..83d4027T}
Tamaki T, Sakai N (2011{\natexlab{b}}) How does gravity save or kill {Q}-balls?
  Phys Rev D 83:044027. \doi{10.1103/PhysRevD.83.044027}.
  {\href{https://arxiv.org/abs/1105.2932}{{arXiv:1105.2932}}} {[gr-qc]}

\bibitem[{Tamaki and Sakai(2011{\natexlab{c}})}]{2011PhRvD..84d4054T}
Tamaki T, Sakai N (2011{\natexlab{c}}) What are universal features of
  gravitating {Q}-balls? Phys Rev D 84:044054.
  \doi{10.1103/PhysRevD.84.044054}.
  {\href{https://arxiv.org/abs/1108.3902}{{arXiv:1108.3902}}} {[gr-qc]}

\bibitem[{Teodoro et~al.(2021)Teodoro, Collodel, and Kunz}]{Teodoro:2020gps}
Teodoro MC, Collodel LG, Kunz J (2021) {Tidal effects in the motion of gas
  clouds around boson stars}. Phys Rev D 103(10):104064.
  \doi{10.1103/PhysRevD.103.104064}.
  {\href{https://arxiv.org/abs/2003.05220}{{arXiv:2003.05220}}} {[astro-ph.HE]}

\bibitem[{Thorne(1972)}]{thorne_hoop}
Thorne KS (1972) Nonspherical gravitational collapse: A short review. In:
  Klauder JR (ed) Magic Without Magic: John Archibald Wheeler. A Collection of
  Essays in Honor of his Sixtieth Birthday. W.H. Freeman, San Francisco, pp
  231--258

\bibitem[{Torres et~al.(2000)Torres, Capozziello, and
  Lambiase}]{2000PhRvD..62j4012T}
Torres DF, Capozziello S, Lambiase G (2000) Supermassive boson star at the
  {G}alactic center? Phys Rev D 62:104012. \doi{10.1103/PhysRevD.62.104012}.
  {\href{https://arxiv.org/abs/astro-ph/0004064}{{arXiv:astro-ph/0004064}}}

\bibitem[{Unruh(2014)}]{Unruh:2014hua}
Unruh WG (2014) Has hawking radiation been measured? Found Phys 44:532--545.
  \doi{10.1007/s10701-014-9778-0}, proceedings, Horizons of Quantum Physics:
  Taipei, Taiwan, October 14--18, 2012.
  {\href{https://arxiv.org/abs/1401.6612}{{arXiv:1401.6612}}} {[gr-qc]}

\bibitem[{Urbano and Veerm\"ae(2019)}]{Urbano:2018nrs}
Urbano A, Veerm\"ae H (2019) {On gravitational echoes from ultracompact exotic
  stars}. JCAP 04:011. \doi{10.1088/1475-7516/2019/04/011}.
  {\href{https://arxiv.org/abs/1810.07137}{{arXiv:1810.07137}}} {[gr-qc]}

\bibitem[{Ure{\~{n}}a-L{\'{o}}pez and Bernal(2010)}]{2010PhRvD..82l3535U}
Ure{\~{n}}a-L{\'{o}}pez LA, Bernal A (2010) Bosonic gas as a galactic dark
  matter halo. Phys Rev D 82:123535. \doi{10.1103/PhysRevD.82.123535}.
  {\href{https://arxiv.org/abs/1008.1231}{{arXiv:1008.1231}}} {[gr-qc]}

\bibitem[{Ure{\~{n}}a-L{\'{o}}pez et~al.(2002)Ure{\~{n}}a-L{\'{o}}pez, Matos,
  and Becerril}]{2002CQGra..19.6259U}
Ure{\~{n}}a-L{\'{o}}pez LA, Matos T, Becerril R (2002) Inside oscillatons.
  Class Quantum Grav 19:6259--6277. \doi{10.1088/0264-9381/19/23/320}

\bibitem[{Vaglio et~al.(2022)Vaglio, Pacilio, Maselli, and
  Pani}]{Vaglio:2022flq}
Vaglio M, Pacilio C, Maselli A, Pani P (2022) {Multipolar structure of rotating
  boson stars}. Phys Rev D 105:124020. \doi{10.1103/PhysRevD.105.124020}.
  {\href{https://arxiv.org/abs/2203.07442}{{arXiv:2203.07442}}} {[gr-qc]}

\bibitem[{Valdez-Alvarado et~al.(2011)Valdez-Alvarado, Becerril, and
  Ure{\~{n}}a-L{\'{o}}pez}]{ValdezAlvarado:2011dd}
Valdez-Alvarado S, Becerril R, Ure{\~{n}}a-L{\'{o}}pez LA (2011) $\phi^{4}$
  oscillatons. ArXiv e-prints
  {\href{https://arxiv.org/abs/1107.3135}{{arXiv:1107.3135}}} {[gr-qc]}

\bibitem[{{Valdez-Alvarado} et~al.(2013){Valdez-Alvarado}, {Palenzuela},
  {Alic}, and {Ure{\~n}a-L{\'o}pez}}]{2013PhRvD..87h4040V}
{Valdez-Alvarado} S, {Palenzuela} C, {Alic} D, {Ure{\~n}a-L{\'o}pez} LA (2013)
  Dynamical evolution of fermion-boson stars. Phys Rev D 87:084040.
  \doi{10.1103/PhysRevD.87.084040}.
  {\href{https://arxiv.org/abs/1210.2299}{{arXiv:1210.2299}}} {[gr-qc]}

\bibitem[{Valdez-Alvarado et~al.(2020)Valdez-Alvarado, Becerril, and Ure\~na
  L\'opez}]{Valdez-Alvarado:2020vqa}
Valdez-Alvarado S, Becerril R, Ure\~na L\'opez LA (2020) {Fermion-boson stars
  with a quartic self-interaction in the boson sector}. Phys Rev D
  102(6):064038. \doi{10.1103/PhysRevD.102.064038}.
  {\href{https://arxiv.org/abs/2001.11009}{{arXiv:2001.11009}}} {[gr-qc]}

\bibitem[{V\'asquez~Flores et~al.(2019)V\'asquez~Flores, Parisi, Chen, and
  Lugones}]{VasquezFlores:2019eht}
V\'asquez~Flores C, Parisi A, Chen CS, Lugones G (2019) {Fundamental
  oscillation modes of self-interacting bosonic dark stars}. JCAP 06:051.
  \doi{10.1088/1475-7516/2019/06/051}.
  {\href{https://arxiv.org/abs/1901.07157}{{arXiv:1901.07157}}} {[hep-ph]}

\bibitem[{Vilenkin and Shellard(1994)}]{vilenkinbook}
Vilenkin A, Shellard EPS (1994) Cosmic Strings and Other Topological Defects.
  Cambridge Monographs on Mathematical Physics, Cambridge University Press,
  Cambridge; New York

\bibitem[{Vincent et~al.(2016{\natexlab{a}})Vincent, Gourgoulhon, Herdeiro, and
  Radu}]{Vincent:2016sjq}
Vincent FH, Gourgoulhon E, Herdeiro C, Radu E (2016{\natexlab{a}})
  Astrophysical imaging of {K}err black holes with scalar hair. Phys Rev D
  94:084045. \doi{10.1103/PhysRevD.94.084045}.
  {\href{https://arxiv.org/abs/1606.04246}{{arXiv:1606.04246}}} {[gr-qc]}

\bibitem[{Vincent et~al.(2016{\natexlab{b}})Vincent, Meliani, Grandcl{\'e}ment,
  Gourgoulhon, and Straub}]{Vincent:2015xta}
Vincent FH, Meliani Z, Grandcl{\'e}ment P, Gourgoulhon E, Straub O
  (2016{\natexlab{b}}) Imaging a boson star at the {G}alactic center. Class
  Quantum Grav 33:105015. \doi{10.1088/0264-9381/33/10/105015}.
  {\href{https://arxiv.org/abs/1510.04170}{{arXiv:1510.04170}}} {[gr-qc]}

\bibitem[{Visinelli(2021)}]{Visinelli:2021uve}
Visinelli L (2021) {Boson stars and oscillatons: A review}. Int J Mod Phys D
  30:2130006. \doi{10.1142/S0218271821300068}.
  {\href{https://arxiv.org/abs/2109.05481}{{arXiv:2109.05481}}} {[gr-qc]}

\bibitem[{Wald(1984)}]{1984ucp..book.....W}
Wald RM (1984) General Relativity. University of Chicago Press, Chicago

\bibitem[{Wheeler(1955)}]{Wheeler:1955zz}
Wheeler JA (1955) Geons. Phys Rev 97:511--536. \doi{10.1103/PhysRev.97.511}

\bibitem[{Will(2014)}]{will}
Will CM (2014) The confrontation between general relativity and experiment.
  Living Rev Relativity 17:4. \doi{10.12942/lrr-2014-4}.
  {\href{https://arxiv.org/abs/1403.7377}{{arXiv:1403.7377}}} {[gr-qc]}

\bibitem[{Wystub et~al.(2021)Wystub, Dengler, Christian, and
  Schaffner-Bielich}]{Wystub:2021qrn}
Wystub S, Dengler Y, Christian JE, Schaffner-Bielich J (2021) {Constraining
  exotic compact stars composed of bosonic and fermionic dark matter with
  gravitational wave events}. arXiv e-prints
  {\href{https://arxiv.org/abs/2110.12972}{{arXiv:2110.12972}}} {[astro-ph.HE]}

\bibitem[{Yagi and Stein(2016)}]{Yagi:2016jml}
Yagi K, Stein LC (2016) Black hole based tests of general relativity. Class
  Quantum Grav 33:054001. \doi{10.1088/0264-9381/33/5/054001}.
  {\href{https://arxiv.org/abs/1602.02413}{{arXiv:1602.02413}}} {[gr-qc]}

\bibitem[{Yazadjiev and Doneva(2019)}]{Yazadjiev:2019oul}
Yazadjiev SS, Doneva DD (2019) {Dark compact objects in massive
  tensor-multi-scalar theories of gravity}. Phys Rev D 99(8):084011.
  \doi{10.1103/PhysRevD.99.084011}.
  {\href{https://arxiv.org/abs/1901.06379}{{arXiv:1901.06379}}} {[gr-qc]}

\bibitem[{Yoshida and Eriguchi(1997)}]{1997PhRvD..56..762Y}
Yoshida S, Eriguchi Y (1997) Rotating boson stars in general relativity. Phys
  Rev D 56:762--771. \doi{10.1103/PhysRevD.56.762}

\bibitem[{Yuan et~al.(2004)Yuan, Narayan, and Rees}]{2004ApJ...606.1112Y}
Yuan YF, Narayan R, Rees MJ (2004) Constraining alternate models of black
  holes: {Type I} x-ray bursts on accreting fermion-fermion and boson-fermion
  stars. Astrophys J 606:1112--1124. \doi{10.1086/383185}.
  {\href{https://arxiv.org/abs/astro-ph/0401549}{{arXiv:astro-ph/0401549}}}

\bibitem[{Yunes et~al.(2016)Yunes, Yagi, and Pretorius}]{Yunes:2016jcc}
Yunes N, Yagi K, Pretorius F (2016) Theoretical physics implications of the
  binary black-hole mergers {GW150914} and {GW151226}. Phys Rev D 94:084002.
  \doi{10.1103/PhysRevD.94.084002}.
  {\href{https://arxiv.org/abs/1603.08955}{{arXiv:1603.08955}}} {[gr-qc]}

\bibitem[{Zhang et~al.(2022{\natexlab{a}})Zhang, Jain, and
  Amin}]{Zhang:2021xxa}
Zhang HY, Jain M, Amin MA (2022{\natexlab{a}}) {Polarized vector oscillons}.
  Phys Rev D 105(9):096037. \doi{10.1103/PhysRevD.105.096037}.
  {\href{https://arxiv.org/abs/2111.08700}{{arXiv:2111.08700}}} {[astro-ph.CO]}

\bibitem[{Zhang et~al.(2021)Zhang, Zeng, Wang, Wei, Seoane, and
  Liu}]{Zhang:2021ojz}
Zhang YP, Zeng YB, Wang YQ, Wei SW, Seoane PA, Liu YX (2021) {Gravitational
  radiation pulses from Extreme-Mass-Ratio-Inspiral system with a supermassive
  boson star}. arXiv e-prints
  {\href{https://arxiv.org/abs/2108.13170}{{arXiv:2108.13170}}} {[gr-qc]}

\bibitem[{Zhang et~al.(2022{\natexlab{b}})Zhang, Zeng, Wang, Wei, and
  Liu}]{Zhang:2021xhp}
Zhang YP, Zeng YB, Wang YQ, Wei SW, Liu YX (2022{\natexlab{b}}) {Motion of test
  particle in rotating boson star}. Phys Rev D 105(4):044021.
  \doi{10.1103/PhysRevD.105.044021}.
  {\href{https://arxiv.org/abs/2107.04848}{{arXiv:2107.04848}}} {[gr-qc]}

\bibitem[{Zhang et~al.(2022{\natexlab{c}})Zhang, Zeng, Wang, Wei, and
  Liu}]{Zhang:2022qzw}
Zhang YP, Zeng YB, Wang YQ, Wei SW, Liu YX (2022{\natexlab{c}}) {Stable
  circular orbit of a spinning test particle in rotating boson star}. arXiv
  e-prints {\href{https://arxiv.org/abs/2201.01498}{{arXiv:2201.01498}}}
  {[gr-qc]}

\end{thebibliography}

\end{document}